\documentclass[11pt]{article}
\pdfoutput=1
\usepackage{graphicx}
\usepackage{graphics}
\usepackage{dsfont}
\usepackage{epsfig}
\usepackage{amsmath,amssymb,amsthm,amscd}
\usepackage[numbers,sort&compress]{natbib}
\usepackage{tikz}
\usepackage{float}
\usepackage{arydshln}

\usepackage{multirow}

\usepackage{xcolor}
\newcommand{\Tr}{\textrm{Tr}}

\newcommand{\be}{\begin{equation}}
\newcommand{\ee}{\end{equation}}
\newcommand{\ba}{\begin{aligned}}
\newcommand{\ea}{\end{aligned}}

\newcommand{\rank}{\,\mathrm{rank}\,}

\newcommand{\n}{d_E}
\newcommand{\C}{\widehat{C}}
\newcommand{\q}{\hat{q}}
\newcommand{\ri}{i}

\numberwithin{equation}{section}

\usepackage{jheppub}

\begin{document}

\begin{flushright}
    USTC-ICTS/PCFT-25-03
\end{flushright}

\baselineskip=16pt

\title{Refined BPS numbers on compact Calabi-Yau threefolds from Wilson loops}
\author[a,b]{Min-xin Huang}
\author[c]{Sheldon Katz}
\author[d,e]{Albrecht Klemm}
\author[a,b]{Xin Wang}
\affiliation[a]{Interdisciplinary Center for Theoretical Study,\\
University of Science and Technology of China,\\
96 Jinzhai Road, Hefei, Anhui 230026, China}
\affiliation[b]{Peng Huanwu Center for Fundamental Theory,\\
96 Jinzhai Road, Hefei, Anhui 230026, China}
\affiliation[c]{Department of Mathematics, University of Illinois Urbana-Champaign,\\
1409 W. Green St., Urbana, IL, 61801, USA}
\affiliation[d]{Bethe Center for Theoretical Physics, Universit\"at Bonn,\\
Nußallee 12, 53115 Bonn, Germany}
\affiliation[e]{Department of Mathematical and Physical Sciences, University of Sheffield,\\
S3 7RH Sheffield, UK}

\abstract{We relate the counting of  refined BPS numbers  on compact elliptically fibred 
Calabi-Yau threefolds $X$ to Wilson loop expectations values in the gauge theories that emerge in 
various rigid local limits  of the 5d supergravity theory defined by M-theory compactification 
on $X$. In these local limits $X_*$ the volumes of curves in certain classes go to infinity, the corresponding very massive M2-brane states can be treated as Wilson loop particles and the refined topological string partition 
function on $X$ becomes a sum of terms proportional to associated refined Wilson loop expectation values. The resulting ansatz for the complete refined topological partition function on $X$ is written in terms  of the proportionality coefficients which depend only on the $\epsilon$ deformations and the Wilson loop expectations values which 
satisfy holomorphic anomaly equations. Since the ansatz is quite restrictive 
and can be further constrained by the one-form symmetries and E-string type limits for large base curves, 
we can efficiently evaluate the refined BPS numbers on $X$,  which we  do explicitly for local gauge groups up to rank three and $h_{11}(X)=5$. These refined BPS numbers pass an impressive number of consistency checks imposed by the direct counting of these numbers using the moduli space of one dimensional stable sheaves on $X$ and give us numerical predictions for the complex structure dependency of the refined BPS numbers.}

\emailAdd{minxin@ustc.edu.cn}
\emailAdd{katzs@illinois.edu}
\emailAdd{aoklemm@th.physik.uni-bonn.de}
\emailAdd{wxin@ustc.edu.cn}

\maketitle
\tableofcontents
\section{Introduction and Summary}
\label{sec:Intro} 
Wilson loop operators are gauge invariant operators whose expectation value gives a 
measure of quark confinement introduced in~\cite{Wilson:1974sk}, where the relation of the string 
approach to strong interactions was already pointed out. In the AdS$_4\times S^5$ 
conformal gauge string theory correspondence this expectation value has been calculated to leading order in the 't Hooft
coupling as the string world-sheet volume in the AdS space bounding the loop of the gauge theory on the boundary~\cite{Maldacena:1998im}. Direct localisation calculations 
of the expectation value of circular supersymmetric Wilson loops  
in 4d $\mathcal{N}=2$ supersymmetric gauge theory were performed in~\cite{Pestun:2007rz} 
to prove that the latter are related to a Gaussian matrix model. 

It has been a very fruitful approach to  study lower 
dimensional rigid, i.e. non-gravitational, supersymmetric theories to start with string, M-- or 
F--theory compactifications on compact Calabi-Yau spaces $X$ and 
take a local limit of $X$ in which the gravity sector decouples, an approach that is known as \emph{geometric engineering} of rigid 
(gauge) theories. Since the refinement can be  clearly defined  
for these rigid theories  we propose in this paper to go in the opposite direction and start from local rigid gauge theories and reconstruct 
properties of the refined spectrum of  supergravity completions in the topological sector of the corresponding string theory. As 
in the above case of Gauge theory/Gravity duality, Wilson loop expectation values in the gauge theory play a key role in this program. Of course we do not claim that these theories are dual, however the use of Wilson line operators and of the one-form symmetries put severe constraints on the structure of the gravity completion. Moreover if several local  
limit exists we get consistency conditions on the structure of 
our ansatz.   More concretely  we study the enumeration of spinning massive BPS states in 5d $\mathcal{N}=1$ supergravity theories on $\mathbb{R}^4\times S^1$ from M-theory compactification on compact Calabi-Yau threefolds. Our approach connects the local and the global case using Wilson loop operators, wrapping along the time circle $S^1$, of the gauge theories in rigid limits of the 5d supergravity theory, addressing the questions raised in the weak gravity conjecture~\cite{Harlow:2022ich}. 
Since all our Calabi-Yau manifolds are elliptically fibred, F-theory compactifications yield  6d $\mathcal{N}=(1,0)$ theories and we can in particular focus on the Kaluza-Klein (KK) theories arising from the circle compactification. In the weak gravity limit of the 5d KK theory, where the gauge force dominates, the BPS spectrum is captured by gauge theories in rigid limits of the 5d supergravity theory. In this limit, the KK particles become very massive, freezing all their dynamical degrees of freedom, so that they effectively act as sources for Wilson loop operators along the time direction. Based on this argument, we propose an ansatz \eqref{eq:ansatz_struc} for the BPS partition function, which states that it can be expressed as a linear combination of Wilson loops in \emph{all} representations. This ansatz also meets the requirement by the \emph{Completeness Hypothesis} \cite{Polchinski:2003bq}, which states that any gauge theory coupled to gravity, there must exist charged matter in \emph{every}
representation of the gauge group, and suggests the broken of one-form symmetry for the supergravity theory. Furthermore, we argue in Section~\ref{sec:3.4} that the structure of the expansion is constrained by one-form symmetries of the 5d gauge theories as well as by the positivity and integrality of the K\"ahler parameter expansion of the compact Calabi-Yau threefold. 

A benefit of our approach is that the gauge theory partition functions, as well as the expectation values of Wilson loop operators, can be exactly calculated, even on the Omega-deformed background. For instance, one can in principle apply localization methods \cite{Nekrasov:2002qd,Nekrasov:2004vw,Shadchin:2004yx,Nekrasov:2015wsu,Kim:2016qqs,Chang:2016iji,Haouzi:2020yxy,Nawata:2023wnk}, introduce additional very heavy hypermultiplets \cite{Gaiotto:2015una,Grassi:2018bci,Kim:2021gyj,Huang:2022hdo}, utilize blowup equations \cite{Kim:2021gyj,Wang:2023zcb} as well as refined holomorphic anomaly equations \cite{Huang:2022hdo,Wang:2023zcb} to calculate the refined BPS partition functions or the refined BPS numbers for Wilson loops in arbitrary 5d $\mathcal{N}=1$ quantum field theories and in arbitrary representations. By expanding the partition functions of gravity theories in the form of \eqref{eq:ansatz_struc} and using these expressions for the Omega-deformed Wilson loops, up to a few additional coefficients that only depend on the Omega-deformed parameters $\epsilon_{1,2}$, we formally define refined BPS partition functions for 5d supergravity theories, leading to refined BPS counting for the spinning BPS states in 5d supergravity theories.

In fact, by employing the refined holomorphic anomaly equations for Wilson loops studied in \cite{Huang:2022hdo,Wang:2023zcb} to the ansatz \eqref{eq:ansatz_struc}, we obtain a novel refined holomorphic anomaly equations for compact Calabi-Yau threefolds \eqref{eq:HAEgeneral1}. The undetermined coefficients in the ansatz are holomorphic ambiguities, which can be fixed with a few additional inputs of refined BPS numbers. For the first few orders, these inputs can be obtained from one-loop contributions from the matter content of 6d supergravity theories and the higher instanton string contributions of their 6d SCFT limits.  
As a concrete test of our proposal, we explicitly fix the first few undetermined coefficients in the ansatz \eqref{eq:ansatz_struc} for a few compact elliptically fibered Calabi-Yau threefolds ${X}$ up to $h^{1,1}({X})=5$. These calculations are also consistent with the refined results in \cite{Huang:2020dbh} and the mathematical calculations performed in Sections~\ref{subsec:rank1}--\ref{subsec:rank3}.  Some of these geometric calculations have been used to fix boundary conditions, while others serve as checks.  We also observe in Section~\ref{subsubsec:ellipticP2} that even if one is only interested in the unrefined GV invariants, our proposal gives a new way to use the refinements to compute GV invariants. All  refined
BPS numbers turn out to be non-negative. This is compatible with 
their interpretations as dimensions of vector spaces. The latter might be finite dimensional representations of hidden geometrical symmetries.

\section{Refinement on local and global Calabi-Yau threefolds} 
\label{sec:refinementlocalglobal} 
M-theory compactified on a Calabi-Yau threefold $X$ yields an 
$\mathcal{N}=1$ supergravity theory with eight conserved supercharges. Its 
BPS states come from wrapping the M-theory membrane or  
fivebrane on holomorphic curves or divisors in $X$ respectively. 
The Poincar\'e representations of these BPS states are fixed by their representations $(j_L,j_R)$ with respect to the little group \be 
{\mathrm{SU}}(2)_L\times {\mathrm{SU}}(2)_R\subset {\mathrm{Sp}}(4)
\ee 
of the 5d Lorentz group and their non-vanishing 
masses, which are proportional to their charges $\gamma$ in a lattice $\Gamma$.
The multiplicities $N^\gamma_{j_L,j_R}\in \mathbb{N}$ of these BPS states depend 
on the particular fibre $X$ in the family $X\hookrightarrow  {\cal X}\rightarrow {\cal M}$ over the deformation space ${\cal M}$ of the Calabi-Yau threefold within which eight supercharges  are preserved. These multiplicities are of great physical interest as they count light states in the effective 
action including those which are proposed to account for the microscopic entropy of supersymmetric black holes in the uncompactified dimensions. 
Mathematically they are related to the counting of perverse sheaves on $X$. 

The geometric description~\cite{Gopakumar:1998jq} of at least some of these  BPS states becomes  
clearer in the 4d Type IIA compactification on $X$ obtained from 5d by circle compactification.
This 4d $\mathcal{N}=2$ supergravity theory also has eight supercharges and is related to the 5d theory by the exact 
4d/5d correspondence \cite{Gopakumar:1998jq}\cite{Gaiotto:2005gf}. Here 
the charge lattice can be identified with the K-theory charge lattice of $\mathrm{D}(2K)$-branes of the type IIA theory 
\begin{equation} 
\Gamma=(q_0,q_A,p^A,p^0) \
\end{equation} 
in the even cohomology group $\oplus_{k=0}^3 H^{2k}(X)$ of $X$.  
In particular the BPS states coming from the  M-theory membrane become bound states of $\mathrm{D}0$ and $\mathrm{D}2$ branes labelled by the $\mathrm{D}0$ brane charge $q_0\in \mathbb{Z}$ 
and the $\mathrm{D}2$-brane charge $q_A$ identified with the curve class $\beta\in H_2(X,\mathbb{Z})$. The 5d/4d 
correspondence \cite{Gaiotto:2005gf} identifies the  
left spin with the $\mathrm{D}0$ brane charge  
\begin{equation} 
q_0=2 \frac{j_L}{(p^0)^2} \ . 
\end{equation}
The holomorphic topological string free energies 
${\cal F}_g(t)$ to genus $g$ encodes completely 
the non-vanishing unrefined BPS indices $n_g^\beta$ with charge D$6$ brane charge one\footnote{The charge 
multiplicity of the highest dimensional brane determines the rank of the gauge group 
with maximal dimensional support and is called therefore the rank $r$.} 
and no D$4$ brane   
\begin{equation}  
(\mathrm{D}0,\mathrm{D}2,\mathrm{D}4,\mathrm{D}6)_{q,p}=(q_0,\beta,0,1)
\label{eq:relstates} 
\end{equation} 
\cite{Gopakumar:1998jq}\cite{KKV}, where 
$\beta\le \beta(g)$ has to be below the 
Castelnuovo bound \cite{MR2596635}. These BPS indices 
$n_g^\beta$ are invariant under complex 
structure deformations within the family 
${\cal X}$ \cite{Gopakumar:1998jq}. They are related to the actual 
BPS multiplicities by a weighted summation over the right 
spins and a triangular change in the left spin basis
$I^n_*=\left(2 [0]_*+\left[\frac{1}{2}\right]_*\right)^{\otimes n}=\sum_{j\ge 0} \left(\left(\substack{ 2n\\[1 mm] n-j }\right) -\left(\substack{2n \\[1 mm] n-2-j}\right)\right)\left[\frac{j}{2}\right]_*$ as
\begin{equation}
\sum_{g=0} n_g^\beta I^g_L=\sum_{j_R} (-1)^{2 j_R}(2 j_R+1)N_{j_L,j_R}^\beta \left[\frac{j_L}{2}\right]_L \ .
\label{eq:BPSindex} 
\end{equation} 
In contrast to the cases  of local toric Calabi-Yau spaces $X$, the complete 
solution of even the unrefined topological  string on compact Calabi-Yau spaces $X $  is 
an open problem due to incomplete knowledge of the boundary conditions 
in the direct integration of the holomorphic anomaly equations, which 
prevents the holomorphic ambiguity from being completely fixed \cite{MR2596635}. Recently 
using the wall crossing to $(\mathrm{D}0,\mathrm{D}2,\mathrm{D}4,\mathrm{D}6)_{p,q}=(n,\beta,r,0)$ BPS 
states and the modular gauge indices of supersymmetric gauge theories of ranks $r=1$ 
\cite{Alexandrov:2023Jan} and $r=2$ \cite{Alexandrov:2023Dec}, 
these  boundary conditions have been improved \cite{Alexandrov:2023Jan,Alexandrov:2023Dec}. 
The data currently available for hypergeometric
one parameter Calabi-Yau threefold families are accessible  here  \cite{link}. 
If $X$ is a local Calabi-Yau space, supergravity decouples and in this 
so-called rigid limit one gets an ${\cal R}$ symmetry $SU(2)_{\cal R}$ 
acting  on the supersymmetry algebra which in this case do not depend on the complex structure.

\subsection{Refinement on local Calabi-Yau spaces}

Refined holomorphic anomaly equations for local Calabi-Yau and geometrical engineered gauge theories were studied in ~\cite{Huang:2010kf}\cite{Krefl:2010fm}. The boundary conditions on local Calabi-Yau are completely determined and the results in the large radius limit coincide with topological vertex~\cite{Iqbal:2007ii} and the localization calculation using the Bialiniki-Birula 
decomposition~\cite{Choi:2012jz}. 

In the gauge theory context~\cite{Nekrasov:2002qd} defines a five dimensional index using the global $U(1)_{\cal R}\subset SU(2)_{\cal R}$ charge that is 
manifest in any  rigid $\mathcal{N}=1$ supersymmetric theory in 5d as well as in its circle compactification to 4d 
\begin{equation}
Z_{BPS}(\epsilon_L,\epsilon_R,t)={\mathrm{Tr}}_{{\cal H}_{BPS}} (-1)^{2(j_L+j_R)} e^{-2 \epsilon_L j_L} e^{-2 \epsilon_R j_{R}} e^{-2 \epsilon_R j_{\cal R}} e^{\beta H} \ .
\label{eq:refinedBPSindex}
\end{equation}
Here the $j_{L/R}$ denote the operators that correspond to the Cartan generators of the $SU(2)_{L/R}$ above, while $j_{{\cal R}}$ denotes the 
charge operator of the $U(1)_{\cal R}$ global symmetry.  First 
note that due to supersymmetry and the insertion of the Fermion number operator $(-1)^{2 (j_L+j_R)}$ the trace receives only contributions from the BPS groundstate of the Hamiltonian $H\sim (Q_L^2+Q_R^2)$. 
The mass of the BPS states is related to its charge and for the 
bound states between D$2$ and D$0$ branes under considerations is 
therefore ultimately to the varying  $\beta,q_0$ in \eqref{eq:relstates}. The twisting of the $j_R$ by the global $U(1)_{\cal R}$ symmetry is essential so that the expansion coefficients in the expansion of  \eqref{eq:refinedBPSindex} with respect to 
\begin{equation}
q_{L}=e^{\epsilon_L}=e^{\frac{1}{2}(\epsilon_1-\epsilon_2)}\ \ {\mathrm{ and}} \ \ q_{R}=e^{\epsilon_R}=e^{\frac{1}{2}(\epsilon_1+\epsilon_2)}
\end{equation} 
are invariant BPS indices. The latter are  related to geometrical perverse sheaf counting problems and in particular to the positive integral invariants $N^\beta_{j_L,j_R}\in \mathbb{N}$. 

First note that geometrically the $U(1)_{\cal R}$ is directly related to the $\mathbb{C}^*$ isometry of the toric local Calabi-Yau spaces $X_{\text{tor}}$~\cite{Choi:2012jz}. The latter can be pulled back to the moduli space of Pandharipande Thomas invariants \cite{PT1}  of stable  pairs of sheaves in order to define a virtual Bialinicki-Birula decomposition on the latter, which refines the localisation calculation~\cite{Choi:2012jz} to evaluate their individual coefficients in an $\epsilon_1$, $\epsilon_2$ expansion. Using the latter and the refined Castelnuovo bounds enables 
one to calculates the index  $Z_{BPS}(\epsilon_L,\epsilon_R,H)$ to arbitrary orders in $q_{L/R}$ and the K\"ahler parameter $e^{\beta\cdot t}$. After identifying ${\cal F}(\epsilon_L,\epsilon_R,t)=\log(Z_{BPS}(\epsilon_L,\epsilon_R,H))$, $H=t$ and reorganizing the result in form the suggested by~\cite{Gopakumar:1998jq}
\begin{equation}\label{eq:BPSexpansion}
    \mathcal{F}(\epsilon_L,\epsilon_R,t)= \sum_{\substack{\beta\in H_2(X,\mathbb{Z})\\[.5 mm]   \beta \neq 0} }\sum_{\substack{k=1\\[.5mm] j_L,j_R=0}}^{\infty}(-1)^{2(j_L+j_R)}  N_{j_L,j_R}^{\beta} \frac{\chi_{j_L}(q_{L}^k)\, \chi_{j_R}(q_{R}^k)}{k\,  {\cal I}(k \epsilon_1,k \epsilon_2)}\, e^{k \, \beta \cdot t }, 
\end{equation}
where 
\be
{\cal I}(\epsilon_1,\epsilon_2)=2 \sinh\left(\frac{\epsilon_1}{2}\right) 2 \sinh\left(\frac{\epsilon_2}{2}\right) \qquad {\mathrm{and}} \qquad 
\chi_j(x)=\left(\sum_{m=-j}^jx^{2m}\right)\, , 
\label{eq:Ifactor}
\ee
the refined BPS invariants $N^\beta_{j_L,J_R}$ are obtained. This lead to the first systematic geometric approach to define and calculate of the $N^\beta_{j_L,j_R}$ for toric local Calabi-Yau threefolds  in and outside the gauge theory context and in particular directly for local  ${\cal O}(-3) \rightarrow \mathbb{P}^2$~\cite{Choi:2012jz}.

On local toric Calabi-Yau manifolds the refined topological vertex~\cite{Iqbal:2007ii}, the refined 
holomorphic anomaly equations~\cite{Huang:2010kf}\cite{Krefl:2010fm}, direct localisation calculations~\cite{Choi:2012jz}, the extension of the blowup equations from four dimensional $\mathcal{N}=2$ 
supersymmetric gauge theories to five dimensions~\cite{Gu:2018gmy} allow the in-principle calculation of all 
refined BPS invariants.  The refined holomorphic anomaly equations~\cite{Huang:2013yta}  
and further extended blowup equations~\cite{Gu:2019dan,Gu:2019pqj,Gu:2020fem,Duan:2021ges,Lee:2022uiq,Kim:2020hhh} apply in great generality 
to local Calabi-Yau manifolds based on elliptically fibered surfaces. They are confirmed by the elliptic index 
calculations in $(2,0)$ and $(1,0)$  supersymmetric theories from F-theory in six dimensions~\cite{Haghighat:2013,Haghighat:2014vxa,Gu:2017ccq,DelZotto:2017mee}. Many relevant example calculations are 
performed in the cited literature and will be used as boundary conditions in the local limits of  
our compact elliptically fibred Calabi-Yau manifolds below.

In~\cite{MR3504535} an index calculation was proposed related to geometrisation of 
D6 branes arising from an eleven dimensional M-theory background $Z_{11}=({\cal V}\times R^1_{\rm time}) \rightarrow X$, a fibration of a Taub-Nut geometry~\cite{Hull:1997kt}\cite{Sen:1997js}. The base $X$ is a six dimensional manifold, which has to have sufficiently many (toric) 
isometries in order to allow localisation calculations, and in the most relevant case is simply a toric  
Calabi-Yau manifold as above. In special cases the refined topological vertex results~\cite{Iqbal:2007ii} have been reproduced~\cite{MR3504535}. ${\cal V}$ is in general a  rank two holomorphic vector bundle equipped with a (k multi-centered) Taub-Nut metric ${\mathrm d}s^2_{TN}$, $\det({\cal V})=K_X$ and $U(1)\subset SU(2) \subset U(2)$ isometries. For the latter to exist the authors \cite{MR3504535} assume a global splitting of ${\cal V}={\cal L}_1\oplus {\cal L}_2$ into line 
bundles. Using the ideas of Kaluza-Klein monopoles from the Taub-Nut 
metric  \cite{Hull:1997kt}\cite{Sen:1997js} one concludes that one can pick the M-theory circle within the $U(1)$ isometry and thereby reduce 
M-theory on $Z_{11}$ to  a type IIA compactification $Z_{10}=(K_X \times R^{1,1})\rightarrow X$ with $k$ D6 branes wrapped on $X$. Further~\cite{MR3504535} argue that for $k=1$ one gets a quite general
prescription. On the question of the existence of a global Calabi-Yau orientation, i.e. a compatible choice 
of the square root of $(K^{\rm vir}_{\widehat{ {M_\beta}}})^{1/2}$ see section \ref{Sec:GeometriccalculationsofrefBPS}, 
is speculated from the physics point of view in~\cite{MR3504535}, but whether the approach can contribute to actual calculation of refined BPS numbers on compact Calabi-Yau manifolds is an open question.   A mathematical proof of the existence of a canonical orientation is given in \cite{ju}.

\subsection{Refinement proposal for compact Calabi-Yau threefolds} 
Since in supergravity  one expects no global symmetries~\cite{Banks:2010zn}, one cannot use the 
twisting by the global $U(1)_{\cal R}$ charges to define a protected index for the refined BPS states 
as in \eqref{eq:refinedBPSindex}. However in this case  one still has a  protected 
supersymmetric index called $n_g^\beta$ defined in \eqref{eq:BPSindex}  that is conjectured to be  
invariant under complex structure deformations~\cite{Gopakumar:1998jq}. There are two related 
mathematical formulations to calculate the unrefined invariants. Directly inspired from 
physical model of the D2-D0 brane moduli space~\cite{Gopakumar:1998ii,Gopakumar:1998jq} is the attempt to reconstruct the ${\rm SU}_L(2)\times {\rm SU}_R(2)$
Lefshetz actions on the cohomology of this moduli space of a D2-brane  supported on a curve $C$. Mathematically
this moduli space is identified with a degenerate Jacobian fibration Jac$_g(C)$ over the family of image curves $C$ of genus $g$. The Lefshetz action on its cohomology can be reconstructed  by the Abel Jacobi map from ${\rm Hilb}^p(C)$ to Jac$_g(C)$ and at least if the degenerations are mild many calculations can be 
performed~\cite{KKV}. As reviewed in section \ref{Sec:GeometriccalculationsofrefBPS} 
we extend there the approach of~\cite{Maulik:2016rip} of reconstructing the cohomology on the moduli of 1-dimensional stable sheaves $F$ supported on $C$ with ${\rm ch}_2=\beta$ and $\chi(F)=n$.

An elaborate example for the complex structure dependence of the $N_{j_R,j_L}^\beta$ is provided in Section~\ref{sec:Model3_deform}.
It is based on the simple geometry of smooth ruled surfaces $S\in  X$. The projection map $\rho:S\rightarrow C$ with $\mathbb{P}^1$ fibres maps to a curve $C_g$  of genus $g$. Such geometries were studied in terms of $N=2$ Higgs transitions where geometrically the $\mathbb{P}^1$ fibre shrinks~\cite{Klemm:1996kv,Katz:1996ht}. They occur for example in hypersurfaces $p=0$ in toric varieties $\mathbb{P}_\Delta$~\cite{Batyrev:1994hm} defined by the 
4d reflexive lattice polytope $\Delta$ which together with 
$\hat \Delta $ form a dual pair $(\Delta,\hat \Delta)$. If a 
codimension two face $\theta_2$ with inner points $l(\theta_2)$ 
is dual to a codimension three edge $\hat \theta_3$ in $\hat \Delta$ with inner points $l(\hat \theta_3)$ then some monomial deformations of $p$ that in general  represent complex structure deformation of $ X$ cannot occur, instead there exist $K=l(\theta_2)\cdot l(\hat \theta_3)$ additional  independent Beltrami differentials $\mu_k$, $k=1,\ldots K$ in $H^1(\hat M,T_{\hat M})$ which correspond to \emph{non-polynomial} complex structure deformations. The latter  are 
frozen to particular values in the toric embedding of $ X$. If 
$g=l(\theta_2)\neq 0$ and $l(\hat \theta_3)=1$ then for these frozen values  of the moduli a rule surface $S\subset  X$ over a genus $g$ curve $C_g$ is realized. 
If $\hat l(\theta_3)>1$ a surface $S\subset  X$ with several 
such components $S_i$ exist, where over each point 
in the base the $l(\hat \theta_3)=n$ $\mathbb{P}^1$'s intersect with 
negative  Cartan matrix of $A_n$. Let us consider a genus zero
curve in the class $\beta$ represented by a fibre $\mathbb{P}^1$. 
The moduli space of each  $\mathbb{P}^1$ ${\cal M}_\beta$ 
is identified with $C_g$ and the ${\rm SU}(2)_L\times {\rm SU}_R(2)$ Lefshetz decompositions yields $R^\beta_B= 2 g \,  [0,0]+\left[\frac{1}{2},\frac{1}{2}\right]$. As explained in 
section~\ref{sec:Model3_deform}, if the geometry is 
deformed w.r.t. to $\mu_k$ deformations, the holomorphically 
embedded curve $C_g$ disappears and the $\mathbb{P}^1$ are 
fixed to $2g-2$ points, which corresponds to the representation $R^\beta_A=(2g-2)[0,0]$. Clearly  the weighted trace  \eqref{eq:BPSindex}
over $j_R$ yields the same $n_b^\beta$ while the $N_{j_L,j_R}^\beta$
change before and after the complex structure deformation.

The point is that the ansatz \eqref{eq:ansatz_struc} for 
the refined invariants of $X$ with Wilson loops is 
so restrictive that one we fix the boundary condition according 
to the representations $R^\beta_B$ or $R^\beta_A$ the predicted refined numbers $N^\gamma_{A/B,\, j_R,j_L}$ in many other 
classes $\gamma\in H_2(X,\mathbb{Z})$ change in a way we 
can at least in part confirm by the geometric methods as explained in 
section ~\ref{sec:Model3_deform}.    

For $K3$ fibred Calabi-Yau threefolds proposals  for the refined in where made in \cite{KKP} based on modularity considerations and for elliptically fibred Calabi-Yau threefolds in~\cite{Huang:2020dbh} 
based on an ansatz for the refined holomorphic anomaly equations. These proposals provide eventually
additional boundary conditions for the ansatz made in \eqref{eq:ansatz_struc}. Once they are fixed again we can make 
the consistency checks. 

Mathematical definitions of GV invariants based on moduli spaces of D2-D0 branes require the notion of an orientation introduced in \cite{KS}.   We review several issues regarding orientations at the beginning of Section~\ref{Sec:GeometriccalculationsofrefBPS}. Each moduli space has a virtual canonical bundle. An orientation is a choice of square root of this bundle, and these choices must satisfy a compatibility condition.  While there can be more than one square root, a canonical orientation can be constructed \cite{ju}. Given an orientation, a perverse sheaf of vanishing cycles on the moduli space can be constructed \cite{BBDJS}, which is then used to define the (unrefined) GV invariants \cite{Maulik:2016rip}.  It is observed in \cite{Maulik:2016rip} that different orientations can lead to different GV invariants.  However, if the orientation is Calabi-Yau, which means that it is trivial on the fibers of the Hilbert-Chow morphism mapping a D2-D0 brane to its support curve, then the resulting GV invariant is independent of the choice of Calabi-Yau orientation \cite{Maulik:2016rip}.  Calabi-Yau orientations are conjectured to exist \cite{Maulik:2016rip}.  The conjecture is open as of this writing.  In particular, it is not know if the canonical orientation of \cite{ju} is necessarily Calabi-Yau.   

Orientations can also be used to define the refined GV numbers, at least under certain conditions.  The dimensions of cohomologies of perverse pushforwards of the perverse sheaf of vanishing cycles gives rise a generating function, a Laurent polynomial in two variables.  In trying to match to \cite{Gopakumar:1998jq}, one can hope that this Laurent polynomial is the character of a finite-dimensional representation of $SU(2)\times SU(2)$, but there is no proof of that fact, or even a reason from mathematics to suspect that this is the case in general.   If the perverse sheaf of vanishing cycles supports a pure Hodge module, then we do have a representation of $SU(2)\times SU(2)$, arriving at a mathematical definition of refined GV numbers which produces the correct unrefined GV invariants in all known cases.  In the general case where the perverse sheaf only supports a mixed Hodge module, a definition of refined GV numbers was proposed in \cite{Kiem:2012fs} by using the associated graded (pure) Hodge module of the mixed Hodge module.  However, this proposal has been shown to produce the incorrect GV invariants in an example \cite{Maulik:2016rip}. In all of our computations used to identify boundary conditions, we will be in the situation of pure Hodge modules. Indeed, most of the moduli spaces we study are smooth, in which case the associated perverse sheaf of vanishing cycles supports a pure Hodge module.  Also, we know of no counterexamples to the existence of an $SU(2)\times SU(2)$ representation in the general case.  

\section{Wilson loop calculation and topological string} \label{sec:Wilsonloop}
This section reviews the calculation of expectation values for half-BPS Wilson loop operators on the Coulomb branch of five-dimensional $\mathcal{N}=1$ supersymmetric quantum field theories. Furthermore, we review the realization of Wilson loops in M-theory and topological string theory, and explain their connection to the BPS spectrum of compact elliptically fibered Calabi-Yau threefolds.

\subsection{Wilson loops in gauge theories}
\begin{sloppypar} 
In the study of topological string theory on a \emph{local Calabi-Yau threefold} $X$~\cite{Chiang:1999tz,Klemm:1999gm,Aganagic:2003db}, 
the 5d ${\mathcal{N}=1}$ supersymmetric quantum field theory (SQFT) on $\mathbb{R}^4_{\epsilon_1,\epsilon_2}\times S^1$ is naturally engineered from M-theory compactification on $X$. In particular the partition function of refined topological string theory on $X$~\cite{Iqbal:2007ii} encodes the degeneracies $N^\beta_{j_L,j_R}$ of the BPS particles with charge $\beta \in H_2(X,\mathbb{Z})$ in the left, right $(j_L,j_R)$ spin representations of the ${\mathrm{ SU}}(2)_L\times {\mathrm{SU}}(2)_R$ little group of the 5d massive particles.
\end{sloppypar}

In gauge theory, another important observable is the correlation function of the Wilson loop operator encoding further BPS degeneracies $\widetilde N_{j_L,j_r}^\beta$.  Wilson loop operators are gauge invariant operators that arise from the parallel transport of the gauge field $A_{\mu}$ around closed loops.
These operators also have a supersymmetric version in supersymmetric quantum field theory. For instance, in a 5d $\mathcal{N}=1$ gauge theory, we can define a half-BPS Wilson loop operator as \cite{Young:2011aa,Assel:2012nf}
\begin{equation}
    W_{\mathbf{r}}=\Tr_{\mathbf{r}} \mathcal{T}\left(i\oint_{S^1}dt\left(A_0(t)-\phi(t)\right)\right),
\end{equation}
  which is located at the origin of the 4d space $\mathbb{R}^4$ and winding around the Euclidean time circle $S^1$. This type of Wilson loop is also called Polyakov loop. Here $\mathcal{T}$ denotes the time ordering operator, $\mathbf{r}$ is a representation for the gauge field $A_{\mu}$, $A_0(t)=A_0(\vec{x}=0,t)$ is the zero component of the gauge field and $\phi(\vec{x}=0,t)$ is the scalar field in the vector multiplet which accompanies the gauge field to preserve half of the supersymmetry. The insertion of the half-BPS Wilson loop operator can be also realized by introducing a half-BPS static, heavy and electrically charged particle at the origin of $\mathbb{R}^4$. We will refer to such a particle as a {\it{Wilson loop particle}}.

Consider a 5d SQFT that is described by a gauge theory with gauge group $G$. On the Coulomb branch of the moduli space, the scalar field $\phi$ in the vector multiplet has non-trivial expectation values in the Cartan subalgebra of the gauge group $G$. The gauge group $G$ is broken to its Abelian subgroup $U(1)^{r}$ with $r=\rank G$ the rank of the gauge group. The representation $\mathbf{r}$ becomes the non-negative electric charge $\q_i$, also referred to as gauge charges, of the Wilson loop particle under the $i$-th Abelian gauge subgroup $U(1)$, and is denoted by $\mathbf{r}=[\q_{1},\cdots,\q_{r}]$. 

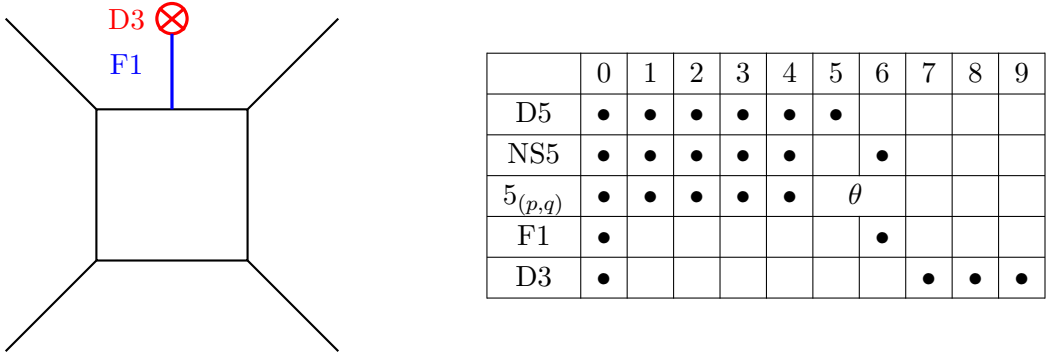
\begin{figure}[h!]
\begin{minipage}{0.36\linewidth}
    \centering
\begin{tikzpicture}
\def\x{1.2};
\def\y{2};
\def\radius{0.17};
\draw[thick] (1,1) -- (1,-1) -- (-1,-1) -- (-1,1) -- (1,1);
\draw[thick](1,1) -- (1+\x,1+\x);
\draw[thick](-1,1) -- (-1-\x,1+\x);
\draw[thick](-1,-1) -- (-1-\x,-1-\x);
\draw[thick](1,-1) -- (1+\x,-1-\x);
\draw[red,very thick] (0,\y+0.2) circle (0.2);
\draw[red,very thick] (-\radius,\y+\radius+\radius) -- (\radius,\y+0.05);
\draw[red,very thick] (-\radius,\y+0.05) -- (\radius,\y+\radius+\radius);
\node at (-0.6,\y+0.2) {\color{red}{D3}};
\draw[very thick,blue](0,\y)--(0,1);
\node at (-0.6,\y-0.4) {\color{blue}{F1}};
\end{tikzpicture}
\end{minipage}
\begin{minipage}{0.64\linewidth}
\centering
\begin{tabular}{|c|c|c|c|c|c|c|c|c|c|c|}
\hline
 &0 &1  &2  &3  &4  &5  &6  &7  &8  &9  \\
\hline
D5 & $\bullet$ & $\bullet$ &$\bullet$  &$\bullet$  &$\bullet$  &$\bullet$  &  &  &  &  \\
\hline
NS5 &$\bullet$  &$\bullet$  &$\bullet$  &$\bullet$  &$\bullet$  &  &$\bullet$  &  &  &  \\
\hline
$5_{(p,q)}$ &$\bullet$  &$\bullet$  &$\bullet$  &$\bullet$  &$\bullet$    &\multicolumn{1}{r}{\,\,\,\,$\theta$\!\!\!\!} & &  &  &\\
\hline 
F1 &$\bullet$  &  &  &  &  &  &$\bullet$  &  &  &  \\
\hline
D3 &$\bullet$  &  &  &  &  &  &  &$\bullet$  &$\bullet$  &$\bullet$  \\
\hline
\end{tabular}

\end{minipage}
\caption{IIB description of the 5d Wilson loops in the $SU(2)$ theory. The $(p,q)$ five-brane web diagram on the left illustrates the $(p,q)$ five-brane configuration in the $x^{5,6}$ directions. It is the dual diagram of the corresponding toric description of the local Calabi-Yau $X={\cal O}(-2,-2)\rightarrow \mathbb{P}^1\times \mathbb{P}^1$. The half-BPS Wilson loop in the fundamental representation of $SU(2)$ is realized by the fundamental string F1 stretched between a D5 brane and the D3 brane. In the description on the right, the brane configurations are detailed, with $\tan \theta=\frac{p}{q}$ for a five-brane with charge $(p,q)$.}
\label{table:IIBofcodimension4defect}
\end{figure}

We are interested in the BPS partition functions of 5d theories in the presence of half-BPS Wilson loop operators. These partition functions can be computed as a sum of the $k$-instanton supersymmetric index of the ADHM quantum mechanics which can be read off from the IIB brane realization \cite{Tong:2014cha,Tong:2014yna,Nekrasov:2015wsu,Assel:2018rcw}. An illustration of the IIB $(p,q)$ five-brane web description for 5d $SU(2)$ theory can be found in Figure~\ref{table:IIBofcodimension4defect}. In this case, the rank of the gauge group is 1 so the representation of the Wilson loop can be labeled by a single positive integer $\hat{q}$. The half-BPS Wilson loop operators are realized by adding semi-infinite F1 strings with charge 1, stretched between D3 branes and D5 branes. The lowest energy modes on such F1 strings are fermionic, so there can be at most one F1 string stretched between a D3 brane and a D5 brane. To obtain the Wilson loop partition function in the representation $\mathbf{r}=[\hat{q}]$, we need to insert $\hat{q}$ D3 branes along the $x^{0,7,8,9}$ directions.

\subsection{Connection to topological string theory}
In M-theory, dynamic electric particles in the 5d quantum field theory are obtained from M2-branes wrapping holomorphic 2-cycles in the Calabi-Yau threefold $X$. The rank $r$ of the  gauge group  is determined by the number of compact divisors in $X$ given by Betti number $r=b^c_4(X)$. It was proposed in \cite{Kim:2021gyj} that the insertion of a Wilson loop particle with charge $\mathbf{r}=[\q_1,\cdots,\q_r]$ can be realized by inserting a wrapped non-dynamic M2-brane over a non-compact curve $\C$, ending on boundary branes. See \cite{Heckman:2022muc} for a more extensive study of the boundary branes and the related generalized symmetry operators. This geometric definition can be used to define the loop operators in 5d SQFTs without gauge theory description. The electric charges of the Wilson loop particles are computed as the intersection numbers of the compact divisors $D_i,i=1,\cdots,r,$ and the non-compact curve $\C=\q_1\C_1+\cdots+\q_r\C_r$, where
\begin{equation}
    D_i\cdot \C_j=\delta_{i,j},\quad i,j=1,\cdots,r.
\end{equation}
It follows that the electric charges $\q_i$ are computed as
\begin{equation}
    \q_i=D_i\cdot \C,\quad i=1,\cdots,r.
\end{equation}
In the presence of Wilson loop operators, the corresponding refined BPS degeneracies $\widetilde{N}_{j_L,j_R}^{\beta}$ for Wilson loops describe the number of BPS states coming from M2-branes wrapping curves $C+\C$, with $[C]=\beta\in H_2(X,\mathbb{Z})$.
It was proposed in \cite{Huang:2022hdo,Kim:2021gyj} that the BPS sector has the BPS expansion
\begin{equation}\label{eq:BPSsector}
    \mathcal{F}_{\text{W},\q}(\epsilon_1,\epsilon_2,t,\hat{t})= \mathcal{I}^{|\q|-1}(\epsilon_1,\epsilon_2)\sum_{\beta\in H_2(X;
    \mathbb{Z})}\sum_{j_L,j_R}(-1)^{2j_L+2j_R}{\widetilde{N}_{j_L,j_R}^{\beta}}\chi_{j_L}(q_L)\chi_{j_R}(q_R)e^{\beta \cdot t+\q\cdot \hat{t}}\,,
\end{equation}
where the notation is like in (\ref{eq:BPSexpansion},\ref{eq:Ifactor}) and  $\q=[\q_1,\cdots,\q_r]$ is the charge vector for the Wilson loop with total charge $|\q|=\q_1+\cdots+\q_r$, $t_i,\ i=1,\cdots,b_2$ are K\"ahler moduli of the Calabi-Yau threefold $X$, $\hat{t}_i,\ i=1,\cdots,b_4$ are K\"ahler moduli related to the non-compact curves $\C_i$. Note that
there is no multi covering sum and  $\mathcal{I}(\epsilon_1,\epsilon_2)$ is the momentum factor encoding the dynamical contributions of the Wilson lines.
The expectation values of the Wilson loop operators are denoted by $\langle W_{\q}\rangle$ and can be computed from the generating function
\begin{align}\label{eq:Zgen}
    Z_{\text{gen}}&=\exp\left(\sum_{\q_i\geq 0}\frac{1}{\prod_{i=1}^r\q_i!}\mathcal{F}_{\text{W},\q}(\epsilon_1,\epsilon_2,t,\hat{t}\,)\right)=e^{\mathcal{F}(\epsilon_1,\epsilon_2,t)}\left(1+\sum_{|\q|>0}\frac{1}{\prod_{i=1}^r\q_i!}\langle W_{\q}\rangle e^{\q\cdot\hat{m}}\right),
\end{align}
where $\mathcal{F}(\epsilon_1,\epsilon_2,t)=\mathcal{F}_{\text{W},[0]}(\epsilon_1,\epsilon_2,t)$ is the refined topological string free energy for the local geometry $X$ and $\hat{m}\sim\hat{t}$ represent the masses of the Wilson loop particles, which will be defined in \eqref{eq:effectivemass} in the context of their connection to compact Calabi-Yau threefolds. 
More precisely, the subscript $\q$ in the Wilson loop operator $W_{\q}$ indicates the Wilson loop is in the tensor product representation $\mathbf{r}_1^{\otimes \q_1}\otimes\cdots\otimes \mathbf{r}_r^{\otimes \q_r}$, where $\mathbf{r}_i=[0,\cdots,0,\q_i=1,0,\cdots,0]$ is the representation with minimal charge in the $i$-th factor of the gauge group. The aspects of the generating functions in Gromov-Witten theory are currently under investigation in \cite{GWW}.

In \cite{Huang:2022hdo}, it was proposed that the BPS spectrum of the Wilson loops can be computed by adding a hypermultiplet with mass $m_i$ in the minimal representation $\mathbf{r}_i$,
providing an alternative way to understand the BPS expansion of the Wilson loops. 
In the large mass limit $m_i\rightarrow  -\infty$ \footnote{The minus sign arises from the notation for the K\"ahler parameters used in this paper.}, the particle in the hypermultiplet corresponds to the Wilson loop particle in the 5d gauge theory. However, the heavy half-BPS particle obtained in this way is not precisely the Wilson loop particle, as the dynamic term $\mathcal{I}(\epsilon_1,\epsilon_2)$ defined in \eqref{eq:Ifactor} must be absorbed into the mass $m_i$ to define the effective mass of the Wilson loop particle:
\begin{align}\label{eq:effectiveM}
    M_i=\mathcal{I}^{-1}(\epsilon_1,\epsilon_2)\,e^{ m_i}.
\end{align}

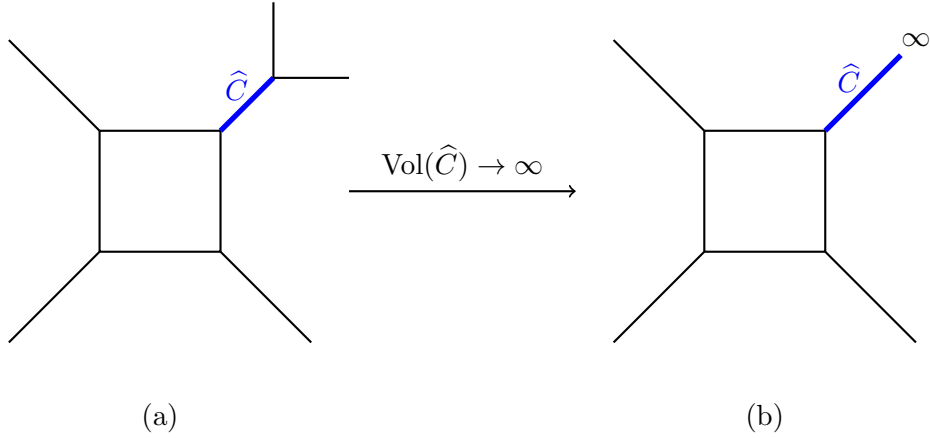
\begin{figure}[ht]
\begin{center}
\begin{tikzpicture}
\def\x{0.8}
\draw[ thick] (-\x,\x) -- (\x,\x) -- (\x,-\x) -- (-\x,-\x) -- (-\x,\x);
\draw[ thick] (-\x,\x) -- (-2,2);
\draw[line width=2pt, blue] (\x,\x) -- (1.5,1.5);
\draw[ thick] (1.5,1.5) -- (2.5,1.5);
\draw[ thick] (1.5,1.5) -- (1.5,2.5);
\draw[ thick] (\x,-\x) -- (2,-2);
\draw[ thick] (-\x,-\x) -- (-2,-2);

\draw[blue] (1.3,1.4) node[anchor=east]{$\C$};
\draw (0,-3) node{(a)};
\def\shift{4}
\draw[ thick,->] (-1.5+\shift,0) -- (1.5+\shift,0) ;
\draw[] (0+\shift,0) node[anchor=south]{$\mathrm{Vol}(\C)\rightarrow \infty$};
\def\s{8}
\def\x{0.8}
\draw[ thick] (-\x+\s,\x) -- (\x+\s,\x) -- (\x+\s,-\x) -- (-\x+\s,-\x) -- (-\x+\s,\x);
\draw[ thick] (-\x+\s,\x) -- (-2+\s,2);
\draw[line width=2pt, blue] (\x+\s,\x) -- (1.8+\s,1.8);
\draw[ thick] (\x+\s,-\x) -- (2+\s,-2);
\draw[ thick] (-\x+\s,-\x) -- (-2+\s,-2);
\draw[blue] (1.4+\s,1.5) node[anchor=east]{$\C$};
\draw[] (2+\s,2) node{$\infty$};
\draw (0+\s,-3) node{(b)};
\end{tikzpicture}
\end{center}
\caption{Brane diagrams in the $x^{5,6}$ directions. (a) The brane diagram for $SU(2)$ gauge theory with one flavor. $\C$ is the curve class related to the edge in the diagram. (b) The brane diagram with an infinitely long curve $\C$.}
\label{fig:su2braneC}
\end{figure}

For example, in the $SU(2)$ case, we can introduce a hypermultiplet in the fundamental representation, where the IIB brane diagram is described in Figure~\ref{fig:su2braneC}, with additional $(1,1)$ strings lying on the $(1,1)$ five-branes along $\C$.  The curve $\C$ is obtained from the one point blowup of $\mathbb{P}^1\times\mathbb{P}^1$ 
and then flopping the exceptional curve $e_1$ of the blowup to obtain $\C$ as the flopped $\mathbb{P}^1$. In the large mass limit where the volume of $\C$ is taken to infinity, only one semi-infinite $(1,1)$ string can survive.  This $(1,1)$ string corresponds to the half-BPS Wilson loop in 5d in the fundamental representation, a bosonic cousin of the fermionic Wilson loop depicted in Figure~\ref{table:IIBofcodimension4defect}. 
Higher representations can be introduced by adding more fundamental matter, which can be done in the geometry by blowing-up local $\mathbb{P}^1\times\mathbb{P}^1$ $\q$ times, with exceptional curves $e_i,i=1,\cdots, \q$. Denoting the curve classes for the two $\mathbb{P}^1$ factors by $h_1,h_2$, we deduce that the refined BPS number for the Wilson loop with charge $\q$ can be identified with the refined BPS number of the blown-up geometry via
\begin{align}\label{eq:BPSconnections}
    \widetilde{N}_{j_L,j_R}^{d_1 h_1+d_2 h_2}={N}_{j_L,j_R}^{d_1 h_1+d_2 h_2-e_1-\cdots -e_{\q}},\qquad \text{if }\q\leq 8,
\end{align}
which has been proposed and verified in \cite{Huang:2022hdo}.\footnote{The flop of $e_i$ has class $-e_i$, so in the flopped geometry, the right hand side of (\ref{eq:BPSconnections}) is just a rephrasing of our previous description of Wilson loops. We are asserting that this formula is still valid in the unflopped local geometry, since flopping a curve corresponds to an analytic continuation of the function described in Appendix~\ref{subsec:analcont}.}
The additional momentum factor $\mathcal{I}$ in the generating function \eqref{eq:BPSsector} comes from the normalization of the mass parameter \eqref{eq:effectiveM}.

It was proposed in \cite{Huang:2022hdo} that the Wilson loop partition functions satisfy refined holomorphic anomaly equations, which are reformulated in \cite{Wang:2023zcb} to the refined holomorphic anomaly equations for the BPS sectors as
\begin{align}\label{eq:HAE_Wilson}
   \bar{\partial}_{\bar i} \mathcal{F}^{(n,g)}_{\q}=\frac{1}{2}\bar{C}^{jk}_i\left[D_jD_k\mathcal{F}^{(n,g-1)}_{\q}+{\sum_{n^{\prime},g^{\prime},\q^{\prime}}}^{\prime}\prod_{i=1}^r\binom{\q_i}{\q_i^{\prime}}\,
   D_j\mathcal{F}^{(n^{\prime},g^{\prime})}_{\q^{\prime}}D_k\mathcal{F}^{(n-n^{\prime},g-g^{\prime})}_{\q-\q^{\prime}}\right].
\end{align}
The prime sum means the sum is over $0\leq n^{\prime} \leq n$, $0\leq g^{\prime} \leq g$ and $0\leq \q_i^{\prime} \leq \q_i$ by excluding $(n^{\prime},g^{\prime},\q_i^{\prime})=(0,0,0)$ and $(n^{\prime},g^{\prime},\q_i^{\prime})=(n,g,\q_i)$. The charge $\q^{\prime}$ is defined as $\q^{\prime}=[\q_1^{\prime},\cdots,\q_r^{\prime}]$.
The binomial coefficients in the last summation come from the permutation of $\q_i$ strings with charge 1.
In the holomorphic limit, $\mathcal{F}^{(n,g)}_{\q}$ is the genus $(n,g)$ free energy of the BPS sector defined from the genus expansion
\begin{align}
    \mathcal{F}_{\text{W},\q}=\sum_{n,g=0}^{\infty}(\epsilon_1+\epsilon_2)^{2n}(\epsilon_1\epsilon_2)^{g+|\q|-1}\mathcal{F}^{(n,g)}_{\q}.
\end{align}
In the special case $|\hat{q}|=0$, the refined holomorphic anomaly equations \eqref{eq:HAE_Wilson} reduce to the conventional refined holomorphic anomaly equations for local Calabi-Yau threefolds \cite{Huang:2010kf,Krefl:2010fm}.

\subsection{Connection to compact Calabi-Yau threefolds}\label{sec:compactCY3viaWilson}
In the last section, we reviewed the connection between the Wilson loops and topological strings on non-compact Calabi-Yau threefolds. The key consequence is that if a Calabi-Yau threefold $X$ can reduce to a non-compact Calabi-Yau threefold $X_{*}$ by taking the volumes of some of the curve classes $\C_1,\cdots$ to infinity, one can treat the BPS particles from M2-branes wrapping over $\C_1,\cdots$ as Wilson loop particles and the topological string partition functions on $X$ in some curve class degrees are proportional to the partition functions of Wilson loops in the 5d gauge theory coming from M-theory compactified on $X_{*}$. In this section, we provide evidence and physical arguments that the correspondence to Wilson loops can be generalized to a compact Calabi-Yau threefold $X$.

Consider a 5d $\mathcal{N}=1$ supergravity theory obtained from M-theory compactified on $X$. Denote a compact curve in $X$ as $\C$. In the large volume limit of $\C$, the volume of $X$ becomes infinite and the gravity in the 5d theory is decoupled. Under this limit, the half-BPS particles arising from M2 branes wrapping $\C$ become heavy and their dynamic degrees of freedom are frozen. Consequently, we can effectively treat these particles as Wilson loop particles in the local theory. Thus, we expect the topological string partition function on $X$, with at least $\C_1,\cdots$ degree one, can be calculated from the BPS partition functions of the Wilson loops. Furthermore, since the calculations for the partition functions of Wilson loops can be refined, we expect that this approach provides a framework for calculating \emph{refined BPS numbers} for compact Calabi-Yau threefolds.

More precisely, let $C,\C\in H_2({X};\mathbb{Z})$ denote curves in the compact Calabi-Yau threefold ${X}$, with $t,\hat{t}$ as the corresponding K\"ahler parameters. In a local limit by taking $\hat{t}$ to $-\infty$, we obtain a non-compact Calabi-Yau threefold $X_*$ with K\"ahler parameters $t$. Let $D_i\in H^4(X_*;\mathbb{Z})$ represent the compact divisors in $X_*$ and let $\phi_i$ denote the dual K\"ahler parameters associated with $D_i$. The gauge charge neutral combination
\begin{align}\label{eq:effectivemass}
    \hat{m}=\hat{t}-\sum_i \q_{i} \phi_i,\qquad \q_{i}=D_i\cdot \C,
\end{align}
defines the effective masses for the Wilson loop particles, where the parameters $\phi_i$ are Coulomb parameters in the effective 5d local theory.
If all the intersection numbers satisfy $\q_{i}\leq 1$, in the following expansion of the compact Calabi-Yau free energy
\begin{align}
    \mathcal{F}(t,\hat{t};\epsilon_1,\epsilon_2)=\sum_{\hat{\beta}}
    \mathcal{F}_{\hat{\beta}}(t;\epsilon_1,\epsilon_2)e^{ \hat{\beta}\cdot \hat{m}},
\end{align}
we claim $\mathcal{F}_{\hat{\beta}}(t;\epsilon_1,\epsilon_2)$ with $\hat{\beta}\cdot\q=1$ is proportional to the refined BPS sector \eqref{eq:BPSsector} of the Wilson loop in the local theory with charge $\hat{\beta}$. In the following, we study examples to test our statement.

\paragraph{Elliptic $\mathbb{P}^2$}
Our first example is the smooth elliptic fibration over $\mathbb{P}^2$, also known as the Calabi-Yau hypersurface denoted by $X_{18}(1,1,1,6,9)$. This case is a two parameter model, as discussed in \cite{Hosono:1993qy,Candelas:1994}, with the following Mori cone charges:
\begin{equation}\label{eq:MC_ellP2}
\begin{tabular}{cccccccccccc}
\\
$l^{(1)}$& $=$ &$($&$-6$ & ; & $2$ & $3$ & $1$ & $0$ & $0$ & $0$ &$)$  \\
$l^{(2)}$& $=$ &$($& $0$ & ; & $0$ & $0$ & $-3$ & $1$ & $1$ & $1$ & $)$ \\
&&&&&&&$\downarrow$ & & & \\
&&&&&&& $D$ & & &  \\
\end{tabular}
\end{equation}
This geometry has a local limit obtained by taking $t_1\rightarrow -\infty$. In this limit the global elliptically fibered Calabi-Yau threefold reduces to a local geometry $\mathcal{O}(-3)\rightarrow \mathbb{P}^2$, where the compact divisor $D$ is the base $\mathbb{P}^2$. In this case, the dual parameter associated with $D$ is given by $\phi=-\frac{t_2}{3}$ and the effective mass for the Wilson loop particle is $\hat{m}=t_1+\frac{t_2}{3}$. In the expansion of the refined free energy
\begin{equation}
    \mathcal{F}(t_1,t_2;\epsilon_1,\epsilon_2)=\sum_{d_1=0}^{\infty}
    \mathcal{F}_{d_1}(t_2;\epsilon_1,\epsilon_2) e^{d_1\hat{m}}\,,
\end{equation}
$\mathcal{F}_{d_1}$ captures the contributions from the curve classes $(d_1,d_2\geq 0)$. The leading term $\mathcal{F}_{0}$ is the refined free energy for local $\mathbb{P}^2$, calculated in \cite{Huang:2010kf}, and  
the subleading term $\mathcal{F}_{1}$ is factorized as
\begin{equation}
    \mathcal{F}_1(t_2;\epsilon_1,\epsilon_2)=f(\epsilon_1,\epsilon_2)\mathcal{F}^{\mathbb{P}^2}_{\text{W},[1]}(\phi)\,,
\end{equation}
where $\mathcal{F}^{\mathbb{P}^2}_{\text{W},[1]}=\left\langle W_{[1]}^{\mathbb{P}^2}\right\rangle$, calculated in \cite{Kim:2021gyj,Huang:2022hdo,Wang:2023zcb}, is the VEV of the Wilson loop for local $\mathbb{P}^2$ in the representation with minimal gauge charge $1$. The factor 
\begin{align}\label{eq:f_ellP2}
    f(\epsilon_1,\epsilon_2)=\frac{546-(q_-^{-1}+q_-)(q_+^{-2}+1+q_+^2)}{2\sinh(\epsilon_1/2)\cdot 2\sinh(\epsilon_2/2)}\,,\qquad q_{\pm}=e^{\frac{1}{2}(\epsilon_1\pm\epsilon_2)}\,,
\end{align}
is the generating function for the refined BPS numbers of the class of the elliptic fiber, which has degree $(d_1,d_2)=(1,0)$. Physically, the factor $f(\epsilon_1,\epsilon_2)$ is the `frozen' contribution from the M2-brane wrapping on the elliptic fiber in the massive limit $\hat{m}\rightarrow -\infty$. At genus zero, we find
\begin{align}
    \epsilon_1\epsilon_2\mathcal{F}_1(t_2;\epsilon_1,\epsilon_2)|_{\epsilon_{1,2}\rightarrow 0}=540e^{-\frac{1}{3}t_2}\,(1-2e^{t_2}+5e^{2t_2}-32e^{3t_2}+286e^{4t_2}+\cdots)\,,
\end{align}
which indeed agrees with the genus zero GV invariants at $d_E=1$.
\paragraph{Elliptic $\mathbb{F}_0$}
Our second example is the elliptic fibration over the Hirzebruch surface $\mathbb{F}_0$. This is a three-parameter model whose Mori cone generators are given by
\begin{align}\label{eq:ellF0Mori}
\begin{tabular}{ccccccccccccc}
\\
$l^{(1)}$& $=$ &$($&$-6$ & ; & $2$ & $3$ & $1$ & $0$ & $0$ & $0$ & $0$ &$)$  \\
$l^{(2)}$& $=$ &$($& $0$ & ; & $0$ & $0$ & $-2$ & $1$ & $0$ & $0$ & $1$ & $)$ \\
$l^{(3)}$& $=$ &$($& $0$ & ; & $0$ & $0$ & $-2$ & $0$ & $1$ & $1$ & $0$ & $)$ \\
&&&&&&&$\downarrow$ &  & && \\
&&&&&&& $D$ & & &&  \\
\end{tabular}
\end{align}
In the limit $t_1\rightarrow -\infty$, we obtain the local $\mathbb{F}_0$ geometry with the compact divisor $D=\mathbb{F}_0$. 
The 5d low-energy effective theory corresponding to the local geometry is the 5d $SU(2)_0$ theory, where the subscript $0$ denotes the theta angle $0$. The dual parameter $\phi$ associated with $D$ is defined as $\phi=-\frac{t_2}{2}$, which is the Coulomb parameter for $SU(2)_0$, and the effective mass for the Wilson loop particle is given by $\hat{m}=t_1+\frac{1}{2}t_2$.
The refined free energy has the expansion
\begin{equation}
    \mathcal{F}(t_1,t_2,t_3;\epsilon_1,\epsilon_2)=\sum_{d_1=0}^{\infty}
    \mathcal{F}_{d_1}(t_2,t_3;\epsilon_1,\epsilon_2) e^{d_1\hat{m}}\,,
\end{equation}
where the leading term $\mathcal{F}_0$ is
the refined free energy for local $\mathbb{F}_0$ and the subleading term $\mathcal{F}_1$ is factorized
\begin{align}
    \mathcal{F}_{1}(t_2,t_3;\epsilon_1,\epsilon_2)=f(\epsilon_1,\epsilon_2)\mathcal{F}^{\mathbb{F}_0}_{\text{W},[1]}(\phi,m)\,,
\end{align}
where $\mathcal{F}^{\mathbb{F}_0}_{\text{W},[1]}=\left\langle W_{[1]}^{SU(2)_0}\right\rangle$ is the VEV for the Wilson loop in the fundamental representation of the gauge algebra $\mathfrak{su}(2)$ and $m$ is the mass parameter defined from the instanton counting parameter $\mathfrak{q}=e^{m}$. Here $f(\epsilon_1,\epsilon_2)$ is the frozen contribution of the Wilson loop particle, which can be calculated from the BPS spectrum of the M2-brane wrapping the curve class $l^{(1)}$ as 
\begin{align}
    f(\epsilon_1,\epsilon_2)=\frac{488-(q_-^{-1}+q_-)(q_+^{-1}+q_+)^2}{2\sinh(\epsilon_1/2)\cdot 2\sinh(\epsilon_2/2)}\,.
\end{align}

\paragraph{Elliptic $\mathbb{F}_1$, phase I}
The third example is the elliptic fibration over the Hirzebruch surface $\mathbb{F}_1$, whose Mori cone generators are
\begin{align}\label{eq:Mori_EllF1_PhaseI}
\begin{tabular}{ccccccccccccc}
\\
$l^{(1)}$& $=$ &$($&$-6$ & ; & $2$ & $3$ & $1$ & $0$ & $0$ & $0$ & $0$ &$)$  \\
$l^{(2)}$& $=$ &$($& $0$ & ; & $0$ & $0$ & $-2$ & $1$ & $0$ & $0$ & $1$ & $)$ \\
$l^{(3)}$& $=$ &$($& $0$ & ; & $0$ & $0$ & $-1$ & $-1$ & $1$ & $1$ & $0$ & $)$ \\
&&&&&&&$\downarrow$ & $\downarrow$ & && \\
&&&&&&& $D_1$ & $D_2$& &&  \\
\end{tabular}
\end{align}
This geometry has two local limits which are local $\mathbb{F}_1$ and local half K3. 
In the limit $t_1\rightarrow -\infty$, we obtain local $\mathbb{F}_1$ with the compact divisor $D_1=\mathbb{F}_1$. 
The 5d low-energy effective theory corresponding to this local geometry is the 5d $SU(2)_{\pi}$ theory with theta angle $\pi$. The dual parameter $\phi$ associated with $D_1$ is defined as $\phi=-\frac{t_2}{2}$, which is the Coulomb parameter for the $SU(2)_{\pi}$ theory, and the effective mass for the Wilson loop particle is given by $\hat{m}=t_1+\frac{1}{2}t_2$.
The refined free energy has the expansion
\begin{equation}
    \mathcal{F}(t_1,t_2,t_3;\epsilon_1,\epsilon_2)=\sum_{d_1=0}^{\infty}
    \mathcal{F}_{d_1}(t_2,t_3;\epsilon_1,\epsilon_2) e^{d_1\hat{m}}.
\end{equation}
where
\begin{align}
    \mathcal{F}_{1}(t_2,t_3;\epsilon_1,\epsilon_2)=f_{(1,0,0)}(\epsilon_1,\epsilon_2)\left(\mathcal{F}^{\mathbb{F}_1}_{\text{W},[1]}+\cdots\right).
\label{eqn:localF1dE1}
\end{align}
In (\ref{eqn:localF1dE1}), $\mathcal{F}^{\mathbb{F}_1}_{\text{W},[1]}=\left\langle W_{[1]}^{SU(2)_{\pi}}\right\rangle$ is the VEV for the Wilson loop in the fundamental representation of the gauge algebra $\mathfrak{su}(2)$ and $m$ is the mass parameter defined from the instanton couting parameter $\mathfrak{q}=e^{m}$. Here $\cdots$ is the contribution from the curve class $\beta=(d_1,d_2,d_3)=(1,0,1)$, where the corresponding BPS particle has $0$ gauge charge and is therefore not captured by the Wilson loop calculations. We will return to these omitted contributions in Section~\ref{subsubsec:g0basecurve}. The factor $f(\epsilon_1,\epsilon_2)$ is the frozen contribution of the Wilson loop particle, which can be calculated from the BPS spectrum of M2-branes wrapping the curve associated with $l^{(1)}$ and takes the form 
\begin{align}
    f(\epsilon_1,\epsilon_2)=\frac{488-(q_-^{-1}+q_-)(q_+^{-1}+q_+)^2}{2\sinh(\epsilon_1/2)\cdot 2\sinh(\epsilon_2/2)}.
\end{align}

Another local limit is obtained by taking $t_2\rightarrow  -\infty$. The corresponding local geometry is the local half K3, which is related to the circle compactification of the 6d E-string theory. The 6d E-string theory has a tensor multiplet.  On the tensor branch, a non-zero tensor parameter $\phi_0$ is turned on, which corresponds to the K\"ahler parameters as $\phi_0=t_3+\frac{1}{2}t_1$. The shift $\frac{1}{2}t_1$ is added to make the $\mathrm{SL}(2,\mathbb{Z})$ duality in the fiber direction more manifest as has been discussed in \cite{Huang:2015sta,Klemm:2012sx}. See Appendix \ref{app:E-string} for more details about the partition function of the E-string theory.

Consider the expansion of the refined free energy
\begin{equation}\label{eq:ellF1_Estring}
    \mathcal{F}(t_1,t_2,t_3;\epsilon_1,\epsilon_2)=\sum_{d_2=0}^{\infty}
    \mathcal{F}_{d_2}^{\prime}(t_1,t_3;\epsilon_1,\epsilon_2) e^{d_2(t_2+t_3)}.
\end{equation}
We expect that the subleading contribution of the expansion \eqref{eq:ellF1_Estring} at $d_2=1$ is factorized as
\begin{align}\label{eq:ellF1_Estr}
    \mathcal{F}_{1}^{\prime}(t_1,t_3;\epsilon_1,\epsilon_2) =Z^{K3}(\epsilon_1,\epsilon_2,\tau) \left(\left \langle W_{\text{E-str},[1]}\right\rangle +\cdots\right),
\end{align}
where $\tau=\frac{t_1}{2\pi \ri}$ and $\left \langle W_{\text{E-str},[1]}\right\rangle=\frac{Z_{\text{E-str},[1]}}{Z_{\text{E-str},[0]}}$ is the VEV for the Wilson surface defect calculated in \cite{Chen:2021ivd} and reviewed in Appendix \ref{app:E-string}.
Wilson surfaces of a 6d theory are surface operators that carrying tensor charges \footnote{Under circle compactification, the Wilson surface in E-string theory reduces to the corresponding Wilson loop of the 5d Kaluza-Klein (KK) theory. Then the tensor charges are identified as the gauge charges of the 5d KK theory.}. Here $\cdots$ corresponds to contributions at the curve classes $\beta=(d_1,d_2,d_3)=(d_1,1,1),\,d_1\geq 0$ which have zero tensor charge and are therefore not captured by the Wilson surface calculations. The factor $Z^{K3}(\epsilon_1,\epsilon_2,\tau)$ arises because all the curve classes $(d_1\geq 0,1,0)$ have tensor charge 1. These contributions combine to give the effective tension of the Wilson surface defect, which is the refined partition function of the K3 fibration over $\mathbb{P}^1$. A possible refinement for this partition function was proposed in  \cite{KKP} as
\begin{align}
    Z^{K3}(\epsilon_1,\epsilon_2,\tau)=\frac{2E_4(\tau)E_6(\tau)}{\eta(\tau)^{18}\theta_1(\epsilon_1,\tau)\theta_1(\epsilon_2,\tau)}+\frac{(-2+q_++q_+^{-1})\eta(\tau)^2}{q^{\frac{5}{6}}\theta_1(\epsilon_1,\tau)\theta_1(\epsilon_2,\tau)},
    \label{eqn:kkp}
\end{align}
where $\theta_i(z,\tau)$ are the Jacobi theta functions, $\eta(\tau)$ is the Dedekind eta function and $E_4(\tau),E_6(\tau)$ are the weight 4 and the weight 6 Eisenstein series.
\paragraph{Elliptic $\mathbb{F}_1$, phase II}
There exists another Calabi-Yau phase corresponding to the flopping of the curve $l^{(3)}$ in \eqref{eq:Mori_EllF1_PhaseI}. The Mori cone generators for this phase are given by
\begin{align}\label{eq:Mori_EllF1_PhaseII}
\begin{tabular}{ccccccccccccc}
\\
$l^{(1)}$& $=$ &$($&$-6$ & ; & $2$ & $3$ & $0$ & $-1$ & $1$ & $1$ & $0$ &$)$  \\
$l^{(2)}$& $=$ &$($& $0$ & ; & $0$ & $0$ & $-3$ & $0$ & $1$ & $1$ & $1$ & $)$ \\
$l^{(3)}$& $=$ &$($& $0$ & ; & $0$ & $0$ & $1$ & $1$ & $-1$ & $-1$ & $0$ & $)$ \\
&&&&&&&$\downarrow$ & $\downarrow$& && \\
&&&&&&& $D_1$ & $D_2$& &&  \\
\end{tabular}
\end{align}
In the limit $t_1,t_3\rightarrow -\infty$, we obtain local $\mathbb{P}^2$ with the compact divisor $D_1$. Similarly, in the limit $t_2,t_3\rightarrow -\infty$, we obtain local $E_8$ del Pezzo surface $dP_8$ with the compact divisor $D_2$.
Define $\hat{m}=t_3+t_1+\frac{1}{3}t_2$ as the effective mass for Wilson loop particles in both of these local theories. Consider the expansion of the refined free energy
\begin{equation}
    \mathcal{F}(t_1,t_2,t_3;\epsilon_1,\epsilon_2)=\sum_{d_3=0}^{\infty}
    \mathcal{F}_{d_3}(t_1,t_2;\epsilon_1,\epsilon_2) e^{d_3\hat{m}}\,.
\end{equation}
At $d_3=1$, the M2-brane over $l^{(3)}$ provides the Wilson loop particle for both local theories simultaneously. Therefore $\mathcal{F}_{1}$ should be factorized as
\begin{align}\label{eq:ellF1_d3=1}
    \mathcal{F}_{1}=f(\epsilon_1,\epsilon_2)\mathcal{F}^{\mathbb{P}^2}_{\text{W},[1]}(\phi)\left(\mathcal{F}^{dP_8}_{\text{W},[1]}(\phi^{\prime})+\cdots\right)
\end{align}
where $\phi=-\frac{1}{3}t_2,\,\phi^{\prime}=-t_1$ are the Coulomb parameters for the 5d quantum field theories corresponding to local $\mathbb{P}^2$ and local $dP_8$ respectively. The factor $f(\epsilon_1,\epsilon_2)$ is calculated from the BPS spectrum of M2-branes wrapping the curve associated with $l^{(3)}$ and takes the form
\begin{align}
    f(\epsilon_1,\epsilon_2)=\frac{1}{2\sinh(\epsilon_1/2)\cdot 2\sinh(\epsilon_2/2)}.
\end{align}
The terms $\cdots$ in \eqref{eq:ellF1_d3=1} correspond to contributions of the curve class $\beta=(d_1,d_2,d_3)=(1,0,1)$ which have zero gauge charge and are therefore not captured by the Wilson loop calculations.

\subsection{Proposal for the general case}\label{sec:3.4}
In the previous subsection, we demonstrated that the refined free energy of a compact elliptically fibered Calabi-Yau threefold ${X}$ can be related to the Wilson loop amplitudes of a local Calabi-Yau $X_*$, which is obtained from the local limit of ${X}$ by taking the volume of a specific curve class $\widehat{C}$ to infinity.  In this section, we extend this analysis to more general cases, exploring how the refined BPS partition function of compact Calabi-Yau threefolds can similarly be connected to Wilson loop amplitudes in their corresponding local limits.

In the most general case, we can find a curve class $\widehat{C}\in H_2({X};\mathbb{Z})$ in the compact Calabi-Yau threefold ${X}$ whose large volume limit consists of neighborhoods $X_i$ of mutually disjoint connected compact divisors $D^{(1)},D^{(2)},\ldots,D^{(n)}$, each defining a local theory $\mathcal{T}_{X_i}$.  Decomposing $D^{(i)}$ into its components $D^{(i)}_j$, we have the Wilson loop charges 
\begin{align}
     q_j^{(i)}:=D_j^{(i)}\cdot\widehat{C}\geq 0,\quad   D_j^{(i)}\in H^4(X_i;\mathbb{Z}).
\end{align}
In terms of $D_k^{(i)}\in H_4(X_i;\mathbb{Z})$ and a basis $C_l^{(i)}$ for $H_2(X_i;\mathbb{Z})$, the charge matrix $Q_{G,kl}^{(i)}$ is defined as
\begin{align}\label{eq:QG}
    D_k^{(i)}\cdot C_l^{(j)} =Q_{G,kl}^{(i)}\,\delta_{ij}, \quad k=1,\cdots,b_4(X_i);\,l=1,\cdots,b_2(X_j).
\end{align}
The partition function for each local theory $Z_{X_i}(t^{(i)})=Z_{X_i}(\epsilon_1,\epsilon_2; t^{(i)})$  depends on the K\"ahler parameters $t^{(i)}_j,j=1,\cdots,b_2(X_i)$, which can be reformulated in the 5d gauge theory in terms of Coulomb parameters $\phi_j^{(i)},j=1,\cdots,r_i$ and mass parameters $m_l^{(i)},l=1,\cdots,f_i$ by
\begin{align}\label{eq:t2um}
    t_j^{(i)}=\sum_{l=1}^{r_i}\phi_l^{(i)}Q_{G,lj}^{(i)} +\sum_{l=1}^{f_i}m_l^{(i)}Q_{F,lj}^{(i)}\,,
\end{align}
where $r_i=b_4(X_i)$ is the rank of the gauge group and $f_i=b_2(X_i)-b_4(X_i)$ is the rank of the flavor group. The charge matrix $Q_G^{(i)}$ was defined in \eqref{eq:QG} and $Q^{(i)}_F$ is the intersection matrix of selected non-compact divisors and the compact curves $C_j^{(i)}$.  These matrices respectively compute the gauge charges and flavor charges of the BPS particles arising from wrapping an M2-brane on $C_j^{(i)}$.

We claim that the partition function of the compact Calabi-Yau threefold can be written as a linear combination of Wilson loop partition functions in the local theory $\mathcal{T}_{X_i}$ in {\emph{all}} representations. Physically, it can be explained as follows. Consider the 5d $\mathcal{N}=1$ supergravity theory on $\mathbb{R}^4\times S^1$ obtained from M-theory compactification on the compact Calabi-Yau threefold ${X}$. In the large volume limit of the curve class $\widehat{C}$, the supergravity theory is effectively described by a sequence of 5d local gauge theories $\mathcal{T}_{X_i}$. The half-BPS particles from M2-branes on the $\widehat{C}$ direction become heavy, carry electric charges $\hat{q}^{(i)}$, and become the sources of the half-BPS Wilson loops along the time direction $S^1$ in each local theory $\mathcal{T}_{X_i}$. Thus, from the viewpoint of each local theory, the partition function of ${X}$ should be expressed in terms of the partition functions of Wilson loops in $\mathcal{T}_{X_i}$. We arrive at the structure
\begin{align} \label{eq:ansatz_struc}
    Z_{{X}}(t)=\prod_{i=1}^nZ_{X_i}(t^{(i)})\cdot\left[1+\sum_{k=1}^{\infty}e^{ \,k \hat{m}}Z_k\right],
\end{align}
where 
\begin{equation}\label{eq:mhat}
\hat{m}=t_{\widehat{C}}-\sum_{i=1}^n\sum_{j=1}^{b_4(X_i)}q_j^{(i)}\phi_j^{(i)}
\end{equation}
is the effective mass for the Wilson loop particle, $\phi_j^{(i)}$ are K\"ahler parameters with respect to bases in $H_2(X_i;\mathbb{Q})$ that are dual to $D_{j}^{(i)}$ which are also the Coulomb parameters in each local theory $\mathcal{T}_{X_i}$ and $Z_k$ is a product of linear combinations of Wilson loops satisfying the ansatz
\begin{align}\label{eq:general_ansatz}
    Z_k(t)= \prod_{i=1}^n\,\,\sum_{h_1^{(i)},\cdots,h_{r_i}^{(i)}\geq 0}\,\,\sum_{\beta_m^{(i)}\in\mathbb{Z}}\, P^{(i)}_{[h_1^{(i)},\cdots,h_{r_{i}}^{(i)}],k;\beta_m^{(i)}}\left\langle W^{(i)}_{[h_1^{(i)},\cdots,h_{r_{i}}^{(i)}]}\right\rangle\,e^{ \beta_m^{(i)}\cdot m^{(i)}},\quad 
\end{align}
up to coefficients
\begin{align}
    \prod_{i=1}^{n}P^{(i)}_{[h_1^{(i)},\cdots,h_{r_{i}}^{(i)}],k;\beta_m^{(i)}}(\epsilon_1,\epsilon_2)=\frac{\mathcal{P}[e^{\epsilon_1},e^{\epsilon_2}]}{\prod_{l=1}^{k}4\sinh(l\epsilon_1/2)\sinh(l\epsilon_2/2)},
\end{align}
where $\mathcal{P}[e^{\epsilon_1},e^{\epsilon_2}]$ is a Laurent polynomial of $e^{\epsilon_{1,2}}$ which can be determined from additional input for refined BPS numbers.

The expectation values of the Wilson loops have the asymptotic behavior when the Coulomb parameters $\phi^{(i)}$ are large :
\begin{align}\label{eq:W_asymptotic}
    \left\langle W^{(i)}_{[h_1^{(i)},\cdots,h_{r_{i}}^{(i)}]}\right\rangle=e^{ h_1^{(i)}\phi_{1}^{(i)}+\cdots +h_{r_i}^{(i)}\phi_{r_i}^{(i)}}\left(1+\mathcal{O}(e^{t^{(i)}})\right).
\end{align}
Moreover, in the bracket on the right hand side of \eqref{eq:W_asymptotic}, the series expansion can be reformulated in terms of the bases $t^{(i)}$ with integral degrees.
The summation in \eqref{eq:general_ansatz} can be reduced via two additional constraints \eqref{eq:constraint1} and \eqref{eq:constraint2}. These constraints can be derived from the one-form symmetry for the local gauge theories, along with the positivity and integrality conditions as follows:
\begin{itemize}
    \item {\bf One-form symmetry}:
        \subitem Denote the one-form symmetry of the 5d local theory $\mathcal{T}_{X_i}$ by $\Gamma^{(1)}_i=\prod_j Z_{p_j^{(i)}}$ . As pointed out in \cite{Morrison:2020ool}, $\Gamma^{(1)}_i$ can be calculated from the Smith normal form of the charge matrix 
        \begin{equation}\label{eq:SNF}
            \mathrm{SNF}(Q_G^{(i)})=U^{(i)}\cdot Q_G^{(i)}\cdot V^{(i)}=
            \begin{pmatrix}
                p_{1}^{(i)} & 0 & \cdots & 0 & \cdots &0 \\
                0 & p_{2}^{(i)} & \cdots & 0 & \cdots & 0 \\
                \vdots & & \ddots & \vdots & & \vdots \\
                0 & \cdots &  & p_{r_i}^{(i)} & \cdots & 0
            \end{pmatrix}\,,
        \end{equation}
        where the main diagonal entries are positive integers satisfying $p_1^{(i)}\leq \cdots \leq p_{r_i}^{(i)}$.  Here $U^{(i)}$ and $V^{(i)}$ are invertible unimodular matrices satisfying $\det U^{(i)}=\det V^{(i)}=1$, and both their entries and those of their inverse matrices are integers. Let $c^{(i)}_{j,1},\cdots,c^{(i)}_{j,r_i}$ be the integral one-form symmetry charges for the Wilson loops under $Z_{p_j^{(i)}}$. Then the charge of $\left\langle W^{(i)}_{[h_1^{(i)},\cdots,h_{r_{i}}^{(i)}]}\right\rangle$ under $\mathbb{Z}_{p_j^{(i)}}$ is $c_{j,1}^{(i)}h_1^{(i)}+\cdots +c_{j,r_i}^{(i)}h_{r_i}^{(i)} \mod p_j^{(i)}$.  The first constraint on \eqref{eq:general_ansatz} is that the Wilson loops must have the same charge under the one-form symmetry $\Gamma^{(i)}$
        \begin{align}\label{eq:constraint1}
        \boxed{
            c_{j,1}^{(i)}h_1^{(i)}+\cdots +c_{j,r_i}^{(i)}h_{r_i}^{(i)} =k(c_{j,1}^{(i)}q_1^{(i)}+\cdots +c_{j,r_i}^{(i)}q_{r_i}^{(i)}) \mod p_j^{(i)}, \quad i=1,\cdots,n.
        }\end{align}
        Note that if $p_j^{(i)}=1$, the condition \eqref{eq:constraint1} is trivial, hence we only need to consider the non-trivial one-form symmetries with $p_j^{(i)}>1$.
    \item  {\bf Positivity and integrality}:
        \subitem If we replace the Coulomb parameters and mass parameters of the local gauge theory with the K\"ahler parameters of the compact Calabi-Yau threefold, their degrees in the expansion of the partition function must be  non-negative and integral. Define the total charge matrix $Q^{(i)}_{\text{Tot}}=\left(
        \begin{array}{c}
            Q_G^{(i)}\\
            Q_F^{(i)}
        \end{array}\right)$ which is a $b_2(X_i)\times b_2(X_i)$ matrix. For a 5d gauge theory, the charge matrix is an invertible matrix, hence according to \eqref{eq:t2um} we have
        \begin{align}\label{eq:phi_m}
            (\phi^{(i)},m^{(i)})_j=\sum_{l=1}^{b_2(X_i)}t_l^{(i)}\left(Q^{(i)}_{\text{Tot}}\right)^{-1}_{lj},\quad j=1,\cdots,b_2(X_i).
        \end{align}
        For any given $k$, from \eqref{eq:ansatz_struc}, \eqref{eq:general_ansatz} and \eqref{eq:W_asymptotic}, the relevant part we need to consider is  
        \begin{align}\label{eq:constraint2_pre1}
            \sum_{j=1}^{r_i}(h_j^{(i)}-kq_j^{(i)})\phi_j^{(i)}+\sum_{l=1}^{f_i}\beta_{m,l}^{(i)}m^{(i)}_l=\sum_{l=1}^{b_2(X_i)}\beta^{(i)}_lt_l^{(i)}.
        \end{align} 
        The second constraint is that the coefficient $\beta^{(i)}_l$ must be a non-negative integer for any $l=1,\cdots,b_2(X_i)$ and $i=1,\cdots,n$:
        \begin{align}\label{eq:constraint2}
        \boxed{
            \beta^{(i)}_l=\sum_{j=1}^{r_i}\left(Q^{(i)}_{\text{Tot}}\right)^{-1}_{lj}(h_j^{(i)}-kq_j^{(i)})+\sum_{j=r_i+1}^{b_2(X_i)}\left(Q^{(i)}_{\text{Tot}}\right)^{-1}_{lj}\beta_{m,j-r_i}^{(i)} \quad\in \mathbb{Z}_{\geq0}
        }\end{align}
       
\end{itemize}

\paragraph{Comment} In the following, we demonstrate that  condition \eqref{eq:constraint1} can be derived from condition \eqref{eq:constraint2} by borrowing ideas from~\cite{Tian:2025yrj}. More precisely, for any $p_j^{(i)}$ defined in \eqref{eq:SNF}, if there exists a set of integers $n_1^{(i)},\cdots,n_{b_2(X_i)}^{(i)}$, such that the action of the one-form symmetry\footnote{More precisely, this is the effect of the one-form symmetry action on the Coulomb branch partition function.}
\begin{equation}\label{eq:shift2}
    \phi_l^{(i)}\rightarrow \phi_l^{(i)} +2\pi i\, \frac{{{c}}_{j,l}^{(i)}}{p_j^{(i)}}\,,\qquad m_l^{(i)} \rightarrow m_l^{(i)}\,,
\end{equation}
is identical to the $B$-field shift of the complexified K\"ahler parameters
\begin{equation}\label{eq:shift}
    t_l^{(i)}\rightarrow t_l^{(i)} +2\pi i\, n_l^{(i)}\,,
\end{equation}
then according to \eqref{eq:constraint2_pre1},
\begin{align}\label{eq:constraint2_pre2}
    \sum_{j=1}^{r_i}(h_j^{(i)}-kq_j^{(i)})\frac{{{c}}_{j,l}^{(i)}}{p_j^{(i)}}=\sum_{l=1}^{b_2(X_i)}\beta^{(i)}_ln_l^{(i)}.
\end{align} 
If \eqref{eq:constraint2} is satisfied, we conclude that 
\begin{equation}
    \sum_{j=1}^{r_i}(h_j^{(i)}-kq_j^{(i)})\frac{{{c}}_{j,l}^{(i)}}{p_j^{(i)}}\in\mathbb{Z}\,,
\end{equation}
which gives rise to condition \eqref{eq:constraint1}.

In the following, we prove that \eqref{eq:shift2} can be generated by \eqref{eq:shift}. Without loss of generality, we drop the index $i$ which represents the $i$-th non-compact Calabi-Yau threefold $X_i$. 

Define the Smith normal forms of the gauge and flavor charge matrices $Q_G$ and $Q_F$ as
\begin{equation}
    \mathrm{SNF}(Q_G)=U_G\cdot Q_G\cdot V_G=
        \begin{pmatrix}
            P_{G,r\times r}& 0_{r\times f} 
        \end{pmatrix}_{r\times b_2}\,,
\end{equation}
and
\begin{equation}
    \mathrm{SNF}(Q_F)=U_F\cdot Q_F\cdot V_F=
        \begin{pmatrix}
            0_{f\times r} &P_{F,f\times f}
        \end{pmatrix}_{f\times b_2}\,,
\end{equation}
respectively. Here $P_{G,r\times r}=\mathrm{diag}(p_1,\cdots,p_{r})$ and $P_{F,f\times f}=\mathrm{diag}(p_{F,1},\cdots,p_{F,f})$ are diagonal matrices where the diagonal entries are the main diagonal entries for the Smith normal forms of $Q_G$ and $Q_F$ respectively. The inverse total charge matrix can be expressed as
\begin{equation}
    Q_{\text{Tot}}^{-1}=\begin{pmatrix}
    A&B
    \end{pmatrix}
\end{equation}
where
\begin{align}
    A_{}=V_G\cdot \begin{pmatrix} P_G^{-1}\\ C\end{pmatrix}\cdot U_G,\qquad B=V_F\cdot \begin{pmatrix} D\\ P_F^{-1}\end{pmatrix}\cdot U_F,
\end{align}
satisfying
\begin{align}
    Q_F\cdot A=0,\qquad Q_G\cdot B=0\,.
\end{align}
If the shift \eqref{eq:shift2} can be generated from \eqref{eq:shift}, according to \eqref{eq:phi_m},
\begin{align}\label{eq:nA=c}
   n\cdot A=\tilde{c},\qquad n\cdot B=0
\end{align}
where $\tilde{c}_l=\frac{{{c}}_{j,l}}{p_j}$. We find there are $r$ sets of solutions for $n$, given by the rows of the $r\times b_2$ matrix
\begin{align}\label{eq:Nmat}
    N=P_G^{-1}\cdot\begin{pmatrix} P_G& 0_{r\times f}\end{pmatrix}\cdot V_G^{-1}.
\end{align}
It is easy to check all entries of $N$ are integers.
Substituting the solutions \eqref{eq:Nmat} to \eqref{eq:nA=c}, we find the one-form symmetry charges are expressed as
\begin{align}
    \frac{{{c}}_{j,l}}{p_j}=\left(P_G^{-1}\cdot U_G\right)_{jl},
\end{align}
which indeed have the correct property.

\subsubsection{The ansatz for elliptic \texorpdfstring{$\mathbb{P}^2$}{P2}}
\label{subsubsec:ellipticP2}

In this section, we verify \eqref{eq:ansatz_struc} for the model elliptic $\mathbb{P}^2$, where the Mori cone charges are described in \eqref{eq:MC_ellP2}. 
Define $\hat{m}=t_1+\frac{1}{3}t_2$ as the effective mass of the Wilson loop particle in line with \eqref{eq:effectivemass}. In the large mass limit $\hat{m}\rightarrow -\infty$, the geometry of elliptic $\mathbb{P}^2$, which is denoted by ${X}$, reduces to the local Calabi-Yau threefold $X_*=\mathcal{O}(-3)\rightarrow \mathbb{P}^2$, which corresponds to a 5d non-Lagrangian theory $\mathcal{T}_{X_*}$.
Based on the ansatz \eqref{eq:ansatz_struc}, the partition function for the compact Calabi-Yau threefold ${X}$ can be expressed as 
\begin{equation}\label{eq:ansatz}
    Z_{{X}}(t_1,t_2,\epsilon_1,\epsilon_2)=Z_{X_*}(t_2,\epsilon_1,\epsilon_2)\left[1+\sum_{k=1}^{\infty}e^{k\,\hat{m}}Z_k(t_2,\epsilon_1,\epsilon_2)\right]
\end{equation}
where 
\begin{equation}\label{eq:Zk_ansatz}
    Z_k(t_2,\epsilon_1,\epsilon_2)=P_{k}\left\langle W_{[k]}\right\rangle+\sum_{l=1}^{\left\lfloor \frac{k}{3}\right\rfloor}\widetilde{P}_{k,l}\left\langle W_{[k-3l]}\right\rangle.
\end{equation}
Here, $\left\langle W_{[k]} \right\rangle$ represents the Wilson loop\footnote{The 5d SQFT $\mathcal{T}_{X_*}$ corresponding to local $\mathbb{P}^2$ is a non-Langrangian theory without a gauge group. The concept of Wilson loops arises from additional M2-branes on non-compact curves in M-theory.} of charge $k$ in $\mathcal{T}_{X_*}$, which has been computed in \cite{Wang:2023zcb}(see also \cite{Huang:2022hdo,Kim:2021gyj}) using refined holomorphic anomaly equations. The second term on the RHS of \eqref{eq:Zk_ansatz} represents the only possible additional contribution permitted under the constraints of the one-form symmetry $\mathbb{Z}_3$ and the positivity and integrality condition \eqref{eq:constraint2}.

The coefficients $P_k=P_k(\epsilon_1,\epsilon_2)$ can be determined from refined BPS numbers in the fiber direction. In the limit $t_2\rightarrow -\infty$, as studied in \cite{Huang:2022hdo}, the VEVs of the Wilson loop operators have the asymptotic values
\begin{equation}
    \left.\left\langle W_{[k]}\right\rangle\right|_{t_2\rightarrow -\infty}=e^{-\frac{k}{3}t_2}.
\end{equation}
Using $f(\epsilon_1,\epsilon_2)$ defined in \eqref{eq:f_ellP2}, one finds upon defining $P_k$ from the expansion 
\begin{equation}
    \sum_{k=0}^{\infty}P_ke^{k\,t_1}=\exp\left[\sum_{k,l=1}^{\infty}\frac{1}{k}f(k\epsilon_1,k\epsilon_2)e^{k\,l\,t_1}\right],
\end{equation}
we indeed obtain the correct refined BPS expansion in the fiber direction:
\begin{equation}
    \left.Z_{{X}}(t_1,t_2,\epsilon_1,\epsilon_2)\right|_{t_2\rightarrow -\infty}=\exp\left[\sum_{k,l=1}^{\infty}\frac{1}{k}f(k\epsilon_1,k\epsilon_2)e^{k\,l\,t_1}\right].
\end{equation}
The coefficients $\widetilde{P}_{k,l}=\widetilde{P}_{k,l}(\epsilon_{1},\epsilon_2)$, which are rational functions of $q_{1,2}=e^{\epsilon_{1,2}}$, are determined from additional boundary conditions.  In general, they take the form
\begin{equation}
    \widetilde{P}_{k,l}(\epsilon_1,\epsilon_2)=\frac{\widetilde{\mathcal{P}}_{k,l}(q_+,q_-)}{\prod_{s=1}^k 4\sinh(s\epsilon_1/2) \sinh(s\epsilon_2/2)},
\end{equation}
where $\widetilde{\mathcal{P}}_{k,l}$ are Laurent polynomials of $q_{\pm}$.

Consequently, we can solve the refined BPS numbers by considering the expansion
\begin{align}
    Z_{{X}}(t_1,t_2,\epsilon_1,\epsilon_2)=\exp\left[\sum_{k=1}^{\infty}\sum_{d_E=0}^{\infty}\frac{1}{k}\mathcal{F}_{d_E}(k\,t_2,k\epsilon_1,k\epsilon_2)e^{k\,d_E\,\hat{m}}\right],
\end{align}
where the free energy $\mathcal{F}_{d_E}$ has the expansion
\begin{align}
    \mathcal{F}_{d_E}(t_2,\epsilon_1,\epsilon_2)= \mathcal{I}^{-1}\sum_{d_B =0}^{\infty}\sum_{j_L,j_R}(-1)^{2j_L+2j_R}{{N}_{j_L,j_R}^{\beta}}\,\chi_{j_L}(q_-)\chi_{j_R}(q_+)\,e^{(d_B-\frac{1}{3} d_E)t_2}
\end{align}
\begin{table}[t]
    \centering
{\footnotesize\begin{tabular}{|c|cccc|}
 \hline $2j_L \backslash 2j_R$ & 0 & 1 & 2 & 3 \\ \hline
 0 &  & 546 &  &  \\
 1 &  & 1 &  & 1 \\ \hline
\end{tabular} \vskip 3pt  $d_B=1$ \vskip 10pt}

{\footnotesize\begin{tabular}{|c|ccccccc|}
 \hline $2j_L \backslash 2j_R$ & 0 & 1 & 2 & 3 & 4 & 5 & 6 \\ \hline
 0 &  &  &  &  & 546 &  &  \\
 1 &  &  & 1 &  & 1 &  & 1 \\ \hline
\end{tabular} \vskip 3pt  $d_B=2$ \vskip 10pt}

{\footnotesize\begin{tabular}{|c|ccccccccccc|}
 \hline $2j_L \backslash 2j_R$ & 0 & 1 & 2 & 3 & 4 & 5 & 6 & 7 & 8 & 9 & 10 \\ \hline
 0 &  &  &  &  &  & 546 & 1 & 546 & 1 &  & 1 \\
 1 &  &  &  & 1 &  & 2 &  & 2 & 546 & 1 &  \\
 2 &  &  &  &  &  &  & 1 &  & 1 &  & 1 \\ \hline
\end{tabular} \vskip 3pt  $d_B=3$ \vskip 10pt}

{\footnotesize\begin{tabular}{|c|cccccccccccccccc|}
 \hline $2j_L \backslash 2j_R$ & 0 & 1 & 2 & 3 & 4 & 5 & 6 & 7 & 8 & 9 & 10 & 11 & 12 & 13 & 14 & 15 \\ \hline
 0 &  &  &  &  & 546 & 1 & 546 & 3 & 1092 & 5 & 546 & 4 & 546 & 2 &  &  \\
 1 &  &  & 1 &  & 2 &  & 4 & 546 & 5 & 1092 & 6 & 1092 & 4 &  & 2 &  \\
 2 &  &  &  &  &  & 1 &  & 3 &  & 5 & 546 & 5 & 546 & 3 &  & 1 \\
 3 &  &  &  &  &  &  &  &  & 1 &  & 2 &  & 2 & 546 & 1 &  \\
 4 &  &  &  &  &  &  &  &  &  &  &  & 1 &  & 1 &  & 1 \\ \hline
\end{tabular} \vskip 3pt  $d_B=4$ \vskip 10pt}

    \caption{The refined BPS numbers for $d_E=1$}
    \label{tab:my_label}
\end{table}
\begin{table}[t]
    \centering
{\footnotesize\begin{tabular}{|c|ccccc|}
 \hline $2j_L \backslash 2j_R$ & 0 & 1 & 2 & 3 & 4 \\ \hline
 0 & 148785 & 546 & 2 &  &  \\
 1 & 546 & 1 & 546 & 1 &  \\
 2 & 1 &  & 1 &  & 1 \\ \hline
\end{tabular} \vskip 3pt  $d_B=1$ \vskip 10pt}

{\footnotesize\begin{tabular}{|c|cccccccc|}
 \hline $2j_L \backslash 2j_R$ & 0 & 1 & 2 & 3 & 4 & 5 & 6 & 7 \\ \hline
 0 & 546 & 2 &  & 148787 & 546 & 2 &  &  \\
 1 &  & 546 & 2 & 1092 & 1 & 546 & 1 &  \\
 2 &  & 2 &  & 3 &  & 2 &  & 1 \\ \hline
\end{tabular} \vskip 3pt  $d_B=2$ \vskip 10pt}

{\footnotesize\begin{tabular}{|c|cccccccccccc|}
 \hline $2j_L \backslash 2j_R$ & 0 & 1 & 2 & 3 & 4 & 5 & 6 & 7 & 8 & 9 & 10 & 11 \\ \hline
 0 &  &  & 3 & 546 & 148790 & 1092 & 298123 & 1092 & 4 & 546 & 2 &  \\
 1 &  & 1 & 546 & 3 & 2184 & 6 & 2184 & 148791 & 1638 & 4 &  & 1 \\
 2 & 1 &  & 3 &  & 6 & 546 & 7 & 1092 & 5 & 546 & 2 &  \\
 3 &  &  &  & 1 &  & 2 &  & 3 &  & 2 &  & 1 \\ \hline
\end{tabular} \vskip 3pt  $d_B=3$ \vskip 10pt}

    \caption{The refined BPS numbers for $d_E=2$}
    \label{tab:my_label1}
\end{table}
For instance, for $d_E=1$, we have:
\begin{equation}\label{eq:F1ellP2}
    \mathcal{F}_1(t_2;\epsilon_1,\epsilon_2)=f(\epsilon_1,\epsilon_2)\left\langle W_{[1]}\right\rangle.
\end{equation}
For $d_E=2$, we find
\begin{equation}\label{eq:F1ellP2_d2=2}\begin{split}
    \mathcal{F}_2(t_2;\epsilon_1,\epsilon_2)=&\left[f(\epsilon_1,\epsilon_2)+\frac{1}{2}f(2\epsilon_1,2\epsilon_2)+\frac{1}{2}f^2(\epsilon_1,\epsilon_2)\right]\left\langle W_{[2]}\right\rangle\\
    &\qquad\qquad-\frac{1}{2}\mathcal{F}_1(2t_2;2\epsilon_1,2\epsilon_2)-\frac{1}{2}\mathcal{F}_1^2(t_2;\epsilon_1,\epsilon_2).
\end{split}\end{equation}
The refined BPS numbers determined from \eqref{eq:F1ellP2} and \eqref{eq:F1ellP2_d2=2} are listed in Table \ref{tab:my_label} and Table \ref{tab:my_label1} respectively.
For $d_E=3$, we have
\begin{equation}\label{eq:F1ellP2_d2=3}\begin{split}
    \mathcal{F}_3(t_2;\epsilon_1,\epsilon_2)=&P_3\left\langle W_{[3]}\right\rangle-\frac{1}{6}\mathcal{F}_1^3(t_2;\epsilon_1,\epsilon_2)-\frac{1}{2}\mathcal{F}_1(t_2;\epsilon_1,\epsilon_2)\mathcal{F}_1(2t_2;2\epsilon_1,2\epsilon_2)\\
    &\qquad-\frac{1}{3}\mathcal{F}_1(3t_2;3\epsilon_1,3\epsilon_2)-\mathcal{F}_1(t_2;\epsilon_1,\epsilon_2)\mathcal{F}_2(t_2;\epsilon_1,\epsilon_2)+\widetilde{P}_{3,1}(\epsilon_1,\epsilon_2).
\end{split}\end{equation}
where
\begin{equation}
    P_3=f_1+f_1^2+\frac{1}{6}f_1^3+\frac{1}{2}f_1f_2+\frac{1}{3}f_3,\quad f_k=f(k\epsilon_1,k\epsilon_2),
\end{equation}
and $\widetilde{P}_{3,1}(\epsilon_1,\epsilon_2)$ is undetermined. Equation \eqref{eq:F1ellP2_d2=3} shows that the refined BPS numbers in the fiber direction fully determine the refined BPS numbers for $d_E = 3$, except for the curve class degree $(d_B, d_E) = (1,3)$, which depends on the expression for $\widetilde{P}_{3,1}$. Here we list the refined BPS numbers for the first few base degrees with $d_B>1$ in Table \ref{tab:my_label2}. They agree with the calculations at low genera
in \cite[Appendix A]{Huang:2020dbh}.  In Section~\ref{subsubsec:dE3} we will give a geometric explanation for the necessity of a contribution to $Z_3$ beyond $P_3\left\langle W_{[3]}\right\rangle$ .

Although our primary interest lies in the refinement of the topological string on a compact Calabi-Yau threefold, it is still worthwhile to check the ansatz at the unrefined level $\epsilon_1=-\epsilon_2=\lambda$. Using the unrefined BPS invariants that are calculated from the modular bootstrap method in \cite{Huang:2015sta} and the BPS invariants of Wilson loops calculated in \cite{Wang:2023zcb}, we determine the unknown coefficients $\widetilde{P}_{k,l}$ for $k\leq 12$. We confirm that the overlapping BPS invariants from \eqref{eq:ansatz} are consistent with the calculations in \cite{Huang:2015sta}. In particular, we observed that $\widetilde{P}_{k,l}$ has the structure 
\begin{align}
    \widetilde{P}_{k,l}(\lambda,-\lambda)=\frac{(-1)^k\left(2\sinh(\lambda/2)\right)^{4l}}{\prod_{s=1}^k\left(2\sinh(s\lambda/2)\right)^2}\sum_j c_j e^{- j \lambda},
\end{align}
where the summation runs over all the integers $|j|<\frac{1}{2}k(k+1)-\frac{3}{2}l(l+1)+k\,l$ and $c_j$ are integers.

\begin{table}[t]
    \centering
{\footnotesize\begin{tabular}{|c|ccccccccc|}
 \hline $2j_L \backslash 2j_R$ & 0 & 1 & 2 & 3 & 4 & 5 & 6 & 7 & 8 \\ \hline
 0 & 2184 & 298122 & 26981318 & 298123 & 2184 & 4 &  & 1 &  \\
 1 & 148789 & 2730 & 446912 & 2730 & 148793 & 1092 & 4 &  &  \\
 2 & 1092 & 6 & 2730 & 7 & 1638 & 4 & 546 & 1 &  \\
 3 & 3 &  & 5 &  & 5 &  & 2 &  & 1 \\ \hline
\end{tabular} \vskip 3pt  $d_B=2$ \vskip 10pt}

{{\tiny\begin{tabular}{|c|ccccccccccccc|}
 \hline $2j_L \backslash 2j_R$ & 0 & 1 & 2 & 3 & 4 & 5 & 6 & 7 & 8 & 9 & 10 & 11 & 12 \\ \hline
 0 & 4 & 2184 & 298130 & 26985686 & 745037 & 81244800 & 745037 & 5460 & 148797 & 1638 & 5 &  &  \\
 1 & 1092 & 148800 & 4368 & 893831 & 8736 & 1191954 & 26987324 & 745045 & 4914 & 13 & 546 & 3 &  \\
 2 & 5 & 2730 & 16 & 6552 & 148810 & 8190 & 446927 & 6006 & 148802 & 2184 & 7 &  & 1 \\
 3 &  & 8 & 546 & 16 & 1638 & 19 & 2730 & 15 & 1638 & 8 & 546 & 2 &  \\
 4 & 1 &  & 2 &  & 5 &  & 6 &  & 5 &  & 2 &  & 1 \\ \hline
\end{tabular}} \vskip 3pt  {\footnotesize{$d_B=3$}} \vskip 10pt}

    \caption{The refined BPS numbers for $d_E=3$}
    \label{tab:my_label2}
\end{table}

\subsection{Refined holomorphic anomaly equations for compact Calabi-Yau threefolds}\label{sec:3.5}
In this section, we derive refined holomorphic anomaly equations for compact Calabi-Yau threefolds based on the ansatz \eqref{eq:ansatz_struc} and the refined holomorphic anomaly equations \eqref{eq:HAE_Wilson} for Wilson loops.

As proposed in \cite{Huang:2022hdo,Wang:2023zcb}, the refined holomorphic anomaly equations \eqref{eq:HAE_Wilson} for Wilson loop amplitudes can be reformulated as a master equation, in the form of the heat kernel equation:
\begin{equation}\label{eq:HAEgeneral_masterZ}
    \left[\bar{\partial}_{\bar i}-\frac{\epsilon_1\epsilon_2}{2}\bar{C}^{jk}_{\bar{i}}D_jD_k\right] \mathcal{Z}_{\mathbf{r}}(t,\bar{t},\epsilon_1,\epsilon_2)=0,
\end{equation}
where
\begin{equation}
    \mathcal{Z}_{\mathbf{r}}(t,\bar{t},\epsilon_1,\epsilon_2)=\exp\left(\sum_{\substack{n,g\geq 0,\\n+g>0}}\mathcal{F}_0^{(n,g)}(\epsilon_1+\epsilon_2)^{2n}(\epsilon_1\epsilon_2)^{g-1}\right)\left\langle W_{\mathbf{r}}\right\rangle(t,\bar{t},\epsilon_1,\epsilon_2)
\end{equation}
for Wilson loops in arbitrary representations $\mathbf{r}$. By combining the ansatz \eqref{eq:ansatz_struc}\eqref{eq:general_ansatz} and the master equation for Wilson loops \eqref{eq:HAEgeneral_masterZ}, we obtain
\begin{equation}\label{eq:HAEgeneral_masterZ2}
    \left[\bar{\partial}_{\bar{j}}^{(i)}-\frac{\epsilon_1\epsilon_2}{2}\bar{C}^{(i)kl}_{\bar{j}}D_k^{(i)}D_l^{(i)}\right] \mathcal{Z}_{{X}}(t,\bar{t},\hat{m},\epsilon_1,\epsilon_2)=0,
\end{equation}
where the superscript ${}^{(i)}$ in $\bar{\partial}_{\bar{j}}^{(i)}$, $\bar{C}^{(i)kl}_{\bar{j}}$ and $D_k^{(i)}$ indicates they are defined on the moduli spaces for each local Calabi-Yau threefold $X_i$. The variables $t=(t^{(1)},\cdots,t^{(n)})$ and $\bar{t}=(\bar{t}^{(1)},\cdots,\bar{t}^{(n)})$ are collections of the holomorphic and anti-holomorphic coordinates for the moduli spaces of $X_i$. $\mathcal{Z}_{{X}}$ is a non-holomorphic version of the refined partition function by excluding the genus zero free energies of each non-compact Calabi-Yau threefold $X_i$.
Define the genus expansion
\begin{align}
    \mathcal{Z}_{{X}}(t,\bar{t},\hat{m},\epsilon_1,\epsilon_2)=\exp\left[\sum_{\hat{\beta}}
    \mathcal{F}_{k}(t,\bar{t};\epsilon_1,\epsilon_2)e^{ k\,\hat{m}}\right]
\end{align}
where
\begin{align}
    \mathcal{F}_{k}(t,\bar{t};\epsilon_1,\epsilon_2)=\sum_{n,g=0}^{\infty}(\epsilon_1+\epsilon_2)^{2n}(\epsilon_1\epsilon_2)^{g-1}\mathcal{F}_{k}^{(n,g)}(t,\bar{t}).
\end{align}
We propose that there exist refined holomorphic anomaly equations on the union of the moduli spaces of the local Calabi-Yau threefold $X_i$:
\begin{align}\label{eq:HAEgeneral1}
   \bar{\partial}_{\bar{j}}^{(i)} \mathcal{F}^{(n,g)}_{\hat{\beta}}=\frac{1}{2}\bar{C}^{(i)kl}_{\bar{j}}\left[D_k^{(i)}D_l^{(i)}\mathcal{F}^{(n,g-1)}_{\hat{\beta}}+{\sum_{n^{\prime},g^{\prime},\hat{\beta}^{\prime}}}^{\prime}
   D_k^{(i)}\mathcal{F}^{(n^{\prime},g^{\prime})}_{\hat{\beta}^{\prime}}D_l^{(i)}\mathcal{F}^{(n-n^{\prime},g-g^{\prime})}_{\hat{\beta}-\hat{\beta}^{\prime}}\right],
\end{align}
The prime sum means the sum is over $0\leq n^{\prime} \leq n$, $0\leq g^{\prime} \leq g$ and $0\leq \hat{\beta}_i^{\prime} \leq \hat{\beta}_i$ by excluding $(n^{\prime},g^{\prime},\hat{\beta}^{\prime})=(0,0,0)$ and $(n^{\prime},g^{\prime},\hat{\beta}^{\prime})=(n,g,\hat{\beta})$.
In the holomorphic limit, we obtain
\begin{equation}\label{eq:HAEgeneral}
    \frac{\partial}{S^{(i)kl}} \mathcal{F}^{(n,g)}_{\hat{\beta}}=\frac{1}{2}\left[D_k^{(i)}D_l^{(i)}\mathcal{F}^{(n,g-1)}_{\hat{\beta}}+{\sum_{n^{\prime},g^{\prime},\hat{\beta}^{\prime}}}^{\prime}
   D_k^{(i)}\mathcal{F}^{(n^{\prime},g^{\prime})}_{\hat{\beta}^{\prime}}D_l^{(i)}\mathcal{F}^{(n-n^{\prime},g-g^{\prime})}_{\hat{\beta}-\hat{\beta}^{\prime}}\right],
\end{equation}
where $S^{(i)}$ are the propagators for the B-models of $X_i$.

\subsection{Refined holomorphic anomaly equations for elliptic \texorpdfstring{ $\mathbb{P}^2$}{P2}}
In this section, as an example, we study the refined holomorphic anomaly equations (HAE) for the elliptic $\mathbb{P}^2$ geometry.

Define $t=\frac{t_2}{2\pi i}$ in terms of the K\"ahler parameter for the base $\mathbb{P}^2$ and $\tau=\frac{t_1}{2\pi i}$ in terms of the K\"ahler parameter for the elliptic fiber. The free energy near the MUM point can be expanded in terms of $Q=e^{2\pi i t}$ and $q=e^{2\pi i\tau}$ as
\begin{equation}
    \mathcal{F}(Q,q;\epsilon_1,\epsilon_2)=\mathcal{F}_0(Q;\epsilon_1,\epsilon_2)+\sum_{d_E=1}^{\infty}\mathcal{F}_{\n}(Q;\epsilon_1,\epsilon_2) (qQ^{1/3})^{d_E},
\end{equation}
where $\mathcal{F}_0(Q;\epsilon_1,\epsilon_2)$ is precisely the free energy for local $\mathbb{P}^2$. We refer the reader to \cite{Haghighat:2008gw} for the B-model calculations of local $\mathbb{P}^2$, where we will use the same notation therein. For each coefficient of $(qQ^{1/3})^{\n}$, we have the genus expansion
\begin{equation}
    \mathcal{F}_{\n}(Q;\epsilon_1,\epsilon_2) =\sum_{n,g=0}^{\infty}(\epsilon_1+\epsilon_2)^{2n}(\epsilon_1\epsilon_2)^{g-1}\mathcal{F}_{\n}^{(n,g)}(Q).
\end{equation}
The combination $qQ^{1/3}$ corresponds to the effective mass for the Wilson loop particle which comes from the smallest nontrivial curve class which intersects with the base $\mathbb{P}^2$ with intersection number 0. 
In the B-model, suppose we have the complex structure parameters $z$ that are related to the K\"ahler parameters $t$, and denote the propagator for local $\mathbb{P}^2$ by $S=S^{zz}$. \footnote{See \cite[Section 4.4]{Huang:2022hdo} for detailed expressions for $S$ and $z$.}
The following HAEs can be checked recursively (see Section \ref{sec:5})
\begin{equation}
    \frac{\partial}{\partial S}\mathcal{F}_{\n}^{(0,0)}=\frac{1}{2}\sum_{\n^{\prime}=1}^{\n-1}D\mathcal{F}_{\n^{\prime}}^{(0,0)}\cdot D\mathcal{F}_{\n-\n^{\prime}}^{(0,0)},
\end{equation}
where
\begin{equation}
    \mathcal{F}_{\n=1}^{(0,0)}=540{z^{-\frac{1}{3}}}.
\end{equation}
We conjecture for any $\n\geq 1$, $\mathcal{F}_{\n}^{(0,0)}$ is regular at the conifold point and orbifold point, but is singular at the large volume point with $\mathcal{F}_{\n=1}^{(0,0)} \sim Q^{-\n/3}$.

For generic genus $(n,g)$, according to \ref{eq:HAEgeneral}, the following holds
\begin{align}\label{eq:HAEFn}
    \frac{\partial \mathcal{F}_{\n}^{(n,g)}}{\partial S}=\frac{1}{2} \left( D^2 \mathcal{F}_{\n}^{(n,g-1)} +{\sum_{\n^{\prime},n^{\prime},g^{\prime}}}^{\prime}D \mathcal{F}_{\n^{\prime}}^{(n^{\prime},g^{\prime})}\cdot D \mathcal{F}_{\n-\n^{\prime}}^{(n-n^{\prime},g-g^{\prime})} \right),
\end{align}
for any $\n>0$ and $n,g\geq 0$. 
Here the prime sum means we sum over all the integers $0\leq n^{\prime}\leq n$, $0\leq g^{\prime}\leq g$ and $0\leq d_E^{\prime}\leq d_E$ by excluding $(n^{\prime},g^{\prime},d_E^{\prime})=(0,0,0)$ and $(n^{\prime},g^{\prime},d_E^{\prime})=(n,g,d_E)$. 
Here $D$ is the covariant derivative defined as
\begin{align}
    D\mathcal{F}_{\n}^{(n,g)}=\partial_z\mathcal{F}_{\n}^{(n,g)},\quad D^2\mathcal{F}_{\n}^{(n,g)}=(\partial_z+\Gamma_z^{zz})\partial_z\mathcal{F}_{\n}^{(n,g)},
\end{align}
where $\Gamma_z^{zz}$ is the Christoffel symbol on the moduli space. When $d_E=0$, the refined HAE's become the equations for local $\mathbb{P}^2$ proposed in \cite{Huang:2010kf} and they are valid for $n+g\geq 2$. The holomorphic anomaly equation can be solved recursively by directly integrating over the propagator $S$. Up to a holomorphic ambiguity $f^{(n,g)}_{\n}(z)$, the genus $(n,g)$ free energy can be completely solved. We make the ansatz on the holomorphic ambiguity
\begin{align}\label{eq:ansatzP2}
    f^{(n,g)}_{\n}(z)=z^{-\frac{\n}{3}}\left(\sum_{i=1}^{2(n+g-1)+\n}\frac{a_i}{\Delta^i}+\sum_{i=0}^{o}b_i z^i\right)
\end{align}
for suitable constants $a_i,b_i$, where 
\begin{equation}
\Delta=1+27z,
\end{equation}
is the discriminant and
\begin{align}
    o=\left\lfloor \frac{1}{3}(2n+2g+\n)\right\rfloor.
\end{align}
We conjecture that the free energy $\mathcal{F}^{(n,g)}_{\n>0}$ is regular at the conifold point and orbifold point, which provides quite enough boundary conditions to fix most of the coefficients in the holomorphic ambiguity \eqref{eq:ansatzP2}. However, we still need the information for the refined BPS numbers
\begin{equation}
    n_{d_B,d_E}^{g_L,g_R},\quad d_B-d_E/3 \leq 0
\end{equation}
to solve the ambiguity completely. Since we know the complete BPS numbers for $(d_B,d_E)=(0,d_E)$, we can completely determine the refined BPS numbers for $d_E=1,2,$ and $d_B\geq 0$. 
For instance, at genus $(0,0)$:
\begin{align*}
    \mathcal{F}^{(0,0)}_1=\,540z^{-\frac{1}{3}}\,,\qquad
    \mathcal{F}^{(0,0)}_2=\,\frac{405}{2\,z^{8/3}}(80S-37z^2)\,.
\end{align*}
At genus $(0,1)$:
\begin{align*}
    \mathcal{F}^{(0,1)}_1=\,\,&3z^{-{13}/{3}}\Delta^{-1}(10 S^2+5 S z^2+11 z^4+297 z^5) \,,\\
    \mathcal{F}^{(0,1)}_2=\,\,&\frac{9}{4\,z^{26/3}\Delta^2}(800 S^4+200 S^3 z^2+130 S^2 z^4+11610 S^2 z^5+675 S z^6+33345 S z^7\\
    &\qquad\qquad-413 z^8+371790 S
   z^8-23652 z^9-319302 z^{10})\,.\\
\end{align*}
At genus $(1,0)$:
\begin{align*}
    \mathcal{F}^{(1,0)}_1=\,\,&\frac{1}{4}z^{-7/3}\Delta^{-1}(30 S-116 z^2-3537 z^3)\,,\\
    \mathcal{F}^{(1,0)}_2=\,\,&\frac{3}{16\,z^{20/3}\Delta^2}(1200 S^3+600 S^2 z^2+64800 S^2 z^3-9490 S z^4-466290 S z^5\\
    &\qquad\qquad +4243 z^6-7639920 S z^6+236007
   z^7+2732292 z^8)\,.\\
\end{align*}
These calculations match with the calculations from the ansatz \eqref{eq:ansatz}.

\section{Examples}\label{sec:examples}
In this section, based on the ansatz \eqref{eq:ansatz_struc} and the calculations for Wilson loops, we compute the refined BPS numbers for several elliptically fibered Calabi-Yau threefolds with the base $B=\mathbb{F}_n$, where $\mathbb{F}_n$ is a Hirzebruch surface. In the limit where the size of the elliptic fiber becomes infinity, the corresponding 5d theories reduce to 5d gauge theories with gauge groups $SU(N)$. We classify these compact Calabi-Yau threefolds by the ranks of their corresponding 5d gauge groups.

In general, for F-theory compactification on an elliptically fibered Calabi-Yau threefold, the low-energy physics is described by a 6d $\mathcal{N}=(1,0)$ supergravity theory. With further circle compactification of the 6d theory, we obtain a 5d KK theory, which can also be obtained from M-theory compactified on the same elliptically fibered Calabi-Yau threefold. The BPS partition functions are identical for the 6d theory and the 5d KK theory. 
Therefore, the BPS spectrum arising from the one-loop contributions of these 6d supergravity theories as well as their 6d SCFT or little string limits, provides additional input to fix the ansatz \eqref{eq:ansatz_struc}. For this reason, we review 6d $\mathcal{N}=(1,0)$ supergravity theories in Appendix \ref{app:A}.

Consider the compactification of a 6d $\mathcal{N}=(1,0)$ theory on a circle. At the unrefined level $\epsilon_2=-\epsilon_1$, the leading behavior of the genus zero and genus one free energies is fully controlled by the anomaly 4-form polynomial $X_4^{\alpha},\alpha=1,\cdots,b_2(B),$ defined in \eqref{eq:X4_1}.  At the refined level, the 4-form polynomial is modified by an additional term that corresponds to the $SU(2)_{\mathcal{R}}$ $\mathcal{R}$-symmetry, written in the form:
\begin{equation}\label{eq:X4_2}
    X_{4}^{\alpha}=-\frac{a^{\alpha}}{4}p_1(M_6)+\sum_i b_{i}^{\alpha}\mathrm{tr}\,F_i^2+y^{\alpha}c_2(\mathcal{R}).
\end{equation}
Here the coefficients $a^{\alpha},b_{i}^{\alpha}$ are anomaly coefficients determined by the classical geometric data, and $y^{\alpha}$ are undetermined. The term $p_1(M_6)$ is the first Pontryagin class of the tangent bundle of the six dimensional spacetime $M_6$ where the 6d supergravity theory lives, $F_i$ are the field strengths of the non-Abelian
gauge symmetries, and $c_2(\mathcal{R})$ is the second Chern class of the background $SU(2)_{\mathcal{R}}$ $\mathcal{R}$-symmetry bundle. With the anomaly 4-form polynomial \eqref{eq:X4_2} and the matter content of the 6d theory, the leading behavior of the refined genus zero and genus one free energies are derived in \eqref{eq:F6d_00},\eqref{eq:F6d_01} and \eqref{eq:F6d_10}. By considering the 5d $SU(N)$ gauge theory limit and comparing the refined free energies with \eqref{eq:completion}, we can finally determine $y^{\alpha}$.

For the case where the base of the elliptic fibration is a Hirzebruch surface $\mathbb{F}_n$, the anti-canonical class of the base is $-K_B=2E+(n+2)F$, where $E$ and $F$ are divisors in the base with intersection numbers $E^2=-n,F^2=0,E\cdot F=1$. We order the divisor classes as $(F,E)$. The values of $a^{\alpha}$, the intersection matrix $\Omega_{\alpha\beta}$ and its inverse are given by
\begin{align}
    a=(n+2,2),\qquad \Omega=\left(
        \begin{array}{rr}
            0&1\\
            1&-n\\
        \end{array}\right),\qquad \Omega^{-1}=\left(
        \begin{array}{rr}
            n&1\\
            1&0\\
        \end{array}\right).
\end{align}
The values of $b^{\alpha}$ correspond to the locations of  singular fibers.  We will clarify their values and determine $y^{\alpha}$ in the following example sections. 
\paragraph{Comment on Jacobi-form Ansatz}
On the tensor branch, the partition functions of 6d supergravity theories compactified on $T^2$ can be expanded in terms of the tensor parameters $t_{\text{b}}$ as
\begin{align}
    Z_{\text{6d}}=e^{\frac{1}{\epsilon_1\epsilon_2}\mathcal{E}}Z_{\text{1-loop}}\left(1+\sum_{k>0}Z_{k}(z;\tau)e^{k\cdot t_{\text{b}}}\right)
\end{align}
Here, $\tau$ denotes the K\"ahler parameter of the elliptic fiber and $z$ collectively represents $\epsilon_1,\epsilon_2,$ and all other K\"ahler parameters in the fiber direction. The quantities $\mathcal{E}$ and $Z_{\text{1-loop}}$ are defined in \eqref{eq:defE} and \eqref{eq:defZ1loop} respectively.
The functions $Z_{k}(z;\tau)$ are elliptic genera of worldsheet theories describing $k$-strings. They are meromorphic Jacobi-forms of weight zero, whose indices can be computed from the geometries \cite{Huang:2015sta,Schimannek:2019ijf,Cota:2019cjx}. In the 6d SCFT limit, it has been proposed in \cite{DelZotto:2018tcj,DelZotto:2017mee} that the Jacobi-form indices are extracted from the anomaly polynomial via the replacement rule \eqref{eq:replacementrule}. This method has also been applied to the 6d supergravity case \cite{Hayashi:2023hqa}. If one naively applies the replacement rule to the refined anomaly polynomial \eqref{eq:X4_2}, we obtain
\begin{align}
    \mathrm{Index}(Z_{k})=-\frac{1}{2}k^{\alpha}\Omega_{
    \alpha\beta}k^{\beta}(\epsilon_+^2-\epsilon_-^2)+k^{\alpha}\Omega_{
    \alpha\beta}\left(-\frac{a^{\beta}}{4}(\epsilon_1^2+\epsilon_2^2)+\sum_i b_{i}^{\beta}\sum_{i^{\prime},j^{\prime}} K_i^{i^{\prime}j^{\prime}} \phi_{i^{\prime}}\phi_{j^{\prime}}-y^{\beta}\epsilon_+^2\right).
\end{align}
For elliptic $\mathbb{P}^2$,
\begin{align}
    \Omega=1,\quad b_i = 0, \quad a=3,\quad y=-\frac{19}{36}
\end{align}
the resulting index becomes
\begin{align}
    \mathrm{Index}(Z_k)=\frac{1}{2}k(3-k)\epsilon_1\epsilon_2-\frac{89k}{144}(\epsilon_1+\epsilon_2)^2.
\end{align}
The appearance of a large denominator $144$ is unusual compared to other cases. However, this result agrees with the expected index from the refined holomorphic/modular anomaly equation proposed in \cite{Huang:2020dbh}. It would interesting to explore whether a modular expression for $Z_k$ exists at the refined level.
\subsection{Rank 1 examples}
In this section, we compute the refined BPS numbers for elliptic fibrations over the Hirzebruch surfaces $\mathbb{F}_0$ and $\mathbb{F}_1$. Since many details have been addressed in previous works, we focus on the ansatz \eqref{eq:ansatz_struc} for these two models and refer the readers to \cite{Huang:2015sta,Huang:2020dbh} for discussions on other aspects.

\subsubsection{Elliptic \texorpdfstring{$\mathbb{F}_0$}{F0}}
Let's consider the example elliptic $\mathbb{F}_0$, whose Mori cone charges can be found in equation \eqref{eq:ellF0Mori}. The triple intersection ring and the evaluation of the second Chern class $c_2$ on the K\"ahler forms are given by
\begin{align}
    \mathcal{R}=8 J_1^3+2 J_1^2 J_2+2 J_1^2 J_3+J_1 J_2 J_3,
\end{align}
and
\begin{align}
    \{\int c_2J_i\}=\{92,24,24\}\,,
\end{align}
respectively.
We are following the usual conventions here.  The $J_i$ are the K\"ahler generators dual to the $l^{(i)}$, and  a triple intersection evaluates to the coefficient of the corresponding monomial in $\mathcal{R}$, and is zero if the corresponding monomial doesn't occur in $\mathcal{R}$.   

The 6d $\mathcal{N}=(1,0)$ supergravity theory corresponding to this geometry has one tensor multiplet, no vector multiplets, and $244$ hypermultiplets:
\begin{equation}
    H=244,\qquad V=0,\qquad T=1.
\end{equation}
Therefore the spin content in the fiber direction, according to (\ref{eqn:Fnfiberclass}) or Appendix \ref{app:A2}, is given by
\begin{equation}\label{eq:BPSfiber_ellF0}
    \beta=(d_E,d_2,d_3)=(d_E,0,0):\qquad\left[\frac{1}{2},1\right] \oplus \left[\frac{1}{2},0\right] \oplus 488\left[0,0\right].
\end{equation}
The genus zero and genus one free energies are 
\begin{align}
    \mathcal{F}^{(0,0)}&=t_{\text{b}_1}t_{\text{b}_2}\tau+\frac{1}{3}\tau^3+\mathcal{O}(e^t),\\
    \mathcal{F}^{(0,1)}&=-t_{\text{b}_1}-t_{\text{b}_2}-\frac{11}{6}\tau+\mathcal{O}(e^t),\\
    \mathcal{F}^{(1,0)}&=\frac{1}{2}(t_{\text{b}_1}+t_{\text{b}_2})+\frac{1}{4}y^{\alpha}t_{\text{b}_{\alpha}}+\frac{23}{24}\tau+\mathcal{O}(e^t),
\end{align}
where $\tau=t_1$, $t_{\text{b}_1}=t_2-\tau,t_{\text{b}_2}=t_3-\tau$. By considering the exchange symmetry of $t_2$ and $t_3$ arising from the exchange of two $\mathbb{P}^1$'s in the base $\mathbb{F}_0=\mathbb{P}^1\times \mathbb{P}^1$, we further assume $y^{\alpha=1}=y^{\alpha=2}$.

In the large elliptic fiber limit $t_1\rightarrow -\infty$, we obtain local $\mathbb{F}_0$ with the compact surface $D=\mathbb{F}_0$. This local geometry corresponds to 5d pure $SU(2)_0$ theory with theta angle $0$. The curve $l^{(1)}$ corresponds to the Wilson loop particle with charge $1$ in the resulting 5d gauge theory. The rank of the gauge symmetry is $r=1$ and the rank of the flavor symmetry is $f=1$. The total charge matrix for the 5d theory is
\begin{equation}
    Q_{\text{Tot}}=\left(
        \begin{array}{c}
            Q_G\\ \hdashline[1pt/1pt]
            Q_F
        \end{array}\right)=\left(
        \begin{array}{rr}
            -2&-2\\\hdashline[1pt/1pt]
            0&1\\
        \end{array}\right).
\end{equation}
With these preparations, we obtain the relations
\begin{equation}\label{eq:ellF0_tto5d}
    t_1=\hat{m}+\phi,\qquad t_2=-2\phi,\qquad t_3=-2\phi+m,
\end{equation}
where $\phi$ is the Coulomb parameter, $m=\log \mathfrak{q}$ is the instanton parameter and $\hat{m}$ is the effective mass of the Wilson loop particle for the 5d gauge theory. Substituting \eqref{eq:ellF0_tto5d} into the genus zero and genus one free energies and comparing with \eqref{eq:FpolySUN}, we find they are consistent and determine $y=(-\frac{5}{12},-\frac{5}{12})$.

From the positivity and integrality conditions \eqref{eq:constraint2}, we find
\begin{align}
    -\frac{1}{2}(h-k)-\beta_m, \,\beta_m \in \mathbb{Z}_{\geq 0},
\end{align}
such that the ansatz \eqref{eq:ansatz_struc} is expressed as
\begin{equation}\label{eq:ansatz_F0}
    Z_{{X}}(t_1,t_2,t_3,\epsilon_1,\epsilon_2)=Z_{X_*}(t_2,t_3,\epsilon_1,\epsilon_2)\left[1+\sum_{k=1}^{\infty}e^{k\hat{m}}Z_k(t_2,t_3,\epsilon_1,\epsilon_2)\right]\,,
\end{equation}
where
\begin{equation}\label{eq:temp_ellF0_1}
    Z_k(t_2,t_3,\epsilon_1,\epsilon_2)=P_{k}\left\langle W_{[k]}\right\rangle+\sum_{l=1}^{\left\lfloor \frac{k}{2}\right\rfloor}\sum_{\beta_m=0}^l e^{\beta_m m}\widetilde{P}_{k,l,\beta_m}\left\langle W_{[k-2l]}\right\rangle.
\end{equation}
In the second term on the RHS of \eqref{eq:temp_ellF0_1}, one finds the gauge charges of the Wilson loops are decreasing by $2$, which is consistent with the one-form symmetry $\mathbb{Z}_2$ for the 5d $SU(2)_0$ theory.

Denote the refined BPS number contribution at the curve class $\beta$ as 
\begin{equation}
    f_{\beta}(\epsilon_1,\epsilon_2)=\sum_{j_L,j_R}(-1)^{2j_L+2j_R}N_{j_L,j_R}^{\beta}\frac{\chi_{j_L}(q_-)\chi_{j_R}(q_+)}{2\sinh(\epsilon_1/2)\cdot 2\sinh(\epsilon_2/2)}.
\end{equation}
The coefficients $P_k=P_k(\epsilon_1,\epsilon_2)$ can be extracted from the expansion
\begin{equation}\label{eq:Pk_F0}
    \sum_{k=0}^{\infty}P_kq^k=\exp\left[\sum_{k,l=1}^{\infty}\frac{1}{k}f_{1,0,0}(k\epsilon_1,k\epsilon_2)q^{kl}\right],
\end{equation}
where 
\begin{equation}\label{eq:fbeta_100}
    f_{1,0,0}(\epsilon_1,\epsilon_2)=\frac{488-(q_-^{-1}+q_-)(q_+^{-2}+2+q_+^2)}{2\sinh(\epsilon_1/2)\cdot 2\sinh(\epsilon_2/2)},
\end{equation}
is the factor determined from the refined BPS numbers in the fiber direction \eqref{eq:BPSfiber_ellF0}.
For $k=1$, we find
\begin{equation}
    Z_1(t_2,t_3,\epsilon_1,\epsilon_2)=P_{1}\left\langle W_{[k]}\right\rangle.
\end{equation}
For $k=2,3,4$, additional input for the refined BPS numbers of the curve classes\footnote{We note that the refined BPS numbers of the curve classes $\beta=(d_E,d_2,d_3)$ and $\beta=(d_E,d_3,d_2)$ are identical due to the exchange symmetry of two $\mathbb{P}^1$'s in the base.} $\beta=(2,1,0),(3,1,0),(4,1,0),(4,2,0),(4,1,1)$ allows us to fix the undetermined coefficients $\widetilde{P}$. For instance,
\begin{align*}
    &\widetilde{P}_{2,1,0}=\widetilde{P}_{2,1,1}=-2 f_{1,0,0}(\epsilon_1,\epsilon_2)-f_{1,0,0}(2 \epsilon_1,2 \epsilon_2)+f_{2,1,0}(\epsilon_1,\epsilon_2),\\
    &\widetilde{P}_{3,1,0}=\widetilde{P}_{3,0,1}=-2 f_{1,0,0}(\epsilon_1,\epsilon_2){}^2-f_{1,0,0}(3 \epsilon_1,3 \epsilon_2)+f_{3,1,0}(\epsilon_1,\epsilon_2)\nonumber\\
    &\qquad\qquad\qquad\qquad+f_{1,0,0}(\epsilon_1,\epsilon_2)(-f_{1,0,0}(2 \epsilon_1,2 \epsilon_2)+f_{2,1,0}(\epsilon_1,\epsilon_2)-3),\\
    &\qquad \vdots
\end{align*}
Fortunately, the refined BPS content in the curve class direction $\beta=(d_E,d_1,0)$ has been proposed in \cite{KKP} and is summarized in \eqref{eqn:kkp} for $d_1=1$. As has been proposed in \cite{Minahan:1998vr}, for $d_1>1$, the contributions are generated by Hecke transformations of those at $d_1=1$, leading to
\begin{equation}\begin{split}
    Z_{{X}}(t_1,t_2,t_3,\epsilon_1,\epsilon_2)|_{t_3\rightarrow -\infty} =\exp\Bigg[&\sum_{k,l=1}^{\infty}\frac{1}{k}f_{1,0,0}(k\epsilon_1,k\epsilon_2)e^{k\,l\,t_1}\\
    &+\sum_{d_2=1}^{\infty}\frac{1}{d_2}e^{d_2 (t_1+t_2)} \sum_{\substack{ad=d_2\\a,d\in\mathbb{Z}}}\sum_{b(\mathrm{mod}\, d)}Z^{K3}\left(a\epsilon_{1,2};\frac{a\tau+b}{d}\right)\Bigg],\label{eq:ZK3}
\end{split}\end{equation}
for $ \tau=\frac{t_1}{2\pi \ri}$. By using the refined partition function \eqref{eqn:kkp}, equation \eqref{eq:ZK3} completely solves the refined BPS numbers of curve classes $\beta=(d_E,d_2,0)$. In Table \ref{tab:K3direction}, we provide the first few refined BPS numbers. From these results, except for the refined BPS numbers of the curve class $\beta=(4,1,1)$, we completely determine the refined BPS numbers of curve classes $\beta=(d_E,d_2,d_3)$ for $d_E\leq 4$ and any $d_2,d_3\geq0$. We list the refined BPS numbers for the first few degrees in Table \ref{tab:BPSF0dE=1}, Table \ref{tab:BPSF0dE=2}, Table \ref{tab:BPSF0dE=3} and Table \ref{tab:BPSF0dE=4}. Note that the BPS numbers of the curve class $\beta=(3,2,0)$ in Table \ref{tab:K3direction} are not used to determine the ansatz. However, we find that the refined BPS numbers of this curve class, as listed in Table \ref{tab:BPSF0dE=3}, precisely match those in Table \ref{tab:K3direction}, providing an independent consistency check on our calculations.
\begin{table}[t]
\centering
{\footnotesize\begin{tabular}{|c|cc|}
 \hline $2j_L \backslash 2j_R$ & 0 & 1 \\ \hline
 0 &  & 1 \\ \hline
\multicolumn{3}{c}{$(d_E,d_2,d_3)$\,=\,$(0,1,0)$  \rule{0pt}{2.6ex}}\\ 
\end{tabular}}
\hspace{0.5cm}
{\footnotesize\begin{tabular}{|c|ccc|}
 \hline $2j_L \backslash 2j_R$ & 0 & 1 & 2 \\ \hline
 0 & 488 &  &  \\
 1 & 1 &  & 1 \\ \hline
\multicolumn{4}{c}{$(d_E,d_2,d_3)$\,=\,$(d,d,0),d>0$  \rule{0pt}{2.6ex}}\\ 
\end{tabular}}\vskip 6pt

{\footnotesize\begin{tabular}{|c|cccc|}
 \hline $2j_L \backslash 2j_R$ & 0 & 1 & 2 & 3 \\ \hline
 0 & 280964 & 1 &  &  \\
 1 & 1 & 488 & 1 &  \\
 2 &  & 1 &  & 1 \\ \hline
\multicolumn{5}{c}{$(d_E,d_2,d_3)$\,=\,$(2,1,0)$  \rule{0pt}{2.6ex}}\\ 
\end{tabular}}
\hspace{0.5cm}
{\footnotesize\begin{tabular}{|c|ccccc|}
 \hline $2j_L \backslash 2j_R$ & 0 & 1 & 2 & 3 & 4 \\ \hline
 0 & 15928440 & 2 &  & 1 &  \\
 1 & 2 & 281452 & 2 &  &  \\
 2 &  & 2 & 488 & 1 &  \\
 3 &  &  & 1 &  & 1 \\ \hline
\multicolumn{6}{c}{$(d_E,d_2,d_3)$\,=\,$(3,1,0)$  \rule{0pt}{2.6ex}}\\ 
\end{tabular}}\vskip 6pt

{\footnotesize\begin{tabular}{|c|cccccc|}
 \hline $2j_L \backslash 2j_R$ & 0 & 1 & 2 & 3 & 4 & 5 \\ \hline
 0 & 410133618 & 4 & 488 & 1 &  &  \\
 1 & 3 & 16209892 & 4 &  & 1 &  \\
 2 & 488 & 4 & 281452 & 3 &  &  \\
 3 & 1 &  & 2 & 488 & 1 &  \\
 4 &  &  &  & 1 &  & 1 \\ \hline
\multicolumn{7}{c}{$(d_E,d_2,d_3)$\,=\,$(4,1,0)$  \rule{0pt}{2.6ex}}\\ 
\end{tabular}}\vskip 6pt

{\footnotesize\begin{tabular}{|c|ccccc|}
 \hline $2j_L \backslash 2j_R$ & 0 & 1 & 2 & 3 & 4 \\ \hline
 0 & 15928440 & 2 &  & 1 &  \\
 1 & 2 & 281452 & 2 &  &  \\
 2 &  & 2 & 488 & 1 &  \\
 3 &  &  & 1 &  & 1 \\ \hline
\multicolumn{6}{c}{$(d_E,d_2,d_3)$\,=\,$(3,2,0)$  \rule{0pt}{2.6ex}}\\ 
\end{tabular}}\vskip 6pt

{\footnotesize\begin{tabular}{|c|ccccccc|}
 \hline $2j_L \backslash 2j_R$ & 0 & 1 & 2 & 3 & 4 & 5 & 6 \\ \hline
 0 & 6749497860 & 6 & 281452 & 3 &  &  &  \\
 1 & 6 & 426343510 & 8 & 488 & 2 &  &  \\
 2 & 281452 & 7 & 16210380 & 5 &  & 1 &  \\
 3 & 2 & 488 & 5 & 281452 & 3 &  &  \\
 4 &  & 1 &  & 2 & 488 & 1 &  \\
 5 &  &  &  &  & 1 &  & 1 \\ \hline
\multicolumn{8}{c}{$(d_E,d_2,d_3)$\,=\,$(4,2,0)$  \rule{0pt}{2.6ex}}\\ 
\end{tabular}}\vskip 6pt

    \caption{The refined BPS numbers of elliptic $\mathbb{F}_0$ for $d_E\leq 4$, $d_2=1, 2$ and $d_3=0$.}
    \label{tab:K3direction}
\end{table}

\subsubsection{Elliptic \texorpdfstring{${\mathbb{F}_{1}}$}{F1}}
In this section, we compute the refined BPS partition function for the elliptic $\mathbb{F}_1$ model.  As described in Section \ref{sec:compactCY3viaWilson}, the geometry of the elliptic $\mathbb{F}_1$ has two phases. In phase I \eqref{eq:Mori_EllF1_PhaseI}, there are two local limits: local $\mathbb{F}_1$  which is obtained by taking the volume of $l^{(1)}$ to infinity, and local half K3 which is obtained by taking the volume of $l^{(2)}$ to infinity. These two local limits correspond to the 5d $SU(2)_{\pi}$ theory with theta angle $\pi$ and the 6d E-string theory respectively. In phase II \eqref{eq:Mori_EllF1_PhaseII}, there are two local limits, respectively obtained by taking the volume of $l^{(3)}$ to infinity in addition to taking the volume of  $l^{(1)}$ or $l^{(2)}$ to infinity, yielding local $\mathbb{P}^2$ and local $dP_8$, respectively.  
\paragraph{Phase I}

In this phase, The triple intersection ring and the evaluation of the second Chern class $c_2$ on the K\"ahler forms are
\begin{align}
    \mathcal{R}=8 J_1^3+3 J_1^2 J_2+J_1 J_2^2+2 J_1^2 J_3+J_1 J_2 J_3\,,
\end{align}
and
\begin{align}
    \{\int c_2J_i\}=\{92,36,24\}\,,
\end{align}
respectively.

This phase is described by an elliptic fibration over $\mathbb{F}_1$. Its corresponding  6d $\mathcal{N}=(1,0)$ supergravity theory has the same numbers of 6d $(1,0)$ supermultiplets as the elliptic $\mathbb{F}_0$ model, therefore the spin contents in the fiber direction are the same as those in \eqref{eq:BPSfiber_ellF0}. The genus zero and genus one free energies are given by
\begin{align}
    \mathcal{F}^{(0,0)}&=\frac{1}{2}(t_{\text{b}_1}^2+2t_{\text{b}_1}t_{\text{b}_2})\tau+\frac{1}{3}\tau^3+\mathcal{O}(e^t)\,,\\
    \mathcal{F}^{(0,1)}&=-\frac{3}{2}t_{\text{b}_1}-t_{\text{b}_2}-\frac{11}{6}\tau+\mathcal{O}(e^t)\,,\\
    \mathcal{F}^{(1,0)}&=\frac{3}{4}t_{\text{b}_1}+\frac{1}{2}t_{\text{b}_2}+\frac{1}{4}y^{\alpha}t_{\text{b}_{\alpha}}+\frac{23}{24}\tau+\mathcal{O}(e^t)\,,
\end{align}
where $\tau=t_1$, $t_{\text{b}_1}=t_2-\tau,t_{\text{b}_2}=t_3-\frac{1}{2}\tau$. 

By decoupling gravity, the 6d supergravity theory admits a 6d SCFT limit which is the E-string theory. 
For a 6d SCFT with a single tensor branch associated with the $-n$ curve in $\mathbb{F}_n$, the coefficient of $y$ is the dual Coxeter number of the corresponding gauge group \cite{Shimizu:2016lbw}. 
In the case of the 6d supergravity theory, we expect that after subtracting the contribution from the overall form corresponding to the base K\"ahler parameter $\hat{m}_{6\mathrm{d}}=t_{\text{b}_1}+\frac{1}{n}t_{\text{b}_2}$, the anomaly polynomial 4-form of the 6d supergravity theory is identical with the anomaly polynomial 4-form of the 6d SCFT.
This consistency condition imposes a constraint on $y$:
\begin{align}
    y^2-\frac{1}{n}y^1=-h^{\vee}_{G}\,.
\end{align}
For the E-string theory, where $n=1$ and $h^{\vee}_{G}=1$, this constraint implies $y^2=y^1-1$.

In the local limit $t_1\rightarrow -\infty$, we obtain the 5d $SU(2)_{\pi}$ theory with theta angle $\pi$. Physically, this 5d gauge theory can be achieved if we first compactify the 6d theory on a circle and then decouple all the KK modes. 

In this local limit, the total charge matrix for the 5d theory is
\begin{equation}
    Q_{\text{Tot}}=\left(
        \begin{array}{c}
            Q_G\\ \hdashline[1pt/1pt]
            Q_F
        \end{array}\right)=\left(
        \begin{array}{rr}
            -2&-1\\\hdashline[1pt/1pt]
            0&1\\
        \end{array}\right),
\end{equation}
so that the K\"ahler parameters are expressed as the Coulomb parameter $\phi$ and mass parameter $m$ in the form:
\begin{align}\label{eq:ellF1_tto5d}
    t_1=\hat{m}+\phi,\quad t_2=-2\phi,\quad t_3=m-\phi,
\end{align}
where $\hat{m}$ is the effective mass of the Wilson loop particle as before. Substituting \eqref{eq:ellF1_tto5d} into the genus zero and genus one free energies and compare it with \eqref{eq:FpolySUN}, we find they are consistent and determine $y=(-\frac{2}{9},-\frac{11}{9})$.

From the positivity and integrality conditions \eqref{eq:constraint2}, we find
\begin{align}
    \beta_2=-\frac{1}{2}(h-k)-\frac{1}{2}\beta_m, \,\beta_3=\beta_m \in \mathbb{Z}_{\geq 0},
\end{align}
such that the ansatz \eqref{eq:ansatz_struc} is expressed as
\begin{equation}\label{eq:ansatz_F1_I_1}
    Z_{{X},\text{phase I}}(t_1,t_2,t_3,\epsilon_1,\epsilon_2)=Z_{X_*}(t_2,t_3,\epsilon_1,\epsilon_2)\left[1+\sum_{k=1}^{\infty}e^{k\hat{m}}Z_k(t_2,t_3,\epsilon_1,\epsilon_2)\right]\,,
\end{equation}
where
\begin{equation}\label{eq:Zk_SU2_pi}
    Z_k(t_2,t_3,\epsilon_1,\epsilon_2)=P_{k}\left\langle W_{[k]}^{dP_1}\right\rangle+\sum_{l=1}^{k}\sum_{\substack{\beta_2,\beta_3\geq 0,\\2\beta_2+\beta_3=l}}e^{\beta_3 \, m}\widetilde{P}_{k,\beta_2,\beta_3}\left\langle W_{[k-l]}^{dP_1}\right\rangle,
\end{equation}
where $P_k$ is the same as the elliptic $\mathbb{F}_0$ case defined in \eqref{eq:Pk_F0}. The one-form symmetry for the 5d theory is broken due to the theta angle. As a result, both even and odd charges appear in \eqref{eq:Zk_SU2_pi}.

As a check on the ansatz \eqref{eq:ansatz_F1_I_1}, we determine the coefficients $\widetilde{P}_{k,\beta_2,\beta_3}$ up to $k=3$. To fully determine them, we need additional input for refined BPS numbers of the curve classes $\beta=(d_E,d_1,d_2)$ satisfying
\begin{align}\label{eq:ellF1_betalist}
     2d_2+d_3\leq d_E\leq 4,\,d_2,d_3\geq 0,
\end{align}
For $d_3=0$, the BPS spectrum can be obtained under the reasonable assumption that it is also provided by equation \eqref{eq:ZK3}. For $d_2=0$, we expect the BPS spectrum to coincide with that of the E-string theory, which has been calculated in \cite{Huang:2013yta,Kim:2014dza}. The solution to \eqref{eq:ellF1_betalist} also contains the curve class $\beta=(3,1,1)$, where both $d_2$ and $d_3$ are non-zero. This case can not be determined from the local 5d/6d theory calculations, as explained in Section \ref{sec:compactCY3viaWilson}.

After fixing the ansatz, we compare the overlapping BPS spectrum obtained from \eqref{eq:ansatz_F1_I_1} with the BPS spectrum related to the Wilson surface of the E-string theory \eqref{eq:ellF1_Estr}, and find exact agreement.

\paragraph{Phase II} 
The geometry of the elliptic $\mathbb{F}_1$ model has another phase \eqref{eq:Mori_EllF1_PhaseII} obtained by flopping the curve $l^{(3)}_{\text{I}}$ in phase I. In this phase, the elliptic fibration structure is broken. The refined BPS numbers in phase I are related to those  in phase II via \footnote{These relations can be obtained by the analytic continuation of the partition function.} 
\begin{equation}\label{eq:ellF1_relations}
    N_{j_L,j_R}^{\beta_{\text{I}}=(d_E,d_2,d_3)}=
    \begin{cases}
       N_{j_L,j_R}^{\beta_{\text{II}}=(d_E,d_2,d_E+d_2-d_3)}\,,\qquad &\beta_{\text{I}}\neq (0,0,1)\,,\\
       N_{j_L,j_R}^{\beta_{\text{II}}=(0,0,1)}\,,\qquad &\beta_{\text{I}} = (0,0,1)\,.
    \end{cases}
\end{equation}
The calculations for the refined BPS numbers in phase II provide an independent check of our results in phase I.

By taking the volume of the flopped $l^{(3)}_{\text{I}}$ which becomes $l^{(3)}_{\text{II}}$ in  \eqref{eq:Mori_EllF1_PhaseI} to infinity, we obtain two neighboring non-compact Calabi-Yau threefolds: $X_1=\mathcal{O}(-3)\rightarrow \mathbb{P}^2$ and the massless local del Pezzo surface $X_2=dP_8$. They have mutually disjoint
connected compact divisors $D_1$ and $D_2$ respectively. The curve $l^{(3)}_{\text{II}}$ has intersection number $1$ with both $D_1$ and $D_2$, indicating that it corresponds to the charge one Wilson loop particle in both local theories. This leads to the following ansatz:
\begin{equation}\label{eq:ansatz_F1_II_1}
    Z_{{X},\text{phase II}}(t_1,t_2,t_3,\epsilon_1,\epsilon_2)=Z_{X_1}(t_2,\epsilon_1,\epsilon_2)Z_{X_2}(t_1,\epsilon_1,\epsilon_2)\left[1+\sum_{k=1}^{\infty}e^{k\,\hat{m}}Z_k(t_1,t_2,\epsilon_1,\epsilon_2)\right]\,,
\end{equation}
where $\hat{m}=t_3+t_1+\frac{1}{3}t_2$, and
\begin{equation}
    Z_k(t_1,t_2,\epsilon_1,\epsilon_2)=\sum_{l=0}^{\left\lfloor \frac{k}{3}\right\rfloor}\sum_{l^{\prime}=0}^{k}\widetilde{P}_{k,l,l^{\prime}}\left\langle W_{[k-3l]}^{\mathbb{P}^2}\right\rangle\left\langle W_{[k-l^{\prime}]}^{dP_8}\right\rangle\,.
\end{equation}
Here $\left\langle W_{[k]}^{\mathbb{P}^2}\right\rangle$ are VEVs for Wilson loops in the 5d theory $\mathcal{T}_{X_1}$ which only depend on the K\"ahler parameter $t_2$ and $\epsilon_{1,2}$; $\left\langle W_{[k]}^{dP_8}\right\rangle$ are VEVs for Wilson loops for the 5d $SU(2)+7\,\mathbf{F}$ theory\footnote{The refined BPS numbers for $dP_8$ Wilson loops can be found in \cite[Table 25]{Wang:2023zcb}, Table \ref{tab:wilsonBPSdP8} and Table \ref{tab:wilsonBPSdP8_2} for $k=1,2,$ and $3$ respectively.} that only depends on the K\"ahler parameter $t_1$ and $\epsilon_{1,2}$. $\widetilde{P}_{k,l,l^{\prime}}$ are undetermined coefficients as usual, which shall be fixed with additional input for refined BPS numbers. In particular, $P_{k}=\widetilde{P}_{k,0,0}$ can be completely determined from the expansion:
\begin{equation}\label{eq:Pk_F1_phaseII}
    \sum_{k=0}^{\infty}P_kq^k=\exp\left[\sum_{k,l=1}^{\infty}\frac{1}{k}f_{0,0,1}(k\epsilon_1,k\epsilon_2)q^{kl}\right],
\end{equation}
where 
\begin{equation}
    f_{0,0,1}(\epsilon_1,\epsilon_2)=\frac{1}{2\sinh(\epsilon_1/2)\cdot 2\sinh(\epsilon_2/2)},
\end{equation}
is calculated from the BPS spin $[0,0]$ of the curve class $\beta_{\text{II}}=(0,0,1)$. For $k=1$, there is an additional coefficient, which is determined as:
\begin{equation}
    \widetilde{P}_{1,0,1}=f_{1,0,1}(\epsilon_1,\epsilon_2).
\end{equation}
The expression of $f_{1,0,1}(\epsilon_1,\epsilon_2)$ is the same as equation \eqref{eq:fbeta_100} according to \eqref{eq:ellF1_relations}. Following with a similar procedure, by using the minimal input from the overlapping refined BPS numbers with phase I, we solve the ansatz up to $k=3$ and find exact agreement for the other overlapping refined BPS numbers.
\subsection{Rank 2 examples}
In this section, we study two compact elliptically fibered Calabi-Yau hypersurfaces, whose local limits give 5d $SU(3)$ gauge theories. Their dual polytopes and Mori cone charges are: 
\begin{equation} \label{rank2_polytope}
 \begin{array}{ccccc|ccccccc|} 
    &\multicolumn{4}{c}{\nu_i^*}     &l^{(1)} & l^{(2)} & l^{(3)} & l^{(4)} & \\   
    & 0 & 0 & 0 & 0 & 0 & -1 & 0 & -3 &\\
& -1 & 0 & 0 & 0 & 0 & 0 & 0 & 1 & \\
& 0 & -1 & 0 & 0 & 1 & 0 & 0 & 1 &\\
& 2 & 3 & -1 & -n & 0 & 0 & 1 & 0 &\\
& 1 & 2 & 0 & 1 & 0 & 1 & 0 & 0 &\\
& 2 & 3 & 1 & 0 & 0 & 0 & 1 & 0 &\\
& 2 & 3 & 0 & -1 & 0 & 1 & -n & 0 &\\
& 1 & 1 & 0 & 0 & -2 & 1 & 0 & 1 &\\
& 2 & 3 & 0 & 0 & 1 & -2 & -2+n & 0 &\\
\end{array} \  
\end{equation} 
Here, $n$ indicates that the base for the elliptic fibration is the Hirzebruch surface $\mathbb{F}_n$. In the following subsections, we study the cases $n=0,1$, and leave the mathematical analysis for Section \ref{subsec:rank2}. 

\subsubsection{Model 2A: Elliptic fibration over \texorpdfstring{$\mathbb{F}_0$}{F0}}
\label{subsec:4parameter1}
\begin{align}\label{eq:model2A_mori}
\begin{tabular}{cccccccccccccc}
\\
$l^{(1)}$& $=$ &$($&$0$ & ; & $0$ & $1$ & $0$ & $0$ & $0$ & $0$ & $-2$ & $1$ &$)$  \\
$l^{(2)}$& $=$ &$($&$-1$ & ; & $0$ & $0$ & $0$ & $1$ & $0$ & $1$ & $1$ & $-2$ &$)$  \\
$l^{(3)}$& $=$ &$($&$0$ & ; & $0$ & $0$ & $1$ & $0$ & $1$ & $0$ & $0$ & $-2$ &$)$  \\
$l^{(4)}$& $=$ &$($&$-3$ & ; & $1$ & $1$ & $0$ & $0$ & $0$ & $0$ & $1$ & $0$ &$)$  \\
&&&&&&&&&&& $\downarrow$ & $\downarrow$ & \\
&&&&&&&&&&& $D_1$ & $D_2$ &  \\
\end{tabular}
\end{align}
In this section, we study the geometry whose Mori cone charges are described in \eqref{eq:model2A_mori}. We refer to this geometry as Model 2A. It is an elliptically fibered Calabi-Yau threefold with base $\mathbb{F}_0$. Its non-trivial Hodge numbers are $h^{1,1}=4,h^{2,1}=214$. 

The 6d $\mathcal{N} = (1,0)$ supergravity theory corresponding to this geometry is a 6d $Sp(1)$ theory with 16 fundamental hypermultiplets.  Therefore, the numbers of hypermultiplets, vector multiplets and tensor multiplets are given by
\begin{equation}
    H=247,\qquad V=3,\qquad T=1.
\end{equation}
Among the $247$ hypermultiplets, $2\times 16=32$ are charged under the gauge group $Sp(1)$, while the remaining $247-32=215$ are neutral. Here $215=h^{2,1}+1$ which precisely matches the result provided in \eqref{eq:constraint_3}. From the matter content, we conclude that the non-vanishing refined BPS numbers at the curve classes $\beta=(d_1,0,0,d_4)$ are given by
\begin{equation}\label{eq:BPS6d_2A}\begin{split}
    \beta=(n,0,0,2n):\quad &[1/2,1]\oplus[1/2,0]\oplus[0,1/2]\oplus 430[0,0]\\
    \beta=(n,0,0,2n\pm 1):\quad &32[0,0]\\
    \beta=(n,0,0,2n\pm 2):\quad &[0,1/2]
\end{split}\end{equation}
for all possible $n\in \mathbb{Z}$ satisfying $\beta>0$.

In the local limit $t_4\rightarrow -\infty$, the geometry reduces to a non-compact Calabi-Yau threefold, denoted $X$. It has compact surfaces $D_1=\mathbb{F}_2$ and $D_2=\mathbb{F}_0$, corresponding to the 5d $SU(3)_{1}$ theory with Chern-Simons level 1. The total charge matrix for the 5d theory is
\begin{equation}
    Q_{\text{Tot}}=\left(
        \begin{array}{c}
            Q_G\\ \hdashline[1pt/1pt]
            Q_F
        \end{array}\right)=\left(
        \begin{array}{rrr}
            -2&1&0\\
            1&-2&-2\\\hdashline[1pt/1pt]
            0&0&1\\
        \end{array}\right),
\end{equation}
and the Wilson loop particle corresponding to $l^{(4)}$ has gauge charge $(1,0)$. The K\"ahler parameters are expressed in terms of the Coulomb parameters $\phi_i$ and the mass parameter $m$ in the form:
\begin{align}\label{eq:ellF1_tto5d_2A}
    t_1=-2\phi_1+\phi_2,\quad t_2=\phi_1-2\phi_2,\quad t_3=m-2\phi_2,\quad t_4=\hat{m}+\phi_1,
\end{align}
where $\hat{m}$ is the effective mass of the Wilson loop particle.

Using the inverse of the total charge matrix
\begin{equation}
    Q_{\text{Tot}}^{-1}=\left(
        \begin{array}{rrr}
            -\frac{2}{3}&-\frac{1}{3}&-\frac{2}{3}\\
            -\frac{1}{3}&-\frac{2}{3}&-\frac{4}{3}\\
            0&0&1\\
        \end{array}\right)\,,
\end{equation}
the positivity and integrality conditions \eqref{eq:constraint2} impose the constraints
\begin{align}\label{eq:cond2_model2A}
    \beta_1=-\frac{2}{3}(h_1-k)-\frac{1}{3}h_2-\frac{2}{3}\beta_m,\,\beta_2=-\frac{1}{3}(h_1-k)-\frac{2}{3}h_2-\frac{4}{3}\beta_m \,,\beta_3=\beta_m \quad\in \mathbb{Z}_{\geq 0}.
\end{align}
The solution to \eqref{eq:cond2_model2A} gives the ansatz for the partition function:
\begin{equation}\label{eq:ansatz_model2A}
    Z_{{X}}(t_1,t_2,t_3,t_4,\epsilon_1,\epsilon_2)=Z_{X_*}(t_1,t_2,t_3,\epsilon_1,\epsilon_2)\left[1+\sum_{k=1}^{\infty}e^{k\hat{m}}Z_k(t_1,t_2,t_3,\epsilon_1,\epsilon_2)\right],
\end{equation}
where
\begin{equation}\label{eq:Zk_2A}\begin{split}
    Z_k(t_1,t_2,t_3,\epsilon_1,\epsilon_2)=&
    \sum_{\substack{h_1+2h_2=k-3l-4\beta_m\\ h_1,h_2,l,\beta_m\geq 0}}e^{\beta_m\,m}\widetilde{P}_{[h_1,h_2],k;\beta_m}  \left\langle W_{[h_1,h_2]}^{SU(3)_1}\right\rangle\,.
\end{split}\end{equation}
Here
$m=\log\mathfrak{q}=t_3-\frac{2}{3}(t_1+2t_2)$ is the instanton counting parameter of the 5d $SU(3)_1$ theory as defined in \eqref{eq:ellF1_tto5d_2A}. For a 5d $SU(3)$ theory, the one-form symmetry is $\mathbb{Z}_3$. The Wilson loop in the representation $[h_1,h_2]$ has the one-form symmetry charge $h_1+2h_2$. However, adding a Chern-Simons term at level 1 breaks the one-form symmetry completely, making it trivial. Indeed, we find there is no particular constraint on the one-form symmetry charge $h_1+2h_2$ in \eqref{eq:Zk_2A}.

To verify the ansatz \eqref{eq:ansatz_model2A}, we determine the coefficients $\widetilde{P}_{[h_1,h_2],k;\beta_m}$ up to $k=3$. This requires additional input from non-vanishing refined BPS numbers at curve classes
\begin{equation}\begin{split}
    \beta=&(0,0,0,1),(0,0,0,2),(1,0,0,2),(1,0,0,3),(2,1,0,3)\,,
\end{split}\end{equation}
which have been determined in \eqref{eq:BPS6d_2A}, except for the curve class $\beta=(2,1,0,3)$.
In particular, for any given $k$,
$P_{[k,0]}=\widetilde{P}_{[k,0],k;0} $ are extracted from
\begin{align}\label{eq:2A_Pk0}
    \sum_{k=0}^\infty P_{[k,0]}q^k=\exp\left[\sum_{k=1}^{\infty}\frac{1}{k}\left(f_{0,0,0,1}(k\epsilon_1,k\epsilon_2)q^{kl}+f_{0,0,0,2}(k\epsilon_1,k\epsilon_2)q^{2kl}\right)\right].
\end{align}
For $k=1$, the only undetermined coefficient is $P_{[1,0]}$, which is completely determined by \eqref{eq:2A_Pk0}. 
For $k=2$, we find
\begin{equation}
    \widetilde{P}_{[0,1],2;0}=-f_{0,0,0,1}(2 \epsilon _1,2 \epsilon _2)-2 f_{0,0,0,2}(\epsilon _1,\epsilon _2)+f_{1,0,0,2}(\epsilon _1,\epsilon _2)\,.
\end{equation}
For $k=3$, we find
\begin{align}
    \widetilde{P}_{[1,1],3;0}=\,&-f_{0,0,0,1}(\epsilon_1,\epsilon_2) f_{0,0,0,1}(2 \epsilon_1,2 \epsilon_2)-f_{0,0,0,1}(3 \epsilon_1,3 \epsilon_2)-2 f_{0,0,0,1}(\epsilon_1,\epsilon_2) f_{0,0,0,2}(\epsilon_1,\epsilon_2)\nonumber\\
    &\qquad\qquad\qquad\qquad\qquad\qquad+f_{0,0,0,1}(\epsilon_1,\epsilon_2) f_{1,0,0,2}(\epsilon_1,\epsilon_2)+f_{1,0,0,3}(\epsilon_1,\epsilon_2)\,,\\
     \widetilde{P}_{[0,0],3;0}=\,&f_{0,0,0,1}(3 \epsilon_1,3 \epsilon_2)-3 f_{1,0,0,3}(\epsilon_1,\epsilon_2)+f_{2,1,0,3}(\epsilon_1,\epsilon_2)\,.
\end{align}
With the exception of $\beta=(2,1,0,3)$, the refined BPS numbers for all curve classes $\beta=(d_1,d_2,d_3,d_4),\,d_4\leq 3$ are determined. We present the data for the refined BPS numbers at the first few curve class degrees in the supplementary material at the external link \cite{link}.

At the unrefined level $\epsilon_1=-\epsilon_2=\lambda$, the BPS data for the curve class degrees $\beta=(d_1,1,0,d_4)$ can be fixed from a generalization of the $K3$ fibration over $\mathbb{P}^1$ with one mass parameter. We find 
\begin{align}
    Z_{(d_2,d_3)=(1,0)}=\,&\exp\left[\sum_{k=1}^{\infty}\frac{1}{k}\left(f_{d_1,1,0,d_4}(k\lambda,-k\lambda)e^{2\pi i d_1\tau+d_4 z}\right)\right]\nonumber\\
    =\,&-\frac{\varphi_{-2,1}(z)(E_4^3+E_6^2)+2\varphi_{0,1}(z)E_4 E_6}{12\eta^{24}\varphi_{-2,1}(\lambda)},
\end{align}
where $E_4=E_4(\tau),E_6=E_6(\tau)$ are the Eisenstein series, and $\varphi_{-2,1}(z),\varphi_{0,1}(z)$ are the generators of weak Jacobi forms with weights $-2$ and $0$, and index $1$. 

\subsubsection{Model 2B: Elliptic fibration over \texorpdfstring{$\mathbb{F}_1$}{F1}}
\label{subsec:4parameter2}
\begin{align}\label{eq:model2B_mori}
\begin{tabular}{cccccccccccccc}
\\
$l^{(1)}$& $=$ &$($&$0$ & ; & $0$ & $1$ & $0$ & $0$ & $0$ & $0$ & $-2$ & $1$ &$)$  \\
$l^{(2)}$& $=$ &$($&$-1$ & ; & $0$ & $0$ & $0$ & $1$ & $0$ & $1$ & $1$ & $-2$ &$)$  \\
$l^{(3)}$& $=$ &$($&$0$ & ; & $0$ & $0$ & $1$ & $0$ & $1$ & $-1$ & $0$ & $-1$ &$)$  \\
$l^{(4)}$& $=$ &$($&$-3$ & ; & $1$ & $1$ & $0$ & $0$ & $0$ & $0$ & $1$ & $0$ &$)$  \\
&&&&&&&&&&& $\downarrow$ & $\downarrow$ & \\
&&&&&&&&&&& $D_1$ & $D_2$ &  \\
\end{tabular}
\end{align}
In this section, we study the geometry whose Mori cone charges are given in \eqref{eq:model2B_mori}. We refer to this geometry as Model 2B. It is an elliptically fibered Calabi-Yau threefold with base $\mathbb{F}_1$. Its non-trivial Hodge numbers are $h^{1,1}=4,h^{2,1}=202$. 

The 6d $\mathcal{N} = (1,0)$ supergravity theory corresponding to this geometry is a 6d $Sp(1)$ theory. The gauge group is defined along the curve $l^{(2)}+l^{(3)}$, such that the anomaly coefficients $b^{\alpha}$ are $b=(1,1)$. From the anomaly cancellation constraints \eqref{eq:anomalycon1}-\eqref{eq:anomalycon3}, we determine that the theory has 22 hypermultiplets in the fundamental representation.
From the matter content, we conclude that the non-vanishing refined BPS numbers at the curve classes $\beta=(d_1,0,0,d_4)$ are
\begin{equation}\label{eq:BPS6d_2B}\begin{split}
    \beta=(n,0,0,2n):\quad &[1/2,1]\oplus[1/2,0]\oplus[0,1/2]\oplus 406[0,0]\\
    \beta=(n,0,0,2n\pm 1):\quad &44[0,0]\\
    \beta=(n,0,0,2n\pm 2):\quad &[0,1/2]
\end{split}\end{equation}
for all possible $n\in \mathbb{Z}$ satisfying $\beta>0$. Moreover, we expect that if $d_2=0$, all refined BPS numbers at the curve classes $\beta=(d_1,0,d_3,d_4)$ can be determined from the partition function of the 6d E-string theory.

In the local limit $t_4\rightarrow -\infty$, the geometry reduces to a non-compact Calabi-Yau threefold, denoted $X$, with compact surfaces $D_1=\mathbb{F}_3$ and $D_2=\mathbb{F}_1$. It corresponds to the 5d $SU(3)_{2}$ theory with Chern-Simons level 2. The total charge matrix for the 5d theory is
\begin{equation}
    Q_{\text{Tot}}=\left(
        \begin{array}{c}
            Q_G\\ \hdashline[1pt/1pt]
            Q_F
        \end{array}\right)=\left(
        \begin{array}{rrr}
            -2&1&0\\
            1&-2&-1\\\hdashline[1pt/1pt]
            0&0&1\\
        \end{array}\right),
\end{equation}
and the Wilson loop particle corresponding to $l^{(4)}$ has gauge charge $(1,0)$. The K\"ahler parameters are expressed in terms of the Coulomb parameters $\phi_i$ and the mass parameter $m$ in the form:
\begin{align}\label{eq:2B_tto5d}
    t_1=-2\phi_1+\phi_2,\quad t_2=\phi_1-2\phi_2,\quad t_3=m-\phi_2,\quad t_4=\hat{m}+\phi_1,
\end{align}
where $\hat{m}$ represents the effective mass of the Wilson loop particle.

Using the inverse matrix for the total charge matrix
\begin{equation}
    Q_{\text{Tot}}^{-1}=\left(
        \begin{array}{rrr}
            -\frac{2}{3}&-\frac{1}{3}&-\frac{1}{3}\\
            -\frac{1}{3}&-\frac{2}{3}&-\frac{2}{3}\\
            0&0&1\\
        \end{array}\right)\,,
\end{equation}
the positivity and integrality conditions \eqref{eq:constraint2} provide the constraints
\begin{align}\label{eq:cond2_model2B}
    \beta_1=-\frac{2}{3}(h_1-k)-\frac{1}{3}h_2-\frac{1}{3}\beta_m,\,\beta_2=-\frac{1}{3}(h_1-k)-\frac{2}{3}h_2-\frac{2}{3}\beta_m \,,\beta_3=\beta_m \quad\in \mathbb{Z}_{\geq 0}\,,
\end{align}
leading to the ansatz for the partition function:
\begin{equation}\label{eq:ansatz_model2B}
    Z_{{X}}(t_1,t_2,t_3,t_4,\epsilon_1,\epsilon_2)=Z_{X_*}(t_1,t_2,t_3,\epsilon_1,\epsilon_2)\left[1+\sum_{k=1}^{\infty}e^{k\hat{m}}Z_k(t_1,t_2,t_3,\epsilon_1,\epsilon_2)\right],
\end{equation}
where
\begin{equation}\label{eq:Zk_2B}\begin{split}
    Z_k(t_1,t_2,t_3,\epsilon_1,\epsilon_2)=&
    \sum_{\substack{h_1+2h_2=k-3l-2\beta_m\\h_1,h_2,l,\beta_m\geq 0}}e^{\beta_m\,m}\widetilde{P}_{[h_1,h_2],k;\beta_m}  \left\langle W_{[h_1,h_2]}^{SU(3)_2}\right\rangle\,.
\end{split}\end{equation}
To verify the ansatz \eqref{eq:ansatz_model2B}, we determine the coefficients $\widetilde{P}_{[h_1,h_2],k;\beta_m}$ up to $k=3$. This requires additional input from non-vanishing refined BPS numbers for the curve classes
\begin{equation}\label{eq:degrees2B}\begin{split}
    \beta=&(0,0,0,1),(0,0,0,2),(1,0,0,2),(1,0,1,2),(1,0,0,3),(1,0,1,3),(2,1,0,3).
\end{split}\end{equation}
With the exception of $\beta=(2,1,0,3)$, the refined BPS numbers for all of these curve classes \eqref{eq:degrees2B} can be determined from \eqref{eq:BPS6d_2B} and the partition function of E-strings. For instance, those obtained from the E-string partition function are
\begin{align}
    \beta=(1,0,1,2):\quad &[1/2,1/2]\oplus 248[0,0],
\end{align}
and there is no BPS content for $\beta=(1,0,1,3)$.
We present the data for the refined BPS numbers of the curve classes $\beta=(d_1,d_2,d_3,d_4)$ with $d_1+d_2\leq 7$ and $d_3,d_4\leq 3$, except for $\beta=(2,1,0,3)$, in the supplementary material at the external link \cite{link}.

\subsection{Rank 3 examples}
\label{sec:rank3}
In this section, we study the compact elliptically fibered Calabi-Yau hypersurfaces, whose dual polytopes and Mori cone charges are: 
\begin{equation} \label{rank3_polytope}
 \begin{array}{ccccc|ccccccc|} 
    &\multicolumn{4}{c}{\nu_i^*}     &l^{(1)} & l^{(2)} & l^{(3)} & l^{(4)} & l^{(5)} &  \\   
    & 0 & 0 & 0 & 0 & -3 & 0 & 0 & 0 & 0 & \\
& -1 & 0 & 0 & 0 & 1 & 0 & 0 & 0 & 0 & \\
& 0 & -1 & 0 & 0 & 0 & 0 & 0 & 0 & 1 & \\
& 2 & 3 & -1 & 0 & 0 & 0 & 1 & 0 & 0 & \\
& 2 & 3 & 0 & 1 & 0 & 0 & 0 & 1 & 0 & \\
& 2 & 3 & n & -1 & 0 & 0 & 0 & 1 & 0 & \\
& 2 & 3 & 2 & 0 & -2 & 1 & 0 & * & 1 & \\
& 2 & 3 & 1 & 0 & 1 & -2 & 1 & * & 0 & \\
& 2 & 3 & 0 & 0 & 0 & 1 & -2 & * & 0 & \\
& 1 & 1 & 1 & 0 & 3 & 0 & 0 & 0 & -2 & \\
\end{array} \  
\end{equation} 
where $n=1,2,3$ and the detailed expression for $l^{(4)}$ can be found in \eqref{eq:5parameter_n=1}, \eqref{eq:5parameter_n=2} and \eqref{eq:5parameter_n=3} respectively. 

The 6d $\mathcal{N} = (1,0)$ supergravity theories corresponding to these geometries are 6d $G_2$ theories with $(10-3n)$ fundamental hypermultiplets. We determine $V=14$, which is the dimension for the adjoint representation of $G_2$, and $H=273-29\times 1+14=258$. The fundamental representation of $G_2$ has $6$ non-zero weights and one zero weight; therefore, among the $258$ hypermultiplets, there are $H_{\text{neutral}}=258-2\times (10-3n)\times 6=198+18n$ neutral hypermultiplets.

From the matter content, we conclude that the non-vanishing refined BPS numbers for the curve classes $\beta=(d_1,d_2,0,0,d_5)$ are
\begin{equation}\label{eq:BPS6d_3general1}\begin{split}
    \beta=(2n,n,0,0,3n):\quad &[1/2,1]\oplus[1/2,0]\oplus2[0,1/2]\oplus (198+18n)[0,0]\\
    \beta=(2n,n,0,0,3n)\pm \beta_{\Delta_{s}^{+}}:\quad &[0,1/2]\oplus(20-6n)[0,0]\\
    \beta=(2n,n,0,0,3n)\pm \beta_{\Delta_l^{+}}:\quad &[0,1/2]
\end{split}\end{equation}
for all possible $n\in \mathbb{Z}$ satisfying $\beta>0$. Here $\beta_{\Delta_l^{+}}$ and $\beta_{\Delta_s^{+}}$ represent the long and short roots of $G_2$ respectively. In our notation, they are
\begin{align}
    \beta_{\Delta_l^{+}}=&(2,0,0,0,3),(1,0,0,0,3),(1,0,0,0,0),\\
    \beta_{\Delta_s^{+}}=&(1,0,0,0,2),(1,0,0,0,1),(0,0,0,0,1).
\end{align}
If $d_3=0$, all refined BPS numbers for the curve classes $\beta=(d_1,d_2,0,d_4,d_5)$ are determined from the partition functions of the 6d SCFTs $G_2+(10-3n)\mathbf{F}$, which have been calculated for $n=1,2$ with $d_4=1$ in \cite[Appendix D]{Kim:2025} and for $n=3$ with arbitrary $d_4\geq 0$ in \cite{Kim:2018gjo}.

In Section \ref{sec:Model3_deform}, we discuss complex deformations, which correspond to Higgsing processes in physics. For instance, by turning on a special expectation value for one of the mass parameter $m=\epsilon_+$, the gauge group $G_2$ is broken to its subgroup $SU(3)$ with the same rank, and the $(10-3n)$ fundamental hypermultiplets are reduced to $(9-3n)$ fundamental and anti-fundamental hypermultiplets of $SU(3)$. In the unrefined case, the special expectation value of $m$ is still $0$, which doesn't change the unrefined GV invariants.
After Higgsing, the refined BPS numbers \eqref{eq:BPS6d_3general1} become
\begin{equation}\label{eq:BPS6d_3general2}\begin{split}
    \beta=(2n,n,0,0,3n):\quad &[1/2,1]\oplus[1/2,0]\oplus2[0,1/2]\oplus (198+18n)[0,0]\\
    \beta=(2n,n,0,0,3n)\pm \beta_{\Delta_{s}^{+}}:\quad &(18-6n)[0,0]\\
    \beta=(2n,n,0,0,3n)\pm \beta_{\Delta_l^{+}}:\quad &[0,1/2]
\end{split}\end{equation}

In the following subsections, we study the Wilson loop expansion and compute the refined BPS numbers for these models. 

\subsubsection{Model 3A: Elliptic fibration over \texorpdfstring{$\mathbb{F}_2$}{F2}}
\label{subsec:5parameter}
\begin{align}\label{eq:5parameter_n=2}
\begin{tabular}{ccccccccccccccc}
\\
$l^{(1)}$& $=$ &$($&$-3$ & ; & $1$ & $0$ & $0$ & $0$ & $0$ & $-2$ & $1$ & $0$ & $3$ &$)$  \\
$l^{(2)}$& $=$ &$($&$0$ & ; & $0$ & $0$ & $0$ & $0$ & $0$ & $1$ & $-2$ & $1$ & $0$ &$)$  \\
$l^{(3)}$& $=$ &$($&$0$ & ; & $0$ & $0$ & $1$ & $0$ & $0$ & $0$ & $1$ & $-2$ & $0$ &$)$  \\
$l^{(4)}$& $=$ &$($&$0$ & ; & $0$ & $0$ & $0$ & $1$ & $1$ & $0$ & $-2$ & $0$ & $0$ &$)$  \\
$l^{(5)}$& $=$ &$($&$0$ & ; & $0$ & $1$ & $0$ & $0$ & $0$ & $1$ & $0$ & $0$ & $-2$ &$)$  \\
&&&&&&&&&&$\downarrow$ & $\downarrow$ & $\downarrow$ & $\downarrow$ & \\
&&&&&&&&&&$D_1$ & $D_2$ & $D_3$ & $D_4$ &  \\
\end{tabular}
\end{align}
The Moric cone charges for the compact elliptically fibered Calabi-Yau hypersurface ${X}$ that has been described in \eqref{rank3_polytope} with $n=2$, are written in \eqref{eq:5parameter_n=2}. The divisors $D_1,D_2,D_3,D_4$ occur in local limits.
This geometry has non-trivial Hodge numbers $h^{1,1}=5,h^{2,1}=233$.

In the local limit $t_5\rightarrow -\infty$, the geometry reduces to a non-compact Calabi-Yau threefold, denoted by $X$, which contains compact surfaces $D_1=\mathbb{F}_2,D_2=\mathbb{F}_0$ and $D_3=\mathbb{F}_2$. 
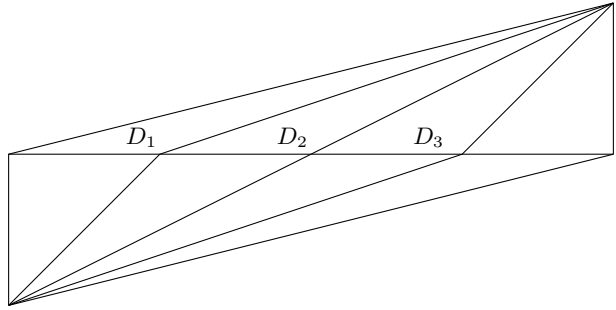
\begin{equation}\label{eq:toric_3A}
\begin{tikzpicture}[scale=2]
\def\xx{1};
\def\yy{1};
\def\x{-3};
\def\y{-1};
\draw[] (-3,0) -- (1,0);
\draw[] (\x,\y) -- (-3,0) -- (\xx,\yy);
\draw[] (\x,\y) -- (1,0) -- (\xx,\yy);
\draw[] (\x,\y) -- (-2,0) -- (\xx,\yy);
\draw[] (\x,\y) -- (-1,0) -- (\xx,\yy);
\draw[] (\x,\y) -- (0,0) -- (\xx,\yy);
\draw[]  (-1.95,-0.02) node[anchor=south east] {\footnotesize{$D_1$}};
\draw[]  (-0.95,-0.02) node[anchor=south east] {\footnotesize{$D_2$}};
\draw[]  (-0.05,-0.02) node[anchor=south east] {\footnotesize{$D_3$}};
\end{tikzpicture}
\end{equation}
The non-compact geometry $X$ is a toric Calabi-Yau threefold, whose toric diagram is depicted in \eqref{eq:toric_3A}. It corresponds to the 5d $SU(4)_{0}$ theory with Chern-Simons level 0, where the one-form symmetry is $\mathbb{Z}_4$. The total charge matrix and its inverse for the 5d theory are
\begin{equation}
    Q_{\text{Tot}}=\left(
        \begin{array}{c}
            Q_G\\ \hdashline[1pt/1pt]
            Q_F
        \end{array}\right)=\left(
        \begin{array}{rrrr}
            -2&1&0&0\\
            1&-2&1&-2\\
            0&1&-2&0\\\hdashline[1pt/1pt]
            0&0&0&1\\
        \end{array}\right),\qquad Q_{\text{Tot}}^{-1}=\left(
        \begin{array}{rrrr}
            -\frac{3}{4}&-\frac{1}{2}&-\frac{1}{4}&-1\\
             -\frac{1}{2}&-1&-\frac{1}{2}&-2\\
             -\frac{1}{4}&-\frac{1}{2}&-\frac{3}{4}&-1\\
            0&0&0&1\\
        \end{array}\right),
\end{equation}
and the Wilson loop particle corresponding to $l^{(5)}$ has gauge charge $(1,0,0)$. The K\"ahler parameters are written in terms of the Coulomb parameters $\phi_i$ and the mass parameter $m$ in the form:
\begin{align}\label{eq:3A_tto5d}
    t_1=-2\phi_1+\phi_2,\quad t_2=\phi_1-2\phi_2+\phi_3,\quad t_3=\phi_2-2\phi_3\quad t_4=m-2\phi_2,\quad t_4=\hat{m}+\phi_1,
\end{align}
where $\hat{m}$ is the effective mass of the Wilson loop particle.

From the inverse of the total charge matrix, we derive the positivity and integrality conditions:
\begin{equation}\label{eq:cond2_model3A}\begin{split}
    \beta_1=&-\frac{3}{4}(h_1-k)-\frac{1}{2}h_2-\frac{1}{4}h_3-\beta_m,\quad\beta_2=-\frac{1}{2}(h_1-k)-h_2-\frac{1}{2}h_3-2\beta_m\,,\\
    \beta_3=&-\frac{1}{4}(h_1-k)-\frac{1}{2}h_2-\frac{3}{4}h_3-\beta_m,\quad \beta_4=\beta_m \quad \in \mathbb{Z}_{\geq 0},
\end{split}\end{equation}
which determine the structure of the partition function for the compact Calabi-Yau threefold:
\begin{equation}\label{eq:ansatz_model3A}
    Z_{{X}}(t,\epsilon_1,\epsilon_2)=Z_{SU(4)_0}(\phi,m,\epsilon_1,\epsilon_2)\left[1+\sum_{k=1}^{\infty}e^{k\hat{m}}Z_k(\phi,m,\epsilon_1,\epsilon_2)\right],
\end{equation}
where
\begin{equation}\label{eq:3Aansatz1}\begin{split}
    Z_k(\phi,m,\epsilon_1,\epsilon_2)=&\sum_{k_1+2k_2+3k_3=k}P_{[k_1,k_2,k_3]}  \left\langle W_{[k_1,k_2,k_3]}^{SU(4)_0}\right\rangle\\
    &+\sum_{l=1}^{
    \lfloor \frac{k}{4}\rfloor} \sum_{\beta_m=0}^{l}e^{\beta_m m}\sum_{h_1+2h_2+3h_3=k-4l}\widetilde{P}_{[h_1,h_2,h_3],k,\beta_m}  \left\langle W_{[h_1,h_2,h_3]}^{SU(4)_0}\right\rangle\,.
\end{split}\end{equation}
Here $m=\log \mathfrak{q}={t_4-t_1-2t_2-t_3}$ is the instanton counting parameter of the 5d $SU(4)_0$ theory defined in \eqref{eq:3A_tto5d}. For a Wilson loop operator with charge $[h_1,h_2,h_3]$ in the $SU(4)$ theory, the charge under the one-form symmetry $\mathbb{Z}_4$ is $h_1+2h_2+3h_3$. The ansatz \eqref{eq:3Aansatz1} also indicates that the one-form symmetry is preserved for the $SU(4)_0$ theory.

To verify the ansatz \eqref{eq:ansatz_model3A}, we determine the coefficients $P$ and $\widetilde{P}$ up to $k=4$. This requires additional input for refined BPS numbers for the curve classes
\begin{align}\label{eq:3A_beta1}
    \beta=(0,0,0,0,1),(1,0,0,0,2),(1,0,0,0,3),(2,1,0,0,3)
\end{align}
and \footnote{We have used the condition that there are no refined BPS numbers for the curve classes $(1,0,0,0,4)$ and $(2,0,0,0,4)$.}
\begin{align}\label{eq:3A_beta2}
    \beta=(2,0,0,1,4),(2,1,0,0,4),(3,2,1,0,4).
\end{align}
Except for $\beta=(3,2,1,0,4)$, all other curve classes in \eqref{eq:3A_beta1} and \eqref{eq:3A_beta2} can be fixed from \eqref{eq:BPS6d_3general1} or \eqref{eq:BPS6d_3general2} and BPS invariants from the 6d SCFT limit:
\begin{align}
    G_2+4\mathbf{F}\qquad \beta=(2,0,0,1,4):\quad& [0,3/2]\oplus 8[0,1]\oplus 29[0,1/2]\oplus 8[0,0]\\
    SU(3)+3(\mathbf{F}+ \overline{\mathbf{F}})\qquad \beta=(2,0,0,1,4):\quad& 15[0,1/2]
\end{align}
For instance, we find $P_{k,0,0}$ can be solved via 
\begin{align}
    \sum_{k=0}^\infty P_{[k,0,0]}Q^k=\exp\left[\sum_{k=1}^{\infty}{f_{0,0,0,0,1}(k\epsilon_1,k\epsilon_2)\frac{Q^k}{k}}\right],
\end{align}
and other coefficients $P$ and $\widetilde{P}$ are determined using a similar method to that used for the rank 2 models. We summarize the first few BPS numbers for Model 3A before and after deformation in \cite{link}.

\subsubsection{Model 3B: Elliptic fibration over \texorpdfstring{$\mathbb{F}_3$}{F3} }
\label{subsec:5parameter1}
\begin{align}\label{eq:5parameter_n=3}
\begin{tabular}{ccccccccccccccc}
\\
$l^{(1)}$& $=$ &$($&$-3$ & ; & $1$ & $0$ & $0$ & $0$ & $0$ & $-2$ & $1$ & $0$ & $3$ &$)$  \\
$l^{(2)}$& $=$ &$($&$0$ & ; & $0$ & $0$ & $0$ & $0$ & $0$ & $1$ & $-2$ & $1$ & $0$ &$)$  \\
$l^{(3)}$& $=$ &$($&$0$ & ; & $0$ & $0$ & $1$ & $0$ & $0$ & $0$ & $1$ & $-2$ & $0$ &$)$  \\
$l^{(4)}$& $=$ &$($&$0$ & ; & $0$ & $0$ & $0$ & $1$ & $1$ & $-1$ & $-1$ & $0$ & $0$ &$)$  \\
$l^{(5)}$& $=$ &$($&$0$ & ; & $0$ & $1$ & $0$ & $0$ & $0$ & $1$ & $0$ & $0$ & $-2$ &$)$  \\
&&&&&&&&&&$\downarrow$ & $\downarrow$ & $\downarrow$ & $\downarrow$ & \\
&&&&&&&&&&$D_1$ & $D_2$ & $D_3$ & $D_4$ &  \\
\end{tabular}
\end{align}

In the local limit $t_5\rightarrow -\infty$, the geometry reduces to a non-compact Calabi-Yau threefold, denoted by $X$, with compact surfaces $D_1=\mathbb{F}_1,D_2=\mathbb{F}_1$ and $D_3=\mathbb{F}_3$. 
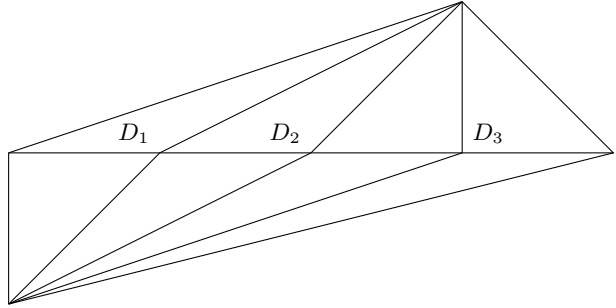
\begin{equation}\label{eq:toric_3B}
\begin{tikzpicture}[scale=2]
\def\xx{0};
\def\yy{1};
\def\x{-3};
\def\y{-1};
\draw[] (-3,0) -- (1,0);
\draw[] (\x,\y) -- (-3,0) -- (\xx,\yy);
\draw[] (\x,\y) -- (1,0) -- (\xx,\yy);
\draw[] (\x,\y) -- (-2,0) -- (\xx,\yy);
\draw[] (\x,\y) -- (-1,0) -- (\xx,\yy);
\draw[] (\x,\y) -- (0,0) -- (\xx,\yy);
\draw[]  (-2,0) node[anchor=south east] {\footnotesize{$D_1$}};
\draw[]  (-1,0) node[anchor=south east] {\footnotesize{$D_2$}};
\draw[]  (0,0) node[anchor=south west] {\footnotesize{$D_3$}};
\end{tikzpicture}
\end{equation}
The non-compact geometry $X$ is a toric Calabi-Yau threefold, whose toric diagram is depicted in \eqref{eq:toric_3B}. It corresponds to the 5d $SU(4)_{1}$ theory with Chern-Simons level 1, where the one-form symmetry is broken. The total charge matrix and its inverse for the 5d theory are
\begin{equation}
    Q_{\text{Tot}}=\left(
        \begin{array}{c}
            Q_G\\ \hdashline[1pt/1pt]
            Q_F
        \end{array}\right)=\left(
        \begin{array}{rrrr}
            -2&1&0&-1\\
            1&-2&1&-1\\
            0&1&-2&0\\\hdashline[1pt/1pt]
            0&0&0&1\\
        \end{array}\right),\qquad Q_{\text{Tot}}^{-1}=\left(
        \begin{array}{rrrr}
            -\frac{3}{4}&-\frac{1}{2}&-\frac{1}{4}&-\frac{5}{4}\\
             -\frac{1}{2}&-1&-\frac{1}{2}&-\frac{3}{2}\\
             -\frac{1}{4}&-\frac{1}{2}&-\frac{3}{4}&-\frac{3}{4}\\
            0&0&0&1\\
        \end{array}\right),
\end{equation}
and the Wilson loop particle corresponding to $l^{(5)}$ has gauge charge $(1,0,0)$. The K\"ahler parameters are expressed in terms of the Coulomb parameters $\phi_i$ and the mass parameter $m$ in the form:
\begin{align}\label{eq:3B_tto5d}
    t_1=-2\phi_1+\phi_2,\quad t_2=\phi_1-2\phi_2+\phi_3,\quad t_3=\phi_2-2\phi_3\quad t_4=m-\phi_1-\phi_2,\quad t_4=\hat{m}+\phi_1,
\end{align}
where $\hat{m}$ is the effective mass of the Wilson loop particle and $m=\log \mathfrak{q}={t_4-\frac{5}{4}t_1-\frac{3}{2}t_2-\frac{3}{4}t_3}$ is the instanton counting parameter for the 5d $SU(4)_{1}$ theory.

The positivity and integrality conditions are given by
\begin{equation}\label{eq:cond2_model3B}\begin{split}
    \beta_1=&-\frac{3}{4}(h_1-k)-\frac{1}{2}h_2-\frac{1}{4}h_3-\frac{5}{4}\beta_m,\quad\beta_2=-\frac{1}{2}(h_1-k)-h_2-\frac{1}{2}h_3-\frac{3}{2}\beta_m\,,\\
    \beta_3=&-\frac{1}{4}(h_1-k)-\frac{1}{2}h_2-\frac{3}{4}h_3-\frac{3}{4}\beta_m,\quad \beta_4=\beta_m \quad \in \mathbb{Z}_{\geq 0}\,,
\end{split}\end{equation}
leading to the ansatz for partition function for the compact Calabi-Yau threefold:
\begin{equation}\label{eq:snsatz_3B}
    Z_{{X}}(t,\epsilon_1,\epsilon_2)=Z_{SU(4)_{1}}(\phi,m,\epsilon_1,\epsilon_2)\left[1+\sum_{k=1}^{\infty}e^{k\hat{m}}Z_k(\phi,m,\epsilon_1,\epsilon_2)\right],
\end{equation}
where
\begin{equation}\begin{split}
    Z_k(\phi,m,\epsilon_1,\epsilon_2)=\sum_{\substack{h_1+2h_2+3h_3=k-4l-3\beta_m\\h_1,h_2,h_3,l,\beta_m\geq 0}}e^{\beta_m\, m}\widetilde{P}_{[h_1,h_2,h_3],k,\beta_m}  \left\langle W_{[h_1,h_2,h_3]}^{SU(4)_{1}}\right\rangle\,.
\end{split}\end{equation}

With the additional input of the refined BPS numbers from \eqref{eq:BPS6d_3general1} or \eqref{eq:BPS6d_3general2}
and those from its 6d SCFT limit:
\begin{align}
    G_2+\mathbf{F}\qquad \beta=(2,0,0,1,3):\quad& [0,2]\oplus 2[0,3/2]\oplus 3[0,1]\oplus 2[0,1/2]\oplus 2[0,0]\\
    \beta=(2,0,0,1,4):\quad& [0,2]\oplus 2[0,3/2]\oplus 2[0,1]\oplus 2[0,1/2]\oplus [0,0]\\
    SU(3)\qquad \beta=(2,0,0,1,3):\quad& [0,1]\oplus[0,0]\\
    \beta=(2,0,0,1,4):\quad& \text{empty}
\end{align}
we determine a closed form expression for \eqref{eq:snsatz_3B} up to $k=4$.
Using these results, we calculate the first few refined BPS numbers for the Model 3B before and after deformation, with the exception of $\beta=(3,2,1,0,4)$, and summarize them in the supplementary material at the external link \cite{link}.

\subsubsection{Model 3C: Elliptic fibration over \texorpdfstring{$\mathbb{F}_1$}{F1}}
\label{subsec:5parameter3}
\begin{align}\label{eq:5parameter_n=1}
\begin{tabular}{ccccccccccccccc}
\\
$l^{(1)}$& $=$ &$($&$-3$ & ; & $1$ & $0$ & $0$ & $0$ & $0$ & $-2$ & $1$ & $0$ & $3$ &$)$  \\
$l^{(2)}$& $=$ &$($&$0$ & ; & $0$ & $0$ & $0$ & $0$ & $0$ & $1$ & $-2$ & $1$ & $0$ &$)$  \\
$l^{(3)}$& $=$ &$($&$0$ & ; & $0$ & $0$ & $1$ & $0$ & $0$ & $0$ & $1$ & $-2$ & $0$ &$)$  \\
$l^{(4)}$& $=$ &$($&$0$ & ; & $0$ & $0$ & $0$ & $1$ & $1$ & $0$ & $-1$ & $-1$ & $0$ &$)$  \\
$l^{(5)}$& $=$ &$($&$0$ & ; & $0$ & $1$ & $0$ & $0$ & $0$ & $1$ & $0$ & $0$ & $-2$ &$)$  \\
&&&&&&&&&&$\downarrow$ & $\downarrow$ & $\downarrow$ & $\downarrow$ & \\
&&&&&&&&&&$D_1$ & $D_2$ & $D_3$ & $D_4$ &  \\
\end{tabular}
\end{align}

In the local limit $t_5\rightarrow -\infty$, the geometry reduces to a non-compact Calabi-Yau threefold, denoted by $X$, with compact surfaces $D_1=\mathbb{F}_3,D_2=\mathbb{F}_1$ and $D_3=\mathbb{F}_1$. 
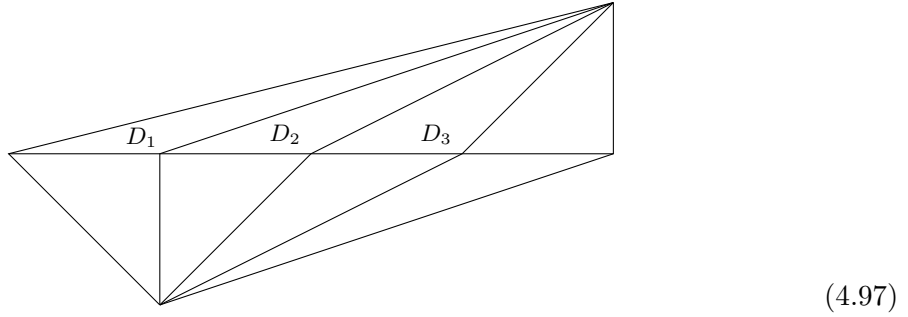
\begin{equation}\label{eq:toric_3C}
\begin{tikzpicture}[scale=2]
\def\xx{1};\def\yy{1};
\def\x{-2};
\def\y{-1};
\draw[] (-3,0) -- (1,0);
\draw[] (\x,\y) -- (-3,0) -- (\xx,\yy);
\draw[] (\x,\y) -- (1,0) -- (\xx,\yy);
\draw[] (\x,\y) -- (-2,0) -- (\xx,\yy);
\draw[] (\x,\y) -- (-1,0) -- (\xx,\yy);
\draw[] (\x,\y) -- (0,0) -- (\xx,\yy);
\draw[]  (-1.95,-0.02) node[anchor=south east] {\footnotesize{$D_1$}};
\draw[]  (-1,0) node[anchor=south east] {\footnotesize{$D_2$}};
\draw[]  (0,0) node[anchor=south east] {\footnotesize{$D_3$}};
\end{tikzpicture}
\end{equation}
The non-compact geometry $X$ is a toric Calabi-Yau threefold, whose toric diagram is depicted in \eqref{eq:toric_3C}. It corresponds to the 5d $SU(4)_{-1}$ theory with Chern-Simons level $-1$, where the one-form symmetry is broken. The total charge matrix and its inverse for the 5d theory are
\begin{equation}
    Q_{\text{Tot}}=\left(
        \begin{array}{c}
            Q_G\\ \hdashline[1pt/1pt]
            Q_F
        \end{array}\right)=\left(
        \begin{array}{rrrr}
            -2&1&0&0\\
            1&-2&1&-1\\
            0&1&-2&-1\\\hdashline[1pt/1pt]
            0&0&0&1\\
        \end{array}\right),\qquad Q_{\text{Tot}}^{-1}=\left(
        \begin{array}{rrrr}
            -\frac{3}{4}&-\frac{1}{2}&-\frac{1}{4}&-\frac{3}{4}\\
             -\frac{1}{2}&-1&-\frac{1}{2}&-\frac{3}{2}\\
             -\frac{1}{4}&-\frac{1}{2}&-\frac{3}{4}&-\frac{5}{4}\\
            0&0&0&1\\
        \end{array}\right),
\end{equation}
and the Wilson loop particle corresponding to $l^{(5)}$ has gauge charge $(1,0,0)$. The K\"ahler parameters are expressed in terms of the Coulomb parameters $\phi_i$ and the mass parameter $m$ in the form:
\begin{align}\label{eq:3C_tto5d}
    t_1=-2\phi_1+\phi_2,\quad t_2=\phi_1-2\phi_2+\phi_3,\quad t_3=\phi_2-2\phi_3\quad t_4=m-\phi_2-\phi_3,\quad t_4=\hat{m}+\phi_1,
\end{align}
where $\hat{m}$ is the effective mass of the Wilson loop particle.

The positivity and integrality conditions are given by
\begin{equation}\label{eq:cond2_model3C}\begin{split}
    \beta_1=&-\frac{3}{4}(h_1-k)-\frac{1}{2}h_2-\frac{1}{4}h_3-\frac{3}{4}\beta_m,\quad\beta_2=-\frac{1}{2}(h_1-k)-h_2-\frac{1}{2}h_3-\frac{3}{2}\beta_m\,,\\
    \beta_3=&-\frac{1}{4}(h_1-k)-\frac{1}{2}h_2-\frac{3}{4}h_3-\frac{5}{4}\beta_m,\quad \beta_4=\beta_m \quad \in \mathbb{Z}_{\geq 0}.
\end{split}\end{equation}
leading to the ansatz for partition function for the compact Calabi-Yau threefold:
\begin{equation}\label{eq:ansatz_3C}
    Z_{{X}}(t,\epsilon_1,\epsilon_2)=Z_{SU(4)_{-1}}(\phi,m,\epsilon_1,\epsilon_2)\left[1+\sum_{k=1}^{\infty}e^{k\,\hat{m}}Z_k(\phi,m,\epsilon_1,\epsilon_2)\right],
\end{equation}
where
\begin{equation}\begin{split}
    Z_k(\phi,m,\epsilon_1,\epsilon_2)=\sum_{\substack{h_1+2h_2+3h_3=k-4l-5\beta_m\\h_1,h_2,h_3,l,\beta_m\geq 0}}e^{\beta_m\, m}\widetilde{P}_{[h_1,h_2,h_3],k,\beta_m}  \left\langle W_{[h_1,h_2,h_3]}^{SU(4)_{-1}}\right\rangle.
\end{split}\end{equation}

With the additional input of the refined BPS numbers from \eqref{eq:BPS6d_3general1} or \eqref{eq:BPS6d_3general2},
we determine a closed form expression for \eqref{eq:ansatz_3C} up to $k=4$.
Using these results, we calculate the first few BPS numbers for Model 3C before and after deformation, with the exception of $\beta=(3,2,1,0,4)$, and summarize them in the supplementary material at the external link \cite{link}.

\section{Geometric computations of refined BPS numbers}
\label{Sec:GeometriccalculationsofrefBPS}

In this section, we extend the mathematical attempts and methods used to define refined BPS numbers from \cite{Hosono:2001gf,Kiem:2012fs,Maulik:2016rip} and surveyed in \cite{Huang:2020dbh}.  Some of these computations are used as input for solving for boundary conditions used in Section~\ref{sec:examples} to compute refined BPS numbers.  Other computations serve to provide many checks on the validity of the methods of Sections~\ref{sec:Wilsonloop} and~\ref{sec:examples} and the resulting computations of refined BPS numbers.   

We start with a quick summary of the mathematical definitions which we use, referring to the above references for more detail.  Let $X$ be a compact Calabi-Yau threefold and let $\beta\in H_2(X)$ be a curve class.  Let $\widehat{M}_\beta$ denote the moduli space of 1-dimensional stable sheaves $F$ with $\mathrm{ch}_2(F)=\beta$ and $\chi(F)=1$, and let $M_\beta$ be the Chow variety of 1-cycles of class $\beta$.  In the physical setup, $\widehat{M}_\beta$ is the space of D2-D0 branes with a fixed unit of D0-brane charge \cite{Gopakumar:1998ii,Gopakumar:1998jq}.  There is the Hilbert-Chow morphism $\pi_\beta:\widehat{M}_\beta^{\mathrm{red}}\to M_\beta$.  

The moduli space $\widehat{M}_\beta$ supports a d-critical locus structure \cite{Joyce:2015dc}, which provides $\widehat{M}_\beta$ with a virtual canonical bundle $K_{\widehat{M}_\beta}^\mathrm{vir}$.  In our situation, $K_{\widehat{M}_\beta}^\mathrm{vir}$ can be described as the line bundle whose fiber at $F\in \widehat{M}_\beta$ is canonically identified with the 1-dimensional vector space 
\begin{equation}
    \left(K_{\widehat{M}_\beta}^\mathrm{vir}\right)_F= \bigotimes_{i=0}^3\left(\det\mathrm{Ext}^i(F,F)\right)^{\otimes(-1)^i}.
\end{equation}

One of the conditions for a space to support a d-critical locus structure is that is it locally isomorphic to the critical locus of a superpotential.  The superpotential then locally gives rise to a perverse sheaf of vanishing cycles.  A global condition is needed to determine how to glue these perverse sheaves together to obtain a perverse sheaf $\Phi_\beta$ on all of $\widehat{M}_\beta$, leading to the notation of an orientation \cite{KS}.  For fixed $\beta$, an orientation is simply a choice of square root $(K_{\widehat{M}_\beta}^{\mathrm{vir}})^{1/2}$ of $K_{\widehat{M}_\beta}^{\mathrm{vir}}$.  There is also a compatibility condition between orientation choices for different $\beta$.  We will return to this point later.  Since $\mathrm{Ext}^*(F,F)$ describes open string states, it is tempting to speculate that orientations are related to a Pfaffian arising from the evaluation of a fermion determinant over the moduli space of sheaves.

If $\widehat{M}_\beta$ is smooth, then $K_{\widehat{M}_\beta}^{\mathrm{vir}}$ is just the square of the ordinary canonical bundle, so a natural choice of orientation is $(K_{\widehat{M}_\beta}^{\mathrm{vir}})^{1/2}=K_{\widehat{M}_\beta}$.  With this orientation choice, the perverse sheaf of vanishing cycles is $\Phi_\beta=\mathbb{C}[\dim \widehat{M}_\beta]$, the shift of the constant sheaf by $\dim \widehat{M}_\beta$ places to the left in the derived category.

At this point we make a technical assumption made in \cite{Maulik:2016rip}: we assume that the orientation is a Calabi-Yau orientation, meaning that it is trivial on the fibers of $\pi_\beta$.  It is conjectured in \cite{Maulik:2016rip} that Calabi-Yau orientations exist.  

A proposal for mathematically defining the refined BPS numbers was made in \cite{Kiem:2012fs}; however a counterexample to this proposal (which even produced the incorrect unrefined BPS invariants) was pointed out in \cite{Maulik:2016rip}.  The authors of this last paper also pointed out the refined BPS numbers can be consistently defined by the definitions in \cite{Kiem:2012fs} for Calabi-Yau orientations when
 $\Phi_\beta$ (with rational coeffients instead of complex coefficients) underlies a polarized pure Hodge module.   We now elaborate on what this means.

We let $\mathcal{D}_Y$ be the sheaf of differential operators on a smooth $Y$, which comes with an increasing filtration $F_p\mathcal{D}_Y\subset \mathcal{D}_Y$ of differential operators of order at most $p$.  Part of the data of a pure Hodge module on $Y$ is a coherent sheaf $\mathcal{M}$ of (left) $\mathcal{D}_Y$-modules,\footnote{There is an equivalent theory using right $\mathcal{D}$-modules, with some ultimately minor differences. See \cite{Schnell:2019ms} for more details.} with an increasing filtration $F_\bullet \mathcal{M}$, satisfying
\begin{equation}
    F_p\mathcal{D}_Y \cdot F_q\mathcal{M}\subset F_{p+q}\mathcal{M},
    \label{eqn:gt}
\end{equation}
with equality for $q\gg0$.

To a $\mathcal{D}_Y$-module $\mathcal{M}$ we associate the de Rham complex
\begin{equation}
    \mathrm{DR}(\mathcal{M})=[\mathcal{M}\to \Omega^1_Y\otimes_{\mathcal{O}_Y}\mathcal{M}\to \cdots \Omega^d_Y\otimes_{\mathcal{O}_Y}\mathcal{M}],
\end{equation}
where $d=\dim Y$ and the rightmost term is in degree 0.  It turns out that $\mathrm{DR}(\mathcal{M})$ is a perverse sheaf on $Y$.  In fact, $\mathrm{DR}$ sets up an equivalence of categories between the category of regular holonomic $\mathcal{D}_Y$ modules and (complex) perverse sheaves on $Y$.  Before giving the next condition, we give a basic example.

Given a family $f:X\to Y$  of smooth projective varieties parametrized by a smooth $Y$, we have the local system of rational cohomologies $\mathcal{H}^k=R^kf_*\mathbb{Q}$, a sheaf of rational vector spaces on $Y$ whose fiber at $y\in Y$ is $H^k(f^{-1}(y),\mathbb{Q})$, each of which underlies a pure polarized Hodge structure of weight $k$ after tensoring with $\mathbb{C}$.  In addition, we have a Hodge filtration on the holomorphic vector bundle $\mathcal{M}=\mathcal{H}^k\otimes_{\mathbb{Q}}\mathcal{O}_Y$, together with the Gauss-Manin connection satisfying Griffiths transversality.  A sign change in the indices turns the decreasing Hodge filtration $F^\bullet\mathcal{M}$ into an increasing filtration $F_\bullet\mathcal{M}$, and then (\ref{eqn:gt}) is just Griffiths transversality.

If $\mathcal{M}$ more generally is any vector bundle with connection, then the de Rham complex is exact except for the first term, and the kernel of the first map is the local system of flat sections of $\mathcal{M}$, placed in degree $-d$.  For example, if $\mathcal{M}$ is $\mathcal{H}^k\otimes_{\mathbb{Q}}\mathcal{O}_Y$ as above, then $\mathrm{DR}(\mathcal{M})$ is quasi-isomorphic to $(\mathcal{H}^k\otimes_{\mathbb{Q}}\mathbb{C})[d]$ in the constructible derived category of $Y$, and is a perverse sheaf, which arises as the complexification of the perverse sheaf $\mathcal{H}^k[d]$ of rational vector spaces.  More generally, the notion of a pure Hodge module requires
\begin{equation}
    \mathrm{DR}(\mathcal{M})\simeq \mathrm{rat}(\mathcal{M})\otimes_\mathbb{Q}\mathbb{C}
    \end{equation}
for some perverse sheaf $\mathrm{rat}(\mathcal{M})$ of rational vector spaces.   As just explained, this condition holds for in our basic example.

Now if $Y=\widehat{M}_\beta$ is smooth and the orientation $(K_{\widehat{M}_\beta}^{\mathrm{vir}})^{1/2}=K_{\widehat{M}_\beta}$ is chosen, the resulting perverse sheaf $\Phi_\beta=\mathbb{C}[\dim \widehat{M}_\beta]$ is the complexification of the rational perverse sheaf $\mathbb{Q}[\dim \widehat{M}_\beta]$ and is the de Rham complex of $\mathcal{O}_Y$ endowed with the natural structure of a $\mathcal{D}_Y$-module by application of differential operators to functions.  
We see that the perverse sheaf $\mathbb{C}[\dim \widehat{M}_\beta]$ on smooth $\widehat{M}_\beta$ associated with the natural orientation support a pure Hodge module and can be reliably used to compute the refined BPS numbers.

While there are additional technical conditions on the notion of a pure Hodge module, we content ourselves with noting that if $Y$ is a point, then $\mathcal{M}$ is just a vector space arising as the complexification of a rational vector space, and the decreasing filtration $\mathcal{F}^\bullet \mathcal{M}$ inferred from $\mathcal{F}_\bullet \mathcal{M}$ is precisely the Hodge filtration of a Hodge structure of pure weight on $\mathcal{M}$.

A thorough treatment would require a deeper dive into Saito's theory of pure Hodge modules and mixed Hodge modules, but we trust that the above comments give the reader a sufficient sense of the theory.  We refer the interested reader to the summary \cite{Saito:2017} for more details.

\smallskip
After choosing a Calabi-Yau orientation, we obtain perverse sheaves of vanishing cycles $\Phi_\beta$ on $\widehat{M}_\beta$ \cite{Maulik:2016rip} as already mentioned. If $\Phi_\beta$ underlies a pure Hodge module (which is the case if $\widehat{M}_\beta$ is smooth, in which case we take $\Phi_\beta=\mathbb{C}[\dim \widehat{M}_\beta]$), then the decomposition theorem \cite{BBD} says that $R(\pi_\beta)_*\Phi_\beta$ decomposes into a  direct sum of shifts of the perverse cohomologies of $R{\pi_\beta}_*\Phi_\beta$:
\begin{equation}
  R{\pi_\beta}_*\Phi_\beta = \bigoplus_i{}^p\mathcal{H}^i\left(R{\pi_\beta}_*\Phi_\beta\right)[-i],
\end{equation}
where $[-i]$ denotes a shift of $i$ places to the right in the derived category of complexes of sheaves of vector spaces on $M_\beta$ with constructible cohomology.  In Section~\ref{subsubsec:g1basecurve} we will express the perverse sheaves ${}^p\mathcal{H}^i\left(R{\pi_\beta}_*\Phi_\beta\right)$ as a direct sum of simple perverse sheaves $\mathrm{IC}(L_i)$ associated with local systems $L_i$ on Zariski open subsets of subvarieties of $M_\beta$.

Cup product with a relatively ample class for $\pi_\beta$ induces maps
\begin{equation}
    {}^p\mathcal{H}^i\left(R{\pi_\beta}_*\Phi_\beta\right) \to {}^p\mathcal{H}^{i+2}\left(R{\pi_\beta}_*\Phi_\beta\right),
\label{eq:SU2L}
\end{equation}
which are identified by hard Lefshetz with the raising operators of an $\mathrm{SU}_2$ action, identified with the $(\mathrm{SU}_2)_L$ action.  Furthermore, given an ample class on $M_\beta$, we similarly get maps on hypercohomologies
\begin{equation}
\mathbb{H}^j\left(M_\beta,{}^p\mathcal{H}^i\left(R{\pi_\beta}_*\Phi_\beta\right)\right)\to 
\mathbb{H}^{j+2}\left(M_\beta,{}^p\mathcal{H}^i\left(R{\pi_\beta}_*\Phi_\beta\right)\right),
\label{eq:SU2R}
\end{equation}
which are identified with the raising operators of an $\mathrm{SU}_2$ action, identified with the $(\mathrm{SU}_2)_R$ action. The unrefined BPS invariants can then be deduced from the $\mathrm{SU}_2\times\mathrm{SU}_2$ representation in the usual way, recovering the definition of \cite{Maulik:2016rip}, where it is proven that the unrefined BPS numbers are independent of the choice of Calabi-Yau orientation (and the purity of the mixed Hodge module is not required). 

\smallskip\noindent
{\bf Conjecture.} The refined BPS numbers defined as above are independent of the choice of Calabi-Yau orientation which leads to a perverse sheaf $\Phi_\beta$ supporting a pure Hodge module. 

\smallskip\noindent
{\bf Remark.}
In \cite{Maulik:2016rip}, an example is given for which there are no orientations which lead to a perverse sheaf $\Phi_\beta$ supporting a pure Hodge module.  So even if the above conjecture is true, we are not making a proposal for a mathematical definition of the refined BPS numbers in the most general case.  However, in all examples that we are aware of where $\Phi_\beta$ does not support a pure Hodge module, the maps (\ref{eq:SU2L}) and (\ref{eq:SU2R}) nevertheless define an action of $\mathrm{SU}_2\times\mathrm{SU}_2$, even though no proof of that fact is available.

\smallskip
Despite this remark, many of our examples will deal with special case where $\widehat{M}_\beta$ is smooth, in which case we have $\Phi_\beta=\mathbb{C}[D]$ with $D=\dim\widehat{M}_\beta$.  For many of our examples, we also have that  $M_\beta$ is smooth and the fibers of $\pi_\beta$ are connected curves $C$ of some fixed arithmetic genus $g$.   Since the space of stable rank~1 sheaves on $C$ has dimension $g$, we see that $\dim M_\beta=D-g$.  The typical stalk of $R^{-D}\pi_*\mathbb{C}[D]$ is $H^0(C,\mathbb{C})$, which is canonically isomorphic to $\mathbb{C}$.  Therefore $R^{-D}\pi_*\mathbb{C}[D]$ is canonically isomorphic to the constant sheaf $\mathbb{C}$ on $M_\beta$.  Since $\mathbb{C}[D-g]$ is a perverse sheaf on $M_\beta$, we can rewrite this result as
\begin{equation}
  {}^p\mathcal{H}^{-g}\left(R{\pi_\beta}_*{\mathbb{C}[D]}\right)=\mathbb{C}[D-g],
\end{equation}
and ${}^p\mathcal{H}^{-h}\left(R{\pi_\beta}_*{\mathbb{C}[D]}\right)=0$ for $h>g$, as $C$ has no negative cohomologies.  Thus the maximum of $2j_L$ is $g$.  For the right spins associated with $2j_L=g$, we look at the Lefschetz of 
$\mathbb{H}^*(M_\beta,\mathbb{C}[D-g])$, which is just the ordinary Lefschetz of $M_\beta$ with a shift of indices.  We arrive at a generalization of an assertion of \cite{Gopakumar:1998jq}.

\smallskip\noindent
{\bf Fact 1.} Assuming moduli spaces are smooth, for families of connected curves of arithmetic genus $g$, the maximum left spin is $g$, and the corresponding right spin content is given by the Lefschetz of the base $M_\beta$.

\smallskip
Since the sum of a relatively ample class for $\pi_\beta$ and the pullback to $\widehat{M}_\beta$ of an ample class on $M_\beta$ is ample on $\widehat{M}_\beta$,
we arrive at a generalization of another assertion of \cite{Gopakumar:1998jq}.

\smallskip\noindent
{\bf Fact 2.} Assuming $\widehat{M}_\beta$ is smooth, the diagonal $\mathrm{SU}_2$ is given by the Lefschetz of $\widehat{M}_\beta$.

\smallskip
An important point is that the refined BPS numbers can depend on the choice of orientation, so some consistency is required between the orientation choices for different $\beta$ if there is any hope of a refined holomorphic anomaly equation being satisfied.  This required consistency is part of the definition of an orientation in \cite{KS}.  We content ourselves with a few comments here.

Given sheaves $F\in\widehat{M}_\beta$ and $F'\in\widehat{M}_{\beta'}$, we compute using Serre duality that the fiber $\otimes(\det\mathrm{Ext}^i(F\oplus F',F\oplus F'))^{\otimes(-1)^i}$ of $K_{\widehat{M}_{\beta+\beta'}}^{\mathrm{vir}}$ at $F\oplus F'$ is
\begin{equation}
\otimes_{i=0}^3(\det\mathrm{Ext}^i(F,F))^{\otimes(-1)^i}\otimes
\otimes_{i=0}^3(\det\mathrm{Ext}^i(F',F'))^{\otimes(-1)^i}\otimes (\otimes_{i=0}^3(\det\mathrm{Ext}^i(F,F'))^{\otimes(-1)^i})^{\otimes2}.
\end{equation}
Thus it makes sense to require the consistency condition of orientations
\begin{equation}
(K_{\widehat{M}_{\beta+\beta'}}^{\mathrm{vir}})^{1/2}\simeq (K_{\widehat{M}_{\beta}}^{\mathrm{vir}})^{1/2}\otimes (K_{\widehat{M}_{\beta'}}^{\mathrm{vir}})^{1/2}\otimes\mathcal{L},
\label{eqn:consistency}
\end{equation}
where $\mathcal{L}$ is a line bundle whose fiber at $F\oplus F'$ is $\otimes_{i=0}^3(\det\mathrm{Ext}^i(F,F'))^{\otimes(-1)^i}$.

We have abused notation in the above since $\chi(F\oplus F')=2$ shows that $F\oplus F'$ is not an element of $\widehat{M}_{\beta+\beta'}$ as written.  However, an analogous compatibility condition is required in \cite[Definition 15]{KS} for the entire derived category and the above abuse of notation is simply the result of oversimplified exposition of the full extent of the required compatibility conditions.

We note that if the $\widehat{M}_{\beta}$ is smooth, then ${Ext}^1(F,F)$ is the fiber of the tangent bundle of $\widehat{M}_{\beta}$ at the point $F$ while ${Ext}^0(F,F)$ is the fiber of the trivial bundle.  It follows that the fiber of $K_{\widehat{M}_\beta}^{\mathrm{vir}}$ at $F$ is $(\det\mathrm{Ext}^1(F,F))^{\otimes{-2}}$, so this bundle is just the square of the ordinary canonicial bundle. It is then natural to choose the orientation $(K_{\widehat{M}_\beta}^{\mathrm{vir}})^{1/2}$ to simply be the ordinary canonical bundle of $\widehat{M}_\beta$.  It is straightforward to check that if $F$ and $F'$ both lie in smooth moduli spaces and these orientation choices are made, then the consistency condition (\ref{eqn:consistency}) follows automatically.  

This discussion shows that the orientation choices associated with Facts 1 and 2 above are consistent for different $\beta$, fulfilling a requirement for a holomorphic anomaly equation to be possible.

\subsection{Rank 1 theories}
\label{subsec:rank1}

We now consider a compact elliptically fibered Calabi-Yau threefold $\pi:X\to B$.  Let $f$ denote the fiber class.  Identifying $B$ with the zero section $S$ of $\pi$, we can write a curve class on $X$ as $\beta+d_Ef$, with $\beta\in H_2(B,\mathbb{Z})$.  

For $B=\mathbb{P}^2$, we can identify $\beta$ with its degree $d_B$ and refer to curves of class $\beta+d_Ef$ as curves of degree $(d_B,d_E)$.  For $B=\mathbb{F}_0$ and $B=\mathbb{F}_1$, we have $H_2(B,\mathbb{Z})\simeq\mathbb{Z}^2$.  In the case of $\mathbb{F}_0$ we choose the two distinct fiber classes as generators, while for $\mathbb{F}_1$ we take generators $\{E,F\}$, where $E$ is the $-1$ curve and $F$ is the class of a $\mathbb{P}^1$ fiber of $\mathbb{F}_1$.  We will describe curve classes on $\mathbb{F}_0$ and $\mathbb{F}_1$ by two degrees $(d_2,d_3)$ relative to the appropriate chosen set of generators.

\subsubsection{\texorpdfstring{$d_E=1$}{dEeq1}}
In this section we compute several refined BPS numbers for $d_E=1$, providing supporting evidence for (\ref{eq:F1ellP2}) and its analogues for the $\mathbb{F}_0$ and $\mathbb{F}_1$ bases.

\subsubsection{Maximum left spin}
\label{subsubsec:mls}

In this brief section, we apply Fact 1 above to all base degrees, finding complete agreement with the bottom rows of Tables~\ref{tab:my_label}, \ref{tab:BPSF0dE=1}, and \ref{tab:BPSF1dE=1}.

Letting $C$ be a connected curve of class $\beta+f$ with $\beta\in H_2(B,\mathbb{Z})$, we have $S\cdot C=K_S\cdot \beta+1$, which is negative unless $B=\mathbb{F}_1$ and $C$ is the $-1$ curve.

In all of the other cases, $S\cdot C<0$ implies that $C$ has a component which is contained within $S$.  Letting $D$ be the union of all components contained in $S$, we see that $C=D\cup f$, where $D$ is identified with a curve in $B$ of class $\beta$ and $f$ is an elliptic fiber.  Since $S\cap f$ is a point $p$ and $C$ is connected, necessarily $D\cap f = p$, and the moduli space $M_{\beta,1}$ of these curves is identified with the universal curve $\{(p,D)\vert p \in D\}$ of degree $\beta$. A standard argument (reviewed in \cite{Huang:2020dbh}) shows that $M_{\beta,1}$ is a projective bundle over $B$: the natural projection $M_{\beta,1}\to\mathbb{P}^2$ has fiber over $p\in B$ the projective space of curves in $B$ degree $\beta$ which contain $p$.  

Explicitly for $B=\mathbb{P}^2$,  $M_{d_B,1}$ is a $\mathbb{P}^{(d_B^2+3d_B-2)/2}$-bundle over $\mathbb{P}^2$, with Lefschetz representation
\begin{equation}
\left[\frac{d_B^2+3d_B-2}4\right]\otimes\Big[1\Big],
\end{equation}
expanding to
\begin{equation}
\left\{
\begin{array}{cl}
\left[\frac32\right]\oplus\left[\frac12\right]&d_B=1\\
\left[\frac{d_B^2+3d_B+2}4\right]\oplus\left[\frac{d_B^2+3d_B-2}4\right]\oplus\left[\frac{d_B^2+3d_B-6}4\right]&d_B>1.
\end{array}\right.
\label{eqn:bottomrows}
\end{equation}
Furthermore, $C$ has arithmetic genus $g(d_B)=(d_B^2-3d_B+4)/2$, corresponding to the rows of Table~\ref{tab:my_label} with $2j_L=(d_B^2-3d_B+4)/2$.  The $j_R$ contents of these rows are
in complete agreement with (\ref{eqn:bottomrows}).

\medskip
We next perform the analogous computations for the $\mathbb{F}_0$ and $\mathbb{F}_1$ bases.  

For $\mathbb{F}_0$, we have that $D$ is a curve of genus $(d_2-1)(d_3-1)$, $M_{(d_2,d_3),1}$ is a $\mathbb{P}^{d_2d_3+d_2+d_3-1}$-bundle over $\mathbb{F}_0$ with Lefschetz
\begin{equation}
    \left[\frac{d_2d_3+d_2+d_3-1}{2}\right]\otimes\left(\Big[1\Big]\oplus\Big[0\Big]\right),
\end{equation}
agreeing with the bottom rows of Table~\ref{tab:BPSF0dE=1} ($2j_L=(d_2-1)(d_3-1)+1$).

For $\mathbb{F}_1$, initially omitting the exceptional case $(d_2,d_3)=(1,0)$ we have that $D$ is a curve of genus $(2d_2d_3-d_2^2-d_2-2d_3+2)/2$, $M_{(d_2,d_3),1}$ is a $\mathbb{P}^{(2d_2d_3-d_2^2+d_2+2d_3-2)/2}$-bundle over $\mathbb{F}_1$ with Lefschetz
\begin{equation}
    \left[\frac{2d_2d_3-d_2^2+d_2+2d_3-2}{4}\right]\otimes\left(\Big[1\Big]\oplus\Big[0\Big]\right),
\end{equation}
agreeing with the bottom rows of Table~\ref{tab:BPSF1dE=1}.

In the exceptional case, the curve class $(d_2,d_3)=(1,0)$ is represented only by the $-1$ curve $E\subset \mathbb{F}_1$.  Let $G=\pi^{-1}(E)$, a half-K3.  The self-intersection of $E$ inside $G$ is also $-1$ by the adjunction formula.  If $C\subset X$ has class $E+f$, then since $C\cdot E=0$ as an intersection in $G$, either $C=E\cup f$ for a fiber $f$ meeting $E$, or $C$ is disjoint from $E$.    If $C$ is disjoint from $E$, then $C$ is irreducible and must be a section of the elliptic fibration $\pi:G\to E$, since $C\cdot f=1$.  Then $\pi|_C:C\to E$ is an isomorphism.  Thus $C$ has genus 0 and does not contribute to $2j_L=1$.  

So we may assume that we are in the first case $C=E\cup f$.  To give a curve $C$ of this form is equivalent to specifying the point $p=E\cap F\in E$.  This gives $M_{(1,0),1}\simeq E\simeq\mathbb{P}^1$ with Lefschetz $[1/2]$, again in agreement with the bottom row of the $(d_2,d_3)=(1,0)$ part of Table~\ref{tab:BPSF1dE=1}.

\subsubsection{Generalities}

We consider stable sheaves $F$ of class $(\beta,1)$.  Equivalently, $F$ has no torsion subsheaves and any nontrivial proper subsheaf $G\subset F$ satisfies $\chi(G)<1$ (which is the same as requiring $\chi(Q)\ge 1$ for all surjections $F\to Q$). We represent $\mathrm{ch}_2(F)$ by a curve $D+f$ ($D$ can be reducible and/or have multiplicities).  We let $p=D\cap f$.
Consider the sheaf $F\vert_f$.  Since this sheaf could have torsion supported at $p$, we put
\begin{equation}
    F_f=\left(F\vert_f\right)/{\mathrm{torsion}},
\end{equation}
and similarly put
\begin{equation}
    F_D=\left(F\vert_D\right)/{\mathrm{torsion}}.
\end{equation}

\smallskip\noindent
{\bf Lemma.} We have a short exact sequence
\begin{equation}
0 \to F \to F_D \oplus F_f \to \mathcal{O}_p\to 0,
\label{eqn:normalization}
\end{equation}
with $F \to F_D$ and $F \to F_f$ both surjective.
Furthermore, $F_D\in\widehat{M}_{\beta,0}$ and $F_f\in\widehat{M}_{0,1}$.  Conversely, given any short exact sequence as above with $F_D\in\widehat{M}_{\beta,0}$, $F_f\in\widehat{M}_{0,1}$, with $F \to F_D$ and $F \to F_f$ both surjective, then $F\in\widehat{M}_{\beta,1}$, $F_D\simeq F\vert_D/{\mathrm{torsion}}$, and $F_f\simeq F\vert_f/{\mathrm{torsion}}$.

\smallskip\noindent
{\bf Remark.} We have already noted that the moduli of the sheaves $F_f$ in (\ref{eqn:normalization}) corresponds to $f(\epsilon_1,\epsilon_2)$ in (\ref{eq:F1ellP2}). The moduli of the sheaves $F_D$ and the point $p$ corresponds to the Wilson loop VEV $\left\langle W_{[1]}\right\rangle$ \cite{Huang:2022hdo}.

\smallskip\noindent
\begin{proof}
The natural surjections $F \to F_D$ and $F \to F_f$ combine to give a map $F \to F_D\oplus F_f$.  This map is clearly an isomorphism away from $p$, so its kernel and cokernel are torsion sheaves supported at $p$. Since $F$ has no torsion subsheaves by stability, the map is injective and we get a short exact sequence
\begin{equation}
0 \to F \to F_D\oplus F_f \to T \to 0,    
\end{equation}
where $T$ is a torsion sheaf supported at $p$.  Stability also implies that 
$s:=\chi(F_f)\ge1$.  

Since the intersection number of the section and the fiber is 1, we see that $p$ must be a smooth point of $f$.  Since $F_f$ is torsion free of rank 1 on $f$, it is locally free at $p$ and $F_f\otimes\mathcal{O}_p\simeq \mathcal{O}_p$.  If the map $F_f\to T$ were zero, then $F_f$ would be identified with a subsheaf of $F$, violating stability.  So $F_f$ maps onto a nontrivial subsheaf $T'\subset T$, necessarily isomorphic to $\mathcal{O}_p$ by $F_f\otimes\mathcal{O}_p\simeq \mathcal{O}_p$.

Writing $\sigma:F_D\to T$, we put $F_D'=\sigma^{-1}(T')$. Recalling $T'\simeq\mathcal{O}_p$, we get a short exact sequence
\begin{equation}
0 \to F \to F_D'\oplus F_f \to \mathcal{O}_p \to 0,    
\end{equation}
so that $\chi(F_D')=2-s$.  Since the surjection $F\to F_D$ factors through $F_D'$, we conclude that $F_D'=F_D$.   Stability then forces $s=1$, and $\chi(F_D)=1$ as well.  

Since a destablizing quotient of $F_D$ or $F_f$ would destablize $F$, we conclude that $F_D$ and $F_f$ are both stable.  The proof of the converse statement is similar and left to the reader.
\end{proof}
The association of $F_f$ to $F$ defines a map $\phi:\widehat{M}_{\beta,1}\to \widehat{M}_{0,1}$.  We will compute the refined BPS numbers for degree $(\beta,1)$ by computing the refined numbers for degree $(0,1)$ and analyzing the fibers of $\phi$.  We will also relate these fibers to Wilson loops in some cases.  

But first, we review the computation of the refined numbers for degree $(0,1)$ \cite{Huang:2020dbh}.  First $\widehat{M}_{0,1}\to {M}_{0,1}$ is identified with $X\to B$ as follows. The sheaves $F_f$ fit into a short exact sequence
\begin{equation}
    0\to \mathcal{O}_f\to F_f\to \mathcal{O}_q\to 0
    \label{eqn:D2D0g1}
\end{equation}
for a point $q\in f$, arising from applying $\underline{\mathrm{Ext}}(-,\mathcal{O}_X)$ to the short exact sequence
\begin{equation}
    0\to \mathcal{I}_{q,f}\to \mathcal{O}_f \to \mathcal{O}_q\to 0,
\end{equation}
where $\mathcal{I}_{q,f}$ is the ideal sheaf of holomorphic functions on $f$ which vanish at $q$.  Since $f$ is determined uniquely by $q\in X$, we see that $\widehat{M}_{\beta,1}\simeq X$. Letting $p=f\cap S \in S\simeq B$, we see that $p\in B$ determines the fiber $f$, and ${M}_{0,1}\simeq B$, the map $\phi$ being identified with the elliptic fibration $\pi:X\to B$ itself. 

Since $g=1$, Fact 1 tells us that the $j_R$ content for $2j_L=1$ is given by the Lefschetz of $B$, so $[1]$ for $B=\mathbb{P}^2$ and $[1]\oplus[0]$ for $B=\mathbb{F}_0$ of $\mathbb{F}_1$. Fact 2 tells us that the diagonal $\mathrm{SU}_2$ is identified with the Lefschetz of $X$, which is $(2+2h^{2,1}(X))[0]\oplus[3/2]\oplus(h^{1,1}(X)-1)[1/2]$.  Combining these two facts we conclude that the refined BPS spectrum in degree $(0,1)$ is
\begin{equation}
    \left[\frac12,1\right]\oplus\left(2+2h^{2,1}(X)\right)\Big[0,0\Big]\oplus\left(h^{1,1}(X)-2\right)\left[0,\frac12\right]
\label{eqn:P2fiberclass}
\end{equation}
for $B=\mathbb{P}^2$ and 
\begin{equation}
    \left[\frac12,1\right]\oplus\left[\frac12,0\right]\oplus\left(2+2h^{2,1}(X)\right]\Big[0,0\Big]\oplus\left(h^{1,1}(X)-3\right)\left[0,\frac12\right]
\label{eqn:Fnfiberclass}
\end{equation}
for $B=\mathbb{F}_0$ or $\mathbb{F}_1$.  

In \cite[Proposition~5]{Katz:2022vwe}, the decomposition theorem was applied to identify $R\pi_*\mathbb{R}$ as a direct sum of shifts of certain real $\mathrm{IC}$ sheaves, some of which are associated to gauge group factors.  To compute $R\pi_*\mathbb{C}$ instead, we simply tensor all terms with $\mathbb{C}$ and arrive at complex $\mathrm{IC}$ sheaves.  It can be checked that for the $\mathbb{P}^2$ and $\mathbb{F}_k$ cases, (\ref{eqn:P2fiberclass}) and (\ref{eqn:Fnfiberclass}) can alternatively be deduced by taking hypercohomologies of the IC sheaves and considering the Lefschetz actions .  As this check is not relevant to the rest of this paper, we say no more here, leaving details to the interested reader.

\medskip

Given $F_D\in \widehat{M}_{\beta,0}$ and $F_f\in \widehat{M}_{0,1}$, we need to classify all surjections $F_D \oplus F_f \to \mathcal{O}_p$.  If the map $F_f\to \mathcal{O}_p$ were zero, then $F_f$ would be identified with a subsheaf of $F$, violating stability. Since $F_f\otimes\mathcal{O}_p\simeq \mathcal{O}_p$, the map $F\to \mathcal{O}_p$ is canonical up to multiplication by a nonzero scalar, so is canonical up to isomorphism.

Similarly, $F_D\to \mathcal{O}_p$ is a surjection, so these maps are classified up to isomorphism by the projectivization of  $\mathrm{Hom}(F\otimes\mathcal{O}_p,\mathcal{O}_p)$.

\subsubsection{Genus 0 base curve}
\label{subsubsec:g0basecurve}
If the curve $D$ in the section $S$ has genus 0, we can determine the moduli spaces  $\widehat{M}_{d_B,1}$ in the $\mathbb{P}^2$ case, and $\widehat{M}_{(d_2,d_3),1}$ in the $\mathbb{F}_0$ and $\mathbb{F}_1$ cases.  These moduli spaces turn out to be smooth, so their Lefschetz actions are identified with the diagonal $SU(2)$-action by Fact 2. The maximum left spin in these cases is $2j_L=1$, since connected curves $D\cup f$ have arithmetic genus 1.  Thus the diagonal $SU(2)$ together with the previously determined $2j_L=1$ contributions completely determine the $SU(2)\times SU(2)$ BPS spectrum in the genus 0 cases.

Given $F\in \widehat{M}_{\beta,1}$ we have the exact sequence (\ref{eqn:normalization}) with sheaves 
$F_D\in \widehat{M}_{\beta,0}$ and $F_f\in \widehat{M}_{0,1}$.  We claim that $F_D\simeq\mathcal{O}_D$.  We assume this claim for the moment and will justify it presently.  It follows that $F_D\otimes\mathcal{O}_p\simeq\mathcal{O}_p$, so there is a unique surjection $F_D\to\mathcal{O}_p$ up to scalar.  Thus given $F_D$ and $F_f$, the sheaf $F$ is uniquely determined up to isomorphism by (\ref{eqn:normalization}).

Returning to the fibration $\widehat{M}_{\beta,1}\to \widehat{M}_{0,1}$ and using $F_D\simeq\mathcal{O}_D$, the fiber is just the projective space of curves $D$ of degree $\beta$ which contain a fixed point $p$.  This determines the diagonal $SU(2)$ as the tensor product of the Lefschetz of $\widehat{M}_{0,1}$ with the Lefschetz of the projective space.  

Before turning to examples, we first justify the claim $F_D\simeq\mathcal{O}_D$ using an argument from \cite{Katz:2008g0gv}.  Since $\chi(F_D)=1$ and $H^i(B,F_D)=0$ for $i>1$, we must have $H^0(B,F_D)\ne 0$. Identify a nonzero global section $s$ of $F_D$ with a map $\mathcal{O}_X\to F_D$.  Since $F_D$ has pure dimension 1, the map must factor through $\mathcal{O}_{D'}$ for some curve $D'\subset D$.  Since $D$ has genus 0, it follows that $D'$ also has genus 0 (as can be seen from the classification of genus zero curves on $\mathbb{P}^2,\ \mathbb{F}_0$, and $\mathbb{F}_1$) and so $\chi(\mathcal{O}_{D'})=1$.  If $D'\ne D$, the inclusion $\mathcal{O}_{D'}\subset F_D$ destabilizes $F$, as these two sheaves both have $\chi=1$.  We conclude that $D'=D$, in which case the inclusion $\mathcal{O}_{D}\hookrightarrow F_D$ must be an isomorphism.

\smallskip
Specializing to the genus 0 case for the $\mathbb{P}^2$ base (i.e.\ $d_B=1$ or $d_B=2$), 
We have $\widehat{M}_{0,1}\simeq X$, with Lefschetz $546\left[0\right]\oplus\left[\frac32\right]\oplus\left[\frac12\right]$ and the fiber of 
$\widehat{M}_{d_B,1}\to \widehat{M}_{0,1}$ is
$\mathbb{P}^{(d_B^2+3d_B-2)/2}$, the moduli space of plane curves of degree $d_B$ passing through a fixed point.
We conclude 
that the diagonal Lefschetz is
\begin{equation}
\left[\frac{d_B^2+3d_B-2}4\right]\otimes\left(546\Big[0\Big]\oplus\left[\frac32\right]\oplus\left[\frac12\right]\right)
\end{equation}
which is
\begin{equation}
\begin{array}{cl}
\left[2\right]\oplus2\left[1\right]\oplus\left[0\right]\oplus546\left[\frac12\right]&d_B=1\\
\left[\frac72\right]\oplus2\left[\frac52\right]\oplus2\left[\frac32\right]\oplus\left[\frac12\right]\oplus546\left[2\right]&d_B=2
\end{array}
\end{equation}
in agreement with the diagonal restrictions in Table~\ref{tab:my_label} and therefore the entire $\mathrm{SU}(2)\times \mathrm{SU}(2)$ representation.

\smallskip
Similarly, the diagonal restrictions in Tables~\ref{tab:BPSF0dE=1} and \ref{tab:BPSF1dE=1} can be checked by the same method for the genus zero cases $d_2=1$ or $d_3=1$ for $\mathbb{F}_0$ and $(d_2,d_3)=(0,1)$, $(2,2)$, or $(1,d_3)$ with $d_3>0$ for $\mathbb{F}_1$.  For both the $\mathbb{F}_0$ and $\mathbb{F}_1$ bases, the Lefschetz of $\widehat{M}_{0,1}\simeq X$ is
\begin{equation}
    488\Big[0\Big] \oplus \left[\frac32\right] \oplus 2\left[\frac12\right].
\end{equation}
and the dimension of the projective spaces are easily worked out.

In the exceptional $\mathbb{F}_1$ case $(d_2,d_3)=(1,0)$, we necessarily have $D=E$, and $\widehat{M}_{(1,0),1}$ is the inverse image $G$ of $E$ in the elliptic fibration $X\to \mathbb{F}_1$, which is identified with half-K3, with Lefschetz $[1]\oplus9[0]$.  Combining this diagonal Lefschetz with the $2j_L=1$ content $(1/2)$, one might conclude that the BPS spectrum is $[1/2,1/2]\oplus 8[0,0]$.  However we have missed 240 BPS states (these are the contributions at degree $(1,0,1)$ omitted in (\ref{eqn:localF1dE1})).  Identifying half K3 with the blowup of $\mathbb{P}^2$ and 9 points, and $E$ with the ninth exceptional curve, we readily check by computing intersections that any of the 240 $(-1)$ curves in $dP_8$ also have base degree $(1,0)$ and $d_E=1$ (recall that in this exceptional case, the argument that $C$ is of the form $D\cup f$ fails).  As each of these curves have genus 0, their spins are $[0,0]$ and we arrive at the complete BPS spectrum $[1/2,1/2]\oplus 248[0,0]$.

In summary, we have completely verified Tables~\ref{tab:BPSF0dE=1} and \ref{tab:BPSF1dE=1} in all cases with genus 0 base curve.

\subsubsection{Genus 1 base curve}
\label{subsubsec:g1basecurve}
Continuing to assume $d_E=1$, if the base curve $D$ has genus 1 (so $d_B=3$ for $\mathbb{P}^2$, $(d_2,d_3)=(2,2)$ for $\mathbb{F}_0$, or $(d_2,d_3)=(2,3)$ or $(3,3)$ for $\mathbb{F}_1$), we now have $g=2$ and the BPS spectrum has contributions from $2j_L=0,1,2$.  Thus the diagonal $SU(2)$ together with the $j_L=2$ content are not sufficient to determine the BPS spectrum.  Below we will describe a new method which gives the $2j_L=g-1$ content directly, allowing us to determine the entire BPS spectrum.  The method applies more generally to give the $j_L=g-1$ content for curves of arbitrary base degree, providing more checks and predictions.  However, for $g>2$ this information is no longer sufficient for determining the entire BPS spectrum.

The moduli space $\widehat{M}_{d_B,1}$ again parametrizes stable sheaves $F$ on $C=D\cup f$ fitting into a short exact sequence (\ref{eqn:normalization}). Since $D$ has genus 1, the argument associated with the fiber class applies and the sheaves $F_D$ are parameterized by $D$ together with a point $r\in D$, corresponding to an extension

\begin{equation}\label{eqn:Dextension}
    0 \to \mathcal{O}_D\to F_D \to \mathcal{O}_r\to 0.
\end{equation}

For $\mathbb{P}^2$ we have the map
\begin{equation}
\phi:\widehat{M}_{3,1} \to \widehat{M}_{0,1}\simeq X.
\label{eqn:projtox}
\end{equation}
sending $F$ to $F_f$.  

We claim that the fibers of $\phi$ are all isomorphic to a $\mathbb{P}^7$-bundle over the blowup of $\mathbb{P}^2$ at a point.  Once the claim has been demonstrated, using the Lefschetz $[1]\oplus[0]$ of the blowup of $\mathbb{P}^2$ together with the fibration structure, we arrive at the Lefschetz of $\widehat{M}_{3,1}$
\begin{equation}
\left[\frac72\right]\otimes\Big(\left[1\Big]\oplus\Big[0\Big]\right)\otimes\left(546\Big[0\Big]\oplus\left[\frac32\right]\oplus\left[\frac12\right]\right)
\end{equation}
which expands to
\begin{equation}
546\left[\frac92\right]\oplus1092\left[\frac72\right]\oplus546\left[\frac52\right]\oplus\Big[6\Big]\oplus4\Big[5\Big]\oplus7\Big[4\Big]\oplus7\Big[3\Big]\oplus4\Big[2\Big]\oplus\Big[1\Big],
\end{equation}
agreeing with the diagonal restriction of the $d_B=3$ row of Table~\ref{tab:my_label}.

To complete the computation of the diagonal restriction, we compute the fiber of $\phi$ over $F_f\in \widehat{M}_{0,1}$.  
Let $p$ be the point where $f$ meets the section. The sheaves $F_D$ are given by extensions (\ref{eqn:Dextension}) with $p\in D$ and $r\in D$.  We distinguish the cases $r\ne p$ and $r=p$ and split the fiber into two contributions:
$\phi^{-1}(F_f)=\phi^{-1}(F_f)_{r\ne p}\cup \phi^{-1}(F_f)_{r=p}$.

If $r\ne p$, then $F_D$ is locally free at $p$ so that $F_D\otimes\mathcal{O}_p\simeq\mathcal{O}_p$. It follows that there is a unique surjection $F_D\to \mathcal{O}_p$ up to scalar multiple, so that the sheaf $F$ defined by (\ref{eqn:normalization}) is uniquely determined up to isomorphism by $F_f$ and $F_D$.  Since the set of degree 3 plane curves passing through distinct points $r$ and $p$ is parametrized by $\mathbb{P}^7$, we conclude that $\phi^{-1}(F_f)_{r\ne p}$ is a $\mathbb{P}^7$ bundle over $\mathbb{P}^2-\{p\}$.

For $r=p$, there are two subcases. If $D$ is smooth at $p$, then $F_D$ is locally free at $p$, so as in the preceding paragraph, $F$ is uniquely determined up to isomorphism by $F_f$ and $F_D$.  If $D$ is singular at $p$, then $F_D$ is not locally free at $p$ and after tensoring (\ref{eqn:Dextension}) with $\mathcal{O}_p$ we see that $F_D\otimes\mathcal{O}_p\simeq \mathcal{O}_p^2$, and the surjections $F_D\to\mathcal{O}_p$ up to scalar are parametrized by $\mathbb{P}^1$, introducing additional moduli in $\phi^{-1}(F_f)_{r=p}$ beyond those corresponding to $F_D$.  We identify this $\mathbb{P}^1$ with the exceptional divisor $E$ of the blowup $\widetilde{\mathbb{P}^2}$ of $\mathbb{P}^2$ at $p$.

We can at last construct a $\mathbb{P}^7$ bundle $\phi^{-1}(F_f)\to \widetilde{\mathbb{P}^2}$, at least set-theoretically.  We have already constructed this fibration on $\phi^{-1}(F_f)_{r\ne p}$ and we turn to completing this by exhibiting a $\mathbb{P}^7$ bundle $\phi^{-1}(F_f)_{r=p}\to E$.

If $F\in \phi^{-1}(F_f)_{r=p}$ with $F_D$ smooth at $p$, we send $F$ to the tangent space to $D$ at $p$, identified with a point of $E$.  If $F_D$ is singular at $p$, we send $F$ to the point of $E$ corresponding to the surjection $F_D\to\mathcal{O}_p$.

Finally, we check that the map $\phi^{-1}(F_f)_{r=p}\to E$ just constructed is a $\mathbb{P}^7$ bundle.  Pick a point in $E$, corresponding to a tangent direction in $\mathbb{P}^2$ at $p$.  The set of all degree 3 curves in $\mathbb{P}^2$ which contain $p$ and a specified tangent direction is parametrized by a $\mathbb{P}^7$.  While these curves can be either smooth at $p$ or singular at $p$, we have completely accounted for all of the moduli in either case and our computation of the diagonal restriction is complete.
 
\smallskip
The analysis for $\mathbb{F}_0$ and the $(d_2,d_3)=(2,3)$ case for $\mathbb{F}_1$ is similar.  For $\mathbb{F}_0$ we have instead of (\ref{eqn:projtox})
\begin{equation}
\phi:\widehat{M}_{(2,2),1} \to X,
\label{eqn:projtox2}
\end{equation}
and the fibers of $\phi$ are $\mathbb{P}^6$ bundles over the blowup of $\mathbb{F}_0$ at a point.  So the Lefschetz of $\widehat{M}_{(2,2),1}$ is
\begin{equation}
\Big[3\Big]\otimes\left(\Big[1\Big]\oplus2\Big[0\Big]\right)\otimes\left(488\Big[0\Big]\oplus\left[\frac32\right]\oplus2\left[\frac12\right]\right)
\label{eqn:f022Lefschetz}
\end{equation}
which expands to 
\begin{equation}
\left[\frac{11}{2}\right]\oplus6\left[\frac92\right]\oplus488\Big[4\Big]\oplus13\left[\frac72\right]\oplus1464\Big[3\Big]\oplus13\left[\frac52\right]\oplus488\Big[2\Big]\oplus6\left[\frac32\right]\oplus\left[\frac12\right],
\label{eqn:f022Lefschetzexp}
\end{equation}
agreeing with the diagonal restriction of the $(d_2,d_3)=(2,2)$ row of Table~\ref{tab:BPSF0dE=1}.  

For the $(d_2,d_3)=(2,3)$ case of $\mathbb{F}_1$ we have 
\begin{equation}
\phi:\widehat{M}_{(2,3),1} \to X
\end{equation}
and the fibers are $\mathbb{P}^6$ bundles over $\widetilde{\mathbb{F}}_1$.
The Lefschetz representation coincides with (\ref{eqn:f022Lefschetz}) and (\ref{eqn:f022Lefschetzexp}), and agrees with diagonal restriction of the $(d_2,d_3)=(2,3)$ part of Table~\ref{tab:BPSF1dE=1}.

There is a subtlety in the $\mathbb{F}_1$ $(d_2,d_3)=(3,3)$ case: if the point $r$ in (\ref{eqn:Dextension}) is contained in the $-1$ curve $E$, then $F_D$ is unstable, requiring additional techniques to describe $\widehat{M}_{(3,3),1}$.  We content ourselves with an outline of the issue.  The $(3,3)$ class is $3E+3F$.  Since $(3E+3F)\cdot E=0$ and $r\in E$, we see that $D $ must contain $E$ as a component, and then (\ref{eqn:Dextension}) implies that $(F_D)|_E\simeq\mathcal{O}_E(1)$ in the generic case where $D$ contains $E$ with multiplicity 1. Write $D=D'+E$, with $D'$ being in the class $2E+3F$. Then the subsheaf of $F_D$ consisting of sections vanishing on $D'$ (forcing the sections to vanish at the point $D'\cap E$) is isomorphic to $\mathcal{O}_E$, destabilizing $F_D$.
Accordingly, a blowdown of the moduli space naively constructed as in the previous cases is required to arrive at $\widehat{M}_{(3,3),1}$.  The result is that the fibers of $\phi:\widehat{M}_{(3,3),1}\to M_{(3,3),1}$ are $\mathbb{P}^7$ bundles over $\mathbb{F}_1$ rather than $\mathbb{P}^7$ bundles over the blowup of $\mathbb{F}_1$ at a point, and again we find agreement using this description of the fibers of $\phi$.  We omit the details.

\medskip
Our final geometric calculation in the rank 1 cases is the $j_R$ content of the rows immediately above the bottom rows of the tables, i.e.\ $2j_L=g-1$.  For a curve class $\beta$ in the base, we have the map
\begin{equation}
    \pi_\beta:\widehat{M}_{\beta,1}\rightarrow M_{\beta,1}.
\end{equation}
The fiber of $\pi_\beta$ over a curve $D\cup f$ with $D$ and $f$ smooth is the product of (compactified) Jacobians $J(D)\times \overline{J(f)}\simeq J(D)\times f$.
The sheaf $R^1{\pi_\beta}_*\mathbb{C}_{M_{\beta,1}}$ on $M_{\beta,1}$ restricts to a local system $L_\beta$ on the dense open subset $M^\circ_{\beta,1}\subset M_{\beta,1}$ of curves $D\cup f$ with $D$ and $f$ smooth, as the dimensions of the cohomology groups $H^1(J(D)\times f,\mathbb{C})$ are independent of the choice of smooth $D$ and intersecting smooth fiber $f$.  This local system determines the perverse sheaf $\mathrm{IC}(L_\beta)$, which restricts to $L_\beta[\dim M_{\beta,1}]$ on $M^\circ_{\beta,1}$.  Then the $j_R$ content of the row above the bottom is given by the Lefschetz representation on the hypercohomology $\mathbb{H}^*(M_{\beta,1},\mathrm{IC}(L_\beta))$.

Since $H^1(J(D)\times f,\mathbb{C})=H^1(D,\mathbb{C})\oplus H^1(f,\mathbb{C})$ we see that the rank 4 local system $L_\beta$ is the direct sum of the two rank 2 local systems $L_{D}$ and $L_{f}$, so that
\begin{equation}
    \mathrm{IC}(L_\beta) = \mathrm{IC}(L_D) \oplus \mathrm{IC}(L_f).
\end{equation}
We compute the Lefschetz representations on $\mathbb{H}^*(M_{\beta,1},\mathrm{IC}(L_D))$ and $\mathbb{H}^*(M_{\beta,1},\mathrm{IC}(L_f))$ and then combine the results to get the desired $\mathrm{SU}(2)_R$ representation.

We begin with $L_f$. Since $H^1(f,\mathbb{C})$ is independent of $D$, we see that via the projection map $M_{\beta,1}\to B$ sending the curve $D\cup f$ to $D\cap f\in S\simeq B$, the local system $L_f$ is pulled back from a local system $B_\beta$ on (a dense open subset of) $B$.  However, $B_\beta$ coincides with the local system $R^1\pi_*\mathbb{C}_X$ with $\pi:X\to B$ the elliptic fibration after restriction to the locus with smooth elliptic fibers.  

The cohomology of $\mathrm{IC}(B_\beta)$ was worked out in \cite[Section~7]{Huang:2020dbh} and we summarize the calculation here.  The decomposition theorem gives
\begin{equation}
    R\pi_*\mathbb{C}_X[3]=\mathbb{C}_B[3]\oplus \mathrm{IC}(B_\beta) \oplus  \mathbb{C}_B[1].
\label{eqn:decomposition}
\end{equation}
Taking hypercohomologies, we get for $B=\mathbb{P}^2$ that the Lefschetz of $\mathbb{H}^*(B,\mathrm{IC}(B_\beta))$ is $546[0]$.  

Returning to the class $(d_B,1)$, since $M_{\beta,1}\to B$ is a $\mathbb{P}^{(d_B^2+3d_B-2)/2}$ bundle, we conclude after pullback that the Lefschetz representation of $\mathbb{H}^*(M_{\beta,1},\mathrm{IC}(L_f))$ is 
\begin{equation}
    546\left[\frac{d_B^2+3d_B-2}4\right],
\label{eqn:fcontribution}
\end{equation}
in agreement with the corresponding entry in Table~\ref{tab:my_label}.  As we will see presently, all of the other nonzero entries in the row corresponding to $2j_L=g(d_B)-1$ exactly match the Lefschetz representation of $\mathbb{H}^*(M_{\beta,1},\mathrm{IC}(L_D))$.

Similarly, for the elliptic fibrations over the bases $B=\mathbb{F}_0$ or $B=\mathbb{F}_1$, we can again deduce from (\ref{eqn:decomposition}) that Lefschetz of $\mathbb{H}^*(B,\mathrm{IC}(B_\beta))$ is $488[0]$.  If $M_{\beta,1}\to B$ is a $\mathbb{P}^N$ bundle, we conclude after pullback that the Lefschetz representation of $\mathbb{H}^*(M_{\beta,1},\mathrm{IC}(L_f))$ is 
\begin{equation}
    488\left[\frac{N}2\right]
\end{equation}
and the placement of the 488's in Tables~\ref{tab:BPSF0dE=1} and \ref{tab:BPSF1dE=1} has been completely verified geometrically in all cases.

\smallskip
We next compute the Lefschetz representations of $\mathbb{H}^*(M_{\beta,1},\mathrm{IC}(L_D))$, beginning with $B=\mathbb{P}^2$.  Consider the moduli space
\begin{equation}
M'_{d_B,1}=\left\{D\cup f \in M_{d_B,1},\ r \in D
\right\}.
\end{equation}
Repeating the argument leading to the description of $\widehat{M}_{3,1}$, we see that $M'_{d_B,1}$ is a $\mathbb{P}^{(d_B^2+3d_B-4)/2}$-bundle over a space which fibers over $\mathbb{P}^2$ with fiber $\widetilde{\mathbb{P}}^2$.  There is a map
\begin{equation}
\pi_\beta': M'_{d_B,1} \to M_{d_B,1},\qquad (D\cup f,\ r)\mapsto D\cup f   
\end{equation}
whose fiber over $D\cup f$ is isomorphic to $D$.  It follows that the local system $L_D$ is identical to the local system associated with $R^1{\pi_\beta'}_*\mathbb{C}$.  The advantage of this description of $L_D$ is that $M'_{d_B,1}$ is much simpler than $\widehat{M}_{d_B,1}$ for $d_B>3$.

Finally, we combine the method of \cite[Section~7]{Huang:2020dbh} with (\ref{eqn:bottomrows}).  Putting $D=\dim M'_{d_B,1}=(d_B^2+3d_B+4)/2$, the decomposition theorem gives
\begin{equation}
R{\pi_\beta'}_*\mathbb{C}_{M'_{dB,1}}[D] = \mathbb{C}_{M_{d_B,1}}[D] \oplus\mathrm{IC}(L_D) \oplus \mathbb{C}_{M_{d_B,1}}[D-2]
\label{eqn:decompdB}
\end{equation}
Computed from the left hand side, the total hypercohomology of (\ref{eqn:decompdB}) has Lefschetz representation
\begin{equation}
    \left[\frac{d_B^2+3d_B-4}4\right]\otimes\left(\Big[1\Big]\oplus\Big[0\Big]\right)\otimes\Big[1\Big]
\end{equation}
For $d_B\ge 3$ this expands to 
\begin{equation}
    \left[\frac{d_B^2+3d_B+4}4\right]\oplus3\left[\frac{d_B^2+3d_B}4\right]\oplus4\left[\frac{d_B^2+3d_B-4}4\right]\oplus3\left[\frac{d_B^2+3d_B-8}4\right]\oplus\left[\frac{d_B^2+3d_B-12}4\right]
\label{eqn:totalspace}
\end{equation}
From this, we have to subtract the cohomology of $\mathbb{C}_{M_{d_B,1}}[D] \oplus \mathbb{C}_{M_{d_B,1}}[D-2]$ which is, using (\ref{eqn:bottomrows})
\begin{equation}
    \left[\frac12\right]\otimes\left(\left[\frac{d_B^2+3d_B+2}4\right]\oplus\left[\frac{d_B^2+3d_B-2}4\right]\oplus\left[\frac{d_B^2+3d_B-6}4\right]\right),
\end{equation}
which expands to
\begin{equation}
    \left[\frac{d_B^2+3d_B+4}4\right]\oplus2\left[\frac{d_B^2+3d_B}4\right]\oplus2\left[\frac{d_B^2+3d_B-4}4\right]\oplus\left[\frac{d_B^2+3d_B-8}4\right].
\label{eqn:fromtopspin}
\end{equation}
Subtracting (\ref{eqn:fromtopspin}) from (\ref{eqn:totalspace}), we are left with
\begin{equation}
\left[\frac{d_B^2+3d_B}4\right]\oplus2\left[\frac{d_B^2+3d_B-4}4\right]\oplus2\left[\frac{d_B^2+3d_B-8}4\right]\oplus\left[\frac{d_B^2+3d_B-12}4\right].
\end{equation}
Combining this summand with (\ref{eqn:fcontribution}), we find complete agreement with the $2j_R=g(d_B)-1$ rows of Table~\ref{tab:my_label} for $d_B=3,4$.  These cases $d_B=1,2$ are handled similarly and also agree.

The same method works just as well for $B=\mathbb{F}_0$ and $B=\mathbb{F}_1$. 
We find complete agreement with Tables~\ref{tab:BPSF0dE=1} and \ref{tab:BPSF1dE=1}.  We leave the details for the interested reader to check.

\subsubsection{\texorpdfstring{$d_E=2$}{dEeq2}}
\label{subsubsec:dE2}
In this section we perform a few simple checks for $d_E=2$ in the case of the $\mathbb{P}^2$ base.  A curve with $d_E=2$ is necessarily the union of plane curve $D$ of degree $d_B$ and two elliptic fibers $f_1$ and $f_2$.  The fibers $f_1$ and $f_2$ may coincide with a multiplicity 2 structure on a single fiber $f$.  We have a map $M_{d_B,2}\to \mathrm{Hilb}^2(\mathbb{P}^2)$ taking $D\cup f_1\cup f_2$ to the two points of the section where the fibers are attached. This map is a $\mathbb{P}^{(d_B^2+3d_B-4)/2}$-bundle, a fiber being the $\mathbb{P}^{(d_B^2+3d_B-4)/2}$ of plane curves of degree $d_B$ containing two fixed points.  Since the Lefschetz of $\mathrm{Hilb}^2(\mathbb{P}^2)$ is $[2]\oplus[1]\oplus[0]$ by G\"ottsche's formula, we conclude that the Lefschetz of $M_{d_B,2}$ is
\begin{equation}
    \left[\frac{d_B^2+3d_B-4}4\right]\otimes\left(\Big[2\Big]\oplus\Big[1\Big]\oplus\Big[0\Big]\right),
\end{equation}
which expands to

\begin{equation}
    \begin{array}{cc}
  \left[2\right]\oplus\left[1\right]\oplus\left[0\right]     &  d_B=1\\
 \left[\frac72\right] \oplus 2\left[\frac52\right]\oplus3\left[\frac32\right]\oplus2\left[\frac12\right]& d_B=2\\
 \left[\frac{d_B^2+3d_B+4}4\right] \oplus2\left[\frac{d_B^2+3d_B}4\right]  \oplus3\left[\frac{d_B^2+3d_B-4}4\right]\oplus2\left[\frac{d_B^2+3d_B-8}4\right]\oplus\left[\frac{d_B^2+3d_B-12}4\right]     & d_B\ge 3,
    \end{array}
\end{equation}
in complete agreement with the bottom rows of Table~\ref{tab:my_label1}.

The above calculation is readily adapted to the  $\mathbb{F}_0$ base, using the Lefschetz $[2]\oplus2[1]\oplus3[0]$ of $\mathrm{Hilb}^2(\mathbb{F}_0)$.   A modification is needed for curves in the base which cannot be made to pass through two general points (degrees $(1,0),(0,1),(2,0)$ and $(0,2)$), but these cases can be analyzed directly.  We find complete agreement with the bottom rows of Table~\ref{tab:BPSF0dE=2}, as well as consistency with (\ref{eqn:kkp}). The $\mathbb{F}_1$ base can be handled similarly.

\subsubsection{\texorpdfstring{$d_E=3$}{dEeq3}}
\label{subsubsec:dE3}

In this brief section, we consider $(d_B, d_E) = (1,3)$.  If $C$ more generally is a connected curve of class $\ell+d_E f$ for a line $\ell\subset S$, we have $S\cdot C = d_E-3$.  In the cases $d_E=1,2$ considered above, this intersection is negative, implying that $C$ is necessarily the union of a line in $S$ and $d_E$ fibers.  However, for $(d_B, d_E) = (1,3)$ we have $S\cdot C=0$, and we have additional curves $C$ which are disjoint from $S$ instead of being the union of a line and three fibers.  

In the partition function $Z_3$, the curves consisting of a line together with three fibers are accounted for by the term $P_3\left\langle W_{[3]}\right\rangle$, as can be verified by the same methods which we used above for $d_B=1$ and $d_E=1,2$.  So we see that an additional term is needed to account for the curves which are disjoint from $S$, corresponding to the term involving $\widetilde{P}_{3,1}$ in the application of Section~\ref{eq:Zk_ansatz} to $Z_3$.

\subsection{Rank 2 examples}
\label{subsec:rank2}

In this section, we provide geometric computations for the rank~2 models (Model 2A in Section~\ref{subsec:4parameter1} and Model 2B in Section~\ref{subsec:4parameter2}).  
We begin by describing these geometries in some detail before computing refined BPS numbers geometrically and comparing to the results of computations using the physical methods of Sections~\ref{sec:Wilsonloop} and~\ref{sec:examples} and catalogued in \cite{link}.

In each of the examples, we assign homogeneous coordinates $(x_1,\ldots,x_8)$ to the last 8 columns of the respective charge matrices, giving a toric description of a fourfold $\mathbb{P}$ containing $X$ as a Calabi-Yau hypersurface .   
The Mori cone of $X$ is generated by curve classes $l^{(1)},\ldots,l^{(4)}$ associated to the rows of the charge matrices in Sections~\ref{subsec:4parameter1} and~\ref{subsec:4parameter2}\footnote{Our computational algorithms for BPS invariants begin with the Mori cone of $\mathbb{P}$. Via the map $H_2(X,\mathbb{Z})\to H_2(\mathbb{P},\mathbb{Z})$, the Mori cone of $X$ is included the Mori cone of $\mathbb{P}$.  As we will see below, the Mori cone of $\mathbb{P}$ is strictly larger than the Mori cone of $X$ in the rank~2 models, necessitating a change of basis for integral $H_2$.} while the columns are associated to divisors $G_j:=X\cap\{x_j=0\}$ in $X$.  The charges forming the last 8 columns of the charge matrix are the intersection numbers $G_j\cdot\ell^{(i)}$.  The first columns consist of the intersection numbers $K_{\mathbb{P}}\cap l^{(i)}$.  

We let $J_1,\ldots J_4$ be the basis of the K\"ahler cone dual to the $\{l^{(i)}\}$.  The divisors $G_j$ can be expressed as a linear combination of the $J_i$ with coefficients given by the charges in the corresponding column.  Thus in both models, we have $G_7$ (which has been relabeled as $D_1$ in the charge matrices; also $G_8$ has been relabled as $D_2$) is given by $D_1=-2J_1+J_2+J_4$.  We consider each example separately in the following sections.

\subsubsection{Geometry of Model 2A}\label{sec:geom_Model2}
The triple intersection numbers of the $J_j$ can be determined algorithmically.  
The intersection ring is given by
\begin{equation}
\begin{split}
 \mathcal{R}=   8 J_1^3+2 J_1^2 J_2+2 J_1^2 J_3+J_1 J_2 J_3+16 J_1^2 J_4+4 J_1 J_2 J_4+4 J_1 J_3 J_4+2
   J_2 J_3 J_4+\\
   28 J_1 J_4^2+8 J_2 J_4^2+6 J_3 J_4^2+48 J_4^3.
   \end{split}
\end{equation}

This convention
indicates that a triple intersection evaluates to its coefficient in the above expression, and is zero if the corresponding monomial doesn't occur.   
For example $J_1^2 J_3=2$.  Then any triple intersection of divisors can be computed by expressing each divisor in terms of the $J_j$.  In the rest of this section, we will simply evaluate triple intersections without giving the details of the calculations.

We also see that $J_3^2=0$, since $J_3^2J_j=0$ for $j=1,\ldots,4$.  Similarly $J_2^2=0$.  More generally, the (multi)degree of any curve $C$ is given by the 4-tuple $(J_1\cdot C,\ldots,J_4\cdot C)$.  In the rest of the paper, we will simply indicate the degree of curves without giving the details of the calculations.

The projection $(x_1,\ldots,x_8)\mapsto(x_2,x_3,x_5,x_6)$ defines a map from $\mathbb{P}$ to a toric surface $B$, producing an elliptic fibration $\pi:X\to B$ of the Calabi-Yau hypersurface $X\subset \mathbb{P}$.

Note that $x_2,x_5$ have class $J_3$ with $J_3^2=0$ and define homogeneous coordinates on $\mathbb{P}^1$.  Similarly $x_3,x_6$ have class $J_2$ with $J_2^2=0$ and define homogeneous coordinates on another $\mathbb{P}^1$.  It follows that $B=\mathbb{F}_0=\mathbb{P}^1\times \mathbb{P}^1$.  The elliptic fiber class is given by the curve class $f=J_2\cdot J_3$, which has degree $(1,0,0,2)$.  Since $D_2\cdot f=1$, we see that $D_2$ is the section of $X\to \mathbb{F}_0$.  When identified with curves in $D_2$, the two fiber classes of $\mathbb{F}_0$ are $D_2\cdot J_3=l^{(2)}$ and $D_2\cdot J_2=l^{(3)}$, respectively.

In passing, we explain why the Mori cone of $X$ is strictly smaller than the Mori cone of $\mathbb{P}$.  
To see this, our initial calculations showed that the Mori cone of $\mathbb{P}$ contains a generator $-l^{(1)}+l^{(4)}$, whose intersection number with $D_2$ is $-1$.  If this class were represented by a curve $C\subset X$, it follows that a component of $C$ of negative self-intersection must be contained in $D_2$.  But $D_2=\mathbb{F}_0$ has no curves of negative self-intersection, and we conclude that $-l^{(1)}+l^{(4)}$ is not in the Mori cone of $X$.\footnote{In Model 2B with $B=\mathbb{F}_1$, a similar analysis initially leaves open the possibility that the class $-l^{(1)}+l^{(4)}$ in the Mori cone of $\mathbb{P}$ might be in the Mori cone of $X$, being represented by the $-1$ curve of the section.  However, as we will see below, the $-1$ curve has class $l^{(3)}$, so this is not possible and again the Mori cone of $X$ is strictly smaller than the Mori cone of $\mathbb{P}$.}

We claim that the divisor $D_1$ is a Hirzebruch surface $\mathbb{F}_2$ over a $\mathbb{P}^1$ in the base, with fibers of class $(1,0,0,0)$.  

To see this, observe that $X\subset\mathbb{P}$ is a hypersurface of degree $(0,1,0,3)$ in the indicated basis.  The only monomials of degree $(0,1,0,3)$ which do not contain $x_7$ (whose vanishing locus is $D_1$), are $x_1^3x_3$ and $x_1^3x_6$.  After setting $x_7=0$, the equation of $X$ simplifies to $x_1^3(ax_3+bx_6)=0$.  Here, the variable $x_1^3$ is an inconsequential factor which can be scaled away by the torus action.  We conclude that $D_1$ projects via $\pi$ to a curve $L\subset B=\mathbb{F}_0$ defined by $ax_3+bx_6=0$.  This is one of the fibers of $\mathbb{F}_0$, with $L\simeq\mathbb{P}^1$.
The fiber of $\pi:D_1\to L$ is given as the intersection $J_3\cap D_1$.  Computing intersections, the fiber has class $(1,0,0,0)$ and has genus 0 by adjunction: $D_1\cdot J_1\cdot (D_1+J_1)=-2$.  Thus $D_1$ is either a Hirzebruch surface or the blowup of a Hirzebruch surface.  A blowup $S$ of a Hirzebruch surface at $n$ points which is contained in a Calabi-Yau threefold satisfies $S^3=K_S^2=8-n$ \cite{Griffiths:1978}.  Since $D_1^3=8$, we see that $D_1$ is a Hirzebruch surface. Since $L=D_1\cap D_2$ has self-intersection 0 in $D_2$, it has self-intersection $-2$ in $D_1$ by adjunction and therefore $D_1\simeq\mathbb{F}_2$.

Now, $\pi^{-1}(L)\subset X$ is a surface containing $D_1$, so $\pi^{-1}(L)=D_1\cup D_1'$ for some surface $D_1'$.  We claim that $D_1'$ is a ruled surface with 16 degenerate fibers consisting of two $\mathbb{P}^1$s.  To see this, the description of $D_1'$ shows that $D_1'\sim J_2-D_1$.  The fiber of $D_1'\to L$ has class $J_3\cdot D_1'$, which we compute has class $(0,0,0,2)$ and genus zero by adjunction.  Thus $D_1'$ is a Hirzebruch surface or the blowup of a Hirzebruch surface.  Next, we compute $(D_1')^3=(J_2-D_1)^3=-8=8-16$, so 16 points in a Hirzebruch surface have been blown up.  Suppose that $p$ is a point of a Hirzebruch surface and let $F_p$ be the fiber containing $p$.  After blowing up $p$, the inverse image of $F_p$ contains two $\mathbb{P}^1$s meeting at a point: the proper transform of $F_p$ and the exceptional divisor, and these are degenerate fibers of the blown-up Hirzebruch surface.  We conclude that 16 of the fibers of $D_1'$ are degenerate, consisting of two $\mathbb{P}^1$s, each of class $(0,0,0,1)$.

From $D_1\cdot 2l^{(4)}=2$, we see that $D_1\cup D_1'\to L$ forms on $I_2$ configuration.

We can now list some refined BPS numbers.

\begin{equation}
    \begin{array}{|c|c|} \hline
    {\mathrm{Classes}}& {\mathrm{Refined\ BPS}}\\ \hline
    (0,0,0,1);(1,0,0,1);(1,1,0,1);(1,0,1,1) & N_{0,0}=32\\ \hline
    (1,0,0,0);(1,0,1,0);(1,1,0,0);(0,0,0,2);&N_{0,1/2}=1\\ \hline
    \end{array}
\label{eqn:RGVN2A}
\end{equation}
all in agreement with \cite{link} (see also \cite{Sun:2016obh}).

We have seen that the class $(0,0,0,1)$ is represented by either $\mathbb{P}^1$ in any of the 16 degenerate fibers of $D_1'$, explaining $N_{0,0}=32$.  Letting $C$ denote any of these curves, we can use $C$ to produce a unique (reducible) rational curve in any of the classes listed in the first row of (\ref{eqn:RGVN2A}) as follows.

The class $(1,0,0,1)$ is realized by the union of $C$ with the unique fiber of $D_1$ which meets it.  This configuration meets a unique fiber of $D_2$ with class $J_2$, which can then be attached to produce a curve in the class $(1,1,0,1)$.  Similarly, the union of $C$ with a fiber of $D_1$ meets a unique fiber of $D_2$ which class $J_3$ (the fiber denoted by $L$ above) resulting in a curve of degree $(1,0,1,1)$.  In each of the indicated degrees, we have exactly one rational curve for each of the 32 curves $C$.

The second row of (\ref{eqn:RGVN2A}) corresponds to families of rational curves parametrized by $\mathbb{P}^1$: the fiber class of $D_1$ ($(1,0,0,0)$), the fiber class of $D_1$ glued to $L$ ($(1,0,1,0)$), the fiber class of $D_1$ glued to the unique fiber of $D_2$ with class $J_2$ that meets it ($(1,1,0,0)$), and the fiber class of $D_1'$ ($(0,0,0,2)$).

\smallskip
The refined BPS numbers for classes of the form $(d_1,d_2,d_3,2d_1)$ are computed by our methods for elliptic $\mathbb{F}_0$ with $d_E=d_1$, as these are the classes of curves of degree $(d_2,d_3)$ in $\mathbb{F}_0$ with $d_1$ fiber classes $(1,0,0,2)$ added.  The only difference is that the refined BPS spectrum for the fiber class changes to 
\begin{equation}
    \left[\frac12,1\right]\oplus\left[\frac12,0\right]\oplus430\Big[0,0\Big]\oplus\left[0,\frac12\right]
    \label{eqn:rank2fiberGVN}
\end{equation}
according to (\ref{eqn:Fnfiberclass}) and $\chi(X)=-420$.  

The methods of Section~\ref{subsubsec:g0basecurve} for genus 0 base curves apply in our situation.  For genus 0 base curves $(d_2,d_3)=(1,d_3)$, $M_{(1,d_2,d_3,2)}$ is the same as $M_{d_2,d_3,1}$ was in the rank 1 case, with the same Lefschetz $[d_3+1]\oplus2[d_3]\oplus[d_3-1]$, and the diagonal Lefschetz calculation becomes $[d_3](\left[\frac32\right]\oplus3\left[\frac12\right]\oplus430\left[0\right])$. Putting these facts together, we conclude that for $d_3\ge 1$ the BPS representation is
\begin{equation}
    \left[\frac12,d_3+1\right]\oplus2\left[\frac12,d_3\right]\oplus\left[\frac12,d_3-1\right]\oplus\left[0,d_3+\frac12\right]\oplus\left[0,d_3-\frac12\right]\oplus430\Big[0,d_3\Big]
\end{equation}
while for $d_3=0$ we get the same representation as for the fiber class.  The calculation has been done by our physical methods described in Section~\ref{subsec:4parameter1} for $d_3\le 6$ and we find complete agreement with \cite{link}.

 Similarly, the methods of Section~\ref{subsubsec:g1basecurve} for genus 1 base curves apply to curves of degree $(1,2,2,2)$, corresponding to a degree $(2,2)$ base curve plus a fiber class, and we find complete agreement with \cite{link}.\footnote{There is an inconsequential extra step needed beyond our calculation in the rank 1 case.  The decomposition theorem for $\pi$ includes an extra summand supported on the curve $L$ \cite{Katz:2022vwe}.  However, this same summand also appears in the local system for $\pi_\beta=\pi_{2,2}$, so the calculation can be completed exactly as before.} 
 
 We can also repeat the method of Section~\ref{subsubsec:dE2} to check the maximum $j_L$ content for the classes $(2,d_2,d_3,4)$, the union of a base curve and two fiber classes, i.e.\ the analogue of $d_E=2$ in the rank~1 cases. Except for the base degrees $(d_2,d_3)=(1,0)$ or $(0,1)$, given any two points of the base $B$, there exist curves of degree $(d_2,d_3)$ containing both points, so the method applies. 
 
 Since the moduli space of these curves is a $\mathbb{P}^D$ with $D=(d_2+1)(d_3+1)-3$, the Lefschetz of the moduli space of curves of class $(2,d_2,d_3,4)$ is the tensor product of $[D/2]$ and $[2]\oplus2[1]\oplus3[0]$, the Lefschetz of $\mathrm{Hilb}^2(\mathbb{F}_0)$, which expands to
 \begin{equation}
     \Big[\frac{D+4}2\Big]\oplus3\Big[\frac{D+2}2\Big]\oplus6\Big[\frac{D}2\Big]\oplus3\Big[\frac{D-2}2\Big]\oplus\Big[\frac{D-4}2\Big]
     \label{eqn:rk2dE2jLmax}
 \end{equation}
if $D\ge4$.  If $D$ is 2 or 3, (\ref{eqn:rk2dE2jLmax}) is still valid after dropping the last term, and is still valid after dropping the last two terms if $D=1$.

Since the genus of a base curve of degree $(d_2,d_3)$ is $(d_2-1)(d_3-1)$ and we are adding two elliptic fibers, we get $(2j_L)_{\mathrm{max}}=(d_2-1)(d_3-1)+2$, with the associated $\mathrm{SU}(2)_R$ representation given by (\ref{eqn:rk2dE2jLmax}), modified as described above for $D<4$.  These computations are all in agreement with the results of our physical methods as far as we have computed \cite{link}.

\subsubsection{Geometry of Model 2B}

The geometric analysis is similar to the analysis of Model 2A. 

The intersection ring is now given by

\begin{equation}
\begin{split}
  \mathcal{R}=  8 J_1^3+3 J_1^2 J_2+J_1 J_2^2+2 J_1^2 J_3+J_1 J_2 J_3+16 J_1^2 J_4+6 J_1 J_2 J_4+2
   J_2^2 J_4+4 J_1 J_3 J_4+\\
  2 J_2 J_3 J_4+26 J_1 J_4^2+ 10 J_2 J_4^2+6 J_3 J_4^2+42 J_4^3
\end{split}
\end{equation}

In this model, $x_2,x_5$  again have class $J_3$ with $J_3^2=0$, giving a further projection $B\to \mathbb{P}^1$ identifying $B$ as a Hirzebruch surface.  Since $x_3$ has class $J_2$ while $x_6$ has class $J_2-J_3$, we see that $B\simeq\mathbb{F}_1$, with $x_6=0$ being the $-1$ curve and $x_3=0$ a section of self-intersection $+1$.  Again, $D_2$ is the section.  
When identified with curves in $D_2$, the fiber class of $\mathbb{F}_1$ is $D_2\cdot J_3=l^{(2)}$, and the exceptional curve is $D_2\cdot (J_2-J_3)=l^{(3)}$.  The fiber class is again $J_2\cdot J_3$, which has class $(1,0,0,2)$.

The monomials of degree $(0,1,0,3)$ not involving $x_7$ are $x_1^3x_3$, $x_1^3x_2x_6$, and $x_1^3x_5x_6$, so $D_1$ projects onto a curve $L\subset B=\mathbb{F}_1$ with equation of the form $ax_3+bx_2x_6+cx_5x_6=0$, a section of $\mathbb{F}_1$ of class $J_2$.  Identifying $L$ with a curve $J_2\cdot D_2$ in the section $D_2$, we see that $L$ has degree $(0,1,1,0)$. The inverse image of $L$ is the union of $D_1$ and another surface $D_1'$, with $D_1'=J_2-D_1$ as in the last example, forming an $I_2$ configuration over $L$.  The fiber class $J_3\cdot D_1$ of $D_1$ is again $l^{(1)}$ and the fiber class $J_3\cdot D_1'$ of $D_1'$ is again $(0,0,0,2)$.

Once again we compute $(D_1)^3=8$ and $D_1$ is a Hirzebruch surface (which can be shown to be $\mathbb{F}_3$), but now $(D_1')^3=-14$, so $D_1'$ is a Hirzebruch blown up at 22 points, and so $D_1'$ has 22 degenerate fibers, each fiber consisting of two curves of degree $(0,0,0,1)$.  We get
\begin{equation}
    \begin{array}{|c|c|} \hline
    {\mathrm{Classes}}& {\mathrm{Refined\ BPS}}\\ \hline
    (0,0,0,1);(1,0,0,1);(1,1,0,1) & N_{0,0}=44\\ \hline
    (1,1,1,1)&N_{0,1/2}=44\\ \hline
    (1,0,0,0);(1,1,0,0);(0,0,0,2);&N_{0,1/2}=1\\ \hline
    \end{array}
\label{eqn:RGVN2B}
\end{equation}
all in agreement with our physical calculations using the methods of Section~\ref{subsec:4parameter2} \cite{link}.  We only explain the geometry of the class $(1,1,1,1)$, as the other cases are similar to what was done before.  Let $C$ be any of the 44 curves of degree $(0,0,0,1)$, which glues to a unique fiber of $D_1$ as before to produce a rational curve of class $(1,0,0,1)$ as before.  The fiber of $D_1$ meets $L$ as before.  Denoting the intersection point of $D_1$ and $L$ by $p$ and recalling that $D_2=\mathbb{F}_1$, we see there is a $\mathbb{P}^1$ moduli space of curves of degree $(0,1,1,0)$ (necessarily contained in $D_2$) which contain $p$.  Gluing these curves to the curves of degree $(1,0,0,1)$, we get a moduli space of rational curves of degree $(1,1,1,1)$ consisting of a disjoint union of 44 copies of $\mathbb{P}^1$.  The middle row of (\ref{eqn:RGVN2B}) follows.

Similarly, the methods of Section~\ref{subsubsec:g0basecurve} for genus 0 base curves and of Section~\ref{subsubsec:g1basecurve} for genus 1 base curves apply.
We can also repeat the method of Section~\ref{subsubsec:dE2} to check the maximum $j_L$ content for the classes $(2,d_2,d_3,4)$, the analogue of $d_E=2$ in the rank~1 cases.

\subsection{Rank 3 examples}
\label{subsec:rank3}

We begin by describing the rank 3 geometries from Section~\ref{sec:rank3} in some detail before computing their refined BPS numbers geometrically and comparing to the results of computations using the physical methods of Sections~\ref{sec:Wilsonloop} and~\ref{sec:examples} and catalogued in \cite{link}.

\subsubsection{Geometry of Model 3A}
\label{subsec:I02}
We start by summarizing the geometry of Model 3A with  explanation to follow.  As we will see, the elliptic fibration has $I_0^*$ fibers over a $-2$ curve.

Referring to the divisors which have been identified in Section~\ref{subsec:5parameter}, the base of the elliptic fibration is the surface $D_3$, which is isomorphic to the Hirzebruch surface $\mathbb{F}_2$.  The fibration degenerates to an $I_0^*$ fibration over the $-2$ curve of $D_3$.  The $I_0^*$ fibers consist of unions of fibers of the ruled surfaces $D_1,D_2$, and $D_4$. $D_2$ is the Hirzebruch surface $\mathbb{F}_0$ fibered over the $-2$ curve of $D_3$, and its fibers represent the curve class $l^{(2)}$.  $D_1$ is the Hirzebruch surface  $\mathbb{F}_2$, and its fibers represent the curve class $l^{(1)}$.  $D_4$ is a ruled surface over the genus 4 curve $C_{14}=D_1\cap D_4$, and its fibers represent the curve class $l^{(5)}$.  The curve $C_{14}$ is a trisection of $D_1$, so any fiber of $D_1$ intersects three fibers of $D_4$.  An $I_0^*$ fiber consists of a fiber of $D_2$, a fiber of $D_1$ with multiplicity~2, and the three fibers of $D_4$ which intersect the fiber of $D_1$.  Thus the fiber class is visibly $2l^{(1)}+l^{(2)}+3l^{(5)}$, as we will also see below by other methods.

\begin{figure}[t]
\begin{center}
\begin{tikzpicture}[>=latex, square/.style={regular polygon,regular polygon sides=4},scale=.75]
\coordinate (D11) at (0,0);
\coordinate (D12) at (6,0);
\coordinate (D13) at (2,2);
\coordinate (D14) at (8,2);
\coordinate (D21) at (2,2);
\coordinate (D22) at (8,2);
\coordinate (D23) at (1,5);
\coordinate (D24) at (7,5);
\coordinate (D31) at (1,5);
\coordinate (D32) at (7,5);
\coordinate (D33) at (1,9);
\coordinate (D34) at (7,9);
\coordinate (S1) at (7,5.5);
\coordinate (S2) at (1,6.5);
\coordinate (S3) at (7,7.5);
\coordinate (S4) at (1,8.5);
\coordinate (sS1) at (9,9);
\coordinate (sS2) at (3,10);
\coordinate (sS3) at (9,11);
\coordinate (sS4) at (3,12);
\coordinate (D1) at (7.5,1.2);
\coordinate (D2) at (8,4);
\coordinate (D3) at (1.5,5.7);
\coordinate (D4) at (6,11.5);
\coordinate (l1) at (.6,7.5);
\coordinate (l2) at (1.2,3.5);
\coordinate (l3) at (.6,1.5);
\coordinate (l4) at (4.5,2.7);
\coordinate (l5) at (1.7,10.8);
\draw[-] (D11) to (D12);
\draw[-] (D11) to (D13);
\foreach \a in {1,...,6} 
\draw[-] (\a,0) to (2+\a,2);
\draw[-] (D12) to (D14);
\draw[-] (D13) to (D14);
\node (D1) at  (D1) [below] {$D_3$};
\node (l3) at  (l3) [below] {$l^{(3)}$};
\draw[-] (D21) to (D22);
\foreach \a in {1,...,6} 
\draw[-,green] (2-\a/6,2+\a/2) to (8-\a/6,2+\a/2);
\draw[-] (D21) to (D23);
\foreach \a in {1,...,6} 
\draw[-] (\a+2,2) to (1+\a,5);
\draw[-] (D22) to (D24);
\draw[-] (D23) to (D24);
\node (D2) at  (D2) [below] {$D_2$};
\node (l4) at  (l4) [below, green] {$l^{(4)}$};
\node (l2) at  (l2) [below] {$l^{(2)}$};
\draw[-] (D31) to (D32);
\draw[-] (D31) to (D33);
\foreach \a in {1,...,6} 
\draw[-] (\a+1,5) to (1+\a,9);
\draw[-] (D32) to (D34);
\draw[-] (D33) to (D34);
\node (D3) at  (D3) [below] {$D_1$};
\node (l1) at  (l1) [below] {$l^{(1)}$};
\node (l5) at  (l5) [below] {$l^{(5)}$};
\draw [cyan] plot [smooth, tension=2] coordinates {(S1) (S2) (S3) (S4) };
\draw [cyan] plot [smooth, tension=2] coordinates { (sS1) (sS2) (sS3) (sS4)};
 \fill[white] (2.9,9) rectangle (11,10.86);
\draw[-,cyan] (S1) to (sS1); 
\draw[-,cyan] (1.05,6.67) to (2.14,8.5);
\draw[-,cyan] (7,7.4) to (8.98,10.88);
\draw[-,cyan] (S4) to (sS4); 
\node (D4) at  (D4) [below,cyan] {$D_4$};
\draw[-,cyan] (2,8.5) to (2+2,8.5+3.5);
\draw[-,cyan] (3,8.46) to (3+2,8.46+3.5);
\draw[-,cyan] (4,8.4) to (4+2,8.4+3.5);
\draw[-,cyan] (5,8.3) to (5+2,8.3+3.5);
\draw[-,cyan] (6,8.1) to (6+2,8.1+3.5);
\draw[-,cyan] (2,6.95) to (2+0.43*2,6.95+0.43*3.5);
\draw[-,cyan] (3,7) to (3+0.41*2,7+.41*3.5);
\draw[-,cyan] (4,7) to (4+0.38*2,7+0.38*3.5);
\draw[-,cyan] (5,6.95) to (5+0.36*2,6.95+0.36*3.5);
\draw[-,cyan] (6,7) to (6+0.27*2,7.+0.27*3.5);
\draw[-,cyan] (2,5.84) to (2+0.33*2,5.84+0.33*3.5);
\draw[-,cyan] (3,5.67) to (3+0.38*2,5.67+0.38*3.5);
\draw[-,cyan] (4,5.57) to (4+0.4*2,5.57+0.4*3.5);
\draw[-,cyan] (5,5.53) to (5+0.42*2,5.53+0.42*3.5);
\draw[-,cyan] (6,5.52) to (6+1*2,5.52+1*3.5);
\node at (2,8.5) [fill=red,scale=.3pt, rotate=45] (s1){};
\node at (s1) [below right, red]{$I_0^*$};
\end{tikzpicture}
\end{center}
\caption{The geometry of the rank 3A model. The unions of all intersecting $D_1$, $D_2$ and $D_4$ fibers
constitutes an $I_0^*$ fiber, one of which is indicated.}
\label{fig:Model3A}
\end{figure}
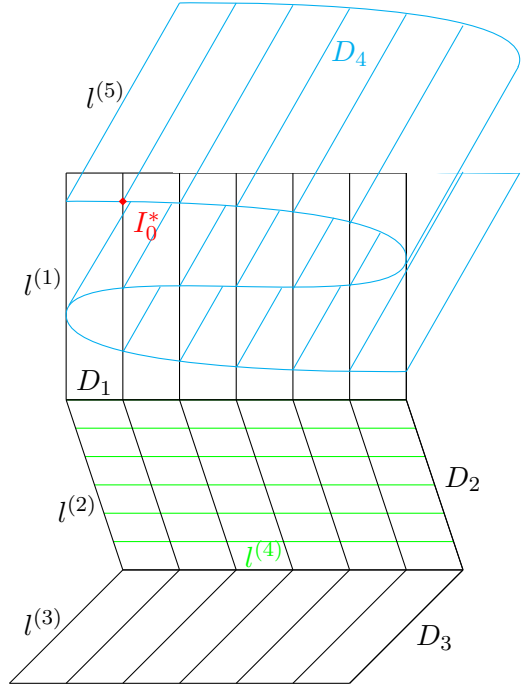

We next describe the geometry in more detail, followed by  explanations.

Denoting the curves $D_i\cap D_j$ (when nonempty) by $C_{ij}$, the $-2$ curve of $D_3$
is identified with $C_{36}$. The surfaces $D_1$ and $D_2$ are Hirzebruch surfaces $\mathbb{F}_2$ and $\mathbb{F}_0$ respectively, $D_1$ being identified with the central vertex of the $\hat{D}_4$ Dynkin diagram.  The curve $C_{23}$ is identified with the $-2$ section of $D_3$ and a section of $D_2$ with self-intersection zero.  The curve $C_{12}$ is identified with the $-2$ section of $D_1$ and a section of $D_2$ with self-intersection zero. The local limit with compact divisors $D_1$ and $D_2$ is $SU(3)$.

The surface $D_4$ is ruled over the genus 4 curve $C_{14}$, which is identified with a section of $D_4$ and a trisection of $D_1$.  The five components of an $I_0^*$ fiber are as follows: a fiber of $D_1$, a fiber of $D_2$, and each of the fibers over the three points of $C_{24}$ which lie on the fiber of $D_1$.  Clearly these fibers undergo monodromy, and the rank~2 local limit with compact divisors $D_1$ and $D_4$ is $G_2$ with four hypermultiplets.

Turing to explanations, we let $J_k$ be the divisor class satisfying $J_k\cdot l^{(i)}=\delta_k^i$. Then the (multi)degree of any curve $C$ can be expanded as
\begin{equation}
    [C]=\sum_{k=1}^5 \left(J_k\cdot C\right)l^{(k)}.
\label{eq:Moribasis}
\end{equation}
We apply (\ref{eq:Moribasis}) to $C_{ij}=D_iD_j$ in the table below.
The arithmetic genus can be found from adjunction: $2g-2=D_iD_j(D_i+D_j)$. 

\begin{equation}
\label{eq:intersectioncurves}
\begin{array}{|c|c|c|}\hline
{\mathrm{curve}}&{\mathrm{class}}&{\mathrm{genus}}\\ \hline
C_{35} & l^{(3)}&0\\ \hline
C_{23} & l^{(4)}&0\\ \hline
C_{25} & l^{(2)}&0\\ \hline
C_{12} &l^{(4)} &0\\ \hline
C_{15} & l^{(1)}&0\\ \hline
C_{14} &6l^{(1)}+3l^{(4)} &4\\ \hline
C_{56}  & 2l^{(1)}+l^{(2)}+3l^{(5)} &1 \\ \hline
C_{45} & 3l^{(5)} &-2\\ \hline
\end{array}
\end{equation}
Since $C_{45}$ has negative arithmetic genus, it must be reducible.  Since it has class $3l^{(5)}$, each component has class either $l^{(5)}$ or $2l^{(45)}$, and $C_{45}$ has two or three irreducible components. For generic hypersurfaces, each component is a smooth curve.  If the $g_i$ denote the genera of the respective components, then the arithmetic genus of the configuration of curves is $\sum_{i=1}^kg_i-k+1$, where $k$ is the number of components.  The only possibility is $k=3$ and each $g_i=0$.  We conclude that $C_{45}$ is the disjoint union of three curves of class $l^{(5)}$, each of genus 0.  Each of these curves will be identified with fibers of $D_4$ presently.  The curve $C_{56}$ will be identified with the elliptic fiber class.  

We next explain why the surfaces $D_1,D_2,D_3$, and $D_4$ are ruled surfaces.  First, they are all smooth for generic hypersurfaces.  We claim that the fiber classes are $l^{(1)},l^{(2)},l^{(3)}$ and $l^{(5)}$, respectively.  The argument for each of $l^{(1)},l^{(2)},l^{(3)}$ is identical and we do those cases first.  From (\ref{eq:intersectioncurves}) we see that these $l^{(i)}$ are represented by the curve $C_{i5}$, which has genus 0.  From $D_i\cdot C_{i5}=-2$ and adjunction, we see that $C_{i5}$ has self-intersection zero in $D_i$, so it must be a fiber class and $D_i$ is a ruled surface for $i=1,2,3$.
For $D_4$, we have seen above that $l^{(5)}$ has genus 0.  Since $D_4\cdot l^{(5)}=-2$, we conclude as before the $l^{(5)}$ is the fiber class of $D_4$.

If a surface $S$ is a $\mathbb{P}^1$ bundle over a curve of genus $g$, then $K_S^2=8(1-g)$ \cite{Griffiths:1978}. Blowing up $S$ to introduce degenerate fibers only lowers $K_S^2$.  Finally, we compute $K_{D_i}^2=D_i^3=8$, so the genus of the base curve is zero and there are no degenerate fibers.  So each $D_i$ is a Hirzebruch surface.  The identification of $D_1,D_2,D_3$ with $\mathbb{F}_2,\mathbb{F}_0,\mathbb{F}_2$ respectively, as well as the identification of the intersection curves $C_{12}$ and $C_{23}$, follow from the intersection numbers $D_1^2D_2=0, D_1 D_2^2=-2,D_2^2D_3=-2$, and $D_2D_3^2=0$.

From the intersection numbers of the $D_i$ with each of the $\mathbb{P}^1$ fiber classes, we see that the surfaces are glued in a chain in the order $D_3-D_2-D_1-D_4$. 
 From $D_3\cdot l^{(1)} = D_3\cdot l^{(5)}=D_4\cdot l^{(2)}=0$ we see that $D_3$ does not intersect $D_1$ or $D_4$, and $D_4$ does not intersect $D_2$.

The intersection curve $C_{14}$ of $D_1$ and $D_4$ is embedded in $D_4$ as a section and in $D_1$ as trisection.  To see this, we count intersections of $C_{14}$ with the respective fibers.  From (\ref{eq:intersectioncurves}) we see that the fiber class of $D_1$ is obtained by intersecting $D_1$ and $G_5$, so the intersection number is $G_5\cdot C_{14}=3$. Similarly, as $G_5\cap D_4$ consists of three fibers of $D_4$, each fiber intersects $C_{14}$ once.  Here, we are using the same notation as in Section~\ref{subsec:rank2}: $G_j=X\cap \{x_j=0\}$.

It will be useful for geometric verification of the refined BPS numbers to relate the geometries of $D_1,D_2,D_3$ to standard notation for Hirzebruch surfaces $\mathbb{F}_n$.  The group of divisors of $\mathbb{F}_n$ is generated by two classes, $E$ and $F$, where $E^2=-n$ and $F$ is the fiber class.  It is sometimes convenient to also define $H=E+nF$, the class of a section with $H^2=n$. We distinguish between divisors on $D_1,D_2,D_3$ by subscripts.

For $D_3$ we have $E_3=C_{23}$, $F_3=C_{35}$, and $H_3=C_{36}$.  For $D_2$ we have $F_2=C_{25}$, $H_2=C_{23}$, and $E_2=C_{12}$ (in this case, $E_2$ and $H_2$ are the same class because $n=0$).  For $D_1$ we have $F_1=C_{15}$ and $E_1=C_{12}$.  In addition, viewed as a divisor on $D_1$, we have $C_{14}=3H_1=3E_1+6F_1$.

\medskip
We can now compute many refined numbers geometrically and find agreement with our methods. Let's start with the class of $l^{(5)}$, which we have seen is the fiber class of $D_4$.  The moduli space of these curves is the base of the fibration, the genus 4 base curve $C_{14}$. Since these curves have genus 0, the left spin is zero and the right spin coincides with the Lefschetz representation on $C_{14}$, resulting in $[0,1/2]\oplus8[0,0]$.

We hasten to point out that this is not a check on the full power of our methods, as this class was already realizable in the local $G_2$ limit.    However, one can arrive at other curve classes by gluing fiber classes of $D_1$ to $l^{(5)}$, and then one can continue to glue fiber classes of $D_2$ and then $D_3$ in succession.  All of these curve classes have moduli space $C_{14}$, so we also get $[0,1/2]\oplus8[0,0]$ for the classes $l^{(1)}+l^{(5)}$, $l^{(1)}+l^{(2)}+l^{(5)}$, and $l^{(1)}+l^{(2)}+l^{(3)}+l^{(5)}$.  These all agree with calculations.  The last two classes are not realizable in either the $G_2$ local limit or the $SU(4)$ local limit, so these cases serve as a nontrivial check of our methods.

The methods of Section~\ref{subsubsec:g0basecurve} apply to genus~0 base curves glued to a fiber.  For example, we can consider base curves of degree $(d,1)$, which after glueing a fiber represent the class $(2,1,d,1,3)$.  We find that for $d>1$ we get the representation
\begin{equation}
    \Big[\frac12,d\Big]\oplus2\Big[\frac12,d-1\Big]\oplus\Big[\frac12,d-2\Big]
    \oplus2\Big[0,\frac{2d-1}2\Big]\oplus2\Big[0,\frac{2d-3}2\Big]\oplus468\Big[0,d-1\Big],
\end{equation}
in agreement with our physical calculations described in Section~\ref{subsec:5parameter} as far as we have computed \cite{link}.

\subsubsection{Geometry of Models 3B and 3C.}

The geometric analyses for models 3B and 3C, the elliptic fibrations introduced in Sections~\ref{subsec:5parameter1} and \ref{subsec:5parameter3} are largely similar to the analysis for model 3A done in Section~\ref{subsec:I02} for the elliptic fibration introduced in Section~\ref{subsec:5parameter}, so we begin by quickly pointing out the similarities and differences without giving the detailed calculations. We let $-n$ be the self-intersection of the curve in the base over which we have an $I_0^*$ fibration, so that we have $n=2$ in Model 3A, $n=3$ in Model 3B, and $n=1$ in Model 3C. 

In each case, the base of the elliptic fibration is $D_3$, a Hirzebruch surface $\mathbb{F}_n$. The surface $D_2$ is the Hirzebruch surface $\mathbb{F}_{|n-2|}$, which is attached to $D_3$ along the curve $C_{23}$, which is identified with the $-n$ section of $D_3$ and a section of $D_2$ of self-intersection $n-2$. The surface $D_1$ is also a Hirzebruch surface, and $D_4$ is ruled over the curve $C_{14}$, which is a section of $D_4$ and a trisection of $D_1$ representing the class $3H_1$. The adjunction formula tells us that the genus of a curve of class $3H$ on a Hirzebruch surface $\mathbb{F}_k$ is $3k-2$.  
In each case, the classes $l^{(1)},l^{(2)},l^{(3)}$, and $l^{(5)}$ are the fiber classes of $D_1,D_2,D_3$, and $D_4$, respectively.

In the case $n=1$, $D_1$ is a copy of $\mathbb{F}_3$, and so $C_{14}$ has genus 7.  The surfaces $D_1$ and $D_2$ are attached along $C_{12}$, which is identified with the $-3$ curve of $D_1$ and a section of $D_2$ with self-intersection 1.  The class $l^{(4)}$ is represented by $C_{23}$ (which is the unique curve in that class).

In the case $n=3$, $D_1$ is a copy of $\mathbb{F}_1$, and so $C_{14}$ has genus 1.  The surfaces $D_1$ and $D_2$ are attached along $C_{12}$, which is identified with the $-1$ curve of each of $D_1$ and $D_2$.  The class $l^{(4)}$ is represented by $C_{12}$ (which is the unique curve in that class).

\smallskip
In either case, the moduli space of $l^{(5)}$ is the base $C_{14}$ of the ruled surface $D_4$. We conclude from the Lefschetz representation of $C_{14}$ that the $SU(2)\times SU(2)$ representation for this class is
\begin{equation}
    \begin{array}{cc}
    \left[0,\frac12\right]\oplus14\left[0,0\right]& n=1\\
        \left[0,\frac12\right]\oplus2\left[0,0\right]& n=3
    \end{array}
\end{equation}
in agreement.  As in the preceding section, in each case $n=1$ and $n=3$ we get the same representations for the classes $l^{(1)}+l^{(5)}$, $l^{(1)}+l^{(2)}+l^{(5)}$, and $l^{(1)}+l^{(2)}+l^{(3)}+l^{(5)}$.  

The methods of Section~\ref{subsubsec:g0basecurve} again apply to genus~0 base curves glued to a fiber.  We again consider base curves of degree $(d,1)$, which after gluing a fiber represent the class $(2,1,d,1,3)$.  We find that for $d>n$ we get the representation
\begin{equation}
\begin{split}
    \Big[\frac12,\frac{2d-n+2}2\Big]\oplus2\Big[\frac12,\frac{2d-n}2\Big]\oplus\Big[\frac12,\frac{2d-n-2}2\Big]\oplus\\
    2\Big[0,\frac{2d-n+1}2\Big]\oplus2\Big[0,\frac{2d-n-1}2\Big]\oplus\left(396+36n\right)\Big[0,\frac{2d-n}2\Big],
    \end{split}
\end{equation}
in agreement with our physical calculations described in Sections~\ref{subsec:5parameter1} and~\ref{subsec:5parameter3} as far as we have computed \cite{link}.

\subsubsection{Dependence of refined BPS numbers on complex moduli}\label{sec:Model3_deform}

Suppose that a Calabi-Yau threefold $X$ contains a smooth surface $S$, and that $S$ is a $\mathbb{P}^1$ bundle $\rho:S\to C$ over a smooth curve $C$ of genus $g$.  Letting $\beta\in H_2(X,\mathbb{Z})$ be the class of any fiber of $\rho$, we see that $C$ is a connected component of $M_\beta$.  It follows that the contribution of this family of curves to the BPS spectrum is $[1/2,0]\oplus2g[0,0]$.  This is a general result which does not require $X$ to be elliptically fibered.

We let $\sigma\in H^1(X,TX)$ be a first-order deformation of the complex structure of $X$,\footnote{It is easy to see by the unobstructedness of Calabi-Yau moduli that a first-order analysis suffices for our purposes.} and let $F\simeq\mathbb{P}^1$ be any fiber of $\rho$. The space of first order deformations of $F$ is given by $H^0(F,N_{F/X})$ and the obstructions lie in $H^1(F,N_{F/X})$ (this is the case for submanifolds $F$ of any dimension).  Then $F$ deforms holomorphically along with the deformation $\sigma$ precisely when $r(\sigma)=0$, where $r$ is the restriction map
\begin{equation}
    H^1(X,TX)\stackrel{r}{\to} H^1(F,N_{F/X}).
    \label{eqn:obstruction}
\end{equation}
From the short exact sequence
\begin{equation}
    0\to N_{F/S}\to N_{F/X}\to N_{S/X}\vert_F \to 0,
\end{equation}
the triviality of $N_{F/S}$ deduced from $N_{F/S}\simeq \mathcal{O}_F\otimes T_{C,\rho(F)}$, and the Calabi-Yau condition $\Lambda^2 N_{F/X}\simeq K_F\simeq \mathcal{O}_F(-2)$, we conclude that
\begin{equation}
    N_{F/X}\simeq\mathcal{O}_F\oplus \mathcal{O}_F(-2),
\end{equation}
so that $H^1(N_{F/X})\simeq\mathbb{C}$ is nonzero and there is potentially an obstruction to deforming $F$.  

Now we consider all fibers simultaneously.  Let $\mathcal{F}\subset S\times C$ be the graph of $\rho$. The projection $\pi_C:\mathcal{F}\to C$ identifies $\mathcal{F}$ with the family parametrized by $C$ of fibers of $\rho$, so that for $p\in C$, $\pi_C^{-1}(p)$ is identified with the fiber $F_p$ of $\rho$ over $p$.  Then the obstruction map (\ref{eqn:obstruction}) globalizes to a map
\begin{equation}
    R:H^1(X,TX) \to H^0(C,R^1{\pi_C}_*N_{\mathcal{F}/S\times C}).
    \label{eqn:globalobstruction}
\end{equation}
In (\ref{eqn:globalobstruction}),  the obstruction bundle $R^1{\pi_C}_*N_{\mathcal{F}/S\times C}$ is a line bundle on $C$ whose fiber at $p\in C$ is canonically identified with the 1-dimensional vector space $H^1(F_p,N_{F_p/X})$. So the fibers which deform along with the deformation $\sigma$ are precisely those parametrized by the zeros of the global section $R(\sigma)$ of the obstruction bundle.  Our next task is to identify the obstruction bundle with the canonical bundle $K_C$ of $C$.  

Returning to the study of a single fiber $F$, we have $H^0(F,N_{F/X})\simeq H^0(F,\mathcal{O}_F\oplus \mathcal{O}_F(-2))\simeq \mathbb{C}$.  By Serre duality we have 
\begin{equation}
    H^1(F,N_{F/X})^*\simeq H^0(F,K_F\otimes N_{F/X}^*)\simeq H^0(F,N_{F/X})^*,
    \label{eqn:fiberwiseduality}
\end{equation}
the last isomorphism arising from the Calabi-Yau condition.  

The key point is that these duality isomorphisms are canonical, so extend to the duality of line bundles
\begin{equation}
 R^1{\pi_C}_*N_{\mathcal{F}/S\times C}\simeq \left(R^0{\pi_C}_*N_{\mathcal{F}/S\times C}
 \right)^*,
 \label{eqn:globalduality}
\end{equation}
a special case of Grothendieck duality.  

Taking global sections of the canonical isomorphisms $N_{F/S}\simeq \mathcal{O}_F\otimes T_{C,\rho(F)}$, we get canonical isomorphisms $H^0(F,N_{F/S})\simeq T_{C,\rho(F)}$, which globalizes to the isomorphism 
\[
R^0{\pi_C}_*N_{\mathcal{F}/S\times C}\simeq TC
\]
of line bundles on $C$.  Then (\ref{eqn:globalduality}) implies that the obstuction bundle is the canonical bundle of $C$, as claimed.

Since $K_C$ has no global sections if $g=0$, it follows that the obstruction section $R(\sigma)$ is identically zero in that case, so all fibers deform.  Rephrasing:

\smallskip\noindent
\emph{If a Calabi-Yau threefold $X$ contains a Hirzebruch surface $S$, then $S$ deforms holomorphically along with any complex structure deformation of $X$.}

\smallskip
If $g>0$, we assume in addition that the fiber class $\beta$ is an edge of the Mori cone of $X$. Recalling that $K_C$ has degree $2g-2$, we will show below that for the generic complex structure deformation $\sigma$, the global section $R(\sigma)$ has $2g-2$ isolated zeroes.  Then the deformed $M_\beta$ consists of $2g-2$ isolated points, each parametrizing a $\mathbb{P}^1$.  So the BPS spectrum changes to $(2g-2)[0,0]$.  As a check, $\mathrm{Tr}(-1)^{F_R}$ is unchanged before and after the deformation.

Using a holomorphic 3 form to identify $TX\simeq\Omega^2_X$, the obstruction map (\ref{eqn:globalobstruction}) can be rewritten as a map $R:H^{2,1}(X)\to H^{1,0}(C)$.  It suffices show that this map is surjective, as a generic holomorphic differential on $C$ has $2g-2$ isolated zeros.  The surjectivity of $R$ is equivalent to the surjectivity of $R+\overline{R}:H^{2,1}(X)\oplus H^{1,2}(X)\to H^1(C)$, which is equivalent to the surjectivity of the map $H^3(X)\to H^1(C)$ obtained by precomposing $R+\overline{R}$ with a projection.  We show surjectivity by showing injectivity of its dual $A:H_1(C)\to H_3(X)$.

The map
$A:H_1(C)\to H_3(X)$ takes a 1-cycle $\gamma$ in $C$ to the 3-cycle $\rho^{-1}(\gamma)$ in $X$. Since $\beta$ is assumed to be an edge of the Mori cone, the ruled surface $S$ can be contracted to a singular Calabi-Yau $\underline{X}$ containing a curve $C$ of transverse $A_1$ singularities, and then $X$ can be recovered as the blowup of $\underline{X}$ along its singular curve $C$.  A Mayer-Vietoris calculation of $H_3(X)$ then shows that $A$ is injective.  

We can also give physical explanation of our assertion about the behavior of the curves in the fiber class under a generic deformation of complex structure.  The contraction to $\underline{X}$ gives rise to an $\mathrm{SU}(2)$ gauge theory with $g$ adjoints.  Partially Higgsing to $U(1)$, we are left with $g$ neutral hypermultiplets and $2g-2$ charged hypermultiplets.  The $g$ neutral hypermultiplets correspond to complex structure deformations $\underline{X}_t$ of $\underline{X}$, and the charged hypermultiplets indicate that $\underline{X}_t$ contains $2g-2$ conifolds for a generic deformation instead of a curve of $A_1$ singularities.  Performing small resolutions, we deduce a complex structure deformation of $X$ with $2g-2$ $\mathbb{P}^1$'s in the class $\beta$.

\smallskip
In the rank 3 cases, the elliptically fibered CY $X$ has complex structure deformations of the above type which are not embeddable in the toric variety $\mathbb{P}$. These deformations provide the first explicit example of a compact elliptically fibered Calabi-Yau for which the refined BPS numbers change under complex structure deformations, yet satisfy the same holomorphic anomaly equations with different boundary conditions.

The intersection curve $C=D_1\cap D_4$ has genus $g=10-3n$ in each of the cases $n=1,2,3$, and parametrizes the $\mathbb{P}^1$ fibers of $D_4$.  So $D_4$ does not survive a generic complex structure deformation but we are left with $2g-2=18-6n$ $\mathbb{P}^1$'s in the original fiber class, and the spins of the class $l^{(5)}$ change from $[0,1/2]\oplus(20-6n)[0,0]$ to $(18-6n)[0,0]$.  

We can solve the refined holomorphic anomaly equation with these new boundary conditions and find that some other refined numbers necessarily change.  We next compute new refined BPS numbers after the deformation, and find agreement.

The key point is that the reducible surface $D_1\cup D_4$ does not survive a generic complex structure deformation, but deforms to an irreducible surface $D_1'$.  
This is because the $I_0^*$ divisor before the deformation was $2D_1+D_2+D_4$, remembering that the central component of an $I_0^*$ configuration has multiplicity 2.  Since $D_1$ and $D_2$ are ruled over curves of genus 0, they both survive the complex structure deformation, and the total divisor over the gauge curve is of the form $D_1+D_2+D_1'$ for some surface $D_1'$.  Comparing the divisors over the gauge curve before and after the deformation, we see that $D_1'$ is in the same class as $D_1+D_4$.  Since $D_4$ does not survive the deformation, we conclude that $D_1'$ is irreducible and is a deformation of $D_1\cup D_4$.

The divisor class $J_2$ restricted to $D_1'$  defines a projection to $\mathbb{P}^1$, with generic fiber $\mathbb{P}^1$ by adjunction or deformation invariance of the arithmetic genus.  These fibers have degree $(1,0,0,0,3)$, the sum of the degrees of the fibers of $D_1$ and three fibers of the original $D_4$. Computing intersections, we see that $D_1\cap D_1'$ is a section of $D_1$  and $D_1'\cap D_2$ is a section of $D_2$.  Thus $D_1\cup D_2\cup D_1'$ is an $I_3$ configuration.  

We have already seen that there are $2g-2$ points of $C$ over which we have a curve $M$ of degree $(0,0,0,0,1)$.  Since $D_1'\cdot l^{(5)}=(D_1+D_4)\cdot l^{(4)}=-1$ in all cases, these curves are contained in $D_1'$.  In other words, we have $2g-2$ fibers of $D_1'$ which split into a pair $M\cup M'$ of intersecting $\mathbb{P}^1$s, with $M$ of degree $(0,0,0,0,1)$, and $M'$ of degree $(1,0,0,0,2)$.  Each of the curves $M$ intersect $D_1$ in one point and is disjoint from $D_2$.  Each of the curves $M'$ intersect $D_2$ in one point and is disjoint from $D_1$.

From the above considerations, we deduce that each of the curve classes in the following table have BPS representation $(18-6n)[0,0]$, in agreement with our solution of the refined holomorphic anomaly equations with the new boundary conditions, recorded in \cite{link}. 

\begin{equation}
    \begin{array}{|c|c|} \hline
    {\mathrm{Classes}}& {\mathrm{Spins}}\\ \hline
 (0,0,0,0,1); (1,0,0,0,2); (1,0,0,0,1); (1,1,0,0,1)&(18-6n)[0,0]\\ 
 (1,1,1,0,1); (1,1,0,0,2); (2,1,0,0,2); (1,1,1,0,2); (2,1,1,0,2)   & \\ \hline
    \end{array}
\label{eqn:RGVN3}
\end{equation}

The classes $(0,0,0,0,1)$ and $(1,0,0,0,2)$ have already been identified as being associated with the $18-6n$ curves $M$ and $M'$ respectively.  Any of the curves $M$ can be glued to a fiber of $D_1$, and then to a fiber of $D_2$, and finally to a fiber of $D_3$, producing $18-6n$ rational curves in each of the classes $(1,0,0,0,1),\ (1,1,0,0,1)$, and $(1,1,1,0,1)$.  Similarly, any of the curves $M'$ can be glued to a fiber of $D_2$, and then to a fiber of $D_1$ and/or $D_3$, producing $18-6n$ rational curves in each of the classes $(1,1,0,0,2),\ (2,1,0,0,2),\ (1,1,1,0,2)$, and $(2,1,1,0,2)$.  All of these spin representations have changed from the $[0,1/2]\oplus(20-6n)[0,0]$ which we had in each case for the same curve class in the toric hypersurface.

\section{Holomorphic anomaly equation: genus zero case}\label{sec:5}
We can hope to use the BCOV holomorphic anomaly equation in the compact elliptic Calabi-Yau geometries to derive the holomorphic anomaly equation for the fiber parameter, similar to the case of base parameters in \cite{Huang:2015sta, Huang:2020dbh}. The refinement for the fiber case may also help to improve the refined BCOV equation for the compact case, which was only partially successful  in \cite{Huang:2020dbh}. 

As a first step, we study the genus zero anomaly equation from the Picard-Fuchs differential equations of the compact elliptic $\mathbb{P}^2$ model.  The Picard-Fuchs equations are 
\be\ba \label{PFP2}
& \mathcal{L}_1 = \theta_1(\theta_1-3\theta_2) -12 z_1 (6\theta_1+1)(6\theta_1+5), \\
& \mathcal{L}_2 = \theta_2^3 +z_2 \prod_{i=0}^2 (3\theta_2 -\theta_1 +i) ,
\ea\ee
where $\theta_i :=z_i \frac{\partial}{\partial z_i}$ and $z_1, z_2$ are the fiber and base complex structure parameters. Suppose the power series and logarithmic solutions are $\omega_0\sim 1, \omega_1\sim \log(z_1),  \omega_2\sim \log(z_2)$.  Then the fiber and base Kahler parameters are identified via the mirror map as $t_1=\frac{\omega_1}{\omega_0}$ and  $t_2=\frac{\omega_2}{\omega_0}$, respectively. 

The instanton part of the free energy of the compact model is expanded as 
\be
\mathcal{F} (Q, q; \epsilon_1, \epsilon_2) = \sum_{d_E=0}^{\infty}  \mathcal{F}_{d_E} (Q; \epsilon_1, \epsilon_2) (q Q^{\frac{1}{3}})^{d_E}, 
\ee
where $q= e^{t_1}, Q=e^{t_2}$ are the exponentiated fiber and Kahler parameters, and $\mathcal{F}_{0} (Q; \epsilon_1, \epsilon_2)$ is simply the free energy of the local $\mathbb{P}^2$ model. The free energy has a further genus expansion 
\be 
 \mathcal{F}_{d_E} (Q; \epsilon_1, \epsilon_2)  =\sum_{n,g=0}^{\infty} (\epsilon_1+\epsilon_2)^{2n} (\epsilon_1\epsilon_2)^{g-1} \mathcal{F}^{(n,g)}_{d_E} (Q) . 
 \ee
 It is proposed that $\mathcal{F}^{(n,g)}_{d_E}$ can be written as a polynomial in $S$ with rational functions of $\tilde{z}^{\frac{1}{3}}$ as coefficients, where $S$ and $\tilde{z}$ are the propagator and complex structure parameters of the local $\mathbb{P}^2$ model, and can be expanded in terms of the exponentiated Kahler parameter $Q$. We should note that $\tilde{z}$ differs from the base complex structure parameter $z_2$ of the compact model, though they agree in the large fiber limit $z_1\sim 0$. Then the following holomorphic anomaly equation is satisfied 
 \be 
 \frac{\partial}{\partial S}  \mathcal{F}_{d_E}^{(0,0)} =\frac{1}{2} \sum_{d_E^{\prime} =1}^{d_E-1} \partial_{\tilde{z}}  \mathcal{F}_{d^{\prime}_E}^{(0,0)}  \cdot  \partial_{\tilde{z}}  \mathcal{F}_{d_E- d^{\prime}_E}^{(0,0)} , 
 \ee
 with the initial condition $ \mathcal{F}_1^{(0,0)} =540 \tilde{z}^{-\frac{1}{3}}$. 
 
 We should derive some low order  formulas for $\mathcal{F}_{d_E}^{(0,0)}$ from the compact Picard-Fuchs equations.  Up to the first few orders in $z_1$, we denote the solutions 
 \be \ba \label{PFsolutions}
& \omega_0 =1+ 60 z_1 +13860 z_1^2 + 4084080(1 + 6z_2)z_1^3 +\mathcal{O}(z_1^4)  ,  \\
& \omega_1 =w_0 [ \log(z_1) + \sum_{n=0}^{\infty} b_n(z_2) z_1^n ], \\
& \omega_2 =w_0 [ \log(z_2) + \sum_{n=0}^{\infty} c_n(z_2) z_1^n ].
 \ea\ee
The coefficient of $z_1^n$ in $\omega_0$ is a polynomial in $z_2$ of degree $[\frac{n}{3}]$, but for the logarithmic solutions, $b_n(z), c_n(z)$'s are power series with $b_0(z)\sim z, c_0(z) \sim z$. Putting the ansatz for $\omega_1, \omega_2$ into the Picard-Fuchs equations, we can derive differential equations for $b_n(z)$'s, $c_n(z)$'s.  For example, the first equation in (\ref{PFP2}) gives a linear relation between $c_0(z), c_1(z), c_1^{\prime}(z)$. The second equation in (\ref{PFP2}) gives a third order linear differential equation for $c_0(z)$, as well as a linear relation between $c_0(z), c_1(z)$ and their derivatives up to the third order.  The differential equation for $c_0(z)$ is 
 \be   \label{relationc0}
 z^2 (1+27z) c_0^{\prime\prime\prime}(z) + 3z (1+36z) c_0^{\prime\prime}(z) + (1+60z) c_0^{\prime}(z)  +6=0. 
 \ee 
 Using the above equation, we can eliminate the derivatives of $c_1(z)$ and find the relation
 \be   \label{relationc1}
 c_1(z) = 180 (1 + 27 z) [1 + 4 z c_0^{\prime}(z) +    3 z^2 c_0^{\prime\prime}(z)] .
 \ee
 Similarly, working up to the order $z_1^3$ we find the equations for $c_2(z), c_3(z)$ as 
 \be\ba  \label{relationc2}
 c_2(z) &= 135 (1 + 27 z) [382 + 1759 z c_0^{\prime} (z)+ 
   1377 z^2 c_0^{\prime\prime}(z)]  , \\
   c_3(z) & = 120 (1 + 27 z) [140612 + 708517 z c_0^{\prime} (z)  + 
   567905 z^2 c_0^{\prime\prime} (z) ]. 
 \ea\ee
 
 The differential relations for $b_i(z)$'s can be similarly derived. By comparing the difference we can check that
 \be\ba    \label{relationbc}
& b_0(z) =-\frac{c_0(z)}{3}, ~~~ b_1(z) = 372 -\frac{c_1(z)}{3}, ~~~ b_2(z) = 76122 - \frac{c_2(z)}{3}, \\
 &b_3(z) = 21249376 + 161331216 z - \frac{c_3(z)}{3}  . 
\ea \ee
 This can also be  confirmed by checking that the following combination is a solution of the Picard-Fuchs equation 
 \be \ba \label{combination}
 \omega_1+\frac{1}{3}  \omega_2 & = \omega_0 [  \log(z_1) + \frac{1}{3} \log(z_2) + 372z_1 +76122 z_1^2 \\ & ~~~ + (21249376 + 161331216 z_2 ) z_1^3 + \mathcal{O}(z_1^4) ].  
\ea \ee
 This combination of logarithmic solutions is similar to the power series solution $\omega_0$ in that the coefficient of $z_1^n$ is a polynomial of $z_2$ of degree $[\frac{n}{3}]$. 
 
 The genus zero free energy is determined by the fact that including the classical cubic term,  $\omega_0 \partial_{t_i} \mathcal{F}^{(0,0)}(t_1, t_2)$ is a double logarithmic solution of the Picard-Fuchs equations. For the compact elliptic $\mathbb{P}^2$ model we have 
 \be 
 \mathcal{F}^{(0,0)}_{classical} = \frac{3}{2} t_1^3 + \frac{3}{2} t_1^2t_2 +\frac{1}{2} t_1 t_2^2. 
 \ee
 We can compute the partial derivative 
 \be
 \partial_{t_1}  \mathcal{F}^{(0,0)}_{classical} =\frac{1}{2} (3t_1+t_2)^2,
 \ee
 where the combination (\ref{combination}) appears. So we can compute $\partial_{t_1}  \mathcal{F}^{(0,0)}_{classical}$ exactly in $z_2$ at each perturbative order of $z_1$, without dependence on the $b_n(z_2), c_n(z_2)$ functions.  It is then also straightforward to compute the actions of Picard-Fuchs operators on the classical contribution.  We find 
 \be \ba \label{PFclassical}
 & \mathcal{L}_1 [ \omega_0 \partial_{t_1}  \mathcal{F}^{(0,0)}_{classical}] = -540 z_1 -309420 z_1^2 -152832960 z_1^3 +\mathcal{O}(z_1^4) ,  \\
  &\mathcal{L}_2 [ \omega_0 \partial_{t_1}  \mathcal{F}^{(0,0)}_{classical}] = -227826432 z_2 z_1^3 +\mathcal{O}(z_1^4). 
\ea  \ee
We then consider the instanton contributions. The $d_E =0 $ contribution is eliminated by the partial derivative $\partial_{t_1}$.  Some low order formulas for $d_E\geq 1$ are 
 \be\ba  \label{Fformulas}
 &\mathcal{F}_{1}^{(0,0)} = 540 \tilde{z}^{-\frac{1}{3}} , ~~~  \mathcal{F}_{2}^{(0,0)} = \frac{ 405(80S -37 \tilde{z}^2)} {2\tilde{z}^{\frac{8}{3}}  },\\   &   \mathcal{F}_{3}^{(0,0)} =   \frac{1}{ \tilde{z}^7 (1 + 27  \tilde{z})} [ 324000 S^3 + 486000 S^2  \tilde{z}^2 - 656100 S  \tilde{z}^4 (1 + 37  \tilde{z}) \\ &  ~~~~~~~ + 
 2  \tilde{z}^6 (83305 + 107785707  \tilde{z} + 2864246994  \tilde{z}^2) ].
\ea \ee
These are regarded as functions of the compact Kahler base parameter $t_2$ when taking the partial derivatives $\partial_{t_i}$. 

In order to prove these formulas, we shall rewrite them in terms of the compact complex structure parameters $z_{1,2}$. More specifically, we shall expand perturbatively in the fiber parameter $z_1$ but keep the exact expressions in the base parameter $z_2$. The local $\tilde{z}$ parameter is related to the Kahler parameter by the
 \be \ba
 t_2 &=  \log(z_2) + c_0(z_2) + c_1(z_2) z_1  + c_2(z_2) z_1^2 +\mathcal{O}(z_1^3) , \\
  & = \log(\tilde{z}) + c_0(\tilde{z}) . 
\ea \ee
 We can use an expansion  ansatz   $\tilde{z} = z_2 + \sum_{i=1}^{\infty} a_n(z_2) z_1^n$, and systematically solve for the coefficients $a_n$'s in terms of $c_n$'s. For example, up to the second order we find 
 \be \label{expandz}
  \tilde{z} = z_2  + \frac{z_2 c_1(z_2)}{ 1+ z_2 c_0^{\prime}(z_2)} z_1+
  \frac{ 2 c_2(z_2) (1 + z_2 c_0^{\prime} (z_2) )^2 + 
    c_1(z_2)^2 (1 - z_2^2 c_0^{\prime\prime} (z_2) )} 
{ 2 (1 + z_2 c_0^{\prime} (z_2)  )^3} z_2 z_1^2    + \mathcal{O}(z_1^3) . 
 \ee 
 We use the local propagator $S$  as in the convention of \cite{Huang:2010kf} defined by 
 \be
  \Gamma^{\tilde{z}}_{\tilde{z} \tilde{z}} = -C_{\tilde{z}\tilde{z}\tilde{z}} S -\frac{7+216 \tilde{z}}{6\tilde{z}(1+27\tilde{z})}, 
 \ee
 where the local three point function and the Christoffel symbol are
 \be
 C_{\tilde{z}\tilde{z}\tilde{z}} =  -\frac{1}{3\tilde{z}^3 (1+27\tilde{z})}, ~~~ \Gamma^{\tilde{z}}_{\tilde{z} \tilde{z}} = \frac{\partial^2 t_2}{\partial \tilde{z}^2}  (\frac{\partial t_2}{\partial \tilde{z}} )^{-1} = \frac{-\tilde{z}^{-2}+  c_0^{\prime\prime} (\tilde{z})} {\tilde{z} ^{- 1}+  c_0^{\prime} (\tilde{z}) }. 
 \ee
 We can solve for $S$ in terms of the local complex structure parameter 
 \be  \label{Sformula}
 S = \frac{ \tilde{z}^2 [1 + 54 \tilde{z} + \tilde{z} (7 + 216 \tilde{z}) c_0^{\prime}(\tilde{z}) +    6 \tilde{z}^2 (1 + 27 \tilde{z}) c_0^{\prime\prime}(\tilde{z}) ]}{ 2(1 +   \tilde{z} c_0^{\prime}(\tilde{z}))} .
 \ee
 With the above formulas, we can straightforwardly compute 
 \be \ba \label{checkformulas}
&  ~~ ~~  \partial_{t_1} (\sum_{d_E=1}^3 \mathcal{F}_{d_E}^{(0,0)}  e^{d_E(t_1+\frac{t_2}{3})}  )  \\
& =    \mathcal{F}_{1}^{(0,0)} z_2^{\frac{1}{3}} z_1e^{372z_1++76122z_1^2} + 2 \mathcal{F}_{2}^{(0,0)}  z_2^{\frac{2}{3}} z_1^2  e^{744z_1} + 3 \mathcal{F}_{3}^{(0,0)}  z_2z_1^3  + \mathcal{O}(z_1^4)  \\
& = 540 z_1 + 169695 z_1^2 + (58866000 + 687019032 z_2) z_1^3 +\mathcal{O}(z_1^4) 
 \ea \ee
 where we need to use the relations (\ref{relationc1}, \ref{relationc2}) as well as the differential equation (\ref{relationc0}), since the higher derivatives of $c_0(z_2)$ appear when we expand the local coordinate $\tilde{z}$ in terms of the global coordinate $z_2$ in (\ref{expandz}). Then we find the function $c_0(z_2)$ and its derivatives cancel out in the result.  Multiplying by the factor $\omega_0$ and acting with the Picard-Fuchs operators, we find that they exactly cancel  the classical contributions (\ref{PFclassical}).  So we confirm that the formulas (\ref{Fformulas}) provide the correct solution of the Picard-Fuchs equation at this order $\mathcal{O}(z_1^4)$. 
 
A better approach which would be more convenient for a derivation of the holomorphic anomaly equation is to expand $z_2$ in terms of the local coordinate $\tilde{z}$ by making an ansatz  $z_2  = \tilde{z} + \sum_{i=1}^{\infty} \tilde{a}_n(\tilde{z} ) z_1^n$ and solving for the coefficients. This is the inverse of the expansion (\ref{expandz}). For example to the first order, one finds 
\be
z_2 =  \tilde{z} -\frac{  \tilde{z} c_1( \tilde{z}) }{ 1 +  \tilde{z} c_0^{\prime} ( \tilde{z}) }z_1 +\mathcal{O}(z_1^2) . 
\ee
Using the formulas (\ref{relationc0}, \ref{relationc1}, \ref{relationc2}), we can convert the coefficients into rational functions of $\tilde{z}, c_0^{\prime} ( \tilde{z}), c_0^{\prime\prime} ( \tilde{z})$. Furthermore, we find that the derivative functions $c_0^{\prime} ( \tilde{z}), c_0^{\prime\prime} ( \tilde{z})$ precisely combine into the local propagator $S$ in (\ref{Sformula}). So the coefficients can be written as polynomials of $S$ with rational functions of $\tilde{z}$ as their coefficients. For the first few orders we have 
\be \ba
z_2 &= \tilde{z}  - \frac{ 90 (2 S +  \tilde{z}^2)}{  \tilde{z} } z_1    - 135 [80 S^3 - 360 S^2  \tilde{z}^2 (1 + 36  \tilde{z}) + 2 S  \tilde{z}^4 (329 + 8073  \tilde{z}) \\
& +      \tilde{z}^6 (275 + 2997  \tilde{z} - 112266  \tilde{z}^2 )] /[ 2  \tilde{z}^5 (1 + 27  \tilde{z} ) ] z_1^2  +\mathcal{O}(z_1^3).
\ea \ee
So from the power series of the double logarithmic solution, one can work in the reverse direction of (\ref{checkformulas}), and expand $z_2$ in the appropriate form. It is not difficult to see that the coefficients $\tilde{z}^{\frac{d_E}{3}} \mathcal{F}_{d_E}^{(0,0)}$ are polynomials in $S$ with rational functions of $\tilde{z}$ as their coefficients. 

 \subsection{Some recursion relations}
 We provide some useful formulas for the coefficients in the solutions of Picard-Fuchs equations in (\ref{PFsolutions}). For the power series solution $\omega_0 =\sum_{m,n=0}^{\infty} a_{m,n} z_1^m z_2^n$, it is not difficult to derive the recursion relations from the Picard-Fuchs equations and find a closed formula  for the coefficients
 \be 
 a_{m,n} = \frac{12^m}{ m! (m-3n)! n!^3 } \prod_{k=1}^m (6k-1)(6k-5), 
 \ee
 which is valid for $ m\geq 3n \geq 0$ and understood to vanish otherwise. 
 
 We then consider the mirror map $t_2 =\frac{\omega_2}{\omega_0} $. It appears from the low order solutions (\ref{relationc1}, \ref{relationc2})  that the coefficients for $n\geq 1$ have the following structure 
 \be 
 c_n(z) = (1+27z) [ \alpha_n + \beta_n z c_0^{\prime} (z) +\gamma_n z^2 c_0^{\prime\prime} (z) ], 
 \ee
 where $\alpha_n, \beta_n, \gamma_n$ are constants. While we are not aware of a simple proof of the structure or a closed formulas for the constants, we can derive a recursion relation for the constant coefficients assuming it to be true. Using the first Picard-Fuchs operator and the fact $ \mathcal{L}_1 (\omega_0 )=0$, we find 
 \be \ba \label{L1t2}
 \mathcal{L}_1 (\omega_0 t_2 ) &= (\theta_1 \omega_0) (\theta_1 -3\theta_2) t_2 + (\theta_1 t_2) (\theta_1 -3\theta_2 )\omega_0 
 +\omega_0 \theta_1(\theta_1-3\theta_2) t_2 \\ &~~~ -432 z_1 [ (\theta_1 t_2) (2\theta_1+1) \omega_0 +\omega_0 \theta_1^2 t_2 ] \\
 &= -3 \sum_{m,n} m a_{m,n} z_1^mz_2^n [1 + z_2 c_0^{\prime} (z_2) ]+ \sum_{m,n} \sum_{l=1}^{\infty} a_{m,n}  z_1^{m+l} z_2^n \{  -3(m+l) z_2 c_l^{\prime}(z_2) \\ &~~~ + l  c_l(z_2)   [2m+l-3n -432(2m+l+1) z_1 ]    \} . 
\ea \ee
Using the relation  (\ref{relationc0}), this can be reduced to an expression depending only on $c_0^{\prime} ( z_2), c_0^{\prime\prime} ( z_2)$. Setting $ \mathcal{L}_1 (\omega_0 t_2 )=0$, and assuming the algebraic independence of $z, c_0^{\prime} ( z), c_0^{\prime\prime} ( z)$, we can obtain many relations which would overdetermine the constant coefficients $\alpha_n, \beta_n, \gamma_n$'s. 

We can consider some simple choices. The coefficient of $z_1^m$ in the equation (\ref{L1t2}) is 
\be 
-3ma_{m,0} +\sum_{k=1}^m k\alpha_k [(2m-k) a_{m-k,0} -432 (2m-k-1) a_{m-k-1,0} ] ,
\ee
 whose vanishings can determine all constants $\alpha_k$'s recursively.  
 
 The coefficient of $z_1^m z_2$ is 
 \be \ba
 & -3ma_{m,1} + \sum_{k=1}^m \{ k\alpha_k [(2m-k-3) a_{m-k,1} +27 (2m-k) a_{m-k,0}  \\ & -432(2m-k-1) (a_{m-k-1,1} +27 a_{m-k-1,0} ) ] -3m a_{m-k,0} (27\alpha_k -6\gamma_k) \}, 
\ea \ee
  whose vanishings can further determine all constants $\gamma_k$'s recursively.  
 
 Finally, the coefficient of $z_1^m z_2c_0^{\prime} (z_2)$ is 
 \be
 -3ma_{m,0} +\sum_{k=1}^m \{ k\beta_k [(2m-k) a_{m-k,0} -432 (2m-k-1) a_{m-k-1,0} ] -3ma_{m-k,0} (\beta_k -\gamma_k) \},
 \ee
   whose vanishings can further determine all constants $\beta_k$'s recursively. 

\section*{Acknowledgements}
SK thanks Davesh Maulik and Yukinobu Toda for discussions about GV invariants, and Hee-Cheol Kim for discussions about his related work. XW thanks Kimyeong Lee, Shuai Guo, Longting Wu and Jiahua Tian for collaborations on related subjects. The research of SK is partially supported by NSF grant DMS-2201203. MH is supported by the National Natural Science Foundation of China Grants No.12325502. MH and XW are supported by the National Natural Science Foundation of China Grants No.12247103. AK is supported by an international Leverhulme Professorship at Sheffield. XW is grateful to Max Planck Institute for Mathematics in Bonn for its hospitality and financial support during the completion of part of this project.

\appendix

\section{Perturbative terms for six-dimensional \texorpdfstring{
$\mathcal{N}=(1,0)$}{Neq(1,0)} supergravities}\label{app:A}

In this appendix, we derive the leading terms of the genus zero and genus one refined free energies based on the analysis of the anomaly polynomial for 6d $\mathcal{N}=(1,0)$ supergravity theories. First, we review the basic concepts of 6d $\mathcal{N}=(1,0)$ supergravity theories in Appendix \ref{app:A1}. Then, we derive the leading behavior of the genus-zero and genus-one free energies in Appendix \ref{app:A2}. Finally, in Appendix \ref{subsec:analcont}, we provide the analytic continuation for the refined Gopakumar-Vafa expansion, which is relevant to our calculations.
\subsection{A brief review of 6d \texorpdfstring{$\mathcal{N}=(1,0)$}{Neq(1,0)} supergravity theory}\label{app:A1}
A six-dimensional $\mathcal{N}=(1,0)$ supergravity theory has an associated unimodular lattice $\Gamma$ with pairing $\Omega$ of signature $(1,T)$ corresponding to the charge lattice of one self-dual and $T$ anti-self-dual 2-form fields $B_{\alpha}=\{B_0,B_{i=1,\cdots,T}\}$ in the supergravity and tensor multiplets respectively. The VEV of the scalars in these multiplets are parameterized by a vector $\mathcal{J}\in \Gamma\otimes\mathbb{R}$ which has unit length under $\Omega$
\begin{align}
    \Omega_{\alpha\beta}\mathcal{J}^{\alpha}\mathcal{J}^{\beta}=\mathcal{J}\cdot \mathcal{J}=1.
\end{align}
We consider 6d $(1,0)$ supergravity theories whose gauge algebras $\mathfrak{g}=\bigoplus_i \mathfrak{g}_i$ consist of a product of simple Lie algebras $\mathfrak{g}_i$ without $\mathfrak{u}(1)$ factors, with hypermultiplet spectrum
\begin{align}
    \bigoplus n_{\mathbf{R}}\times\mathbf{R}=\bigoplus_{i} n_{\mathbf{R}_i}\times\mathbf{R}_i\,\,\bigoplus_{i,j} n_{\mathbf{R}_i,\mathbf{R}_j}\times(\mathbf{R}_i,\mathbf{R}_j)
\end{align}
Here $n_{\mathbf{R}_i}$ and $n_{\mathbf{R}_i,\mathbf{R}_j}$ are the numbers of matter fields that transform in the irreducible representation $\mathbf{R}_i$ of $\mathfrak{g}_i$ and $(\mathbf{R}_i,\mathbf{R}_j)$ of $\mathfrak{g}_i\times \mathfrak{g}_j$ respectively. 

The massless spectrum of the 6d theory consists of a gravity multiplet, $T$ tensor multiplets, $H$  hypermultiplets and $V$ vector multiplets.
Their values and the matter contents are constrained by the Green-Schwarz mechanism \cite{Green:1984sg,Sagnotti:1992qw,Sadov:1996zm}, which requires the factorization of the 8-form anomaly polynomial, derived in \cite{Alvarez-Gaume:1983ihn,Schwarz:1995zw} from gauge anomalies and the gravitation anomaly, as
\begin{equation}\label{eq:anomaly6d}
    I_{\text{GS}}=\frac{1}{2}\Omega_{\alpha\beta}X_{4}^{\alpha}\wedge X_{4}^{\beta},
\end{equation}
with the notations $\Omega^{\alpha\beta}=\Omega_{\alpha\beta}^{-1}, v^{\alpha}=\Omega^{\alpha\beta}v_{\beta}$ and the 4-forms $X_{4,\alpha}$ read
\begin{equation}\label{eq:X4_1}
    X_{4}^{\alpha}=\frac{a^{\alpha}}{4}\mathrm{tr}\,R^2+\sum_i b_{i}^{\alpha}\mathrm{tr}\,F_i^2\,.
\end{equation}
Here $\mathrm{tr}\,=\frac{1}{2h_{\mathfrak{g}}^{\vee}}\mathrm{tr}_{\mathbf{Adj}_{\mathfrak{g}}}\,$ is the normalized trace, $F_i$ are the field strengths of the non-Abelian gauge symmetries, $R$ is the spacetime curvature 2-form and $a,b_i$ are anomaly coefficients.
Then the anomaly cancellation condition \eqref{eq:anomaly6d} gives \cite{Sadov:1996zm}
\begin{align}\label{eq:anomalycon1}
    H-V=273-29 T,\qquad & 0=\sum_{\mathbf{R}_i}n_{\mathbf{R}_i} B_{\mathbf{R}_i}-B_{\mathbf{Adj}_{\mathfrak{g}_i}},\\
    a\cdot a=9-T,\qquad & a\cdot b_i =\frac{1}{6}\left(\sum_{\mathbf{R}_i}n_{\mathbf{R}_i} A_{\mathbf{R}_{i}}-A_{\mathbf{Adj}_{\mathfrak{g}_i}}\right),\label{eq:anomalycon2}
\end{align}
and
\begin{align}\label{eq:anomalycon3}
     b_i\cdot b_i =\frac{1}{3}\left(\sum_{\mathbf{R}_i}n_{\mathbf{R}_i} C_{\mathbf{R}_i}- C_{\mathbf{Adj}_{\mathfrak{g}_i}}\right),\quad \quad b_i\cdot b_j =\sum_{\mathbf{R}_i,\mathbf{R}_j}n_{\mathbf{R}_i,\mathbf{R}_j} A_{\mathbf{R}_i} A_{\mathbf{R}_j},(i\neq j),
\end{align}
where $A_{\mathbf{R}}=\mathrm{index}({\mathbf{R}}),B_{\mathbf{R}},C_{\mathbf{R}}$ are group theory coefficients defined through
\begin{align}
    \mathrm{tr}_{\mathbf{R}}\,F^2=\,\,&A_{\mathbf{R}}\mathrm{tr}\,F^2,\\
    \mathrm{tr}_{\mathbf{R}}\,F^4=\,\,&B_{\mathbf{R}}\mathrm{tr}\,F^4+C_{\mathbf{R}}(\mathrm{tr}\,F^2)^2,
\end{align}
which can be found in \cite[Table 2]{Grassi:2011hq}. In particular, $A_{\mathbf{Adj}}=2h_{\mathfrak{g}}^{\vee}$ is 2 times the dual Coxeter number for $\mathfrak{g}$. For those numbers that are relevant to our paper, we list them in Table \ref{tab:group_coeff}.
\begin{table}[H]
\centering
\begin{tabular}{|c|c|c|c|c|}
\hline
$\mathfrak{g}$ & $\mathbf{R}$  & $A_{\mathbf{R}}$  & $B_{\mathbf{R}}$ & $C_{\mathbf{R}}$\\ \hline
$\mathfrak{su}(N),$ & $\mathbf{Adj}$ &$2N$&$0$&$N+6$\\ \cline{2-5}
$N=2,3$                   & $\mathbf{F}$   &$1$&$0$&$\frac{1}{2}$\\ \hline
$\mathfrak{su}(N),$ & $\mathbf{Adj}$ &$2N$&$2N$&$6$\\ \cline{2-5}
$N\geq 4$                   & $\mathbf{F}$   &$1$&$1$&$0$\\ \hline
\multirow{2}{*}{$\mathfrak{g}_2$} & $\mathbf{Adj}$ &$8$&$0$&$10$\\ \cline{2-5}
                   & $\mathbf{7}$   &$2$&$0$&$1$\\ \hline
\end{tabular}
\caption{Group theory coefficients.}
\label{tab:group_coeff}
\end{table}
When the anomaly cancellation conditions \eqref{eq:anomalycon1},\eqref{eq:anomalycon2},\eqref{eq:anomalycon3} are satisfied, the anomaly is cancelled by adding to the action a Green-Schwarz term
\begin{align}
    S_{\text{GS}}=\int_{M_6} \Omega_{\alpha\beta}B_{2}^{\alpha}\wedge X_{4}^{\beta}\,.
\end{align}
Now we consider the 6d theory coming from F-theory compactification on a compact Calabi-Yau threefold, which is an elliptic fibration over a complex surface $B$. The fiber is a smooth elliptic cuvre over a generic point on the base $B$ and it degenerates over curves $\Sigma_i$ in $B$. The anomaly coefficients $a,b_i$ have a clear geometric correspondence. For instance, the coefficients $a_{\alpha}$ are the decomposition coefficients of the anti-canonical class of the base into a basis $\{h_{\alpha}\}$ for $h^{1,1}(B)$ as \cite{Morrison:1996na}
\begin{align}
    -K_B=\sum_{\alpha}a^{\alpha}h_{\alpha},\qquad \Sigma_{i}=\sum_{\alpha}b_{i}^{\alpha} h_{\alpha}.
\end{align}
So we can perform the following identifications 
\begin{align}
   a\cdot a= K_B\cdot K_B=9-n, \qquad a\cdot b_i= -K_B\cdot \Sigma_i,\qquad b_i\cdot b_j= \Sigma_i\cdot \Sigma_j\,.
\end{align}
We also have
\begin{align}
    T=h^{1,1}(B)-1\,.
\end{align}
Upon further compactification of the 6d theory on $T^2$, the information from the 4d Coulomb branch indicates that the rank of the vector multiplet $\mathrm{rank}(V)$ and the number of neutral hypermultiplets which are uncharged with respect to the Cartan of $V$ are \cite{Morrison:1996pp}
\begin{align}\label{eq:constraint_3}
    \mathrm{rank}(V)=h^{1,1}(X)-h^{1,1}(B)-1\,, \qquad H_{\text{neutral}}=h^{2,1}(X)+1\,.
\end{align}

\subsection{The leading behaviour of  the genus zero and genus one free energies}\label{app:A2}
In this section, we compute the tree-level and 1-loop free energies
\begin{equation}\label{eq:defE}
    \mathcal{E}=\mathcal{E}_{\text{tree}}+\mathcal{E}_{\text{1-loop}}
\end{equation}
for 6d $\mathcal{N}=(1,0)$ theories on a circle $S^1$. The tree-level contribution $\mathcal{E}_{\text{tree}}$ comes from the 6d $(1,0)$ effective action 
\begin{equation}\label{eq:6d_action}
    S=\int_{M_6} \frac{1}{2}g_{\alpha\beta}G^{\alpha}\wedge *G^{\beta}+\Omega_{\alpha\beta}B^{\alpha}\wedge X_4^{\beta},
\end{equation}
where $g_{\alpha\beta}$ is the 6d spacetime metric, $G^{\alpha}$ is the field strength for $B^{\alpha}$ and the 4-form $X_4^{\beta}$ is modified by adding a term $c_2(\mathcal{R})$, which is the second Chern class of the background $SU(2)_{\mathcal{R}}$ $\mathcal{R}$-symmetry bundle, to cancel the mixed anomalies with the $SU(2)_{\mathcal{R}}$ $\mathcal{R}$-symmetry. Up to undetermined coefficients $y^{\alpha}$, we propose
\begin{equation}
    X_{4}^{\alpha}=-\frac{a^{\alpha}}{4}p_1(M_6)+\sum_i b_{i}^{\alpha}\mathrm{tr}\,F_i^2+y^{\alpha}c_2(\mathcal{R}),
\end{equation}
where $p_1(M_6)=-\frac{1}{2}\mathrm{tr}\,R^2$ is the first Pontryagin class of the tangent bundle of the six dimensional spacetime $M_6$.

Under circle compactification on $S^1$, using the replacement rules proposed in \cite{DelZotto:2018tcj,DelZotto:2017mee}
\begin{equation}\label{eq:replacementrule}
    p_1(M_6)\mapsto \epsilon_1^2+\epsilon_2^2,\qquad \mathrm{tr}F_i^2\mapsto \sum_{i^{\prime},j^{\prime}} K_i^{i^{\prime}j^{\prime}} \phi_{i^{\prime}}\phi_{j^{\prime}},\qquad c_2(\mathcal{R})\mapsto -\epsilon_+^2,
\end{equation}
the action \eqref{eq:6d_action} contributes
\begin{align}\label{eq:E_tree}
    \mathcal{E}_{\text{tree}}=\frac{\tau}{2}\Omega_{\alpha\beta}^{-1}t_{\text{b}_{\alpha}}t_{\text{b}_{\beta}}+ t_{\text{b}_{
    \alpha}}\left(-\frac{1}{4}a^{\alpha}(\epsilon_1^2+\epsilon_2^2)+\sum_i b_{i}^{\alpha}\sum_{i^{\prime},j^{\prime}} K_i^{i^{\prime}j^{\prime}} \phi_{i^{\prime}}\phi_{j^{\prime}}-y^{\alpha}\epsilon_+^2\right),
\end{align}
to the refined free energies. Here $\tau$ is the inverse for the radius of the compactified circle $S^1$, $t_{\text{b}_{\alpha}}$ are the tensor parameters which are related to the K\"ahler parameters $t_{B,\alpha}$ for the base $B$ via $t_{B,\alpha}=t_{\text{b}_{\alpha}}-\frac{a_{\alpha}}{2}$. $K_i$ is the Killing form for the Lie algebra $\mathfrak{g}_i$ satisfying 
\begin{equation}
\sum_{i^{\prime},j^{\prime}} K_i^{i^{\prime}j^{\prime}} \phi_{i^{\prime}}\phi_{j^{\prime}}=\frac{1}{2h_{\mathfrak{g}_i}^{\vee}}\sum_{\alpha\in\Delta^+}(\alpha\cdot t)^2=\frac{1}{2A_{\mathbf{R}}}\sum_{w\in\mathbf{R}}(w\cdot t)^2.
\end{equation}
The 1-loop contribution contains regularization from the 6d $(1,0)$ multiplets
\begin{align}\label{eq:1loopE}
    \mathcal{E}_{\text{1-loop}}=\mathcal{E}_{\text{vec}}+\mathcal{E}_{\text{hyper}}+\mathcal{E}_{\text{tensor}}+\mathcal{E}_{\text{grav}}\,.
\end{align}
To determine \eqref{eq:1loopE}, we first study the 1-loop contributions of these supermultiplets to the BPS spectrum
\begin{align}\label{eq:defZ1loop}
    Z_{\text{1-loop}}=Z_{\text{vec}}\,Z_{\text{hyper}}\,Z_{\text{hyper},\text{neutral}}\,Z_{\text{tensor}}\,Z_{\text{grav}}.
\end{align}
The vector multiplet contribution is
\begin{align}
    Z_{\text{vec}}=\mathrm{PE}\left[\frac{-1-q_1q_2}{(1-q_1)(1-q_2)}\sum_i\left(\sum_{\alpha\in\Delta^+_i}e^{\alpha\cdot t}+\sum_{n=1}^{\infty}\sum_{\alpha \in \text{root}_i}e^{\alpha\cdot t+n\tau}\right)\right],
\end{align}
where $\Delta^+_i$ and $\text{root}_i$ are the positive roots and roots for the gauge algebra $\mathfrak{g}_i$.
The hypermultiplet contribution for the charged matter fields is 
\begin{align}
    Z_{\text{hyper}}=\mathrm{PE}\left[\frac{\sqrt{q_1q_2}}{(1-q_1)(1-q_2)}\sum_{\mathbf{R}}2n_{\mathbf{R}}\left(\sum_{w\in\mathbf{R}^+}e^{w\cdot t}+\sum_{n=1}^{\infty}\sum_{\substack{w\in\mathbf{R}\\ w\neq 0}}e^{w\cdot t+n\tau}\right)\right],
\end{align}
where we have used $\mathbf{R}^+$ and $\mathbf{R}$ for the positive weights and weights in the representation $\mathbf{R}$. For the neutral hypermultiplets
\begin{align}
    Z_{\text{hyper},\text{neutral}}=\mathrm{PE}\left[\frac{\sqrt{q_1q_2}}{(1-q_1)(1-q_2)}H_{\text{neutral}}\sum_{n=1}^{\infty}e^{n\tau}\right].
\end{align}
For the tensor multiplets and gravity multiplet, the contribution are
\begin{align}
    Z_{\text{tensor}}=\mathrm{PE}\left[-\frac{\sqrt{q_1q_2}}{(1-q_1)(1-q_2)}T\sum_{n=1}^{\infty}(e^{\epsilon_-}+e^{-\epsilon_-})e^{n\tau}\right]
\end{align}
and
\begin{align}
    Z_{\text{grav}}=\mathrm{PE}\left[-\frac{\sqrt{q_1q_2}}{(1-q_1)(1-q_2)}\sum_{n=1}^{\infty}(e^{\epsilon_-}+e^{-\epsilon_-})(1+e^{\epsilon_1+\epsilon_2}+e^{-\epsilon_1-\epsilon_2})e^{n\tau}\right]
\end{align}
respectively.
The classical part of the genus zero and genus one free energies from these 1-loop contributions can be obtained by adding addition terms according to \eqref{eq:completion} and utilize the Zeta regularization\footnote{The Zeta regularization regularize the divergent summation, we have
\begin{align}
    \sum_{n=1}^{\infty}n^{s}=\zeta(-s)=-\frac{B_{s+1}(1)}{s+1},\quad s\in\mathbb{Z}_{+}\,,
\end{align}
where $B_s(x)$ are the Bernoulli numbers defined from the expansion
\begin{align}
    \frac{t\, e^{xt}}{e^t-1}=\sum_{s=0}^{\infty} B_s(x)\frac{t^n}{n!},
\end{align}
in particular, $\zeta(-s)=-\frac{1}{2},-\frac{1}{12},0,\frac{1}{120}$ for $s=0,1,2,3$.} to regularize the infinite summation. 
\begin{align}\label{eq:F6d_00}
    \mathcal{E}_{\text{vec}}=&\frac{1}{12}\sum_i\left(\sum_{\alpha \in \Delta^+_i}(\alpha\cdot t\pm\epsilon_+)^3+\sum_{n=1}^{\infty}\sum_{\alpha \in \text{root}_i}(\alpha\cdot t+n\tau\pm\epsilon_+)^3\right)\nonumber\\
    &\qquad\qquad\qquad\qquad\qquad-\frac{\epsilon_1^2+\epsilon_2^2}{24}\sum_i\left(\sum_{\alpha \in \Delta^+_i}\alpha\cdot t+\sum_{n=1}^{\infty}\sum_{\alpha \in \text{root}_i}(\alpha\cdot t+n\tau)\right)\nonumber\\
    =&\sum_i\left(\frac{1}{6}\sum_{\alpha \in \Delta^+_i}(\alpha\cdot t)^3-\frac{1}{12}\tau\sum_{\alpha \in \Delta^+_i}(\alpha\cdot t)^2+\frac{\epsilon_+^2}{2}\sum_{\alpha \in \Delta^+_i}\alpha\cdot t-\frac{\epsilon_1^2+\epsilon_2^2}{24}\sum_{\alpha \in \Delta^+_i}\alpha\cdot t\right)\nonumber\\
    &\qquad\qquad\qquad\qquad\qquad\qquad\qquad+\frac{V}{720}\tau^3-\frac{V}{24}\epsilon_+^2\tau-\frac{\epsilon_1^2+\epsilon_2^2}{24}\left(-\frac{V}{12}\tau\right)
\end{align}

\begin{align}\label{eq:F6d_01}
    \mathcal{E}_{\text{hyper}}=&-\frac{1}{12}\left(\sum_{\mathbf{R}}2n_{\mathbf{R}}\left(\sum_{w \in \mathbf{R}^+}(w\cdot t)^3+\sum_{n=1}^{\infty}\sum_{w \in \mathbf{R}^+}(\pm w\cdot t+n\tau)^3\right)+2H_{\text{neutral}}\sum_{n=1}^{\infty}(n\tau)^3\right)\nonumber\\
    &+\frac{\epsilon_1^2+\epsilon_2^2}{48}\left(\sum_{\mathbf{R}}2n_{\mathbf{R}}\sum_{w \in \mathbf{R}^+}w\cdot t+2H\sum_{n=1}^{\infty}n\tau\right)\nonumber\\
    =&-\frac{1}{12}\sum_{\mathbf{R}}n_{\mathbf{R}}\left(2\sum_{w \in \mathbf{R}^+}(w\cdot t)^3-\tau\sum_{w \in \mathbf{R}^+}(w\cdot t)^2\right)-\frac{H}{720}\tau^3\nonumber\\
    &+\frac{\epsilon_1^2+\epsilon_2^2}{48}\left(\sum_{\mathbf{R}}2n_{\mathbf{R}}\sum_{w \in \mathbf{R}^+}w\cdot t-\frac{H}{6}\tau\right)
\end{align}
\begin{align}\label{eq:F6d_10}
    \mathcal{E}_{\text{tensor}}=\frac{T}{12}\left(\sum_{n=1}^{\infty}(n\tau\pm\epsilon_-)^3\right)-\frac{T}{24}(\epsilon_1^2+\epsilon_2^2)\sum_{n=1}^{\infty}n\tau=\frac{T}{720}\tau^3-\frac{T}{24}\epsilon_-^2\tau+\frac{T}{288}(\epsilon_1^2+\epsilon_2^2)\tau
\end{align}
\begin{align}
    \mathcal{E}_{\text{grav}}=\,\,&\frac{1}{12}\left(\sum_{n=1}^{\infty}(n\tau+\epsilon_{1}+\epsilon_2\pm\epsilon_-)^3+(n\tau-\epsilon_{1}-\epsilon_2\pm\epsilon_-)^3+(n\tau\pm\epsilon_-)^3\right)-\frac{1}{8}(\epsilon_1^2+\epsilon_2^2)\sum_{n=1}^{\infty}n\tau\nonumber\\
    =&\frac{1}{240}\tau^3-\frac{5\tau}{48}(\epsilon_1^2+\epsilon_2^2+\epsilon_1\epsilon_2).
\end{align}
By collecting all these contributions, we have
\begin{align}
    \mathcal{E}=\mathcal{E}_{\text{tree}}+\mathcal{E}_{\text{1-loop}}=\mathcal{F}^{(0,0)}+(\epsilon_1+\epsilon_2)^2\mathcal{F}^{(1,0)}+\epsilon_1\epsilon_2\mathcal{F}^{(0,1)}
\end{align}
where
\begin{align}
    \mathcal{F}^{(0,0)}=&-\frac{1}{6}\left(\sum_i\sum_{\alpha\in\Delta^+_i}(\alpha\cdot t)^3-\sum_{\mathbf{R}}{n_{\mathbf{R}}}\sum_{\omega\in \mathbf{R}^{+}}(\omega\cdot t)^3\right)+\frac{9-T}{24}\tau^3\nonumber\\
    &+\sum_i\left(t_{\text{b}_\alpha}b^{\alpha}_i-\frac{a\cdot b_i}{2}\tau\right)\frac{1}{2h^{\vee}_{\mathfrak{g}_i}}\sum_{\alpha\in\Delta^+_i}(\alpha\cdot t)^2+\frac{1}{2}\Omega_{\alpha\beta}^{-1}t_{\text{b}_{\alpha}}t_{\text{b}_{\beta}}\tau\,,
\end{align}
and
\begin{align}
    \mathcal{F}^{(0,1)}=\,\,&-\frac{1}{12}\sum_i\sum_{\alpha\in\Delta^+_i}\alpha\cdot t+\sum_{\mathbf{R}}\frac{n_{\mathbf{R}}}{12}\sum_{\omega\in \mathbf{R}^{+}}\omega\cdot t-(2-\frac{T}{6})\tau-\frac{1}{2}a^{\alpha}t_{\text{b}_{\alpha}}\,,\\
    \mathcal{F}^{(1,0)}=\,\,&-\frac{1}{12}\sum_i\sum_{\alpha\in\Delta^+_i}\alpha\cdot t-\sum_{\mathbf{R}}\frac{n_{\mathbf{R}}}{24}\sum_{\omega\in \mathbf{R}^{+}}\omega\cdot t+\frac{\tau}{96}(101+V-9T)+\frac{1}{4}(a^{\alpha}+y^{\beta})t_{\text{b}_{\alpha}}\,.
\end{align}

\subsubsection{Elliptic \texorpdfstring{$\mathbb{P}^2$}{P2}}
In this section, as an example, we compute the leading contributions to the genus zero and genus one free energies for elliptic $\mathbb{P}^2$. The calculations for the other models studied in this paper are straightforward and are left to the reader.

The 6d supergravity theory corresponds to elliptic $\mathbb{P}^2$ has the number of supermultiplets:
\begin{align}
    H=273,\qquad V=0,\qquad T=0.
\end{align}
Let $h$ denote the hyperplane class of $\mathbb{P}^2$, which has self-intersection $h^2=1$.  The canonical class is $-K_B=3h$. So that $\Omega=1$ and $a=3$. From these data, we obtain the genus zero and genus one free energies
\begin{align}
   \mathcal{F}^{(0,0)}=&\frac{1}{2}\tau t_{\mathrm{b}}^2+\frac{3}{8}\tau^3=\frac{1}{2}(3t_1^3+3t_1^2t_2+t_1t_2^2)\\
   \mathcal{F}^{(0,1)}=&-2\tau-\frac{3}{2}t_{\mathrm{b}}=-\frac{1}{4}(17t_1+3t_2),\\
   \mathcal{F}^{(1,0)}=& \frac{101}{96}\tau+\frac{1}{4}(3+y)t_{\mathrm{b}}=\frac{1}{96}(209+36y)t_1+\frac{1}{4}(3+y)t_2,
\end{align}
where $\tau,t_{\mathrm{b}}$ are related to the K\"ahler parameters via
\begin{align}
    t_1=\tau,\quad t_2=t_{\mathrm{b}}-\frac{3}{2}\tau.
\end{align}
In the limit to local $\mathbb{P}^2$, $t_1,t_2$ are mapping to the local threefold K\"ahler parameter via
\begin{align}
    t_1=\hat{m}-\frac{1}{3}t,\quad t_2=t.
\end{align}
Therefore we have
\begin{align}
   \mathcal{F}^{(0,0)}=&\frac{3}{2}\hat{m}^3-\frac{1}{18}t^3\\
   \mathcal{F}^{(0,1)}=&-\frac{17}{4}\hat{m}-\frac{1}{12}t,\\
   \mathcal{F}^{(1,0)}=& \frac{1}{96}(209+36y)\hat{m}+\frac{1}{288}(7+36y)t.
\end{align}
From the calculation of local $\mathbb{P}^2$, the coefficient of $t$ in the genus $(1,0)$ free energy is $-\frac{1}{24}$, from which we fix the value of $y=-\frac{19}{36}$. 

\subsection{Analytic continuation}
\label{subsec:analcont}
In the refined Gopakumar-Vafa expansion of the partition function, the most important ingredient is \cite{Lockhart:2012vp}
\begin{align}\label{eq:app_S3}
    S_3(t;\epsilon_1,\epsilon_2)=&\exp\left(B_{3,3}(t;\epsilon_1,\epsilon_2)+\sum_{n=1}^{\infty}\frac{e^{-n\,t}}{n\sinh(n\epsilon_1/2)\sinh(n\epsilon_2/2)}\right)\nonumber\\
    &=\exp\left(B_{3,3}(t;\epsilon_1,\epsilon_2)+\int_{\mathbb{R}+\ri 0}\frac{ds}{s}\frac{e^{-2s\, t-2\pi\ri(2b+1)s}}{8\sinh(2\pi\ri l)\sinh(s\epsilon_1)\sinh(s\epsilon_2)}\right),
\end{align}
where $b\in\mathbb{Z}$ and the second line in \eqref{eq:app_S3} is the integral representation of the function $S_3(z)$, it recovers the expression in the first line by performing the contour integral over the right half-plane by excluding the origin.  
Consider the analytic continuation of the function $S_3(t;\epsilon_1,\epsilon_2)$ which correspond to change the contour integral to the other side of the plane, it gives
\begin{align}
     S_3(t;\epsilon_1,\epsilon_2)&=\exp\Bigg(B_{3,3}(t;\epsilon_1,\epsilon_2)+\sum_{n=1}^{\infty}\frac{e^{n\,t}}{n\sinh(n\epsilon_1/2)\sinh(n\epsilon_2/2)}\nonumber\\
     &+\frac{1}{\epsilon_1\epsilon_2}\left(\frac{1}{6}(t+\pi\ri(2 b+1))^3+\frac{1}{24}(4\pi^2-\epsilon_1^2-\epsilon_2^2)(t+\pi\ri(2 b+1))\right)\Bigg).
\end{align}
We expect the analytic continuation gives $S_3(-t;-\epsilon_1,-\epsilon_2)=S_3(t;\epsilon_1,\epsilon_2)$, with a further assumption that $\epsilon_1\epsilon_2 B_{3,3}(t;\epsilon_1,\epsilon_2)$ is a degree 3 homogeneous polynomial for $t,\epsilon_1,\epsilon_2$, we find
\begin{align}\label{eq:completion}
    B_{3,3}(t):=B_{3,3}(t;\epsilon_1,\epsilon_2)=-\frac{1}{12\epsilon_1\epsilon_2}(\tilde{t}\,{}^3+\pi^2\tilde{t})+\frac{\epsilon_1^2+\epsilon_2^2}{48\epsilon_1\epsilon_2}\tilde{t},\quad \tilde{t}=t+\pi\ri(2b+1).
\end{align}

\section{Formulas from the gauge theory approach}
\label{app:Formulas}
In this appendix, we compute expectation values of Wilson loops for 5d $\mathcal{N}=1$ theories on the Omega-deformed background $\mathbb{R}_{\epsilon_1,\epsilon_2}^4\times S^1$ that are relevant for our discussion.
\subsection{\texorpdfstring{5d $SU(N)_{\kappa}$ theories}{5dSU(N)}}
The instanton partition function for the 5d $SU(N)_{\kappa}$ theory with Chern-Simons level $\kappa$ is calculated using the localization method within the ADHM construction of the instanton moduli space, which has been extensively studied in the literature, e.g. \cite{Nekrasov:2002qd,Tachikawa:2004ur,Iqbal:2007ii,Taki:2007dh,Nakajima:2009qjc,Hwang:2014uwa}, both with and without the Omega-deformation on $\mathbb{R}^4$. For the $SU(2)$ theory, the parameter $\kappa$ can be understood as the theta angle of the theory, with $\theta = \kappa , \pi$. For $\kappa\leq N$, the resulting partition function is 
\begin{align}
    Z(t,m;\epsilon_1,\epsilon_2)=e^{F_{\text{poly}}}Z_{\text{pert}}\,Z_{\text{inst}},
\end{align}
where $Z_{\text{inst}}$ is the instanton contribution
\begin{align}
Z_{\text{inst}}=\sum_{\mu}(-1)^{\kappa\,|\mu|}\,\mathfrak{q}^{|\mu|}\,Z_{\mu},\qquad
 Z_{\mu}\,=\,\prod_{i=1}^{N}\frac{ Q_i^{\kappa |\mu_i|}q_1^{-\frac{\kappa}{2}\,{||\mu_i||^2}+\frac{1}{2}}q_2^{-\frac{\kappa}{2}\,{||\mu_i^t||^2}+\frac{1}{2}}}{\prod_{j=1}^N\mathcal{N}_{\mu_i\mu_j}(Q_{ij};q_1,q_2)}
\end{align}
where $q_{1,2}=e^{ \epsilon_{1,2}}$, $\epsilon_+=\frac{1}{2}(\epsilon_1+\epsilon_2)$, $Q_i=e^{\alpha_i},\,Q_{ij}=e^{\alpha_i-\alpha_j}$, $\mathfrak{q}$ is the instanton counting parameter and $\alpha_i$ are the Coulomb parameters for the $SU(N)$ theory with the constraint $\sum_{i}\alpha_i=0$. 
The summation in the instanton contribution $Z_{\text{pert}}$ is over all Young tableau $\mu_i,i=1,\cdots,N$, for each $\mu_i$, it is a partition $\mu_i=\{\mu_{i,m}\}$ and
\begin{align}
    \mathcal{N}_{\mu_i\mu_j}(Q;q_1,q_2)=\prod_{(m,n)\in\mu_i}\left(1-Qq_1^{-\mu_{i,m}+n}
    {q_2^{\mu_{j,n}^t-m+1}}\right)
    \cdot\prod_{(m,n)\in\mu_j}\left(1-Qq_1^{\mu_{j,m}-n+1}q_2^{-\mu_{i,n}^t+m}\right).
\end{align}
We also use the notation  $|\mu_i|=\sum_{(m,n)\in\mu_i}1=\sum_m \mu_{i,m},\,||\mu_i||=\sum_{m}\mu_{i,m}^2$ ,$|\mu|=\sum_{i=1}^N |\mu_i|$ and $\mu^t_i$ is the transpose partition of $\mu_i$.
$Z_{\text{pert}}$ comes from the perturbative or one-loop contribution:
\begin{align}
    Z_{\text{pert}}=\prod_{m,n=0}^{\infty}\prod_{i<j}(1-Q_{ij}q_1^{m}q_2^{n})(1-Q_{ij}q_1^{m+1}q_2^{n+1}).
\end{align}
$F_{\text{poly}}$ obtains contributions from the classical and one-loop part, it has the expression  
\begin{align}\label{eq:FpolySUN}
    \epsilon_1\epsilon_2\, F_{\text{poly}}=-\frac{1}{6}\sum_{i<j}(\alpha_i-\alpha_j)^3-\frac{\kappa}{6}\sum_{i}\alpha_i^3-&\frac{\log\mathfrak{q}}{2N}\sum_{i<j}(\alpha_i-\alpha_j)^2\nonumber\\
    &-\frac{1}{12}((\epsilon_1+\epsilon_2)^2+   \epsilon_1\epsilon_2)\sum_{i<j}(\alpha_i-\alpha_j)
\end{align}

The localization method also applies to the calculations for the vacuum expectation values of operators. For instance, in the KK reduction of the 5d theory to 4d, one can consider the insertion of Chiral operators \cite{Gorsky:1998rp,Losev:2003py,Nakajima:2003uh}, which are exactly the reduction of Wilson loop operators in 5d.

The VEVs for the Wilson loops can be obtained with the insertion of equivariant Chern characters. Define the equivariant Chern character for the universal bundle of the instanton moduli space \cite{Losev:2003py}
\begin{align}
    \mathrm{Ch}_{\mathbf{F},\mu}(Q_i;q_1,q_2)=\sum_{i=1}^NQ_i-\mathcal{I}\,\sum_{i=1}^N \sum_{(m,n)\in \mu_i} Q_iq_1^{-m+\frac{1}{2}}q_2^{-n+\frac{1}{2}}
\end{align}
where $\mathcal{I}=(1-q_1)(1-q_2)q_1^{-1/2}q_2^{-1/2}$ is the factor defined in \eqref{eq:Ifactor}.  Denote $\mathbf{R}_i$ the $i$-th fundamental representation, the equivariant Chern characters for those representations can be obtained from the tensor power of $\mathrm{Ch}_{\mathbf{F},\mu}$, for instance
\begin{equation}\begin{split}
    \mathrm{Ch}_{\mathbf{\Lambda}^2,\mu}(Q_i;q_1,q_2)&=\frac{1}{2}\left(\mathrm{Ch}_{\mathbf{F},\mu}(Q_i;q_1,q_2)^2-\mathrm{Ch}_{\mathbf{F},\mu}(Q_i^2;q_1^2,q_2^2)\right),\\
    \mathrm{Ch}_{\mathbf{\overline{F}},\mu}(Q_i;q_1,q_2)&=\mathrm{Ch}_{\mathbf{F},\mu}(Q_i^{-1};q_1^{-1},q_2^{-1}),
\end{split}\end{equation}
where we have denoted $\mathbf{F}$ as the fundamental representation, $\mathbf{\Lambda}^2$ as the anti-symmetric representation and $\mathbf{\overline{F}}$ as the anti-fundamental representation, their highest weights are $[1,0,\cdots,0]$, $[0,1,0\cdots,0]$ and $[0,\cdots,0,1]$ respectively. The equivariant Chern characters for $\mathbf{R}=\mathbf{R}_1^{\otimes k_1}\otimes\cdots \otimes \mathbf{R}_{N-1}^{\otimes k_{N-1}}$ is
\begin{align}
    \mathrm{Ch}_{\mathbf{R},\mu}=\prod_{i=1}^{N-1}\mathrm{Ch}_{\mathbf{R_i},\mu}^{ k_i}
\end{align}
whose highest weight is $[k_{1},\cdots,k_{N-1}]$. Utilize the highest weight to denote the representation, the Wilson loop expectation value for the representation $\mathbf{R}$ is
\begin{align}
    \left\langle W_{[k_{1},\cdots,k_{N-1}]}^{SU(N)_{\kappa}}\right\rangle=\frac{1}{Z_{\text{inst}}}\,\sum_{\mu}(-1)^{\kappa\,|\mu|}\,\mathfrak{q}^{|\mu|}\,W_{\mathbf{R},\mu}\,Z_{\mu},
\end{align}
where we have defined
\begin{align}\label{eq:WRmu}
    W_{\mathbf{R},\mu}=\mathrm{Ch}_{\mathbf{R},\mu}+\cdots
\end{align}
The $\cdots$ in \eqref{eq:WRmu} are the correction terms which generally exist if the asymptotic freedom condition on the instanton moduli space is not satisfied when we add matters in the representation $\mathbf{R}$. A physical guess for these corrections leads to the ansatz
\begin{align}\label{eq:CHRprime}
    \sum_{k>0}\sum_{\mathbf{R}^{\prime}}f_{\mathbf{R}^{\prime},k}(q_1,q_2)\mathfrak{q}^k\mathrm{Ch}_{\mathbf{R}^{\prime},\mu},
\end{align}
which can be fixed from the requirement of the BPS expansion \eqref{eq:Zgen} that  the curve classes in the expansions intersect with at least one compact surface.\footnote{This is equivalent to there is no single mass parameter term in the expansion.} The summation over the representations $\mathbf{R}^{\prime}$ and $k$ are finite due to the positivity and integrability of the BPS expansion \eqref{eq:BPSsector}. In Appendix \ref{app:Wilson_SU2}, we will fix the correction terms for $SU(2)_{0,\pi}$ theories, but for $SU(N)$ theories with higher ranks, we will ignore these corrections for simplicity as they do not affect our calculations.

\subsection{\texorpdfstring{$SU(2)$ theories with theta angle $0$ or $\pi$}{SU2with zerotheta}}\label{app:Wilson_SU2}
In this section, we provide the explicit expressions of the \eqref{eq:WRmu} for 5d $SU(2)_{0}$ and $SU(2)_{\pi}$ theories based on the description of the last section. 

\paragraph{\underline{$SU(2)_0$}}
\begin{align*}
   & W_{[1],\mu}=\mathrm{Ch}_{1,\mu}\,,\\
   & W_{[2],\mu}=\mathrm{Ch}_{2,\mu}+q_+^{-1}\,\mathcal{I}\,\mathfrak{q}\,,\\
   & W_{[3],\mu}=\mathrm{Ch}_{3,\mu}+q_+^{-1}\,\mathcal{I}(2+q_1^{-1}+q_2^{-1}-q_1^{-1}q_2^{-1})\,\mathfrak{q}\,\mathrm{Ch}_{1,\mu}\,,\\
   & W_{[4],\mu}=\mathrm{Ch}_{4,\mu}+q_+^{-1}\,\mathcal{I}\,(3+q_1^{-2}+q_2^{-2}+q_1^{-2}q_2^{-2}+2q_1^{-1}+2q_2^{-1}-2q_1^{-1}q_2^{-2}-2q_1^{-2}q_2^{-1})\,\mathfrak{q}\,\mathrm{Ch}_{2,\mu}\nonumber\\
   &\qquad\qquad+\mathcal{I}^3q_+^{-1}(1-q_1^{-1}q_2^{-1})\,\mathfrak{q}+\mathcal{I}^2q_+^{-2}(2+q_1^{-1}+q_2^{-1}+q_1^{-2}q_2^{-2}-q_1^{-1}q_2^{-2}-q_1^{-2}q_2^{-1})\,\mathfrak{q}^2\,.
\end{align*}

\paragraph{\underline{$SU(2)_{\pi}$}}
\begin{equation*}\begin{split}
    W_{[1],\mu}=\,&\mathrm{Ch}_{1,\mu}+q_+^{-1}\mathfrak{q}\,,\\
    W_{[2],\mu}=\,&\mathrm{Ch}_{2,\mu}+(1+q_1^{-1}+q_2^{-1}-q_1^{-1}q_2^{-1})\mathfrak{q}\mathrm{Ch}_{2,\mu}+q_+^{-1}\mathfrak{q}^2\,,\\
    W_{[3],\mu}=\,&\mathrm{Ch}_{3,\mu}+q_+^{-5}((-1+q_2)^2+q_1 (-2+q_2+q_2^2)+q_1^2 (1+q_2+q_2^2))\mathfrak{q}\mathrm{Ch}_{2,\mu}-\mathcal{I}^2 q_+^{-3}(1-q_1q_2)\mathfrak{q}^3\\
    &\quad+q_+^{-3}\mathfrak{q}+q_+^{-8}(q_2 (q_2^2+q_2+1) q_1^3+(q_2^3-1) q_1^2+(q_2^3-3 q_2+2) q_1-(q_2-1)^2)\mathfrak{q}^2\mathrm{Ch}_{1,\mu}\,.
\end{split}\end{equation*}

\subsection{\texorpdfstring{$E_8$ del Pezzo surface}{E8delpezzo}}
The local $E_8$ del Pezzo surface $dP_8$ is related to 5d $SU(2)$ theory with 7 fundamental hypermultiplets. The expectation value of the Wilson loop in the fundamental representation was computed in \cite{Wang:2023zcb}, which can be obtained from the refined partition function of E-strings \cite{Kim:2022dbr}. 

The geometry for the massless E-string theory is the elliptic fibration over a $-1$ curve. Denote $Z_{\text{E-str}}(Q_b,Q_f;q_1,q_2)$ the partition function for massless E-strings, where $Q_b$ and $Q_f$ are the parameters for the base and fiber respectively.  In the limit $Q_f,Q_b^{-1}\rightarrow 0$ while keeping $Q=Q_bQ_f$ finite. Substitute $Q_b=\frac{Q}{Q_f}$ in $Z_{\text{E-str}}$, the expansion of the partition function for E-strings in terms with $Q_f$ gives the Wilson loops for the local $E_8$ theory \footnote{Before perform the expansion, one need change the degree of the BPS invariants at $Q_b=QQ_f^{-1}$ to $Q_b^{-1}=Q^{-1}Q_f$, this corresponds to a flop transition 
in the geometry.},
\begin{align}\label{eq:E-str-expand}
    Z_{\text{E-str}}(Q_b,Q_f;q_1,q_2)=Z_{E_8}(Q;q_1,q_2)\left(1+\sum_k Q_f^k Z_k\right),
\end{align}
where
\begin{align}
    Z_k=P_k(q_1,q_2)\left\langle {W_{[-k]}^{E_8}} \right\rangle+\sum_{l=1}^{k}\widetilde{P}_{k,l}(q_1,q_2)\left\langle {W_{[-k+l]}^{E_8}} \right\rangle.
\end{align}
Utilize the BPS expansion \eqref{eq:BPSsector}, we can fix $\widetilde{P}_{k,l}$. For instance, $\widetilde{P}_{1,0}=\widetilde{P}_{2,1}=0$ and
\begin{align}
    \widetilde{P}_{2,0}=\frac{4125+249\chi_{\frac{1}{2}}(q_-)\chi_{\frac{1}{2}}(q_+)+\chi_1(q_-)\chi_1(q_+)}{(1-q_1^2)(1-q_2^2)q_1^{-1}q_2^{-1}},
\end{align}
where
\begin{align}
    \chi_j(q)=\frac{q^{j+\frac{1}{2}}-q^{-j-\frac{1}{2}}}{q^{\frac{1}{2}}-q^{-\frac{1}{2}}}.
\end{align}
\begin{table}[H]
{\centering
{\centering\tiny\begin{tabular}{|c|ccccc|}
 \hline $2j_L \backslash 2j_R$ & 0 & 1 & 2 & 3 & 4 \\ \hline
 0 & 225252 &  & 250 &  &  \\
 1 &  & 43500 &  & 1 &  \\
 2 & 250 &  & 4623 &  &  \\
 3 &  & 1 &  & 250 &  \\
 4 &  &  &  &  & 1 \\ \hline
 \multicolumn{6}{c}{ {\footnotesize{$d$\,=\,$3$\rule{0pt}{2.6ex}}}}
\end{tabular}}\vskip 10pt

{\centering\tiny\begin{tabular}{|c|ccccccccc|}
 \hline $2j_L \backslash 2j_R$ & 0 & 1 & 2 & 3 & 4 & 5 & 6 & 7 & 8 \\ \hline
 0 &  & 16340118 &  & 295752 &  & 250 &  &  &  \\
 1 & 1477380 &  & 6424374 &  & 48371 &  & 1 &  &  \\
 2 &  & 339002 &  & 1587007 &  & 4874 &  &  &  \\
 3 & 4623 &  & 48620 &  & 300375 &  & 251 &  &  \\
 4 &  & 251 &  & 4874 &  & 43998 &  & 1 &  \\
 5 &  &  & 1 &  & 251 &  & 4624 &  &  \\
 6 &  &  &  &  &  & 1 &  & 250 &  \\
 7 &  &  &  &  &  &  &  &  & 1 \\ \hline
 \multicolumn{10}{c}{ {\footnotesize{$d$\,=\,$4$\rule{0pt}{2.6ex}}}} 
\end{tabular}} \vskip 10pt}

{\hspace{-2.3cm}
{\tiny\begin{tabular}{|c|cccccccccccccc|}
 \hline $2j_L \backslash 2j_R$ & 0 & 1 & 2 & 3 & 4 & 5 & 6 & 7 & 8 & 9 & 10 & 11 & 12 & 13 \\ \hline
 0 & 554613637 &  & 1355489880 &  & 102214234 &  & 1662376 &  & 4624 &  &  &  &  &  \\
 1 &  & 394131111 &  & 855496641 &  & 37819756 &  & 349245 &  & 251 &  &  &  &  \\
 2 & 31317499 &  & 144378490 &  & 344206235 &  & 9607509 &  & 49123 &  & 1 &  &  &  \\
 3 &  & 9606263 &  & 40543262 &  & 111910372 &  & 1998750 &  & 4876 &  &  &  &  \\
 4 & 305245 &  & 2042248 &  & 9687004 &  & 31413876 &  & 349496 &  & 251 &  &  &  \\
 5 &  & 48873 &  & 354368 &  & 1999249 &  & 7733750 &  & 48873 &  & 1 &  &  \\
 6 & 250 &  & 4876 &  & 49124 &  & 349497 &  & 1654380 &  & 4875 &  &  &  \\
 7 &  & 1 &  & 251 &  & 4876 &  & 48873 &  & 300874 &  & 251 &  &  \\
 8 &  &  &  &  & 1 &  & 251 &  & 4875 &  & 43999 &  & 1 &  \\
 9 &  &  &  &  &  &  &  & 1 &  & 251 &  & 4624 &  &  \\
 10 &  &  &  &  &  &  &  &  &  &  & 1 &  & 250 &  \\
 11 &  &  &  &  &  &  &  &  &  &  &  &  &  & 1 \\ \hline
  \multicolumn{15}{c}{ {\footnotesize{$d$\,=\,$5$\rule{0pt}{2.6ex}}}}
\end{tabular}} \vskip 10pt}

    \caption{The refined BPS invariants for the Wilson loop of $dP_8$ in the representation $[2]$ with $d\leq 5$. At $d=1,2$, there is no BPS content.}
    \label{tab:wilsonBPSdP8}
\end{table}

\begin{table}[H]
{\centering
{\tiny\begin{tabular}{|c|cccccccc|}
 \hline $2j_L \backslash 2j_R$ & 0 & 1 & 2 & 3 & 4 & 5 & 6 & 7 \\ \hline
 0 & 22830619 &  & 344374 &  & 251 &  &  &  \\
 1 &  & 8312255 &  & 53496 &  & 1 &  &  \\
 2 & 344374 &  & 1936005 &  & 5126 &  &  &  \\
 3 &  & 53496 &  & 349248 &  & 252 &  &  \\
 4 & 251 &  & 5126 &  & 48873 &  & 1 &  \\
 5 &  & 1 &  & 252 &  & 4875 &  &  \\
 6 &  &  &  &  & 1 &  & 251 &  \\
 7 &  &  &  &  &  &  &  & 1 \\ \hline
   \multicolumn{9}{c}{ {\footnotesize{$d$\,=\,$4$\rule{0pt}{2.6ex}}}}
\end{tabular}} \vskip 10pt}

{\hspace{-1.6cm}
{\centering\tiny\begin{tabular}{|c|ccccccccccccc|}
 \hline $2j_L \backslash 2j_R$ & 0 & 1 & 2 & 3 & 4 & 5 & 6 & 7 & 8 & 9 & 10 & 11 & 12 \\ \hline
 0 &  & 2227558639 &  & 141697614 &  & 2016496 &  & 4875 &  &  &  &  &  \\
 1 & 437321613 &  & 1305731372 &  & 49439639 &  & 403244 &  & 252 &  &  &  &  \\
 2 &  & 186660998 &  & 494415484 &  & 12004630 &  & 54251 &  & 1 &  &  &  \\
 3 & 9986885 &  & 52243140 &  & 152981879 &  & 2402245 &  & 5128 &  &  &  &  \\
 4 &  & 2401492 &  & 12084625 &  & 41151254 &  & 403496 &  & 252 &  &  &  \\
 5 & 49125 &  & 408368 &  & 2402745 &  & 9737877 &  & 54000 &  & 1 &  &  \\
 6 &  & 5127 &  & 54252 &  & 403497 &  & 2004128 &  & 5127 &  &  &  \\
 7 & 1 &  & 252 &  & 5128 &  & 54000 &  & 349748 &  & 252 &  &  \\
 8 &  &  &  & 1 &  & 252 &  & 5127 &  & 48874 &  & 1 &  \\
 9 &  &  &  &  &  &  & 1 &  & 252 &  & 4875 &  &  \\
 10 &  &  &  &  &  &  &  &  &  & 1 &  & 251 &  \\
 11 &  &  &  &  &  &  &  &  &  &  &  &  & 1 \\ \hline
    \multicolumn{14}{c}{ {\footnotesize{$d$\,=\,$5$\rule{0pt}{2.6ex}}}}
\end{tabular}} }

    \caption{The refined BPS invariants for the Wilson loop of $dP_8$ in the representation $[3]$ with $d\leq 5$. At $d=1,2,3$, there is no BPS content.}
    \label{tab:wilsonBPSdP8_2}
\end{table}

\subsection{E-string theory}\label{app:E-string}
The partition function of the six-dimensional E-string theory can be calculated from the elliptic genera of 2d $(0,4)$ $O(k)$ gauge theories on $T^2$, where the matter contents of the 2d theories are obtained from the brane bound states in the IIA description of E-strings \cite{Kim:2014dza}. One can introduce half-BPS codimension 4 defects in the E-string theory, they are constructed by adding additional $n$ parallel $\text{D4}^{\prime}$ branes where the brane configuration can be found in \cite[Table 1]{Chen:2021ivd}.

The expectation values for the Wilson surface in the E-string theory are generated from the partition functions of the E-string theory in the presence of codimension 4 defects. 
It can be expressed as a sum over $k$-string elliptic genus contribution
\begin{align}\label{eq:E-string_Wilson}
    Z_{\text{E-str},[n]}=Q_b^{-n}\left(1+\sum_{k=1}^{\infty}Q_b^k Z_{\text{E-str},[n]}^{(k)}\right),\qquad Q_b=e^{\phi_0},
\end{align}
where $\phi_0$ is the tensor parameter and the $k$-string elliptic genus is
\begin{align}\label{eq:E-str-defect-def}
    Z_{\text{E-str},[n]}^{(k)}=\oint[\mathrm{d}u]Z_{\text{1-loop}}^{O(k)}\cdot \prod_{i=1}^{n}\prod_{\rho\in\text{fund}}\frac{\theta_1(\pm\epsilon_-+\rho(u)+x_i)}{\theta_1(\pm\epsilon_++\rho(u)+x_i)}.
\end{align}
Here $\mathrm{d}[u]=\prod_{I=1}^{r_{\rm cont.}}\frac{\mathrm{d}u_I}{2\pi i}$ is the integral measure, $r_{\rm cont.}$ is the rank for the continuous sector of the $O(k)$ group, and $x_i$ are the positions of $\text{D4}^{\prime}$ branes. $Z_{\text{1-loop}}^{O(k)}$ is the 1-loop determinant of the $k$-string elliptic genus
\begin{align}
    Z_{\text{1-loop}}^{O(k)}=&{\theta_1(2\epsilon_+)}^{r_{\rm cont.}}\prod_{\substack{\alpha\in \text{root},\\\alpha\neq0}}\theta_1(\alpha(u))\theta_1(2\epsilon_++\alpha(u))\cdot \frac{\prod_{\rho \in \text{fund}}\prod_{l=1}^{16}\theta_1(\rho(u)\pm m_l)}{\prod_{\rho\in \text{sym}}\theta_1(\rho(u)+\epsilon_1)\theta_1(\rho(u)+\epsilon_2)}.
\end{align}
In integral \eqref{eq:E-str-defect-def} can be evaluated from the Jeffrey-Kirwan (JK) residues by summing over all the JK poles.    In the following, we compute the one-, two- and three-string elliptic genera.
\paragraph{One string}
The one-string elliptic genus is 
\begin{align}
    Z_{\text{E-str},[n]}^{(1)}=\sum_{I=1}^{4}\frac{\prod_{l=1}^8\theta_I(m_l)}{2\theta_1(\epsilon_1)\theta_1(\epsilon_2)}\prod_{i=1}^n\frac{\theta_I(x_i\pm\epsilon_-)}{\theta_I(x_i\pm\epsilon_+)},
\end{align}
which is contributed from four discrete sectors of the $O(1)$ gauge group. 
\paragraph{Two strings}
The two-string elliptic genus of E-strings is calculated from one continuous sector and six discrete sectors of the $O(2)$ flat connections. The continuous sector contribution is 
\begin{align}
    Z^{(2)}_0=\oint [du] \frac{\theta_1(2\epsilon_+)}{\theta_1(\epsilon_{1,2})\theta_1(\epsilon_{1,2}\pm 2u)} \cdot\prod_{i=1}^n\frac{\theta_1(x_i\pm u\pm \epsilon_-)}{\theta_1(x_i\pm u\pm \epsilon_+)}\cdot\prod_{l=1}^{8}\theta_1(m_l\pm u).
\end{align}
The integral is evaluated from the following JK-poles:
\begin{itemize}
    \item $\epsilon_{1,2}+2u=0,1,\tau,\tau+1$.
        \begin{align}
            Z^{(2)}_1=\frac{1}{2}\left(\sum_{I=1}^4\frac{\prod_{l=1}^8\theta_I(m_l\pm\frac{\epsilon_1}{2})}{\theta_1(\epsilon_{1,2})\theta_1(2\epsilon_1)\theta_1(\epsilon_2-\epsilon_1)}\prod_{i=1}^n\frac{\theta_I(x_i\pm \frac{\epsilon_1}{2}\pm \epsilon_-)}{\theta_I(x_i\pm \frac{\epsilon_1}{2}\pm \epsilon_+)}+(\epsilon_1 \leftrightarrow \epsilon_2)\right).
        \end{align}
    \item $x_i+u\pm \epsilon_+=0,\,i=1,\cdots,n$.
        \begin{align}
            Z^{(2)}_2=\sum_{j=1}^n\Bigg(&\frac{1}{\theta_1(2x_j)\theta_1(2x_j-2\epsilon_+)\theta_1(2x_j-2\epsilon_+-\epsilon_{1,2})}\nonumber\\
            &\cdot\prod_{i\neq j}\frac{\theta_1(x_i+x_j-\epsilon_{1,2})\theta_1(-x_i+x_j+\epsilon_{1,2})}{\theta_1(x_i+x_j)\theta_1(-x_i+x_j)\theta_1(x_i+x_j-\epsilon_{+})\theta_1(-x_i+x_j+\epsilon_{+})}\nonumber\\
            &\quad\qquad\qquad\qquad\qquad\qquad\qquad\qquad\qquad+(\epsilon_1\rightarrow-\epsilon_2,\epsilon_2\rightarrow-\epsilon_1)\Bigg).
        \end{align}
\end{itemize}
The contributions coming from the discrete sectors are
\begin{align}
\begin{aligned}
Z^{(2)}_{I,J} &= \frac{\theta_{\sigma(I,J)}(0) \theta_{\sigma(I,J)}(2\epsilon_+)\prod_{l=1}^{8} \theta_I(m_l) \theta_J(m_l)}{ \theta_1(\epsilon_{1,2})^2 \theta_{\sigma(I,J)}(\epsilon_{1,2}) }\prod_{l=1}^n\frac{\theta_I(x_i\pm\epsilon_-)\theta_J(x_i\pm\epsilon_-)}{\theta_I(x_i\pm\epsilon_+)\theta_J(x_i\pm\epsilon_+)},
\end{aligned}
\end{align}
where $\sigma(I,J)=\sigma(J,I)$ is a symmetric function with
\begin{align}
   \sigma(I,I)=1\,,\quad \sigma(1,J)=J\,,\quad \sigma(2,3)=4\,, \quad \sigma(2,4)=3\,,\quad \sigma(3,4)=2\,.
\end{align}
In total, the 2-string elliptic genus is
\begin{align}
    Z_{\text{E-str},[n]}^{(2)}= \frac{1}{2} \sum_{I=1}^2 Z_{I}^{(2)} + \frac{1}{4} \sum_{I < J}^4 Z^{(2)}_{I,J}.
\end{align}
\paragraph{Three strings}
The three-string elliptic genus of E-strings is calculated from 8 sectors, they are given as follows:
\begin{align}\label{eq:E-str-Z3_I}
    Z^{(3)}_I=\oint [du] &\frac{\theta_1(2\epsilon_+)\theta_I(2\epsilon_+\pm u)\theta_I(\pm u)}{\theta_1(\epsilon_{1,2})^2\theta_I(\epsilon_{1,2}\pm u)\theta_1(\epsilon_{1,2}\pm 2u)} \cdot\prod_{l=1}^{8}\theta_I(m_l)\theta_1(m_l\pm u)\nonumber\\
    &\cdot\prod_{i=1}^n\frac{\theta_I(x_i\pm \epsilon_-)\theta_1(x_i\pm u\pm \epsilon_-)}{\theta_I(x_i\pm \epsilon_+)\theta_1(x_i\pm u\pm \epsilon_+)},
\end{align}
and
\begin{align}
    Z^{(3)}_{I}{}^{\prime}=\frac{\theta_2(0)\theta_3(0)\theta_4(0)\theta_2(2\epsilon_+)\theta_3(2\epsilon_+)\theta_4(2\epsilon_+)}{\theta_{1}(\epsilon_{1,2})^3\theta_2(\epsilon_{1,2})\theta_3(\epsilon_{1,2})\theta_4(\epsilon_{1,2})}\prod_{J\neq I}^4\left(\prod_{i=1}^n\frac{\theta_J(x_i\pm \epsilon_-)}{\theta_J(x_i\pm \epsilon_+)}\cdot\prod_{l=1}^8\theta_J(m_l)\right),
\end{align}
for $I=1,2,3,4$.
The integral \eqref{eq:E-str-Z3_I} is evaluated from the JK-poles:
\begin{itemize}
    \item $\epsilon_{1,2}+2u=0,1,\tau,\tau+1$.
        
    \item $\epsilon_{1,2}+u=\omega_I$, where  $\omega_1=0,\omega_2=\frac{1}{2},\omega_3=\frac{1+\tau}{2},\omega_4=\frac{\tau}{2}$ are the half period for Jacobi theta functions.
    \item $x_i\pm\epsilon_++u=0$.
\end{itemize}
They give
    \begin{align}
            Z^{(3)}_{I}=&\frac{1}{\theta_1(\epsilon_{1,2})^2}\Bigg[\frac{\theta_1(\epsilon_{1,2})\cdot\prod_{l=1}^{8}\theta_I(m_l)\theta_1(m_l\pm \epsilon_1)}{\theta_1(2\epsilon_{1})\theta_1(\epsilon_{2}-\epsilon_1)\theta_1(3\epsilon_{1})\theta_1(\epsilon_{2}-2\epsilon_1)} \prod_{i=1}^n\frac{\theta_I(x_i\pm \epsilon_-)\theta_I(x_i\pm \epsilon_1\pm \epsilon_-)}{\theta_I(x_i\pm \epsilon_+)\theta_I(x_i\pm \epsilon_1\pm \epsilon_+)}\nonumber\\
            &+\frac{1}{2}\sum_{J=1}^4\frac{\theta_{\sigma(I,J)}(\frac{3}{2}\epsilon_1+\epsilon_2)\theta_{\sigma(I,J)}(-\frac{\epsilon_1}{2})\prod_{l=1}^{8}\theta_{I}(m_l)\theta_{J}(m_l\pm\frac{\epsilon_1}{2})}{\theta_1(2\epsilon_1)\theta_1(\epsilon_2-\epsilon_1)\theta_{\sigma(I,J)}(\frac{3}{2}\epsilon_1)\theta_{\sigma(I,J)}(\epsilon_2-\frac{\epsilon_1}{2})}\nonumber\\
            &\qquad\qquad\qquad\qquad\qquad\qquad\cdot\prod_{i=1}^n\frac{\theta_I(x_i\pm \epsilon_-)\theta_J(x_i\pm \frac{\epsilon_1}{2}\pm \epsilon_-)}{\theta_I(x_i\pm \epsilon_+)\theta_J(x_i\pm \frac{\epsilon_1}{2}\pm \epsilon_+)}+(\epsilon_1\leftrightarrow \epsilon_2)\Bigg]\nonumber\\
            &+\Bigg[\frac{\theta_I(x_i+\epsilon_+)\theta_I(x_i-3\epsilon_+)\theta_I( x_i-\epsilon_+)^2\theta_1(2x_i-\epsilon_{1,2})}{\theta_1(\epsilon_{1,2})\theta_I(x_i-\epsilon_+-\epsilon_{1,2})\theta_I(x_i\pm\epsilon_-)\theta_1(2x_i-2\epsilon_+\pm\epsilon_{1,2})\theta_1(2x_i)\theta_1(2x_i-2 \epsilon_+)} \nonumber\\
    &\cdot\prod_{l=1}^{8}\theta_I(m_l)\theta_1(x_i-\epsilon_+\pm m_l )\cdot\prod_{j=1}^n\frac{\theta_I(x_j\pm \epsilon_-)}{\theta_I(x_j\pm \epsilon_+)}\cdot \prod_{j\neq i}^n\frac{\theta_1(x_i-\epsilon_+\pm x_j\pm \epsilon_-)}{\theta_1(x_i-\epsilon_+\pm x_j\pm \epsilon_+)}\nonumber\\
    &\quad\qquad\qquad\qquad\qquad\qquad\qquad\qquad\qquad\qquad\qquad+(\epsilon_1\rightarrow-\epsilon_2,\epsilon_2\rightarrow-\epsilon_1)\Bigg].
        \end{align}
In total, the 3-string elliptic genus is
\begin{align}
    Z_{\text{E-str},[n]}^{(3)}= \frac{1}{4} \sum_{I=1}^4 Z_{I}^{(3)} + \frac{1}{8} \sum_{I=1}^4Z^{(3)}_{I}{}^{\prime}.
\end{align}

\section{Refined BPS numbers at higher degree}\label{app:HigherDegree}
\subsection{\texorpdfstring{The refined BPS numbers for elliptic fibration over $\mathbb{F}_0$}{BPSF0}}
\begin{table}[H]
\centering
{\footnotesize\begin{tabular}{|c|ccc|}
 \hline $2j_L \backslash 2j_R$ & 0 & 1 & 2 \\ \hline
 0 & 488 &  &  \\
 1 & 1 &  & 1 \\ \hline
\multicolumn{4}{c}{$(d_2,d_3)$\,=\,$(0,1),(1,0)$  \rule{0pt}{2.6ex}}\\ 
\end{tabular}}
\hspace{0.5cm}
{\footnotesize\begin{tabular}{|c|ccccc|}
 \hline $2j_L \backslash 2j_R$ & 0 & 1 & 2 & 3 & 4 \\ \hline
 0 &  &  & 488 &  &  \\
 1 & 1 &  & 2 &  & 1 \\ \hline
\multicolumn{6}{c}{$(d_2,d_3)$\,=\,$(1,1)$  \rule{0pt}{2.6ex}}\\  
\end{tabular}}\vskip 6pt

{\footnotesize\begin{tabular}{|c|ccccccc|}
 \hline $2j_L \backslash 2j_R$ & 0 & 1 & 2 & 3 & 4 & 5 & 6 \\ \hline
 0 &  &  &  &  & 488 &  &  \\
 1 &  &  & 1 &  & 2 &  & 1 \\ \hline
\multicolumn{8}{c}{$(d_2,d_3)$\,=\,$(1,2),(2,1)$  \rule{0pt}{2.6ex}}\\ 
\end{tabular}}
\hspace{0.5cm}
{\footnotesize\begin{tabular}{|c|ccccccccc|}
 \hline $2j_L \backslash 2j_R$ & 0 & 1 & 2 & 3 & 4 & 5 & 6 & 7 & 8 \\ \hline
 0 &  &  &  &  &  &  & 488 &  &  \\
 1 &  &  &  &  & 1 &  & 2 &  & 1 \\ \hline
\multicolumn{10}{c}{$(d_2,d_3)$\,=\,$(1,3),(3,1)$  \rule{0pt}{2.6ex}}\\  
\end{tabular}}\vskip 6pt

{\footnotesize\begin{tabular}{|c|cccccccccc|}
 \hline $2j_L \backslash 2j_R$ & 0 & 1 & 2 & 3 & 4 & 5 & 6 & 7 & 8 & 9 \\ \hline
 0 &  &  &  &  & 488 & 1 & 976 & 2 &  & 1 \\
 1 &  &  & 1 &  & 4 &  & 5 & 488 & 2 &  \\
 2 &  &  &  &  &  & 1 &  & 2 &  & 1 \\ \hline
\end{tabular} \vskip 3pt  $(d_2,d_3)=(2,2)$ \vskip 6pt}

{\footnotesize\begin{tabular}{|c|ccccccccccc|}
 \hline $2j_L \backslash 2j_R$ & 0 & 1 & 2 & 3 & 4 & 5 & 6 & 7 & 8 & 9 & 10 \\ \hline
 0 &  &  &  &  &  &  &  &  & 488 &  &  \\
 1 &  &  &  &  &  &  & 1 &  & 2 &  & 1 \\ \hline
\end{tabular} \vskip 3pt  $(d_2,d_3)=(1,4),(4,1)$ \vskip 6pt}

{\footnotesize\begin{tabular}{|c|ccccccccccccc|}
 \hline $2j_L \backslash 2j_R$ & 0 & 1 & 2 & 3 & 4 & 5 & 6 & 7 & 8 & 9 & 10 & 11 & 12 \\ \hline
 0 &  &  &  &  & 488 & 1 & 976 & 4 & 1464 & 5 &  & 2 &  \\
 1 &  &  & 1 &  & 4 &  & 8 & 488 & 9 & 976 & 5 &  & 1 \\
 2 &  &  &  &  &  & 1 &  & 4 &  & 5 & 488 & 2 &  \\
 3 &  &  &  &  &  &  &  &  & 1 &  & 2 &  & 1 \\ \hline
\end{tabular} \vskip 3pt  $(d_2,d_3)=(2,3),(3,2)$ \vskip 6pt}

{\footnotesize\begin{tabular}{|c|ccccccccccccc|}
 \hline $2j_L \backslash 2j_R$ & 0 & 1 & 2 & 3 & 4 & 5 & 6 & 7 & 8 & 9 & 10 & 11 & 12 \\ \hline
 0 &  &  &  &  &  &  &  &  &  &  & 488 &  &  \\
 1 &  &  &  &  &  &  &  &  & 1 &  & 2 &  & 1 \\ \hline
\end{tabular} \vskip 3pt  $(d_2,d_3)=(1,5),(5,1)$ \vskip 6pt}

{\footnotesize\begin{tabular}{|c|cccccccccccccccc|}
 \hline $2j_L \backslash 2j_R$ & 0 & 1 & 2 & 3 & 4 & 5 & 6 & 7 & 8 & 9 & 10 & 11 & 12 & 13 & 14 & 15 \\ \hline
 0 &  &  &  &  & 488 & 1 & 976 & 4 & 1464 & 8 & 1952 & 8 &  & 3 &  &  \\
 1 &  &  & 1 &  & 4 &  & 8 & 488 & 13 & 976 & 15 & 1464 & 9 &  & 2 &  \\
 2 &  &  &  &  &  & 1 &  & 4 &  & 8 & 488 & 9 & 976 & 5 &  & 1 \\
 3 &  &  &  &  &  &  &  &  & 1 &  & 4 &  & 5 & 488 & 2 &  \\
 4 &  &  &  &  &  &  &  &  &  &  &  & 1 &  & 2 &  & 1 \\ \hline
\end{tabular} \vskip 3pt  $(d_2,d_3)=(2,4),(4,2)$ \vskip 6pt}

{\footnotesize\begin{tabular}{|c|ccccccccccccccccc|}
 \hline $2j_L \backslash 2j_R$ & 0 & 1 & 2 & 3 & 4 & 5 & 6 & 7 & 8 & 9 & 10 & 11 & 12 & 13 & 14 & 15 & 16 \\ \hline
 0 &  &  & 488 & 1 & 976 & 5 & 2440 & 13 & 2928 & 21 & 3416 & 19 & 488 & 8 &  & 1 &  \\
 1 & 1 &  & 4 &  & 10 & 488 & 19 & 1464 & 29 & 2928 & 33 & 2928 & 22 & 488 & 6 &  &  \\
 2 &  &  &  & 1 &  & 5 &  & 13 & 488 & 22 & 1464 & 23 & 2440 & 13 &  & 3 &  \\
 3 &  &  &  &  &  &  & 1 &  & 5 &  & 12 & 488 & 14 & 976 & 7 &  & 1 \\
 4 &  &  &  &  &  &  &  &  &  & 1 &  & 4 &  & 5 & 488 & 2 &  \\
 5 &  &  &  &  &  &  &  &  &  &  &  &  & 1 &  & 2 &  & 1 \\ \hline
\end{tabular} \vskip 3pt  $(d_2,d_3)=(3,3)$ }

    \caption{The refined BPS numbers of elliptic $\mathbb{F}_0$ for $d_E=1$, $0<d_2+d_3\leq 6$.}
    \label{tab:BPSF0dE=1}
\end{table}
\begin{table}[H]
\centering
{\footnotesize\begin{tabular}{|c|cccc|}
 \hline $2j_L \backslash 2j_R$ & 0 & 1 & 2 & 3 \\ \hline
 0 & 280964 & 1 &  &  \\
 1 & 1 & 488 & 1 &  \\
 2 &  & 1 &  & 1 \\ \hline
\multicolumn{5}{c}{ $(d_2,d_3)$\,=\,$(0,1),(1,0)$ \rule{0pt}{2.6ex}}\\
\end{tabular}}
\hspace{0.5cm}
{\footnotesize\begin{tabular}{|c|ccc|}
 \hline $2j_L \backslash 2j_R$ & 0 & 1 & 2 \\ \hline
 0 & 488 &  &  \\
 1 & 1 &  & 1 \\ \hline
\multicolumn{4}{c}{ $(d_2,d_3)$\,=\,$(0,2),(2,0)$\rule{0pt}{2.6ex}}\\
\end{tabular}}
\vskip 10pt

{\footnotesize\begin{tabular}{|c|cccccc|}
 \hline $2j_L \backslash 2j_R$ & 0 & 1 & 2 & 3 & 4 & 5 \\ \hline
 0 & 488 & 118832 & 488 & 3 &  &  \\
 1 & 2 & 1464 & 3 & 488 & 1 &  \\
 2 &  & 5 &  & 3 &  & 1 \\ \hline
\end{tabular} \vskip 3pt  $(d_2,d_3)=(1,1)$ \vskip 10pt}

{\footnotesize\begin{tabular}{|c|cccccccc|}
 \hline $2j_L \backslash 2j_R$ & 0 & 1 & 2 & 3 & 4 & 5 & 6 & 7 \\ \hline
 0 &  & 3 & 488 & 118832 & 488 & 3 &  &  \\
 1 & 1 & 488 & 3 & 1464 & 3 & 488 & 1 &  \\
 2 &  & 3 &  & 6 &  & 3 &  & 1 \\ \hline
\end{tabular} \vskip 3pt  $(d_2,d_3)=(1,2),(2,1)$ \vskip 10pt}

{\footnotesize\begin{tabular}{|c|cccccccccc|}
 \hline $2j_L \backslash 2j_R$ & 0 & 1 & 2 & 3 & 4 & 5 & 6 & 7 & 8 & 9 \\ \hline
 0 &  &  &  & 3 & 488 & 118832 & 488 & 3 &  &  \\
 1 &  &  & 1 & 488 & 3 & 1464 & 3 & 488 & 1 &  \\
 2 &  & 1 &  & 3 &  & 6 &  & 3 &  & 1 \\ \hline
\end{tabular} \vskip 3pt  $(d_2,d_3)=(1,3),(3,1)$ \vskip 10pt}

{\footnotesize\begin{tabular}{|c|ccccccccccc|}
 \hline $2j_L \backslash 2j_R$ & 0 & 1 & 2 & 3 & 4 & 5 & 6 & 7 & 8 & 9 & 10 \\ \hline
 0 & 488 & 4 &  & 118840 & 1952 & 356988 & 1952 & 10 & 488 & 2 &  \\
 1 & 1 & 488 & 5 & 2928 & 13 & 4392 & 118843 & 1952 & 7 &  & 1 \\
 2 &  & 5 &  & 14 & 488 & 18 & 1464 & 11 & 488 & 3 &  \\
 3 &  &  & 1 &  & 3 &  & 6 &  & 3 &  & 1 \\ \hline
\end{tabular} \vskip 3pt  $(d_2,d_3)=(2,2)$ \vskip 10pt}

{\footnotesize\begin{tabular}{|c|cccccccccccc|}
 \hline $2j_L \backslash 2j_R$ & 0 & 1 & 2 & 3 & 4 & 5 & 6 & 7 & 8 & 9 & 10 & 11 \\ \hline
 0 &  &  &  &  &  & 3 & 488 & 118832 & 488 & 3 &  &  \\
 1 &  &  &  &  & 1 & 488 & 3 & 1464 & 3 & 488 & 1 &  \\
 2 &  &  &  & 1 &  & 3 &  & 6 &  & 3 &  & 1 \\ \hline
\end{tabular} \vskip 3pt  $(d_2,d_3)=(1,4),(4,1)$ \vskip 10pt}

{\footnotesize\begin{tabular}{|c|cccccccccccccc|}
 \hline $2j_L \backslash 2j_R$ & 0 & 1 & 2 & 3 & 4 & 5 & 6 & 7 & 8 & 9 & 10 & 11 & 12 & 13 \\ \hline
 0 &  & 4 & 488 & 118841 & 1464 & 357003 & 4880 & 594662 & 5368 & 24 & 1464 & 5 &  & 1 \\
 1 & 1 & 488 & 5 & 2928 & 18 & 6832 & 118866 & 9272 & 357013 & 4392 & 21 & 488 & 4 &  \\
 2 &  & 5 &  & 18 & 488 & 33 & 2928 & 44 & 4392 & 118857 & 1952 & 11 &  & 1 \\
 3 &  &  & 1 &  & 5 &  & 14 & 488 & 18 & 1464 & 11 & 488 & 3 &  \\
 4 &  &  &  &  &  & 1 &  & 3 &  & 6 &  & 3 &  & 1 \\ \hline
\end{tabular} \vskip 3pt  $(d_2,d_3)=(2,3),(3,2)$ \vskip 10pt}

    \caption{The refined BPS numbers of elliptic $\mathbb{F}_0$ for $d_E=2$, $0<d_2+d_3\leq 6$.}
    \label{tab:BPSF0dE=2}
\end{table}

\begin{table}[H]
{
{\centering\footnotesize\begin{tabular}{|c|ccccc|}
 \hline $2j_L \backslash 2j_R$ & 0 & 1 & 2 & 3 & 4 \\ \hline
 0 & 15928440 & 2 &  & 1 &  \\
 1 & 2 & 281452 & 2 &  &  \\
 2 &  & 2 & 488 & 1 &  \\
 3 &  &  & 1 &  & 1 \\ \hline
\multicolumn{6}{c}{$(d_2,d_3)$\,=\,$(0,1)$ \rule{0pt}{2.6ex}}\\
\end{tabular}}
\hspace{0.5cm}
{\centering\footnotesize\begin{tabular}{|c|ccccc|}
 \hline $2j_L \backslash 2j_R$ & 0 & 1 & 2 & 3 & 4 \\ \hline
 0 & 15928440 & 2 &  & 1 &  \\
 1 & 2 & 281452 & 2 &  &  \\
 2 &  & 2 & 488 & 1 &  \\
 3 &  &  & 1 &  & 1 \\ \hline
\multicolumn{6}{c}{$(d_2,d_3)$\,=\,$(0,2)$  \rule{0pt}{2.6ex}}\\
\end{tabular}} \vskip 10pt

{\centering\footnotesize\begin{tabular}{|c|ccccccc|}
 \hline $2j_L \backslash 2j_R$ & 0 & 1 & 2 & 3 & 4 & 5 & 6 \\ \hline
 0 & 51107504 & 238155 & 2440 & 7 &  & 1 &  \\
 1 & 356982 & 565344 & 118843 & 976 & 5 &  &  \\
 2 & 2440 & 11 & 2440 & 7 & 488 & 1 &  \\
 3 & 5 &  & 8 &  & 4 &  & 1 \\ \hline
\multicolumn{8}{c}{$(d_2,d_3)$\,=\,$(1,1)$ \rule{0pt}{2.6ex}} \\
\end{tabular}}
\hspace{0.5cm}
{\centering\footnotesize\begin{tabular}{|c|ccc|}
 \hline $2j_L \backslash 2j_R$ & 0 & 1 & 2 \\ \hline
 0 & 488 &  &  \\
 1 & 1 &  & 1 \\ \hline
\multicolumn{4}{c}{$(d_2,d_3)$\,=\,$(0,3)$ \rule{0pt}{2.6ex}} \\
\end{tabular}}
\vskip 10pt

{\centering\footnotesize\begin{tabular}{|c|ccccccccc|}
 \hline $2j_L \backslash 2j_R$ & 0 & 1 & 2 & 3 & 4 & 5 & 6 & 7 & 8 \\ \hline
 0 & 2440 & 238159 & 35181504 & 238160 & 2440 & 7 &  & 1 &  \\
 1 & 118841 & 284868 & 475826 & 284868 & 118846 & 976 & 5 &  &  \\
 2 & 1952 & 15 & 4880 & 16 & 2440 & 7 & 488 & 1 &  \\
 3 & 7 &  & 15 &  & 11 &  & 4 &  & 1 \\ \hline
\end{tabular} \vskip 3pt  $(d_2,d_3)=(1,2)$ \vskip 10pt}

{\centering\footnotesize\begin{tabular}{|c|ccccccccccc|}
 \hline $2j_L \backslash 2j_R$ & 0 & 1 & 2 & 3 & 4 & 5 & 6 & 7 & 8 & 9 & 10 \\ \hline
 0 &  & 7 & 2440 & 238160 & 35181504 & 238160 & 2440 & 7 &  & 1 &  \\
 1 & 5 & 976 & 118846 & 284868 & 475826 & 284868 & 118846 & 976 & 5 &  &  \\
 2 & 488 & 7 & 2440 & 16 & 4880 & 16 & 2440 & 7 & 488 & 1 &  \\
 3 & 3 &  & 11 &  & 16 &  & 11 &  & 4 &  & 1 \\ \hline
\end{tabular} \vskip 3pt  $(d_2,d_3)=(1,3)$ \vskip 10pt}

\hspace{-1.5cm}
{\centering\footnotesize\begin{tabular}{|c|cccccccccccc|}
 \hline $2j_L \backslash 2j_R$ & 0 & 1 & 2 & 3 & 4 & 5 & 6 & 7 & 8 & 9 & 10 & 11 \\ \hline
 0 & 2928 & 238174 & 19260384 & 833318 & 109109720 & 952145 & 9760 & 118858 & 1464 & 7 &  &  \\
 1 & 118853 & 5856 & 951677 & 577056 & 1784964 & 19827680 & 714018 & 6344 & 25 & 488 & 3 &  \\
 2 & 2928 & 33 & 12200 & 118895 & 17568 & 475871 & 10736 & 118867 & 2440 & 10 &  & 1 \\
 3 & 15 & 488 & 45 & 1952 & 58 & 4392 & 42 & 1952 & 17 & 488 & 3 &  \\
 4 &  & 3 &  & 9 &  & 14 &  & 9 &  & 3 &  & 1 \\ \hline
\end{tabular} \vskip 3pt  $(d_2,d_3)=(2,2)$ \vskip 10pt}

{\centering\footnotesize\begin{tabular}{|c|ccccccccccccc|}
 \hline $2j_L \backslash 2j_R$ & 0 & 1 & 2 & 3 & 4 & 5 & 6 & 7 & 8 & 9 & 10 & 11 & 12 \\ \hline
 0 &  & 1 &  & 7 & 2440 & 238160 & 35181504 & 238160 & 2440 & 7 &  & 1 &  \\
 1 &  &  & 5 & 976 & 118846 & 284868 & 475826 & 284868 & 118846 & 976 & 5 &  &  \\
 2 &  & 1 & 488 & 7 & 2440 & 16 & 4880 & 16 & 2440 & 7 & 488 & 1 &  \\
 3 & 1 &  & 4 &  & 11 &  & 16 &  & 11 &  & 4 &  & 1 \\ \hline
\end{tabular} \vskip 3pt  $(d_2,d_3)=(1,4)$ \vskip 10pt}}

\hspace{-2.0cm}
{\tiny \begin{tabular}{|c|ccccccccccccccc|}
 \hline $2j_L \backslash 2j_R$ & 0 & 1 & 2 & 3 & 4 & 5 & 6 & 7 & 8 & 9 & 10 & 11 & 12 & 13 & 14 \\ \hline
 0 & 2928 & 238178 & 19261360 & 833354 & 93196408 & 2380620 & 202578796 & 2499447 & 306340 & 475886 & 4880 & 23 & 488 & 2 &  \\
 1 & 118855 & 5856 & 951706 & 301460 & 2736676 & 20416936 & 4165105 & 94053940 & 1903885 & 20496 & 118906 & 2928 & 14 &  & 1 \\
 2 & 2928 & 39 & 14640 & 118941 & 32696 & 951800 & 326348 & 1785071 & 19560868 & 714070 & 9272 & 39 & 488 & 5 &  \\
 3 & 18 & 488 & 65 & 3416 & 126 & 11712 & 118989 & 17080 & 475928 & 10248 & 118887 & 2440 & 14 &  & 1 \\
 4 &  & 5 &  & 19 & 488 & 44 & 1952 & 56 & 4392 & 40 & 1952 & 16 & 488 & 3 &  \\
 5 &  &  & 1 &  & 3 &  & 9 &  & 14 &  & 9 &  & 3 &  & 1 \\ \hline
\multicolumn{16}{c}{ {\footnotesize{$(d_2,d_3)$\,=\,$(2,3)$\rule{0pt}{2.6ex}}}} \\
\end{tabular} \vskip 10pt}

    \caption{The refined BPS numbers of elliptic $\mathbb{F}_0$ for $d_E=3$, $0<d_2+d_3\leq 6$.}
    \label{tab:BPSF0dE=3}
\end{table}

\begin{table}[H]

{\centering\tiny\begin{tabular}{|c|cccccc|}
 \hline $2j_L \backslash 2j_R$ & 0 & 1 & 2 & 3 & 4 & 5 \\ \hline
 0 & 410133618 & 4 & 488 & 1 &  &  \\
 1 & 3 & 16209892 & 4 &  & 1 &  \\
 2 & 488 & 4 & 281452 & 3 &  &  \\
 3 & 1 &  & 2 & 488 & 1 &  \\
 4 &  &  &  & 1 &  & 1 \\ \hline
\end{tabular} \vskip 3pt  $(d_2,d_3)=(0,1)$ \vskip 10pt}

{\centering\tiny\begin{tabular}{|c|ccccccc|}
 \hline $2j_L \backslash 2j_R$ & 0 & 1 & 2 & 3 & 4 & 5 & 6 \\ \hline
 0 & 6749497860 & 6 & 281452 & 3 &  &  &  \\
 1 & 6 & 426343510 & 8 & 488 & 2 &  &  \\
 2 & 281452 & 7 & 16210380 & 5 &  & 1 &  \\
 3 & 2 & 488 & 5 & 281452 & 3 &  &  \\
 4 &  & 1 &  & 2 & 488 & 1 &  \\
 5 &  &  &  &  & 1 &  & 1 \\ \hline
\end{tabular} \vskip 3pt  $(d_2,d_3)=(0,2)$ \vskip 10pt}

\hspace{1.2cm}
{\centering\tiny\begin{tabular}{|c|cccccc|}
 \hline $2j_L \backslash 2j_R$ & 0 & 1 & 2 & 3 & 4 & 5 \\ \hline
 0 & 410133618 & 4 & 488 & 1 &  &  \\
 1 & 3 & 16209892 & 4 &  & 1 &  \\
 2 & 488 & 4 & 281452 & 3 &  &  \\
 3 & 1 &  & 2 & 488 & 1 &  \\
 4 &  &  &  & 1 &  & 1 \\ \hline
\multicolumn{7}{c}{$(d_2,d_3)$\,=\,$(0,3)$  \rule{0pt}{2.6ex}}\\
\end{tabular}}
\hspace{0.5cm}
{\centering\tiny\begin{tabular}{|c|ccc|}
 \hline $2j_L \backslash 2j_R$ & 0 & 1 & 2 \\ \hline
 0 & 488 &  &  \\
 1 & 1 &  & 1 \\ \hline
\multicolumn{4}{c}{$(d_2,d_3)$\,=\,$(0,4)$  \rule{0pt}{2.6ex}}\\ 
\end{tabular}}
\hspace{0.5cm} \vskip 10pt

{\centering\tiny\begin{tabular}{|c|cccccccccc|}
 \hline $2j_L \backslash 2j_R$ & 0 & 1 & 2 & 3 & 4 & 5 & 6 & 7 & 8 & 9 \\ \hline
 0 & 895036392 & 10108466840 & 485191058 & 951660 & 288284 & 24 & 488 & 2 &  &  \\
 1 & 138538871 & 587410946 & 138895893 & 51687488 & 357036 & 4392 & 15 &  & 1 &  \\
 2 & 17064496 & 1665642 & 17914220 & 832868 & 569736 & 118865 & 976 & 7 &  &  \\
 3 & 33 & 293652 & 62 & 291212 & 38 & 3416 & 10 & 488 & 1 &  \\
 4 &  & 31 & 488 & 32 & 488 & 18 &  & 5 &  & 1 \\
 5 &  &  & 1 &  & 2 &  & 1 &  &  &  \\ \hline
\end{tabular} \vskip 3pt  $(d_2,d_3)=(1,2)$ \vskip 10pt}

{\centering\tiny\begin{tabular}{|c|cccccccccccc|}
 \hline $2j_L \backslash 2j_R$ & 0 & 1 & 2 & 3 & 4 & 5 & 6 & 7 & 8 & 9 & 10 & 11 \\ \hline
 0 & 287796 & 951659 & 485191058 & 10108466860 & 485191058 & 951661 & 288284 & 24 & 488 & 2 &  &  \\
 1 & 357021 & 51687488 & 138895903 & 571205446 & 138895904 & 51687488 & 357036 & 4392 & 15 &  & 1 &  \\
 2 & 568760 & 832865 & 17633744 & 1784503 & 17633744 & 832872 & 569736 & 118865 & 976 & 7 &  &  \\
 3 & 28 & 290724 & 69 & 297068 & 70 & 291212 & 38 & 3416 & 10 & 488 & 1 &  \\
 4 & 488 & 31 & 488 & 48 & 488 & 36 & 488 & 18 &  & 5 &  & 1 \\
 5 & 1 &  & 2 &  & 2 &  & 2 &  & 1 &  &  &  \\ \hline
\end{tabular} \vskip 3pt  $(d_2,d_3)=(1,3)$ \vskip 10pt}

\hspace{-1.5cm}
{\tiny\begin{tabular}{|c|ccccccccccccc|}
 \hline $2j_L \backslash 2j_R$ & 0 & 1 & 2 & 3 & 4 & 5 & 6 & 7 & 8 & 9 & 10 & 11 & 12 \\ \hline
 0 & 58816928 & 2338601017 & 1105638636 & 27280967247 & 1125129844 & 4403222 & 19843784 & 475901 & 5856 & 19 &  & 1 &  \\
 1 & 2142498 & 257889928 & 281957323 & 1393908524 & 2615917012 & 219392584 & 2617892 & 24400 & 118909 & 1952 & 11 &  &  \\
 2 & 1151184 & 4521658 & 54558128 & 7497388 & 131797304 & 4403360 & 20414496 & 832937 & 8296 & 37 & 488 & 3 &  \\
 3 & 118933 & 602920 & 594898 & 622440 & 1546516 & 605360 & 594804 & 15616 & 118887 & 2440 & 11 &  & 1 \\
 4 & 2440 & 119 & 7320 & 172 & 12200 & 138 & 7320 & 69 & 1952 & 20 & 488 & 3 &  \\
 5 & 9 &  & 25 &  & 37 &  & 26 &  & 12 &  & 3 &  & 1 \\ \hline
\multicolumn{14}{c}{$(d_2,d_3)$\,=\,$(2,2)$  \rule{0pt}{2.6ex}}\\
\end{tabular}}
\hspace{0.5cm}\vskip 10pt

\hspace{-1.5cm}
{\tiny\begin{tabular}{|c|cccccccccccccc|}
 \hline $2j_L \backslash 2j_R$ & 0 & 1 & 2 & 3 & 4 & 5 & 6 & 7 & 8 & 9 & 10 & 11 & 12 & 13 \\ \hline
 0 & 488 & 24 & 288284 & 951661 & 485191058 & 10108466860 & 485191058 & 951661 & 288284 & 24 & 488 & 2 &  &  \\
 1 & 14 & 4392 & 357036 & 51687488 & 138895904 & 571205446 & 138895904 & 51687488 & 357036 & 4392 & 15 &  & 1 &  \\
 2 & 976 & 118865 & 569736 & 832872 & 17633744 & 1784503 & 17633744 & 832872 & 569736 & 118865 & 976 & 7 &  &  \\
 3 & 9 & 3416 & 38 & 291212 & 70 & 297068 & 70 & 291212 & 38 & 3416 & 10 & 488 & 1 &  \\
 4 &  & 17 & 488 & 36 & 488 & 49 & 488 & 36 & 488 & 18 &  & 5 &  & 1 \\
 5 & 1 &  & 2 &  & 2 &  & 2 &  & 2 &  & 1 &  &  &  \\ \hline
\multicolumn{15}{c}{$(d_2,d_3)$\,=\,$(1,4)$  \rule{0pt}{2.6ex}}\\
\end{tabular}}
\hspace{0.5cm}\vskip 10pt

\hspace{-3.cm}
{\tiny\begin{tabular}{|c|cccccccccccccccc|}
 \hline $2j_L \backslash 2j_R$ & 0 & 1 & 2 & 3 & 4 & 5 & 6 & 7 & 8 & 9 & 10 & $\cdots$\\\hline 
 0 & 468430462 & 2338958116 & 269725512 & 19515028352 & 2441328392 & 59430191270 & 2065808006 & 149487274 & 130127624 & 2023320 & 21472 & \\
 1 & 2380709 & 225498448 & 147108398 & 1248598098 & 3041769570 & 2666356642 & 20071922561 & 1044125850 & 9163710 & 20175012 & 951876 & \\
 2 & 315612 & 5354145 & 38959452 & 14399806 & 280442068 & 158412779 & 957391518 & 2484995076 & 188999084 & 3688588 & 37576 & \\
 3 & 118984 & 340012 & 1070947 & 1537556 & 4402862 & 37886348 & 7259656 & 114802592 & 4165428 & 19871112 & 833006 & \\
 4 & 3416 & 213 & 16592 & 119250 & 322932 & 595175 & 338548 & 1546688 & 321468 & 594883 & 14152 & \\
 5 & 19 & 488 & 68 & 2440 & 134 & 6832 & 167 & 11712 & 130 & 6832 & 63 & \\
 6 &  & 4 &  & 12 &  & 24 &  & 35 &  & 24 &  & \\\hline
\multicolumn{13}{c}{$(d_2,d_3)$\,=\,$(2,3)$  \rule{0pt}{2.6ex}}\\ 
\end{tabular}}
\hspace{0.5cm}\vskip 10pt

    \caption{The refined BPS numbers of elliptic $\mathbb{F}_0$ for $d_E=4$, $0<d_2+d_3\leq 6$. Here we exclude the degree $(d_E,d_2,d_3)=(4,1,1)$ as it is not fixed from the boundary conditions we have.}
    \label{tab:BPSF0dE=4}
\end{table}

\newpage
\subsection{\texorpdfstring{The refined BPS numbers for elliptic fibration over $\mathbb{F}_1$}{BPSF1}}
\begin{table}[H]
\centering
{\footnotesize\begin{tabular}{|c|ccc|}
 \hline $2j_L \backslash 2j_R$ & 0 & 1 & 2 \\ \hline
 0 & 488 &  &  \\
 1 & 1 &  & 1 \\ \hline
\multicolumn{4}{c}{$(d_2,d_3)$\,=\,$(0,1)$  \rule{0pt}{2.6ex}}\\  
\end{tabular}}
\hspace{0.5cm}
{\footnotesize\begin{tabular}{|c|cc|}
 \hline $2j_L \backslash 2j_R$ & 0 & 1  \\ \hline
 0 & {$248$} &    \\
 1 &  & {$1$}   \\ \hline
\multicolumn{3}{c}{$(d_2,d_3)$\,=\,$(1,0)$  \rule{0pt}{2.6ex}}\\  
\end{tabular}}
\hspace{0.5cm}
{\footnotesize\begin{tabular}{|c|cccc|}
 \hline $2j_L \backslash 2j_R$ & 0 & 1 & 2 & 3 \\ \hline
 0 &  & 488 &  &  \\
 1 &  & 2 &  & 1 \\ \hline
\multicolumn{5}{c}{$(d_2,d_3)$\,=\,$(1,1)$  \rule{0pt}{2.6ex}}\\  
\end{tabular}}\vskip 10pt

{\footnotesize\begin{tabular}{|c|cccccc|}
 \hline $2j_L \backslash 2j_R$ & 0 & 1 & 2 & 3 & 4 & 5 \\ \hline
 0 &  &  &  & 488 &  &  \\
 1 &  & 1 &  & 2 &  & 1 \\ \hline
\multicolumn{7}{c}{$(d_2,d_3)$\,=\,$(1,2)$  \rule{0pt}{2.6ex}}\\  
\end{tabular}}
\hspace{0.5cm}
{\footnotesize\begin{tabular}{|c|cccccccc|}
 \hline $2j_L \backslash 2j_R$ & 0 & 1 & 2 & 3 & 4 & 5 & 6 & 7 \\ \hline
 0 &  &  &  &  &  & 488 &  &  \\
 1 &  &  &  & 1 &  & 2 &  & 1 \\ \hline
\multicolumn{9}{c}{$(d_2,d_3)$\,=\,$(1,3)$  \rule{0pt}{2.6ex}}\\  
\end{tabular}}\vskip 10pt

{\footnotesize\begin{tabular}{|c|ccccccc|}
 \hline $2j_L \backslash 2j_R$ & 0 & 1 & 2 & 3 & 4 & 5 & 6 \\ \hline
 0 &  &  &  &  & 488 &  &  \\
 1 &  &  & 1 &  & 2 &  & 1 \\ \hline
\end{tabular} \vskip 3pt  $(d_2,d_3)=(2,2)$ \vskip 10pt }

{\footnotesize\begin{tabular}{|c|cccccccccc|}
 \hline $2j_L \backslash 2j_R$ & 0 & 1 & 2 & 3 & 4 & 5 & 6 & 7 & 8 & 9 \\ \hline
 0 &  &  &  &  &  &  &  & 488 &  &  \\
 1 &  &  &  &  &  & 1 &  & 2 &  & 1 \\ \hline
\end{tabular} \vskip 3pt  $(d_2,d_3)=(1,4)$ \vskip 10pt }

{\footnotesize\begin{tabular}{|c|cccccccccc|}
 \hline $2j_L \backslash 2j_R$ & 0 & 1 & 2 & 3 & 4 & 5 & 6 & 7 & 8 & 9 \\ \hline
 0 &  &  &  &  & 488 & 1 & 976 & 2 &  & 1 \\
 1 &  &  & 1 &  & 4 &  & 5 & 488 & 2 &  \\
 2 &  &  &  &  &  & 1 &  & 2 &  & 1 \\ \hline
\end{tabular} \vskip 3pt  $(d_2,d_3)=(2,3)$ \vskip 10pt }

{\footnotesize\begin{tabular}{|c|cccccccccccc|}
 \hline $2j_L \backslash 2j_R$ & 0 & 1 & 2 & 3 & 4 & 5 & 6 & 7 & 8 & 9 & 10 & 11 \\ \hline
 0 &  &  &  &  &  &  &  &  &  & 488 &  &  \\
 1 &  &  &  &  &  &  &  & 1 &  & 2 &  & 1 \\ \hline
\end{tabular} \vskip 3pt  $(d_2,d_3)=(1,5)$ \vskip 10pt }

{\footnotesize\begin{tabular}{|c|ccccccccccccc|}
 \hline $2j_L \backslash 2j_R$ & 0 & 1 & 2 & 3 & 4 & 5 & 6 & 7 & 8 & 9 & 10 & 11 & 12 \\ \hline
 0 &  &  &  &  & 488 & 1 & 976 & 4 & 1464 & 5 &  & 2 &  \\
 1 &  &  & 1 &  & 4 &  & 8 & 488 & 9 & 976 & 5 &  & 1 \\
 2 &  &  &  &  &  & 1 &  & 4 &  & 5 & 488 & 2 &  \\
 3 &  &  &  &  &  &  &  &  & 1 &  & 2 &  & 1 \\ \hline
\end{tabular} \vskip 3pt  $(d_2,d_3)=(2,4)$ \vskip 10pt }

{\footnotesize\begin{tabular}{|c|ccccccccccc|}
 \hline $2j_L \backslash 2j_R$ & 0 & 1 & 2 & 3 & 4 & 5 & 6 & 7 & 8 & 9 & 10 \\ \hline
 0 &  &  &  &  &  & 488 & 1 & 488 & 2 &  & 1 \\
 1 &  &  &  & 1 &  & 3 &  & 3 & 488 & 1 &  \\
 2 &  &  &  &  &  &  & 1 &  & 2 &  & 1 \\ \hline
\end{tabular} \vskip 3pt  $(d_2,d_3)=(3,3)$ \vskip 10pt }

    \caption{The refined BPS numbers of elliptic $\mathbb{F}_1$ for $d_E=1$, $0<d_2+d_3\leq 6$. The degree $(d_2,d_3)=(1,0)$ is not captured by our calculations. The refined BPS numbers are those for E-strings. }
    \label{tab:BPSF1dE=1}
\end{table}

\clearpage
\addcontentsline{toc}{section}{References}

\bibliographystyle{utphys} 
\bibliography{Reference}

\end{document}